\documentclass[iop,revtex4,numberedappendix,appendixfloats]{emulateapj}

\usepackage{longtable}

\usepackage{natbib}

\usepackage{lscape}
\usepackage{amsmath}
\usepackage{color}
\usepackage{url}
\usepackage{ulem}
\usepackage{multirow}

\usepackage{wasysym}
\usepackage{amssymb}% http://ctan.org/pkg/amssymb
\usepackage{pifont}% http://ctan.org/pkg/pifont
\newcommand{\cmark}{\ding{51}}%

\bibpunct{(}{)}{;}{a}{}{,}

\def\apjl{ApJ}%
\def\apjs{ApJS}%
\def\ao{Appl.~Opt.}%
\def\aap{A\&A}%
\def\arcmin{\hbox{$^\prime$}}
\def\arcsec{\hbox{$^{\prime\prime}$}}

\slugcomment{To appear in The Astrophysical Journal}

\shorttitle{BASS--V: X-ray properties of the {\it Swift}/BAT 70-month AGN catalog}
\shortauthors{Ricci et al.}

\begin{document}

\title{BAT AGN Spectroscopic Survey -- V. 

X-ray properties of the {\it Swift}/BAT 70-month AGN catalog}

\author{C. Ricci\altaffilmark{1,2,3,4*}, B. Trakhtenbrot\altaffilmark{5,6}, M. J. Koss\altaffilmark{5}, Y. Ueda\altaffilmark{2}, I. Delvecchio\altaffilmark{7}, E. Treister\altaffilmark{1}, K. Schawinski\altaffilmark{5}, S. Paltani\altaffilmark{8}, K. Oh\altaffilmark{5},  I. Lamperti\altaffilmark{5}, S. Berney\altaffilmark{5}, P. Gandhi\altaffilmark{9}, K. Ichikawa\altaffilmark{10,11,12}, F. E. Bauer\altaffilmark{1,13,14}, L. C. Ho\altaffilmark{3,15}, D. Asmus\altaffilmark{16}, V. Beckmann\altaffilmark{17}, S. Soldi\altaffilmark{18}, M. Balokovi\'{c}\altaffilmark{19}, N. Gehrels\altaffilmark{20},  C. B. Markwardt\altaffilmark{21,22}}

\altaffiltext{1}{Instituto de Astrof\'{\i}sica, Facultad de F\'{i}sica, Pontificia Universidad Cat\'{o}lica de Chile, Casilla 306, Santiago 22, Chile.}
\altaffiltext{2}{Department of Astronomy, Kyoto University, Oiwake-cho, Sakyo-ku, Kyoto 606-8502, Japan.}
\altaffiltext{3}{Kavli Institute for Astronomy and Astrophysics, Peking University, Beijing 100871, China}
\altaffiltext{4}{Chinese Academy of Sciences South America Center for Astronomy and China-Chile Joint Center for Astronomy, Camino El Observatorio 1515, Las Condes, Santiago, Chile}
\altaffiltext{5}{Institute for Astronomy, Department of Physics, ETH Zurich, Wolfgang-Pauli-Strasse 27, CH-8093 Zurich, Switzerland}
\altaffiltext{6}{Zwicky fellow}
\altaffiltext{7}{Department of Physics, University of Zagreb, Bijeni\v{c}ka cesta 32, HR-10002 Zagreb, Croatia}
\altaffiltext{8}{Department of Astronomy, University of Geneva, ch. d'Ecogia 16, CH-1290 Versoix, Switzerland}
\altaffiltext{9}{School of Physics \& Astronomy, University of Southampton, Highfield, Southampton, SO17 1BJ}
\altaffiltext{10}{National Astronomical Observatory of Japan, 2-21-1 Osawa, Mitaka, Tokyo 181-8588, Japan}
\altaffiltext{11}{Department of Physics and Astronomy, University of Texas at San Antonio, One UTSA Circle, San Antonio, TX 78249, USA}
\altaffiltext{12}{Department of Astronomy, Columbia University, 550 West 120th Street, New York, NY 10027, USA}
\altaffiltext{13}{Space Science Institute, 4750 Walnut Street, Suite 205, Boulder, Colorado 80301, USA}
\altaffiltext{14}{Millenium Institute of Astrophysics, Santiago, Chile}
\altaffiltext{15}{Department of Astronomy, School of Physics, Peking University, Beijing 100871, China}
\altaffiltext{16}{European Southern Observatory, Casilla 19001, Santiago 19, Chile}
\altaffiltext{17}{CNRS / IN2P3, 3 rue Michel Ange, 75794 Paris Cedex 16, France}
\altaffiltext{18}{Centre Fran\c{c}ois Arago, APC, UniversitŽ Paris Diderot, CNRS/IN2P3, 10 rue Alice Domon et L\'eonie Duquet, 75205 Paris Cedex 13, France}
\altaffiltext{19}{Cahill Center for Astronomy and Astrophysics, California Institute of Technology, Pasadena, CA 91125, USA }
\altaffiltext{20}{NASA Goddard Space Flight Center, Mail Code 661, Greenbelt, MD 20771, USA}
\altaffiltext{21}{Department of Astronomy, University of Maryland, College Park, MD 20742, USA}
\altaffiltext{22}{Astroparticle Physics Laboratory, Mail Code 661, NASA Goddard Space Flight Center, Greenbelt, MD 20771, USA}

\altaffiltext{*}{cricci@astro.puc.cl}

\begin{abstract}
Hard X-ray ($\geq 10$\,keV) observations of Active Galactic Nuclei (AGN) can shed light on some of the most obscured episodes of accretion onto supermassive black holes. The 70-month {\it Swift}/BAT all-sky survey, which probes the 14--195\,keV energy range, has currently detected 838\,AGN. We report here on the broad-band X-ray (0.3--150\,keV) characteristics of these AGN, obtained by combining {\it XMM-Newton}, {\it Swift/XRT}, {\it ASCA}, {\it Chandra}, and {\it Suzaku} observations in the soft X-ray band ($\leq 10$\,keV) with 70-month averaged {\it Swift}/BAT data. The non-blazar AGN of our sample are almost equally divided into unobscured ($N_{\rm H}< 10^{22}\rm\,cm^{-2}$) and obscured ($N_{\rm H}\geq 10^{22}\rm\,cm^{-2}$) AGN, and their {\it Swift}/BAT continuum is systematically steeper than the 0.3--10\,keV emission, which suggests that the presence of a high-energy cutoff is almost ubiquitous. We discuss the main X-ray spectral parameters obtained, such as the photon index, the reflection parameter, the energy of the cutoff, neutral and ionized absorbers, and the soft excess for both obscured and unobscured AGN.

\end{abstract}


\section{Introduction}

Active Galactic Nuclei\footnote{See \cite{Beckmann:2012kq,Netzer:2013fj,Netzer:2015qv,Brandt:2015ty,Ramos-Almeida:2017rm} for recent reviews on the subject.} (AGN) are among the most energetic phenomena in the Universe, and are believed to play a significant role in the evolution of galaxies (e.g., \citealp{Kormendy:1995fk,Ferrarese:2000uq,Gebhardt:2000kx,Tremaine:2002vn,Schawinski:2006fj,Kormendy:2013uf}). One of the most distinctive features of AGN is their strong emission in the X-ray regime, which is produced by Comptonization of optical and UV photons (e.g., \citealp{Haardt:1991qr}) in a hot plasma located very close to the accreting supermassive black hole (SMBH). X-ray emission is therefore an important tracer of the physical properties of the accreting system, and can constrain the amount of matter along the line of sight, typically parameterized as the neutral hydrogen column density ($N_{\rm H}$). X-ray emission can also be used to shed light on the structure of the circumnuclear material, by studying the spectral features created by the reprocessing of the primary X-ray radiation on the material surrounding the SMBH. The two main features produced by reprocessing of X-ray radiation in neutral material are the iron K$\alpha$ line at $6.4$\,keV and a broad Compton ``hump" peaking at $\sim 30$\,keV (e.g., \citealp{Matt:1991ys}, \citealp{Murphy:2009uq}).

\begin{figure*}[t!]
\centering
 %% 1st image
 %% 2nd image
\includegraphics[width=\textwidth,angle=0]{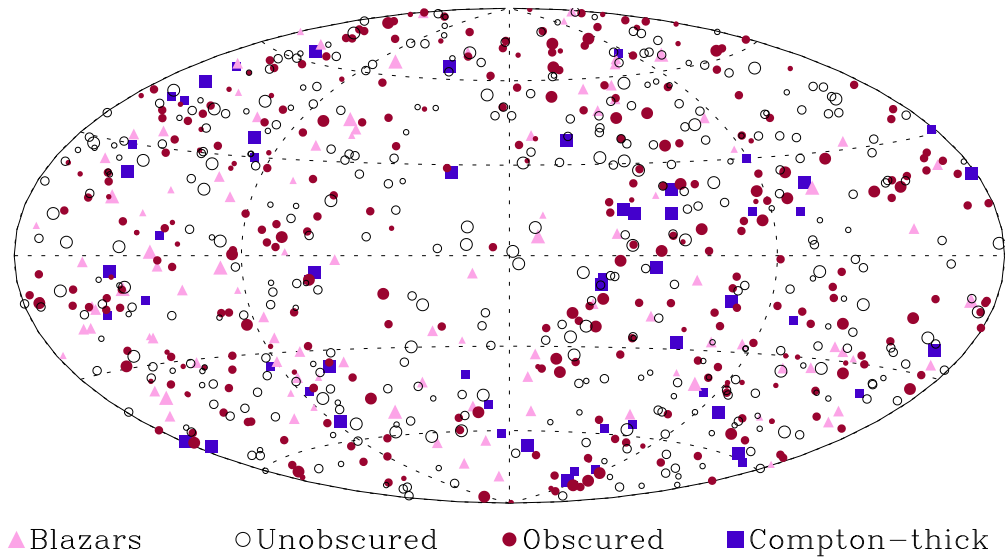}
% %% caption
% \begin{minipage}[t]{1\textwidth}
  \caption{AGN from the 70-month {\it Swift}/BAT catalog (Aitoff projection). Sources are divided into non-blazar AGN and blazars as discussed in $\S$\ref{sect:sample}, and different sizes imply different intrinsic fluxes. Non-blazar AGN are divided, depending on their line-of-sight column density, into unobscured ($N_{\rm H}<10^{22}\rm\,cm^{-2}$), obscured ($10^{22}\leq N_{\rm H}<10^{24}\rm\,cm^{-2}$) and CT ($N_{\rm H}\geq10^{24}\rm\,cm^{-2}$, see $\S$\ref{sect:NeutralAbsorption} for details on the absorption properties of the sample).}
\label{fig:plot_allsky}
% \end{minipage}
\end{figure*}

The integrated emission of unresolved AGN gives rises to the Cosmic X-ray background (CXB; e.g., \citealp{Giacconi:1962cr,Bauer:2004lk,Harrison:2016bf}). Studies carried out below 10\,keV have shown that the shape of the CXB is significantly flatter (with a photon index $\Gamma\sim 1.4$, e.g., \citealp{De-Luca:2004zr}) than the typical X-ray spectrum of unobscured AGN ($\Gamma\simeq 1.9$, e.g., \citealp{Nandra:1994ly}). This, together with the fact that the CXB shows a clear peak at $\sim 30\rm\,keV$, where the bulk of the reprocessed X-ray radiation is emitted, suggests that heavily obscured AGN contribute significantly to the CXB. Synthesis models of the CXB (e.g., \citealp{Ueda:2003nx,Ueda:2014ly}, \citealp{Gilli:2007qf}, \citealp{Treister:2005ve,Treister:2009uq}, \citealp{Draper:2010oq}, \citealp{Akylas:2012uq}) have shown that a fraction of $10-30\%$ of Compton-thick [CT, $\log (N_{\rm H}/\rm cm^{-2})\geq 24$] AGN are needed to reproduce the CXB.  The fraction of CT AGN inferred from synthesis models of the CXB is, however, strongly dependent on the assumptions made on the fraction of reprocessed X-ray emission, with stronger reflection components resulting in smaller fractions of CT AGN (e.g., \citealp{Gandhi:2007fk,Treister:2009uq,Ricci:2011zr,Vasudevan:2013ys,Vasudevan:2016yg,Ueda:2014ly}).

Radiation at hard X-rays ($E \gtrsim 10$\,keV) is less affected by the obscuring material, at least up to $N_{\rm H}\sim 10^{23.5}-10^{24}\rm\,cm^{-2}$ (see Figure 1 of \citealp{Ricci:2015kx}), due to the decline of the photoelectric cross-section with increasing energy. Hard X-ray observations are therefore very well suited to detect heavily obscured AGN, and allow us to obtain the least biased X-ray sample of local AGN, and to directly study the X-ray emission responsible for the peak of the CXB. Currently, there are four operating hard X-ray observatories in-orbit. IBIS/ISGRI on board {\it INTEGRAL} \citep{Winkler:2003fk} was launched in 2002, and has detected so far more than 200\,AGN \citep{Beckmann:2006lh,Beckmann:2009fk,Paltani:2008oq,Panessa:2008wj,de-Rosa:2008ys,de-Rosa:2012qv,Ricci:2011zr,Malizia:2012to,Bottacini:2012qf}. {\it NuSTAR} \citep{Harrison:2013zr}, launched in 2012, is the first focussing hard X-ray telescope on-orbit, and its serendipitous survey has detected 497 sources in the first 40 months of observations (\citealp{Lansbury:2017kq}, see also \citealp{Chen:2017nr}). Thanks to its revolutionary characteristics, {\it NuSTAR} has been very efficient in constraining the properties of heavily obscured AGN (e.g., \citealp{Balokovic:2014dq,Gandhi:2014bh,Stern:2014eq,Koss:2015qf,Koss:2016ys,Brightman:2015cr,Lansbury:2015vn,Lansbury:2017fj,Annuar:2015ve,Annuar:2017rr,Boorman:2016hl,Ricci:2016kq,Ricci:2016bh,Ricci:2017kq,Ricci:2017fj}). The recently launched mission {\it AstroSat} \citep{Singh:2014ai} carries on board two hard X-ray instruments: the Large Area Xenon Proportional Counters (LAXPC; 3--80\,keV) and the Cadmium-Zinc-Telluride coded-mask imager (CZTI, 10--150\,keV). Finally, the NASA mission {\it Swift} \citep{Gehrels:2004kx}, launched in 2005, carries on board the Burst Alert Telescope (BAT, \citealp{Barthelmy:2005uq,Krimm:2013ys}). BAT is a hard X-ray detector that operates in the 14--195\,keV energy range and has proved to be an extremely valuable tool for studying AGN in the local Universe, since it is the only hard X-ray instrument to continuously survey the whole sky.

\begin{figure*}[t!]
\centering
 %% 1st image
 %% 2nd image
\begin{minipage}[!b]{.48\textwidth}
\centering
\includegraphics[width=8.8cm]{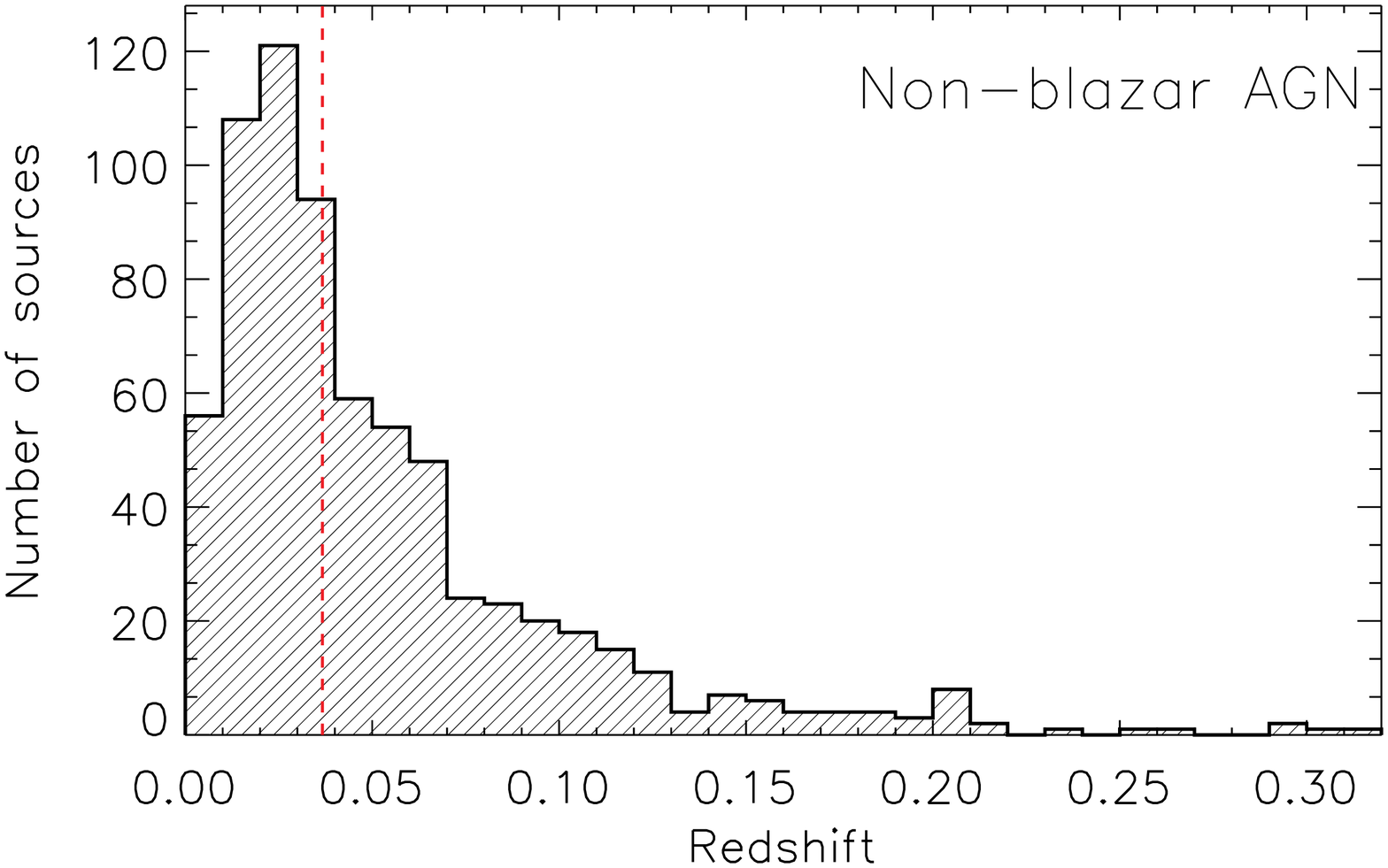}\end{minipage}
\begin{minipage}[!b]{.48\textwidth}
\centering
\includegraphics[width=8.8cm]{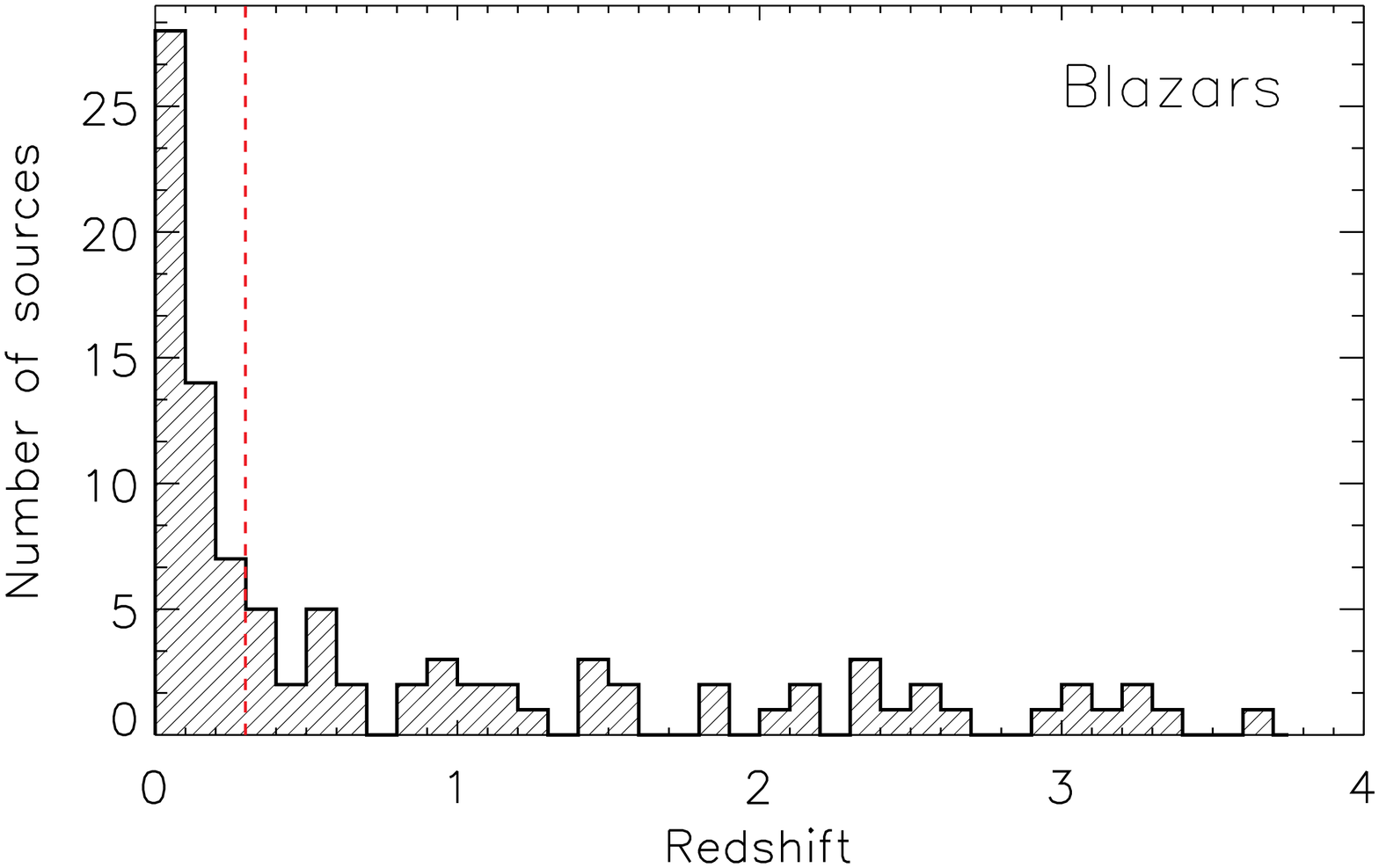}\end{minipage}
\par\bigskip
\begin{minipage}[!b]{.48\textwidth}
\centering
\includegraphics[width=8.8cm]{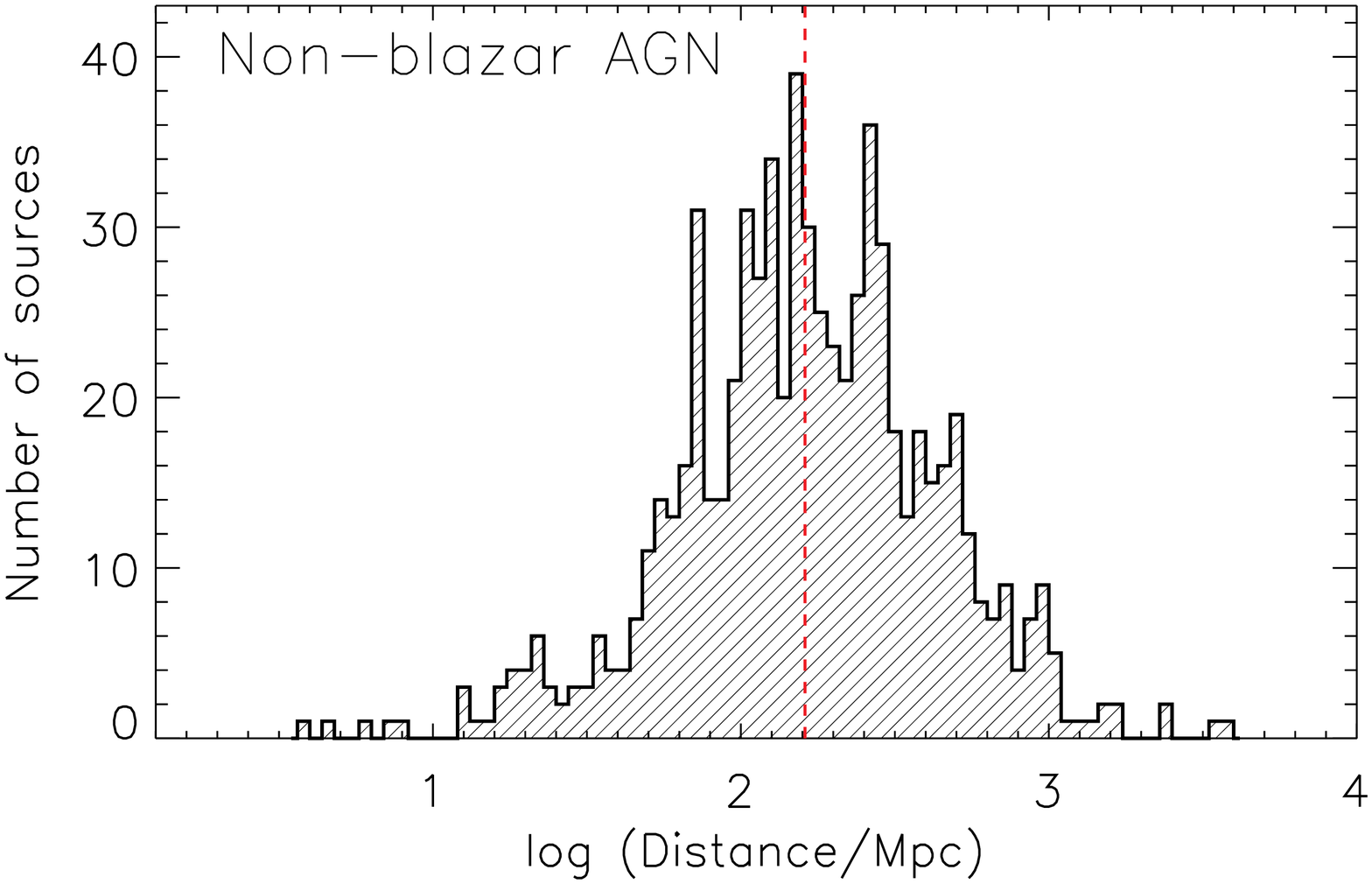}\end{minipage}
\begin{minipage}[!b]{.48\textwidth}
\centering
\includegraphics[width=8.8cm]{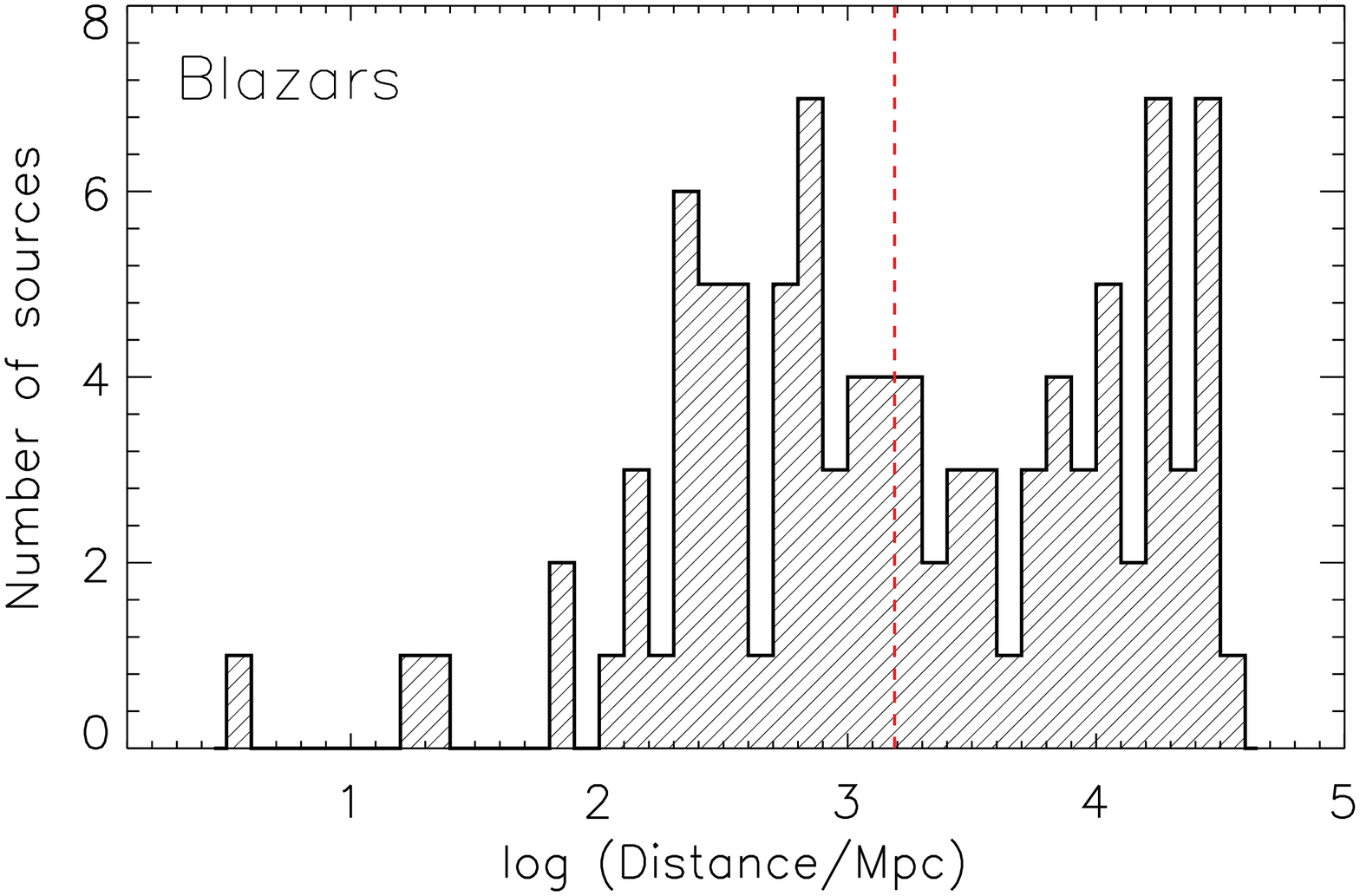}\end{minipage}
% %% caption
 \begin{minipage}[t]{1\textwidth}
  \caption{{\it Top left panel:} zoom in the $z=0-0.32$ range of the redshift distribution of the non-blazar AGN of our sample. There are four non-blazar AGN at $z\geq 0.4$ which are excluded from this plot: SWIFT\,J1131.9$-$1233 ($z=0.6540$), SWIFT\,J2344.6$-$4246 ($z=0.5975$), SWIFT\,J1159.7$-$2002 ($z=0.4500$), and SWIFT\,J0216.3+5128 ($z=0.422$). {\it Top right panel:} redshift distribution of the blazars. {\it Bottom left panel:} Distribution of the distances of the non-blazar AGN of our sample (including the redshift-independent distances). {\it Bottom right panel:} same as the bottom left panel for the blazars. For all panels the red dashed lines show the median for each sample. Among the 70-month catalog sources, non-blazar AGN lie at much lower redshifts and distances (median values: $z=0.0367$, $D=161.6$\,Mpc) compared to the strongly beamed and more luminous (see $\S$\ref{sect:Luminosities} and Fig.\,\ref{fig:Lumdistr_obsunobs}) blazar population ($z=0.299$, $D=1565.3$\,Mpc).}
\label{fig:hist_redshift}
 \end{minipage}
\end{figure*}

Early studies of {\it Swift}/BAT AGN (e.g., \citealp{Tueller:2008qf,Ajello:2008nx,Winter:2008fk,Winter:2009qf,Burlon:2011uq,Ajello:2012fk}) were focussed on the first releases of the {\it Swift}/BAT catalog \citep{Markwardt:2005kq,Tueller:2010qq,Segreto:2010rr,Cusumano:2010hl,Cusumano:2010qf} or on relatively small subsamples (e.g., \citealp{Vasudevan:2013ya}). The latest release of the {\it Swift}/BAT catalog (70-month, \citealp{Baumgartner:2013uq}) has however considerably increased the number of hard X-ray selected AGN, detecting more than 800 extragalactic sources. The all-sky coverage of {\it Swift}/BAT allows to detect very luminous and rare sources, and, being one of the least biased samples of AGN available, it allows to study a growing number of local heavily obscured AGN.  A large number of works have already been carried out studying pointed X-ray observations (e.g., \citealp{Winter:2009fy,Ricci:2010vn,Tazaki:2011lq,Tazaki:2013ys,Kawamuro:2016qv,Tanimoto:2016lh,Marchesi:2017xw,Oda:2017oq}) and the multi-wavelength properties of {\it Swift}/BAT AGN. This includes radio (e.g., \citealp{Burlon:2013la,Wong:2016sf}), far-IR (e.g., \citealp{Melendez:2014la,Mushotzky:2014it,Shimizu:2015xr,Shimizu:2016hc}), mid-IR (e.g., \citealp{Melendez:2008ef,Weaver:2010nx,Ichikawa:2012fk,Ichikawa:2017rw}), near-IR (e.g., \citealp{Mushotzky:2008dn,Lamperti:2017fj,Onori:2017rm,Onori:2017ys,Ricci:2017qv}), optical (e.g., \citealp{Vasudevan:2009jt,Koss:2010rc,Winter:2010kl,Ueda:2015mz}) and $\gamma$-ray (e.g., \citealp{Sambruna:2010ao,Teng:2011gf}) studies, as well as works focussed on the host galaxy properties (e.g., \citealp{Koss:2011qf}), variability (e.g., \citealp{Shimizu:2013uk,Soldi:2014kq}) and on peculiar sources (e.g., \citealp{Koss:2012fj,Hogg:2012rm,Smith:2014bf,Schawinski:2015ef}). Moreover, {\it NuSTAR} has been observing {\it Swift}/BAT AGN since its launch in the framework of a legacy survey (Balokovi\'{c} et al. in prep.), providing high-quality data in the 3--80\,keV energy range. 

Our group has been working on  a systematic study of the multi-wavelength properties of a very large number of {\it Swift}/BAT selected AGN from the currently available 70-month catalog \citep{Baumgartner:2013uq} and a forthcoming extension (105 months, \citealp{Oh:2017aa}). A large effort has been made to collect optical spectroscopy for most of the sources reported in the 70-month {\it Swift}/BAT catalog, which allowed to infer black hole masses for both obscured and unobscured objects \citep{Koss:2017wo}. The first results of the {\it Swift}/BAT AGN Spectroscopical Survey (BASS\footnote{www.bass-survey.com}) include the study of the CT AGN detected by BAT (\citealp{Ricci:2015kx}, see also \citealp{Koss:2016qo} and \citealp{Akylas:2016si}), the analysis of the correlation between high-ionisation optical emission lines and AGN X-ray emission \citep{Berney:2015uq}, the study of the relationship between optical narrow emission lines and the physical parameters of the accreting SMBH \citep{Oh:2017nx}, a near-IR spectroscopic study \citep{Lamperti:2017fj}, and the analysis of the relationship between the X-ray photon index and the mass-normalized accretion rate \citep{Trakhtenbrot:2017kq}. The detailed multi-wavelength analysis of a large sample of local AGN\footnote{See also \citealp{She:2017fk,She:2017qf} for a soft X-ray study of lower luminosity local AGN.} will be a very important benchmark for studies of AGN at higher redshifts, where the typical fluxes are significantly lower.

In this paper we present a compilation and analysis of the X-ray data available for the AGN of the 70-month {\it Swift}/BAT catalog.
This paper is structured as follows. In $\S$\ref{sect:sample} we present our sample, in $\S$\ref{sect:softxrayDA} we describe the data analysis of the soft X-ray data, while in $\S$\ref{sect:Xrayspecanalysis} we illustrate the procedure adopted for the broad-band X-ray spectral analysis of the sources, and the models used. In $\S$\ref{sect:Luminosities} we discuss the luminosity and flux distributions of our sample, in $\S$\ref{sect:Xraycontinuum} we examine the characteristics of the X-ray continuum, in $\S$\ref{sect:Absorption} we report on the results obtained for the neutral and ionized absorbing material, while in $\S$\ref{sect:SoftExcess} we discuss about the properties of the soft excess of obscured and unobscured AGN. Finally, in $\S$\ref{sect:Summary} we summarize our findings and present our conclusions. Throughout the paper we adopt standard cosmological parameters ($H_{0}=70\rm\,km\,s^{-1}\,Mpc^{-1}$, $\Omega_{\mathrm{m}}=0.3$, $\Omega_{\Lambda}=0.7$). Unless otherwise stated, uncertainties are quoted at the 90\% confidence level.

\section{Sample}\label{sect:sample}
Our sample consists of the 838 AGN detected within the 70-month {\it Swift}/BAT catalog\footnote{http://swift.gsfc.nasa.gov/results/bs70mon/} (\citealp{Baumgartner:2013uq}, Fig.\,\ref{fig:plot_allsky}). We flagged all blazars in our sample according to the latest release (5.0.0, \citealp{Massaro:2015bh}) of the Roma BZCAT\footnote{http://www.asdc.asi.it/bzcat/} catalog \citep{Massaro:2009ve}, and using the results of recent works on BAT detected blazars \citep{Ajello:2009ly,Maselli:2013fk}. Overall 105 objects are classified as blazars. Of these 26 are  BL Lacs (BZB), 53 are Flat Spectrum Radio Quasars (BZQ) and 26 are of uncertain type (BZU). This is a different terminology than that used in the optical catalog of \cite{Koss:2017wo} which refers to these sources as beamed AGN. Several sources have been identified as possible blazars by \cite{Koss:2017wo} using optical spectroscopy, and are not treated as blazars here. In Table\,\ref{tbl-list} we report the list of sources in our sample, together with their counterparts, coordinates, redshifts and blazar classification. For completeness, we also report the results obtained for the only non-AGN extragalactic source detected by {\it Swift}/BAT, M\,82 (a nearby starburst with X-ray emission produced by star formation, e.g., \citealp{Ranalli:2008ve}), although we do not include it in our statistical analysis.

\subsection{Counterpart identification}

The counterparts of the {\it Swift}/BAT sources were mostly taken from \cite{Baumgartner:2013uq} and from recent follow-up studies (e.g., \citealp{Parisi:2009zr,Masetti:2010bh,Lutovinov:2012bh,Masetti:2012ys,Parisi:2012uq}). In order to confirm the counterpart association, for all sources we studied the 2--10\,keV images of the fields using {\it XMM-Newton}/EPIC, {\it Swift}/XRT and {\it Chandra}/ACIS (we provide additional information on these data-sets in $\S$\ref{sect:softxrayDA}). Furthermore, the object coordinates were cross-checked with the \textit{Two Micron All Sky Survey} (2MASS, \citealp{Skrutskie:2006gd}) and \textit{Wide-field Infrared Survey Explorer} ({\it WISE}, \citealp{Wright:2010fk}) point source catalogs, and deviations larger than 3\arcsec were investigated individually. In 27 cases, the coordinates of the associated counterparts in the original BAT catalog do not accurately point to the nuclei of the systems, which in all cases are identified with a relatively bright {\it WISE} and 2MASS source. In further five cases the situation is more complex, because the original counterpart does not point to an individual galaxy, but to a pair or triple. In these cases, the closest galaxy to the X-ray source with {\it WISE} colours consistent with an AGN was selected. In Appendix\,\ref{app:new counterparts} we discuss the objects for which new counterparts were found, while in Appendix\,\ref{app:dualAGN} we discuss the {\it Swift}/BAT sources which host dual AGN. In three cases, where both dual AGN contributed significantly to the BAT flux, we report the spectral parameters of the two AGN (named D1 and D2).

\subsection{Redshifts and distances}

\subsubsection{Spectroscopic redshifts and redshift-independent distances}\label{sect:redshift}

Spectroscopic redshifts are available for most of the sources of our sample (803, i.e. $\sim 96$\%). The redshifts were taken from the first release (DR1) of the BASS optical catalog \citep{Koss:2017wo} and from the literature. For the closest objects in our sample (at $z< 0.01$), whenever available, we used redshift-independent measurements of the distance, using the mean reported in the NASA/IPAC Extragalactic Database (NED). Redshift-independent distances were considered for 44\,objects.

The redshift and distance distribution of the non-blazar AGN (left panels) and the blazars (right panels) in our sample are presented in Fig.\,\ref{fig:hist_redshift}, respectively. The median redshift and distance of non-blazar AGN ($z=0.0367$, $D=161.6$\,Mpc) is significantly lower than that of blazars ($z=0.302$, $D=1565$\,Mpc), consistent with the very different luminosity distributions of these two classes of objects (see $\S$\ref{sect:Luminosities} and Fig.\,\ref{fig:Lumdistr_obsunobs}).

\begin{figure}[t!]
\centering
 %% 1st image
 %% 2nd image
%\begin{minipage}[!b]{.48\textwidth}
\centering
\includegraphics[width=9cm]{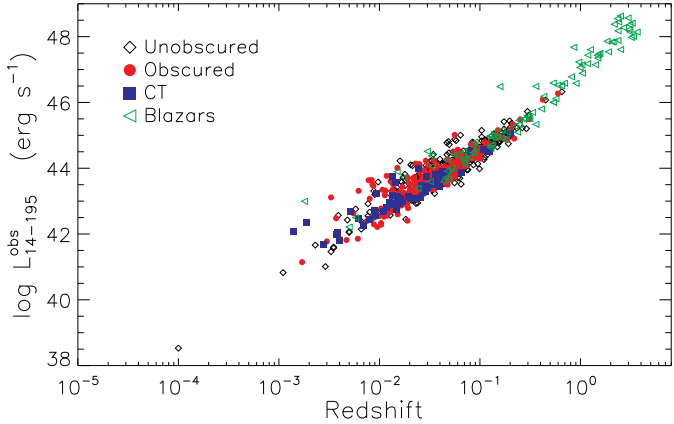}%\end{minipage}
% %% caption
% \begin{minipage}[t]{1\textwidth}
  \caption{Observed 14--195\,keV {\it Swift}/BAT luminosity versus redshift for the AGN of the 70-month {\it Swift}/BAT catalog. The sources are divided into blazars and non-blazar AGN, with the latter being further classified into unobscured ($N_{\rm H}<10^{22}\rm\,cm^{-2}$), obscured ($10^{22}\leq N_{\rm H}<10^{24}\rm\,cm^{-2}$) and CT AGN ($N_{\rm H}\geq10^{24}\rm\,cm^{-2}$, see $\S$\ref{sect:NeutralAbsorption} for details on the absorption properties of the sample).}
\label{fig:loglvsredshift}
% \end{minipage}
\end{figure}

Fig.\,\ref{fig:loglvsredshift} presents the observed 14--195\,keV {\it Swift}/BAT luminosity versus redshift for unobscured ($N_{\rm\,H}<10^{22}\rm\,cm^{-2}$), obscured [$10^{22}\leq (N_{\rm\,H}/\rm cm^{-2})<10^{24}$], CT ($N_{\rm\,H}\geq 10^{24}\rm\,cm^{-2}$) AGN and blazars in the sample.

\tabletypesize{\normalsize}
\begin{deluxetable}{lc} 
\tablecaption{Summary of the soft X-ray spectra used.  \label{tab:instruments}}
%\tablewidth{0pt}
\tablehead{
%\colhead{(1)} & \colhead{(2)} & \colhead{(3)} & \colhead{(4)}  & \colhead{(5)}  \\
%
 \colhead{Facility/Instrument } & \colhead{Sources}  
}
\startdata
%\noalign{\smallskip}
\noalign{\smallskip}
{\it Swift}/XRT	& 	588   \\
\noalign{\smallskip}
{\it XMM-Newton} EPIC/PN	& 	220   \\
\noalign{\smallskip}
{\it Chandra}/ACIS	& 14	   \\
\noalign{\smallskip}
{\it Suzaku}/XIS	& 10	   \\
\noalign{\smallskip}
{\it ASCA} GIS/SIS	& 	3   \\
\noalign{\smallskip}
{\it BeppoSAX}/MECS	& 	 1 
\enddata

\end{deluxetable}

\subsubsection{Photometric redshifts}
The {\it Swift}/BAT 70-month sample includes 28 non-blazar AGN and 7 blazars with no redshift measurement. For the sub-sample of non-blazar AGN we calculated photometric redshifts using the \texttt{LePHARE}\footnote{http://www.cfht.hawaii.edu/~arnouts/LEPHARE/lephare.html} code \citep{Arnouts:1999vn,Ilbert:2006ys}, which is a Spectral Energy Distribution (SED) fitting code based on $\chi^2$ minimization. We adopted a set of templates from \cite{Salvato:2009zr,Salvato:2011ly}, which includes some AGN models from \cite{Polletta:2007ve} and hybrid templates combining AGN and host-galaxy emission. This library has been optimized and extensively tested for SED-fitting of AGN-dominated sources (see \citealp{Salvato:2009zr} for further details). Dust extinction was added to each template as a free parameter in the fit, by assuming the \citet{Calzetti:2000qf} attenuation law. 

To perform the SED-fitting, we collected multi-wavelength photometry in the ultraviolet, optical and infrared regimes. We made use of the publicly available data from the \textit{Galaxy Evolution Explorer} (GALEX, \citealp{Martin:2005iw}) in the far-ultraviolet ($\lambda\sim$ 1550\AA); the \textit{Sloan Digital Sky Survey} Data Release 10 (SDSS-DR10, \citealp{Ahn:2014so}) in the optical ($u$, $g$, $r$, $i$ and $z$ bands); the 2MASS catalog in the near-infrared ($J$, $H$ and $K_{\rm\,s}$ bands);  the {\it WISE} and \textit{AKARI} catalogues in the mid-infrared ($\lambda~\sim$ 3.4, 4.6, 12, 18 and 22$~\mu$m); the \textit{Infra-Red
Astronomical Satellite} (IRAS, \citealp{Neugebauer:1984cy}) in the far-infrared ($\lambda~\sim$ 60 and 100$~\mu$m). We collected broad-band photometry for 27 out of the 28 sources without a listed redshift\footnote{For one source, SWIFT\,J1535.8-5749, we did not retrieve enough data to perform SED-fitting, therefore no redshift is available for this object.}. 

The \texttt{LePHARE} code also builds a probability distribution function (PDF) through the comparison of the observed SED with all the models in the library. This allows us to quantify the uncertainty of the resulting photometric redshift. We finally note that, since they represent a very small fraction ($\sim 4$\%) of our sample, sources with photometric redshifts were only listed out of completeness, and were not used for any study which used the X-ray luminosity.

\section{X-ray data and data analysis}\label{sect:softxrayDA}

The spectral analysis was carried out combining the 70-month time-averaged {\it Swift}/BAT spectra with data obtained by several X-ray facilities: {\it ASCA} ($\S$\ref{sect:ASCA}), {\it Chandra} ($\S$\ref{sect:Chandra}), {\it Suzaku} ($\S$\ref{sect:Suzaku}), {\it Swift}/XRT ($\S$\ref{sect:XRT}), and {\it XMM-Newton} ($\S$\ref{sect:XMM}). Only two AGN (SWIFT\,J1119.5+5132 and SWIFT\,J1313.6+3650A) were not observed by any X-ray facility in the 0.3--10\,keV range, implying a completeness rate of $\sim 99.8\%$. 
The highest energy bin (i.e., 150--195\,keV) of the {\it Swift}/BAT spectrum was not used due to its poor response, such that it has a signal-to-noise ratio a factor of $\sim100-1000$ lower than the other seven BAT energy bins (\citealp{Koss:2013fp}, see top panel of Fig.\,2 of their paper).

The core of our analysis is the spectral decomposition of all the X-ray data available for the {\it Swift}/BAT AGN, to provide measurements of key physical properties related to the X-ray emission, including the intrinsic X-ray luminosity and the column density of matter along the line-of-sight. Therefore we first checked the results obtained by fitting the {\it Swift}/XRT spectrum with a power-law model (see $\S$\ref{sect:Xrayspecanalysis}), visually inspecting the resulting best-fit models and the residuals. We then used X-ray data from {\it Swift}/XRT for the spectral analysis of unobscured sources, unless: i) we found evidence of ionized absorption or peculiar features; ii) {\it Swift}/XRT data had low signal-to-noise ratio or were not available. For these objects we used {\it XMM-Newton} EPIC/PN data, or if no {\it XMM-Newton} observation was publicly available {\it Suzaku}/XIS, {\it Chandra}/ACIS or {\it ASCA} SIS0/SIS1 and GIS2/GIS3 data were used. For obscured sources we used {\it XMM-Newton} EPIC/PN, {\it Suzaku}/XIS, {\it Chandra}/ACIS or {\it ASCA} SIS0/SIS1 and GIS2/GIS3 data. In case none of those were available we used {\it Swift}/XRT observations. For blazars we used {\it Swift}/XRT data, unless none were available. Whenever more than one observation was available, we used the deepest (after accounting for data filtering). We privileged {\it XMM-Newton} EPIC/PN observations over {\it Suzaku}/XIS, {\it Chandra}/ACIS, and {\it ASCA} SIS0/SIS1 and GIS2/GIS3 because of its larger collecting area in the 0.3--10\,keV region, and due to the fact that {\it XMM-Newton} observed a larger number of sources compared to the other satellites. In one case (SWIFT\,J2234.8-2542) the source was only observed by {\it BeppoSAX} below 10\,keV, and we report the results of study of \cite{Malizia:2000yq} combined with the analysis of the {\it Swift}/BAT spectrum. 

\begin{figure}[t!]
\centering
 %% 1st image
 %% 2nd image
%\begin{minipage}[!b]{.48\textwidth}
\centering
\includegraphics[width=9cm,angle=0]{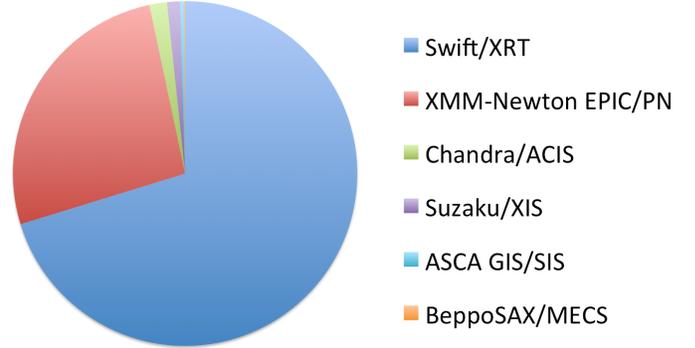}%\end{minipage}
% %% caption
% \begin{minipage}[t]{1\textwidth}
  \caption{Facilities and instruments used for the analysis of the soft X-ray spectra (see Table\,\ref{tab:instruments}).}
\label{fig:pieobs}
% \end{minipage}
\end{figure}

In the following we briefly describe the X-ray instruments we used, and the procedure we adopted, for the spectral extraction. Details of the soft X-ray observation used for the broad-band X-ray spectral analysis of each source are reported in Table\,\ref{tbl-2}. Fig.\,\ref{fig:pieobs} shows the distribution of different instruments used for the analysis of the soft X-ray emission of {\it Swift}/BAT AGN (see also Table\,\ref{tab:instruments}).

\subsection{ASCA}\label{sect:ASCA}

We used {\it ASCA} SIS0/SIS1 and GIS2/GIS3 data for three sources. The reduced spectra were obtained from the Tartarus database\footnote{http://heasarc.gsfc.nasa.gov/FTP/asca/data/tartarus/} \citep{Turner:2001bh}, which collects the products obtained for 611 {\it ASCA} observations of AGN.  

\subsection{Chandra}\label{sect:Chandra}

{\it Chandra} \citep{Weisskopf:2000vn} ACIS \citep{Garmire:2003kx} data were used for 14 sources. The reduction of {\it Chandra}/ACIS data was performed using \textsc{CIAO}\,v.4.6 \citep{Fruscione:2006jl} following the standard procedures. The data were first reprocessed using \textsc{chandra\_repro}, and the source spectra were then extracted using circular apertures of $10\arcsec$ radius, centered on the optical counterpart of each source. Background spectra were extracted using circular regions with identical apertures, centered on regions where no other source was present. Both spectra were extracted using the \textsc{specextract} tool. Sources with significant pileup were modelled with the addition of the \textsc{pileup} model in \textsc{xspec}.

\subsection{Suzaku}\label{sect:Suzaku}

{\it Suzaku} \citep{Mitsuda:2007dq} X-ray Imaging Spectrometer (XIS, \citealp{Koyama:2007cr}) data were used to complement {\it Swift}/BAT spectra for ten source. For most of its operating time XIS was composed of three cameras, the front-illuminated (FI) XIS\,0 and XIS\,3, and the back-illuminated (BI) XIS\,1 (hereafter BI-XIS).

For each of the three XIS cameras, we reprocessed the data and extracted the spectra from the cleaned event files using a circular aperture with a radius of $1.7\,\arcmin$ centred on the source. The background was taken from a source-free annulus centred at the source peak, with an internal and external radius of $3.5\arcmin$ and $5.7\arcmin$, respectively.  We generated the ancillary response matrices (ARFs) and the detector response matrices (RMFs) using the \textsc{xisrmfgen} and \textsc{xissimarfgen} tasks \citep{Ishisaki:2007wd}, respectively. The spectra obtained by XIS\,0 and XIS\,3 were merged using \textsc{mathpha}, \textsc{addrmf} and \textsc{addarf}.

\begin{figure}[t!]
\centering
 %% 1st image
 %% 2nd image
\includegraphics[width=8.8cm]{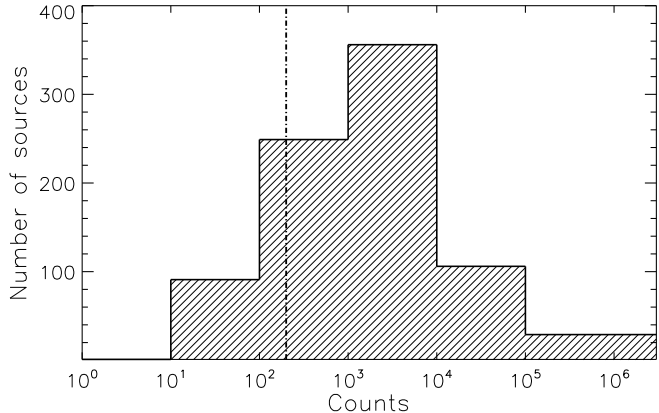}
% %% caption
% \begin{minipage}[t]{1\textwidth}
  \caption{Distribution of the $\leq 10$\,keV counts for the AGN of our sample with soft X-ray observations available. The vertical dot-dashed line shows the threshold used to separate objects fitted with Cash ($< 200$ counts) and $\chi^{2}$ ($\geq 200$ counts) statistics (see $\S$\ref{sect:Xrayspecanalysis} for details). The counts from different instruments were summed for observations carried out but {\it Suzaku}/XIS and {\it ASCA} GIS/SIS. The median of the number of counts of our sample is $\sim 1,600$.}
\label{fig:hist_ctrate}
% \end{minipage}
\end{figure}

\subsection{Swift/XRT}\label{sect:XRT}

The X-ray telescope (XRT, \citealp{Burrows:2005vn}) on board {\it Swift} followed up nearly all of the sources detected by BAT in the first 70-months of operations. {\it Swift}/XRT data analysis was performed using the \textsc{xrtpipeline} following the standard guidelines \citep{Evans:2009fk}. {\it Swift}/XRT observations were used for a total of 588 sources.

\subsection{XMM-Newton}\label{sect:XMM}

We used {\it XMM-Newton} \citep{Jansen:2001ve}  EPIC/PN \citep{Struder:2001uq} observations for 220 sources. The original data files were reduced using the {\it XMM-Newton} Standard Analysis Software version 12.0.1 \citep{Gabriel:2004fk}, and the raw PN data files were then processed using the \texttt{epchain} task.

For every observation, we inspected the background light curve in the 10--12 keV energy band, in order to filter the exposures for periods of high-background activity. Only patterns corresponding to single and double events (PATTERN~$\leq 4$) were selected. We extracted the source spectra from the final filtered event list using circular apertures centred on the object, with a typical radius of $20\arcsec$. Regions with smaller radii were used for the sources detected with a low signal-to-noise ratio. The background was extracted from circular regions of $40\arcsec$ radius, located on the same CCD as the source, where no other X-ray source was detected. We checked for the presence of pile-up using the \textsc{epatplot} task. For observations with significant pile-up we used an annular region with an inner radius set such that no pile-up was present. Finally, we created ARFs and RMFs using the \textsc{arfgen} and \textsc{rmfgen} tasks, respectively.

\begin{figure}[t!]
\centering
 %% 1st image
 %% 2nd image
\begin{minipage}[!b]{.48\textwidth}
\includegraphics[width=9cm]{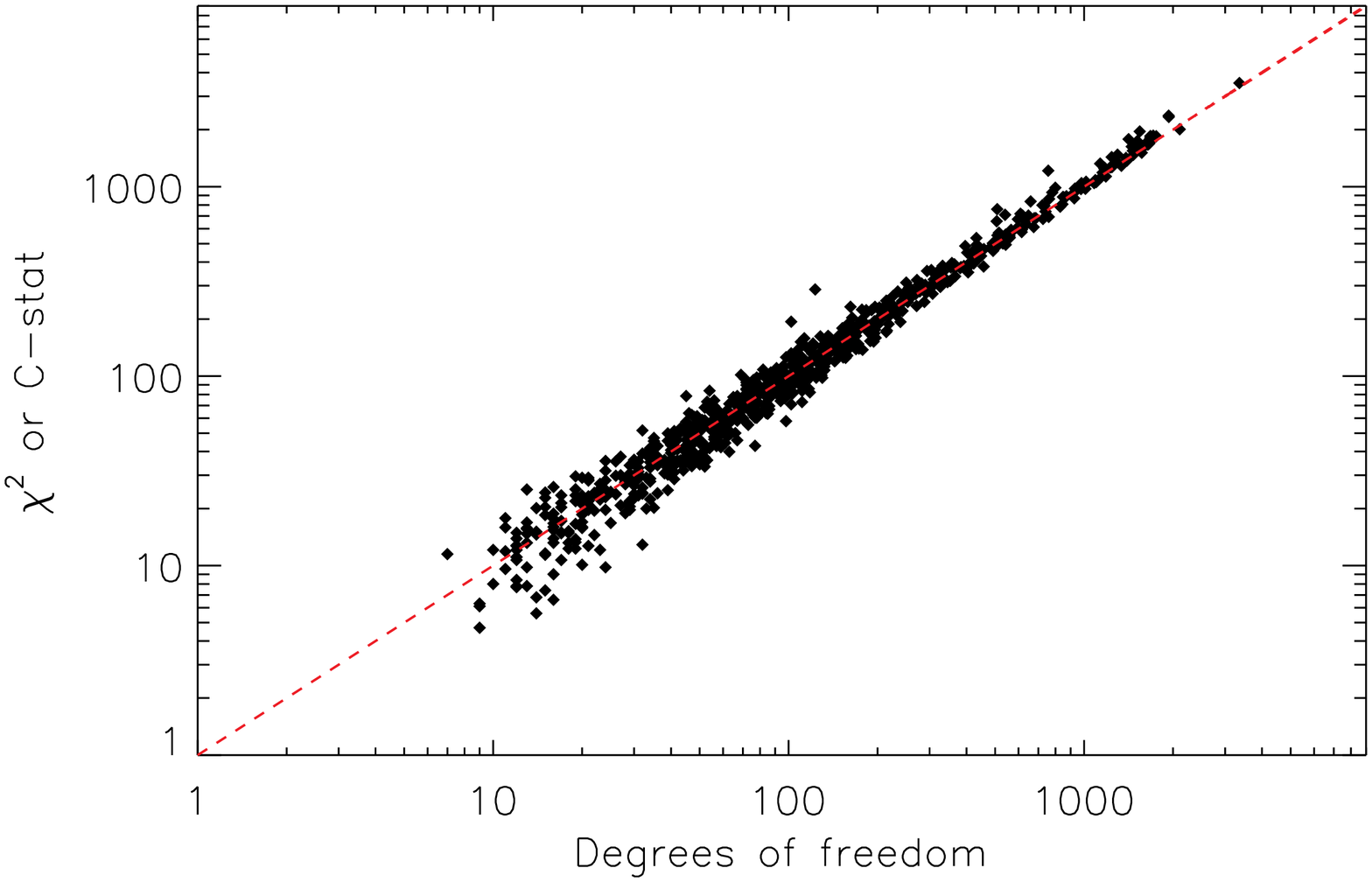}\end{minipage}
\par\medskip
\begin{minipage}[!b]{.48\textwidth}
\includegraphics[width=9cm]{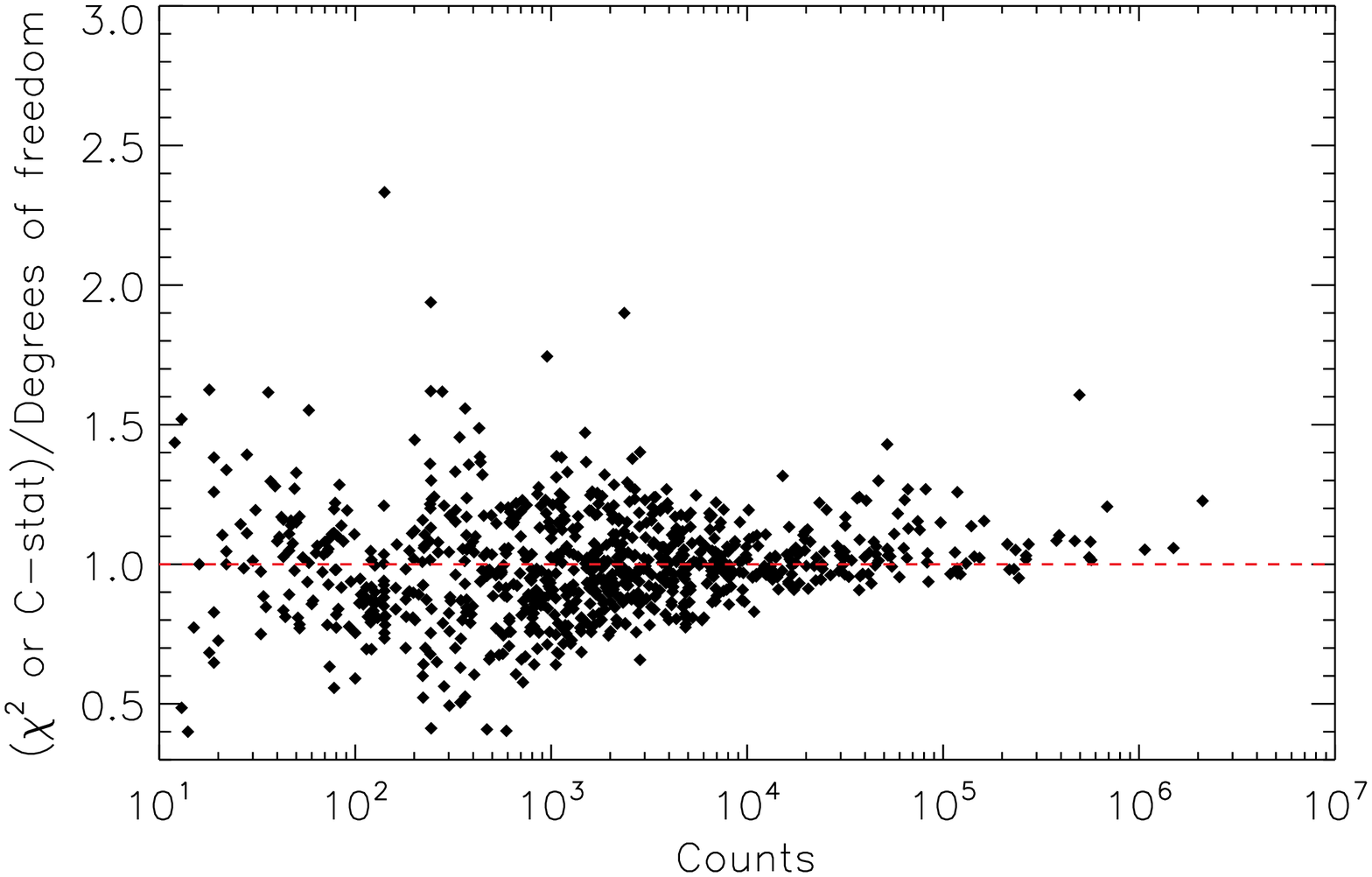}\end{minipage}
% %% caption
 \begin{minipage}[t]{.48\textwidth}
\caption{{\it Top panel:} values of the $\chi^2$ and the C-stat obtained by fitting the whole sample of AGN versus the DOF. {\it Bottom panel:} ratio between $\chi^2$ (or C-stat) and the DOF versus the number of counts. In both panels the red dashed line represents $\chi^2$/DOF=1 or C-Stat/DOF=1.}
\label{fig:ChiDOF}
 \end{minipage}
\end{figure}

\medskip

\section{X-ray spectral analysis}\label{sect:Xrayspecanalysis}

The X-ray spectral analysis was carried out using \textsc{xspec}\,v.12.7.1b \citep{Arnaud:1996kx}. For all models reported in what follows, we took into account Galactic absorption in the direction of the source, by adding photoelectric absorption (i.e., using the \textsc{tbabs} component in \textsc{xspec}, \citealp{Wilms:2000vn}), fixed to the value from the HI maps of \cite{Kalberla:2005fk}, assuming Solar metallicity. The combined X-ray spectra were then analyzed using a series of models of successive complexity, which are listed in Table\,\ref{tab:modelcomponents}. First, all sources were fitted using a simple power-law model [\textsc{tbabs$_{\rm\,Gal}\cdot$(zpow)} in \textsc{xspec}], and the residuals were then visually inspected to assess whether the X-ray spectrum showed signatures of neutral or ionized absorption. A cross-calibration constant ($C_{\rm\,BAT}$) was added to all models to take into account possible variability between the 70-month averaged {\it Swift}/BAT spectra and the considerably shorter soft X-ray observations, as well as cross-calibration uncertainties. It should be remarked that this factor does not take into account possible spectral variability between the hard and soft X-ray spectra, which might accompany flux variability.

Sources were divided into two main categories depending on their BZCAT classification: non-blazar AGN ($\S$\ref{sect:fitUnob} and $\S$\ref{sect:fitOb}) and blazars ($\S$\ref{sect:fitBlaz}). The spectra of eight sources originally classified as blazars show signatures of reprocessed X-ray emission, or are heavily contaminated by other components, and were therefore fitted using non-blazar models. These sources are 3C\,120 \citep{Kataoka:2007fk}, 3C\,273 (\citealp{Haardt:1998jk}, \citealp{Soldi:2008kq}), Cen\,A  \citep{Evans:2004mz}, Mrk\,348 \citep{Marchese:2014vn}, NGC\,1052 \citep{Brenneman:2009sf} and NGC\,7213 \citep{Bianchi:2008ua}. Besides these sources, the X-ray spectrum of Mrk\,1501 also shows evidence of reprocessed X-ray radiation, in the form of a Fe K$\alpha$ feature. NGC\,1275 was also fitted using a different model, since its X-ray spectrum shows peculiar features due to the fact that the source is located at the center of the Perseus cluster. Non-blazar AGN were then further divided, based on the initial power-law fit, into two categories: those showing relatively weak intrinsic absorption from neutral material ($\S$\ref{sect:fitUnob}) and those showing clear signatures of obscuration ($\S$\ref{sect:fitOb}). Different sets of models were used for sources in different categories. In all cases we started with the simplest models and, after visual inspection of the residuals, we increased their complexity, adding components if the fit was significantly improved.

We used $\chi^{2}$ statistics to fit the soft X-ray spectra when the number of counts was $\geq 200$, and Cash statistics (C-stat, \citealp{Cash:1979ys}) when it was below 200. For the 692 sources for which more than 200 source counts were available, we rebinned the spectral data to have 20 counts per bin and used $\chi^2$ statistics. For the remaining 144 objects we rebinned the soft X-ray spectra to have one count per bin and adopted Cash statistics, while we still used $\chi^2$ statistics for the {\it Swift}/BAT spectra. The median number of counts across the entire sample is 1,600. Fig.\,\ref{fig:hist_ctrate} presents the distribution of counts below 10\,keV. Fits were considered to be significantly improved by the addition of a component if $\Delta \chi^2> 2.71$ (or  $\Delta \rm C-stat^2> 2.71$) for each extra free parameter.

\begin{figure}[t!]
\centering
\begin{minipage}[!b]{.48\textwidth}
\centering
\includegraphics[width=9cm]{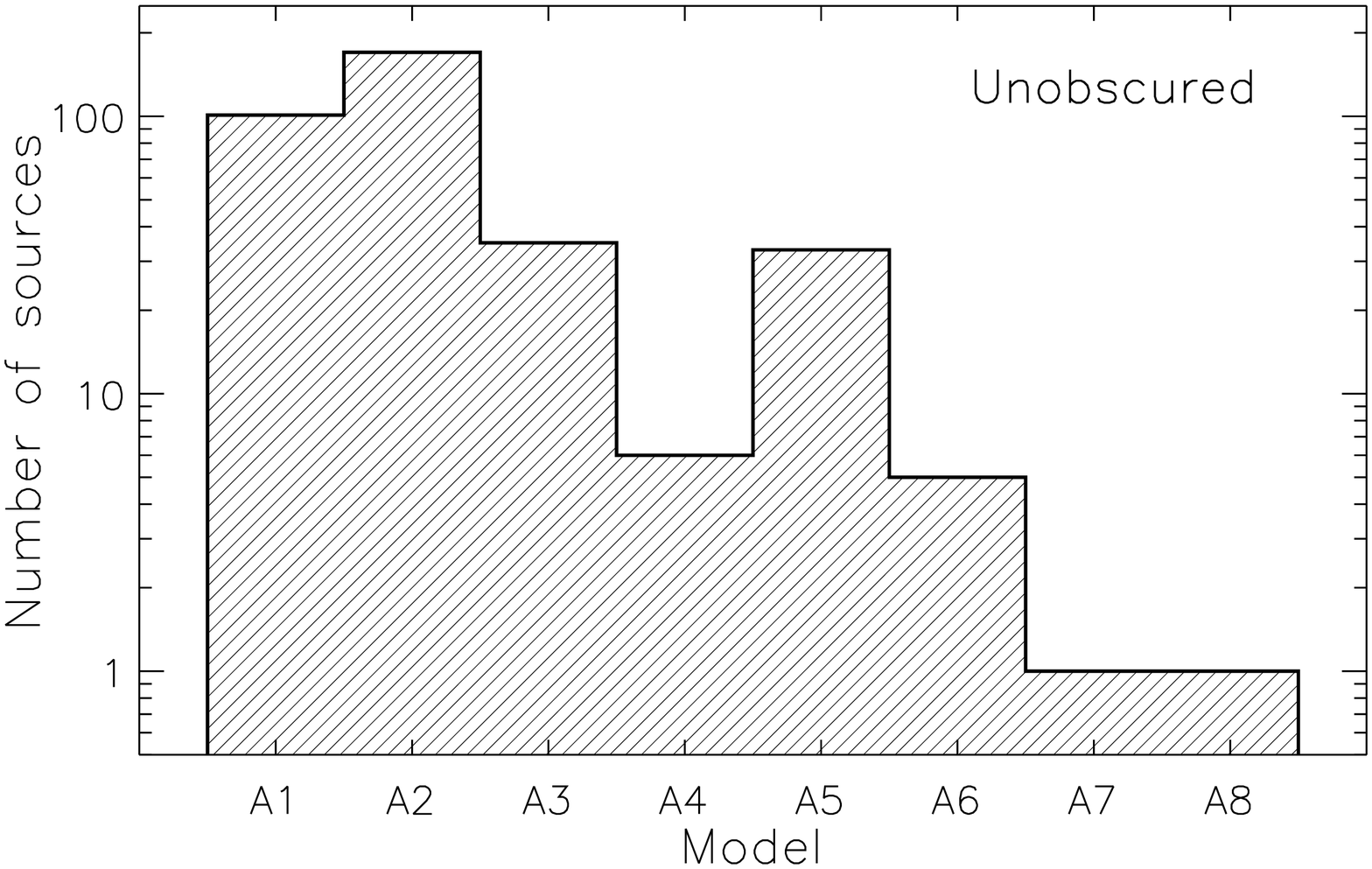}\end{minipage}
\par\smallskip

\begin{minipage}[!b]{.48\textwidth}
\centering
\includegraphics[width=9cm]{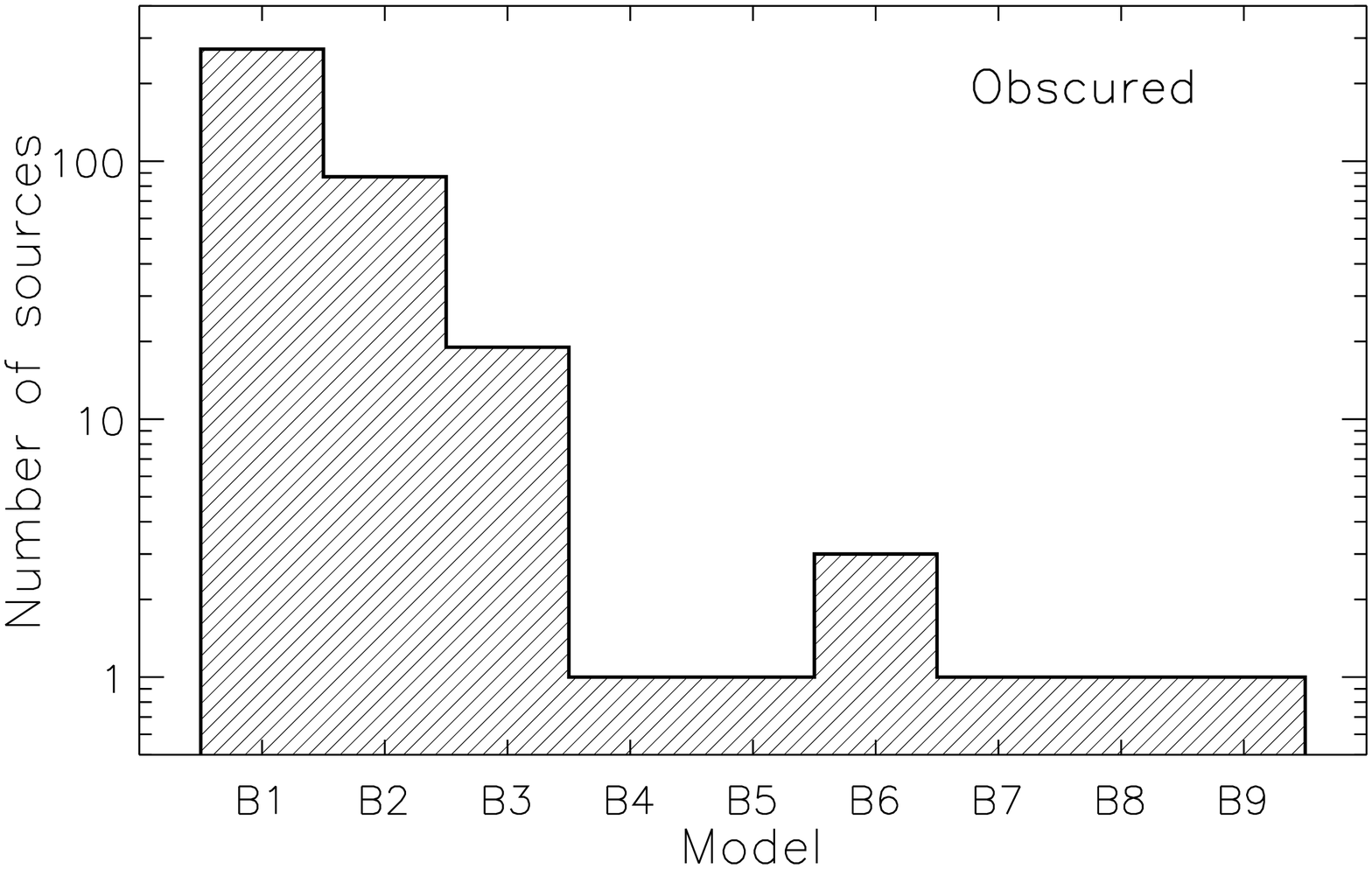}\end{minipage}
\par\smallskip

\begin{minipage}[!b]{.48\textwidth}
\centering
\includegraphics[width=9cm]{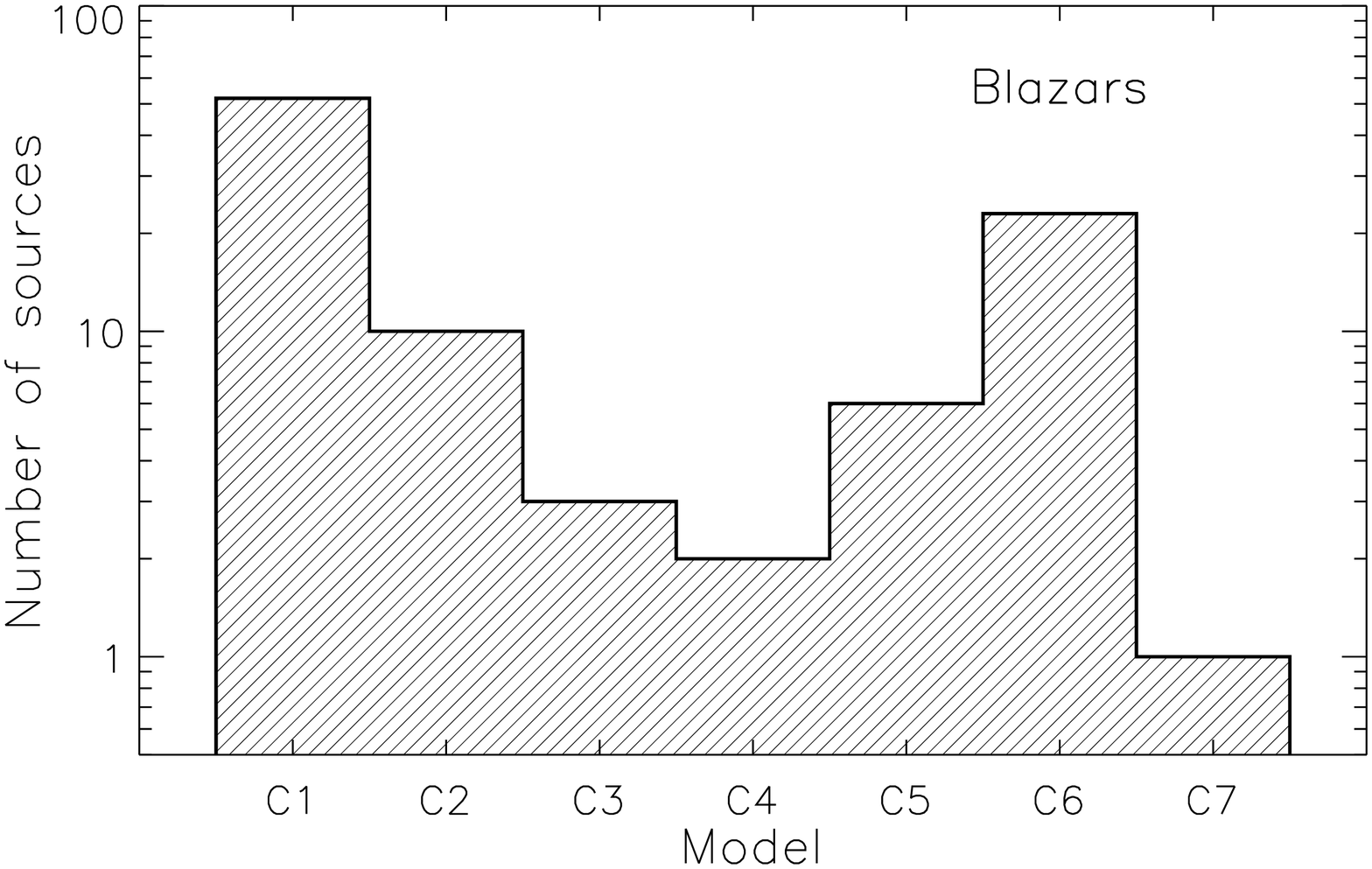}\end{minipage}

% %% caption
 \begin{minipage}[t]{.48\textwidth}
  \caption{Distribution of the different X-ray spectral models used for unobscured AGN (top panel, see $\S$\ref{sect:fitUnob} for details), obscured AGN (middle panel, $\S$\ref{sect:fitOb}) and blazars (bottom panel, $\S$\ref{sect:fitBlaz}). Overall 26 spectral models were used to fit the broad-band X-ray spectra of the AGN in the 70-month {\it Swift}/BAT catalog. The different components used in the models are listed in Table\,\ref{tab:modelcomponents}, and the models are illustrated in Figs.\,\ref{fig:spectra_model1}--\ref{fig:spectra_model4} (with the exception of models D1 and D2, see $\S$\ref{sec:othermodels}).}
\label{fig:distribution_models}
 \end{minipage}
\end{figure}

\tabletypesize{\normalsize}
\begin{deluxetable*}{lccccccccccc} 
\tablecaption{Summary of the components included in the different X-ray spectral models.  \label{tab:modelcomponents}}
\tablewidth{0pt}
\tablehead{
%\colhead{(1)} & \colhead{(2)} & \colhead{(3)} & \colhead{(4)}  & \colhead{(5)}  \\
%
\\
 \colhead{Model } & \colhead{\textsc{tbabs$_{\rm\,Gal}$}} & \colhead{\textsc{$C_{\rm\,BAT}$}} & \colhead{\textsc{zphabs}} & \colhead{\textsc{cabs}}& \colhead{\textsc{zxipcf}} & \colhead{\textsc{zpcfabs}} & \colhead{\textsc{pexrav$^{a}$}} & \colhead{\textsc{bb}}  &  \colhead{\textsc{apec}}  & & \colhead{Sources}  
}
\startdata
%\noalign{\smallskip}
\noalign{\smallskip}
 \multicolumn{1}{c}{A1}  & \cmark   &  \cmark    & \cmark   &   \cmark &   \nodata   & \nodata &  \cmark  &   \nodata   &   \nodata  & & 101   \\
\noalign{\smallskip}
\multicolumn{1}{c}{A2}  & \cmark   &  \cmark    & \cmark   &   \cmark &   \nodata  &  \nodata &  \cmark  &   \cmark   &   \nodata  & & 170   \\
\noalign{\smallskip}
\multicolumn{1}{c}{A3}  & \cmark   &  \cmark    & \cmark   &   \cmark &   \cmark  & \nodata &  \cmark  &   \nodata    &   \nodata   & & 35  \\
\noalign{\smallskip}
\multicolumn{1}{c}{A4}  & \cmark   &  \cmark    & \cmark   &   \cmark &   \cmark \cmark & \nodata &  \cmark  &   \nodata    &   \nodata   & & 6   \\
\noalign{\smallskip}
\multicolumn{1}{c}{A5}  & \cmark   &  \cmark    & \cmark   &   \cmark  &   \cmark & \nodata &  \cmark  &   \cmark    &   \nodata   &  &33  \\
\noalign{\smallskip}
\multicolumn{1}{c}{A6}  & \cmark   &  \cmark    & \cmark   &   \cmark  &   \cmark \cmark & \nodata &  \cmark  &   \cmark   &   \nodata    & & 5  \\
\noalign{\smallskip}
\multicolumn{1}{c}{A7}  & \cmark   &  \cmark    & \cmark   &   \cmark  &   \nodata &\nodata   & \cmark  &   \nodata    &   \cmark \cmark  & &  1   \\
\noalign{\smallskip}
\multicolumn{1}{c}{A8}  & \cmark   &  \cmark    & \cmark   &   \cmark  &   \cmark & \cmark &  \cmark  &   \cmark    &   \nodata   &  & 1  \\
\noalign{\smallskip}
\noalign{\smallskip}
\hline
 \multicolumn{11}{c}{ } \\
\noalign{\smallskip}
Model  & \textsc{tbabs$_{\rm\,Gal}$} & \textsc{$C_{\rm\,BAT}$} & \textsc{zphabs} & \textsc{cabs} & \textsc{zxipcf} & \textsc{zpcfabs}  & \textsc{cutoffpl} & \textsc{pexrav$^{b}$} & \textsc{apec}  &  \textsc{scatt} &   \\
\noalign{\smallskip}
% \multicolumn{11}{c}{ } \\
\hline
\noalign{\smallskip}
\noalign{\smallskip}
\multicolumn{1}{c}{B1}  & \cmark   &  \cmark    & \cmark   &   \cmark  &   \nodata  &   \nodata &   \cmark  &   \cmark   &   \nodata    &  \cmark  &   272  \\
\noalign{\smallskip}
\multicolumn{1}{c}{B2}  & \cmark   &  \cmark    & \cmark   &   \cmark  &   \nodata  &   \nodata &   \cmark  &   \cmark   &   \cmark   &  \cmark  & 87   \\
\noalign{\smallskip}
\multicolumn{1}{c}{B3}  & \cmark   &  \cmark    & \cmark   &   \cmark  &   \nodata  &   \nodata &   \cmark  &   \cmark   &   \cmark \cmark    &  \cmark  &  19  \\
\noalign{\smallskip}
\multicolumn{1}{c}{B4}  & \cmark   &  \cmark    & \cmark   &   \cmark  &   \nodata  &   \nodata &   \cmark  &   \cmark   &   \cmark \cmark \cmark    &  \cmark  &  1  \\
\noalign{\smallskip}
\multicolumn{1}{c}{B5}  & \cmark   &  \cmark    & \cmark   &   \cmark  &   \cmark  &   \nodata &   \cmark  &   \cmark   &   \cmark    &  \cmark  & 1   \\
\noalign{\smallskip}
\multicolumn{1}{c}{B6}  & \cmark   &  \cmark    & \cmark   &   \cmark  &   \nodata  &   \cmark \cmark &   \nodata  &   \cmark$^{a}$   &   \nodata    &  \nodata  & 3   \\
\noalign{\smallskip}
\multicolumn{1}{c}{B7}  & \cmark   &  \cmark    & \cmark   &   \cmark  &   \nodata  &   \nodata &   \cmark  &   \cmark   &   \cmark \cmark$^{c}$    &  \cmark  & 1   \\
\noalign{\smallskip}
\multicolumn{1}{c}{B8}  & \cmark   &  \cmark    & \cmark   &   \cmark  &   \nodata  &   \nodata &   \cmark  &   \cmark   &   \cmark$^{c}$ \cmark$^{c}$    &  \cmark  & 1    \\
\noalign{\smallskip}
\multicolumn{1}{c}{B9}  & \cmark   &  \cmark    & \cmark   &   \cmark  &   \nodata  &   \nodata &   \cmark  &   \cmark   &   \cmark \cmark    &  \nodata  &  1  \\
\noalign{\smallskip}
\noalign{\smallskip}
\hline
 \multicolumn{11}{c}{ } \\
\noalign{\smallskip}
Model  & \textsc{tbabs$_{\rm\,Gal}$} & \textsc{$C_{\rm\,BAT}$} & \textsc{zphabs} & \textsc{cabs} & \textsc{zxipcf} & \textsc{pow}  & \textsc{bkn} & \textsc{bkn2} & \textsc{bb}  &  \textsc{scatt} &   \\
% \multicolumn{11}{c}{ } \\
\noalign{\smallskip}
\hline
\noalign{\smallskip}
\multicolumn{1}{c}{C1}  & \cmark   &  \cmark    & \cmark   &   \cmark  &   \nodata  &   \cmark &   \nodata  &   \nodata   &   \nodata     &  \nodata  &   52 \\
\noalign{\smallskip}
\multicolumn{1}{c}{C2}  & \cmark   &  \cmark    & \cmark   &   \cmark  &   \nodata  &   \cmark &   \nodata  &   \nodata   &   \cmark     &  \nodata  &  10  \\
\noalign{\smallskip}
\multicolumn{1}{c}{C3}  & \cmark   &  \cmark    & \cmark   &   \cmark  &   \cmark  &   \cmark &   \nodata  &   \nodata   &   \nodata     &  \nodata  & 3   \\
\noalign{\smallskip}
\multicolumn{1}{c}{C4}  & \cmark   &  \cmark    & \cmark   &   \cmark  &   \cmark \cmark &   \cmark &   \nodata  &   \nodata   &   \nodata     &  \nodata  &  2  \\ 
\noalign{\smallskip}
\multicolumn{1}{c}{C5}  & \cmark   &  \cmark    & \cmark   &   \cmark  &   \nodata  &   \cmark &   \nodata  &   \nodata   &   \nodata     &  \cmark  & 6   \\
\noalign{\smallskip}
\multicolumn{1}{c}{C6}  & \cmark   &  \cmark    & \cmark   &   \cmark  &   \nodata  &   \nodata &   \cmark  &   \nodata   &   \nodata     &  \nodata  &  23  \\
\noalign{\smallskip}
\multicolumn{1}{c}{C7}  & \cmark   &  \cmark    & \cmark   &   \cmark  &   \nodata  &   \nodata   &   \nodata &   \cmark  &   \cmark     &  \nodata  & 1   \\
\enddata

\tablecomments{The table lists the different components that were used by the models, and the number of sources for which each model was adopted. When more than one checkmark is reported, the component was used more than once. \textsc{scatt} is the scattered component (\textsc{$f_{\rm\,scat}\times$cutoffpl} in \textsc{xspec}), while $C_{\rm\,BAT}$ is the cross-calibration constant (\textsc{cons} in \textsc{xspec}).
Details about the spectral components can be found in $\S$\ref{sect:modelcomponents}, while the accurate syntax used in \textsc{xspec} is reported in $\S$\ref{sect:fitUnob}--\ref{sect:fitBlaz}.}
\tablenotetext{a}{The reflection parameter was set to be $R\geq 0$, i.e. the component takes into account the primary X-ray emission (in the form of a cutoff power-law) and reprocessed radiation at the same time.}
\tablenotetext{b}{The reflection parameter was set to be negative, i.e. the reflection component is disconnected from the primary X-ray emission and assumed to be unobscured.}
\tablenotetext{c}{Absorption by neutral material was considered for the thermal component.}

%\tablenotetext{b}{Another sample footnote for table~\ref{tbl-1}}

\end{deluxetable*}

\begin{figure*}[h!]
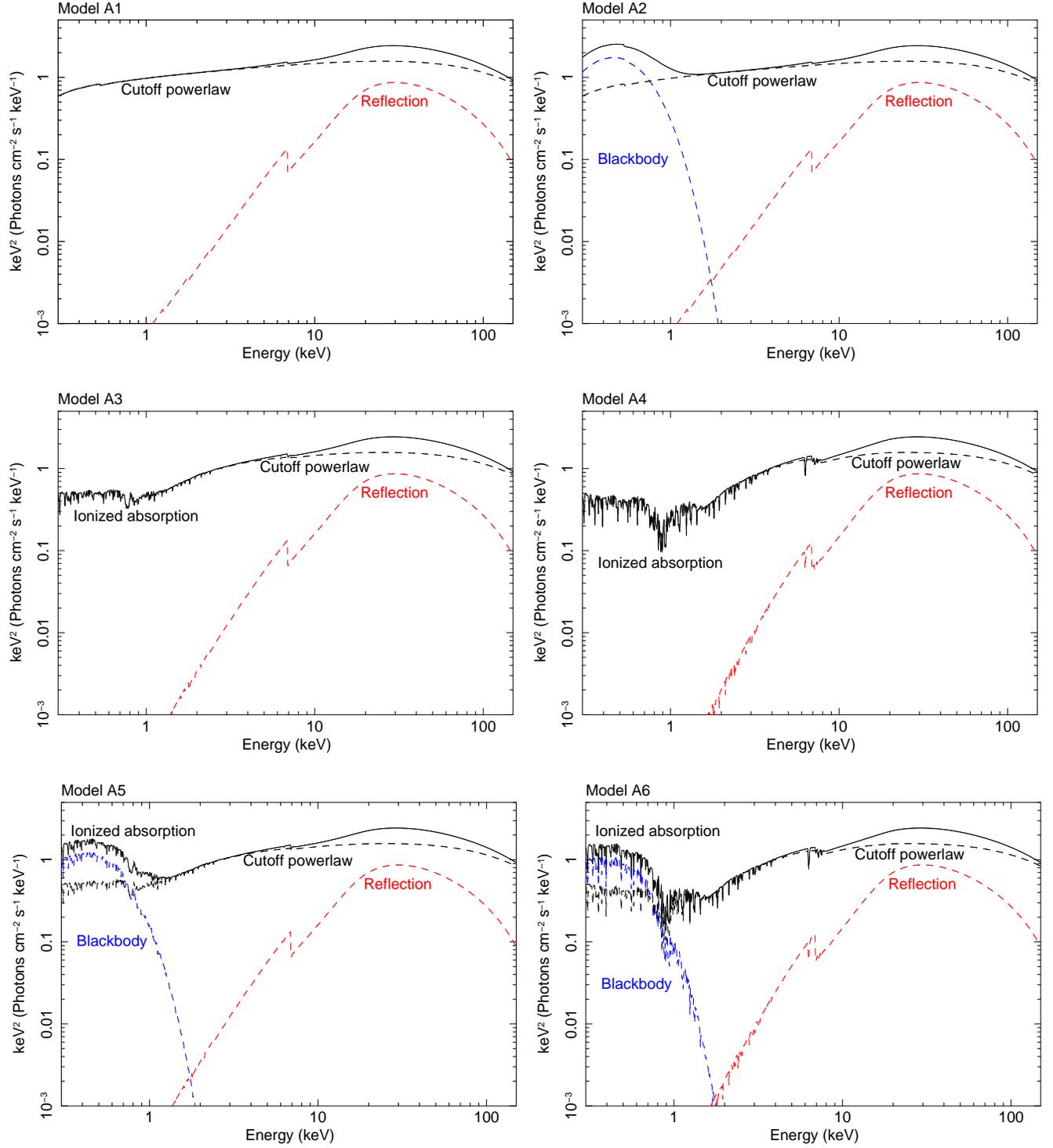

\centering
 %% 1st image
 %% 2nd image
\begin{minipage}[!b]{.48\textwidth}
\centering
\includegraphics[height=9cm,angle=270]{modelA1.ps}\end{minipage}
\begin{minipage}[!b]{.48\textwidth}
\centering
\includegraphics[height=9cm,angle=270]{modelA2.ps}\end{minipage}
\par\bigskip
\begin{minipage}[!b]{.48\textwidth}
\centering
\includegraphics[height=9cm,angle=270]{modelA3.ps}\end{minipage}
\begin{minipage}[!b]{.48\textwidth}
\centering
\includegraphics[height=9cm,angle=270]{modelA4.ps}\end{minipage}
\par\bigskip
\begin{minipage}[!b]{.48\textwidth}
\centering
\includegraphics[height=9cm,angle=270]{modelA5.ps}\end{minipage}
\begin{minipage}[!b]{.48\textwidth}
\centering
\includegraphics[height=9cm,angle=270]{modelA6.ps}\end{minipage}
%
% %% caption
 \begin{minipage}[t]{1\textwidth}
  \caption{Models used for the X-ray spectral analysis (A1 to A6). In the models A1 to A6 the reflection and primary components were decoupled in \textsc{pexrav} for the purpose of visual clarity. See Table\,\ref{tab:modelcomponents} and $\S$\ref{sect:spectralModels} for details. }
\label{fig:spectra_model1}
 \end{minipage}
\end{figure*}

\begin{figure*}[h!]
\centering
 %% 1st image
 %% 2nd image
%
\begin{minipage}[!b]{.48\textwidth}
\centering
\includegraphics[height=9cm,angle=270]{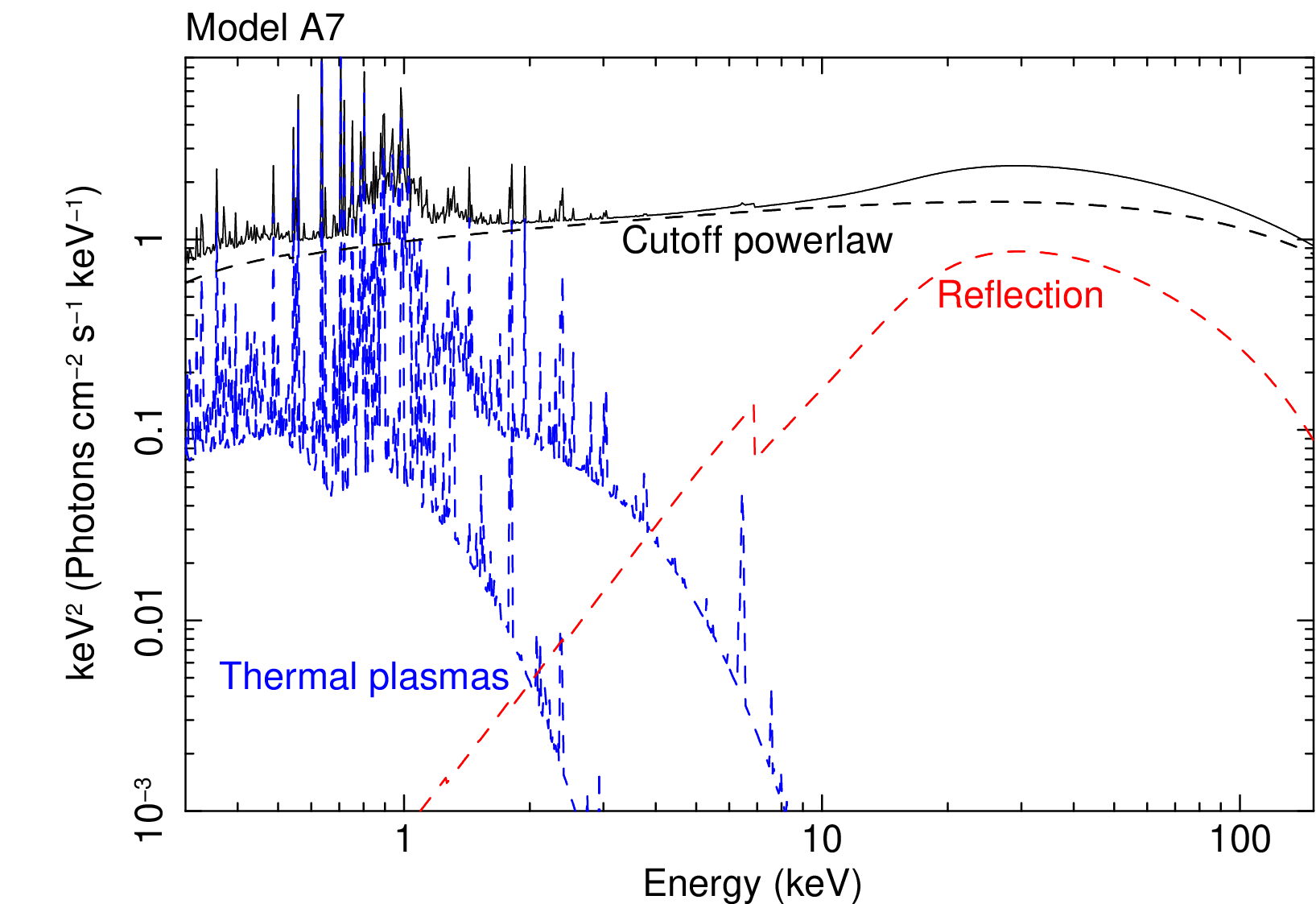}\end{minipage}
\begin{minipage}[!b]{.48\textwidth}
\centering
\includegraphics[height=9cm,angle=270]{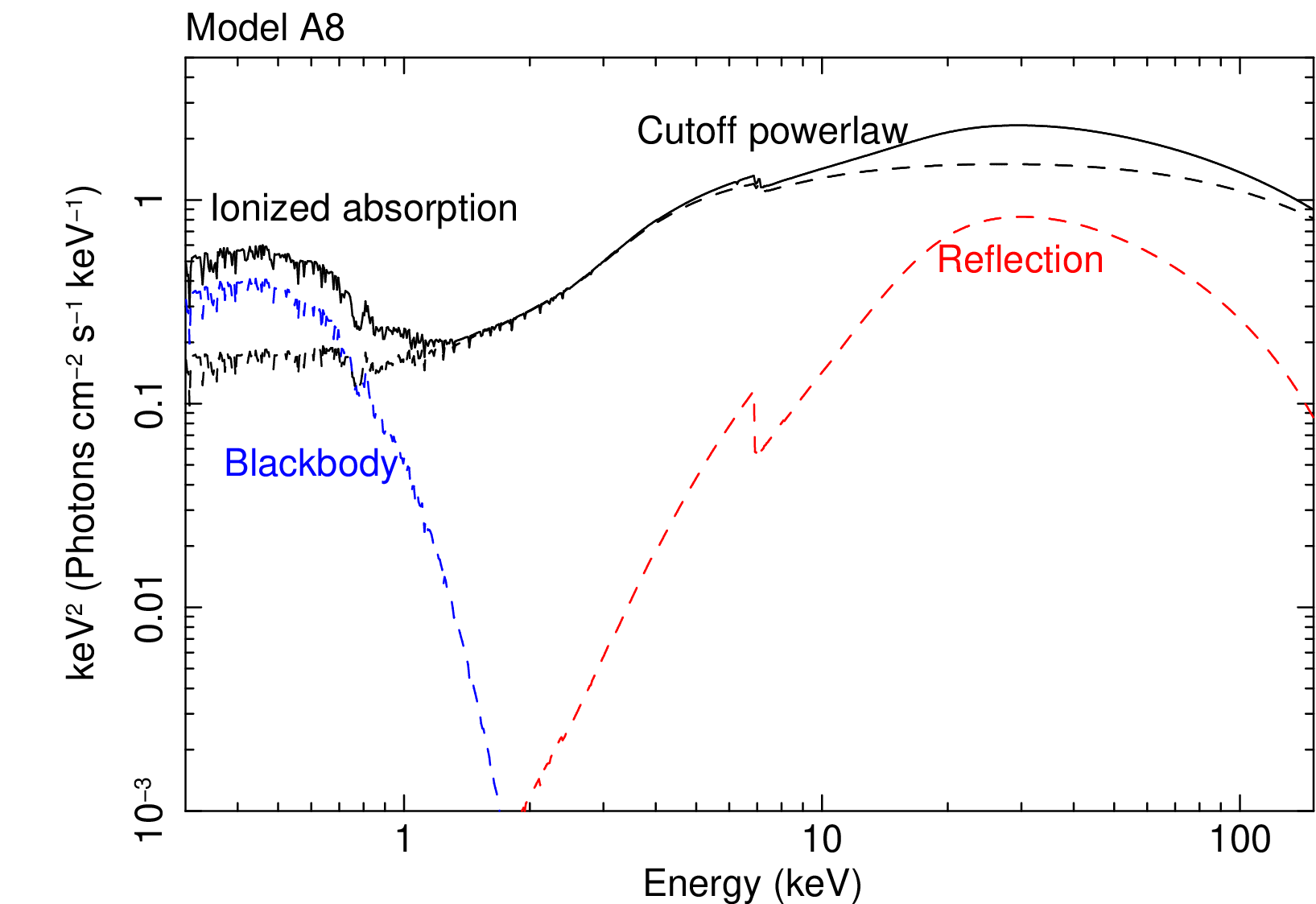}\end{minipage}
\par\bigskip
\begin{minipage}[!b]{.48\textwidth}
\centering
\includegraphics[height=9cm,angle=270]{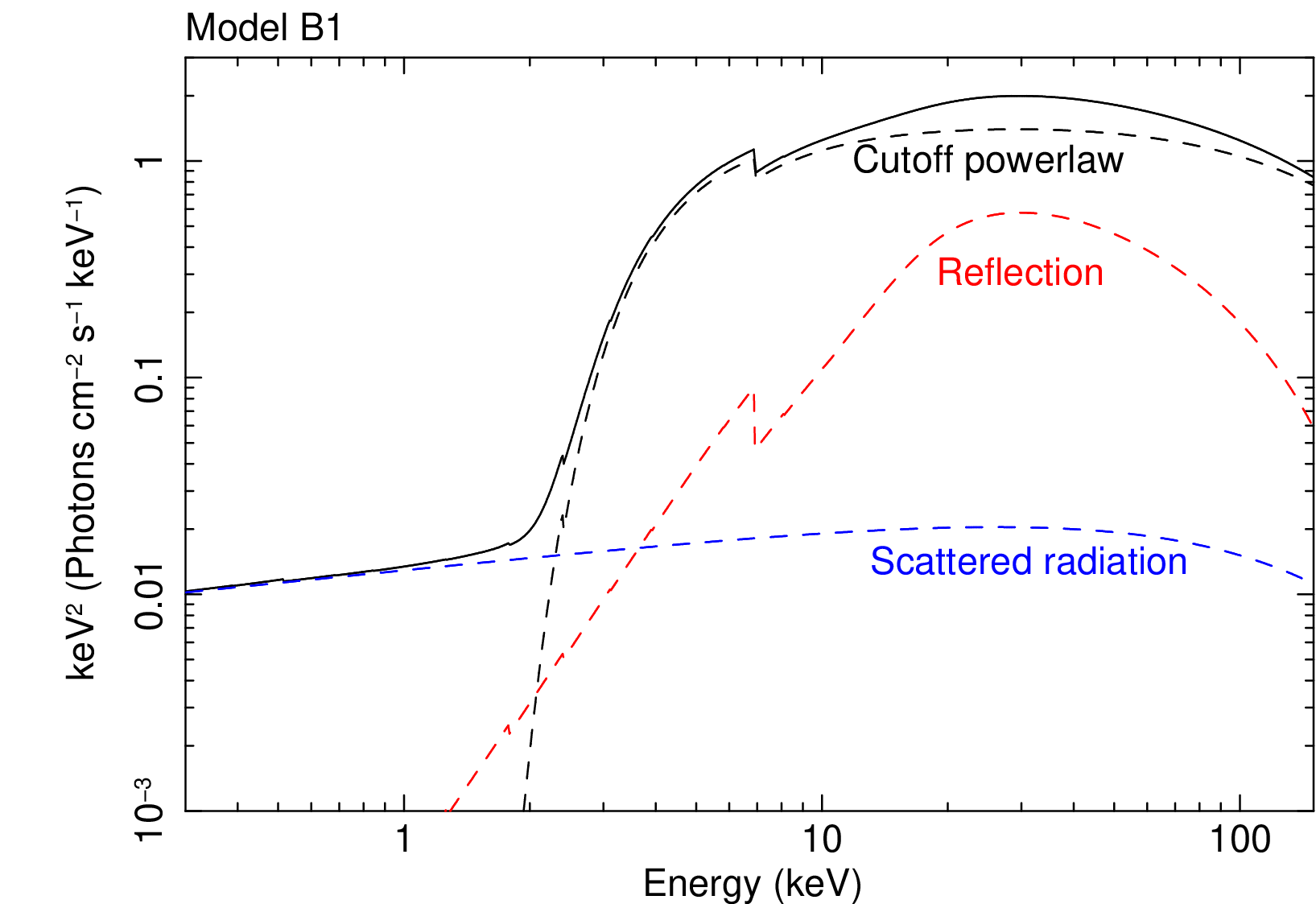}\end{minipage}
\begin{minipage}[!b]{.48\textwidth}
\centering
\includegraphics[height=9cm,angle=270]{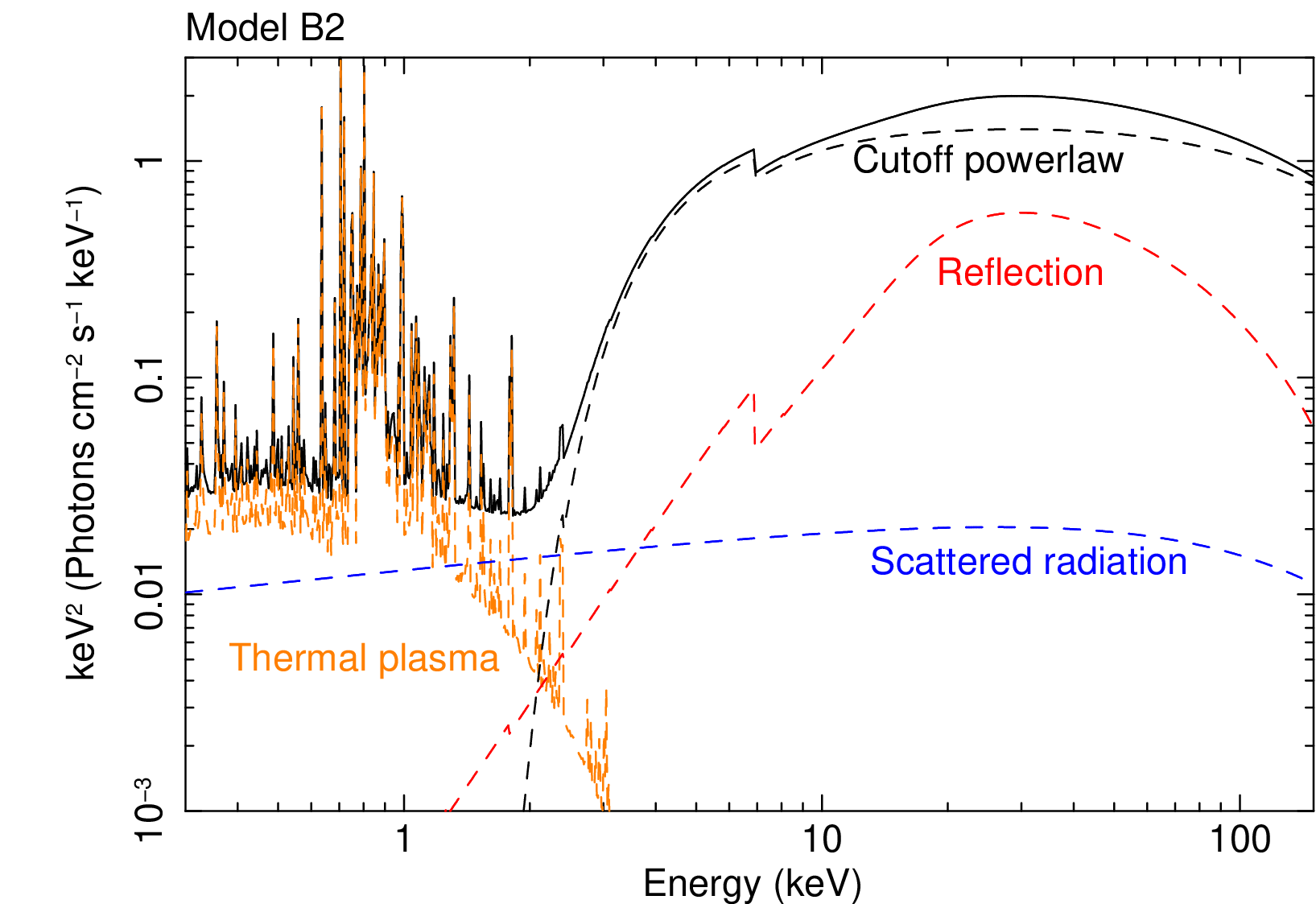}\end{minipage}
\par\bigskip
\begin{minipage}[!b]{.48\textwidth}
\centering
\includegraphics[height=9cm,angle=270]{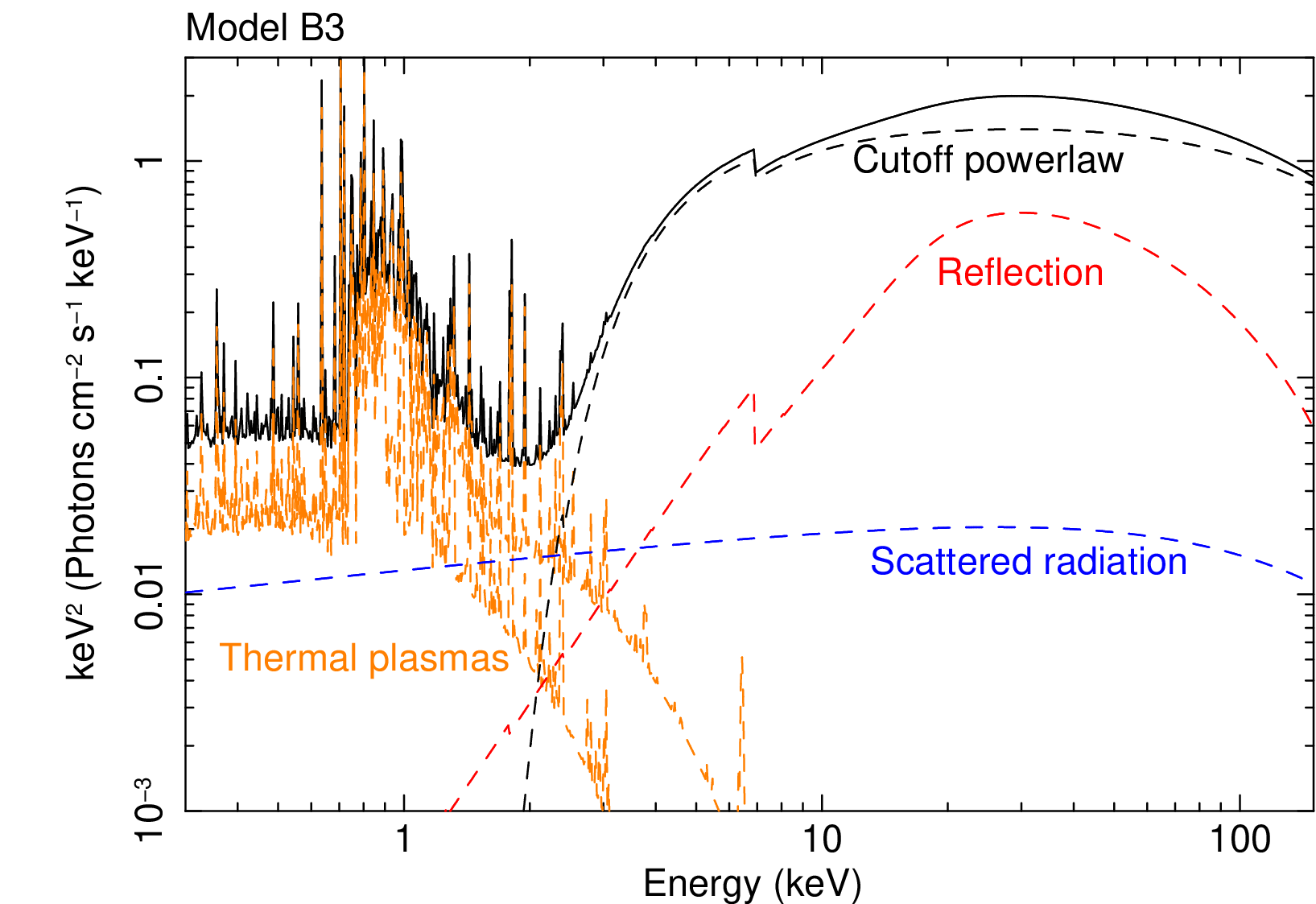}\end{minipage}
\begin{minipage}[!b]{.48\textwidth}
\centering
\includegraphics[height=9cm,angle=270]{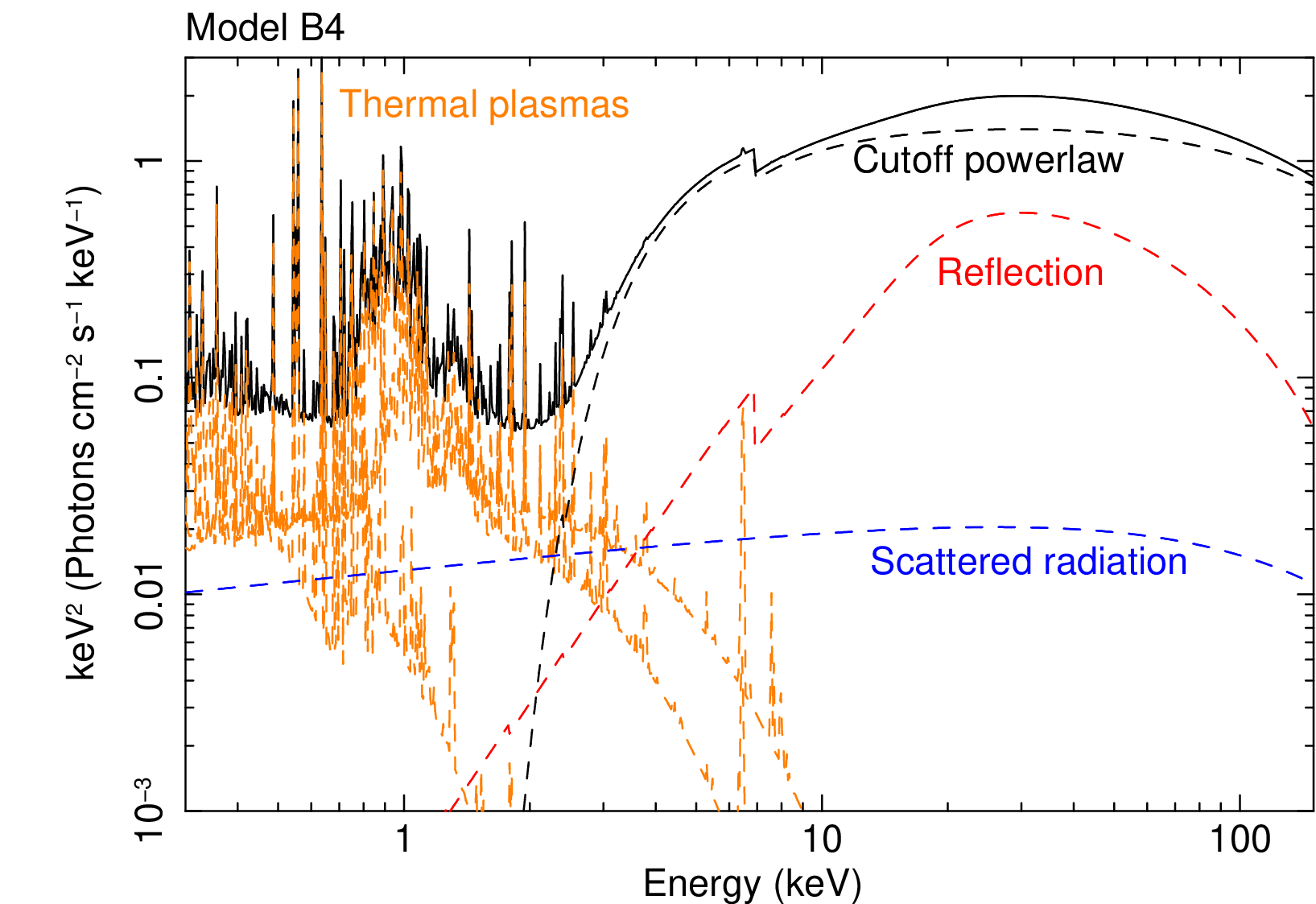}\end{minipage}
%
% %% caption
 \begin{minipage}[t]{1\textwidth}
  \caption{Models used for the X-ray spectral analysis (A7 to B4).  In model A7 and A8 the reflection and primary components were decoupled in \textsc{pexrav} for the purpose of visual clarity. See Table\,\ref{tab:modelcomponents} and $\S$\ref{sect:spectralModels} for details.}
\label{fig:spectra_model2}
 \end{minipage}
\end{figure*}

\begin{figure*}[h!]
\centering
\begin{minipage}[!b]{.48\textwidth}
\centering
\includegraphics[height=9cm,angle=270]{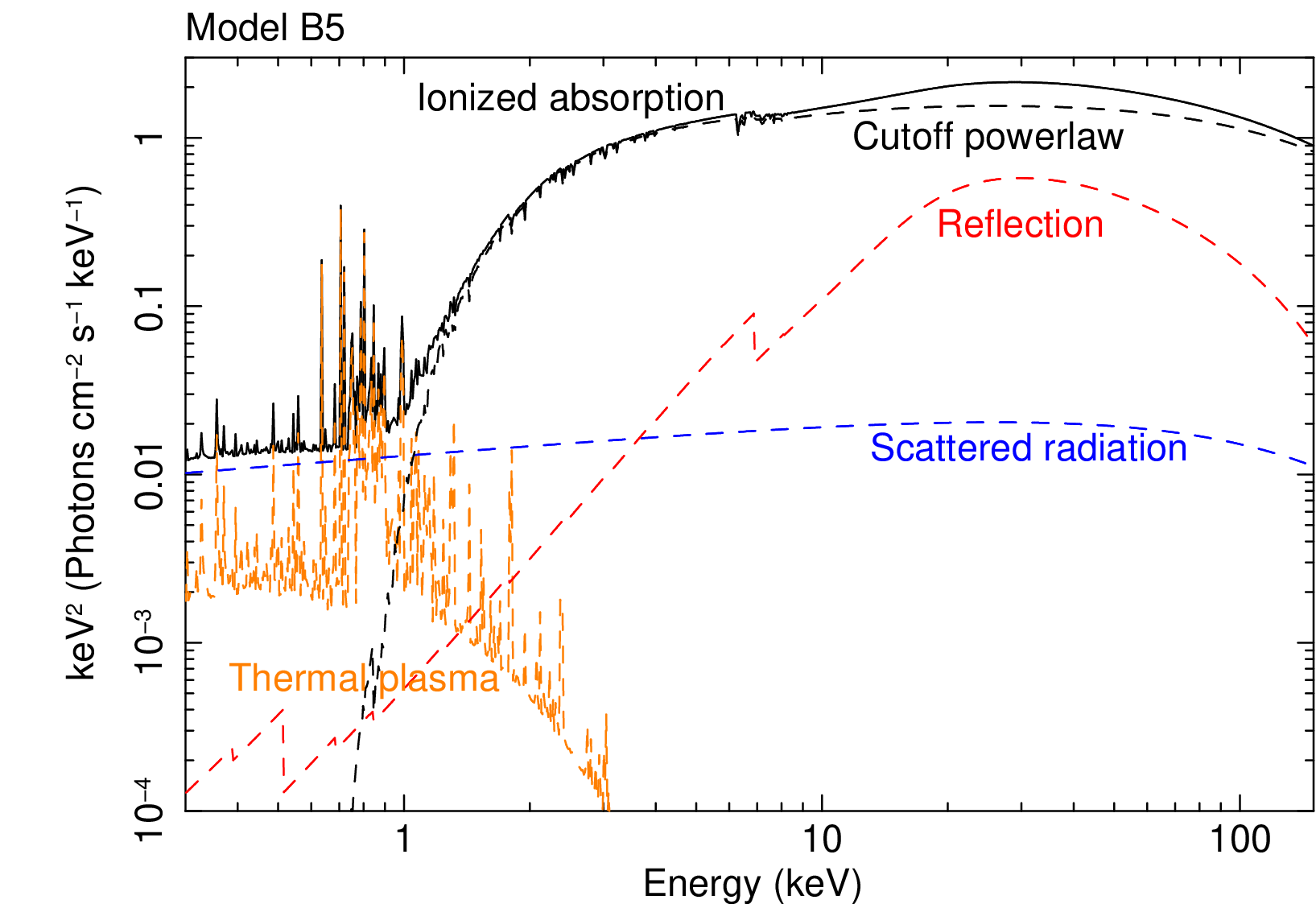}\end{minipage}
\begin{minipage}[!b]{.48\textwidth}
\centering
\includegraphics[height=9cm,angle=270]{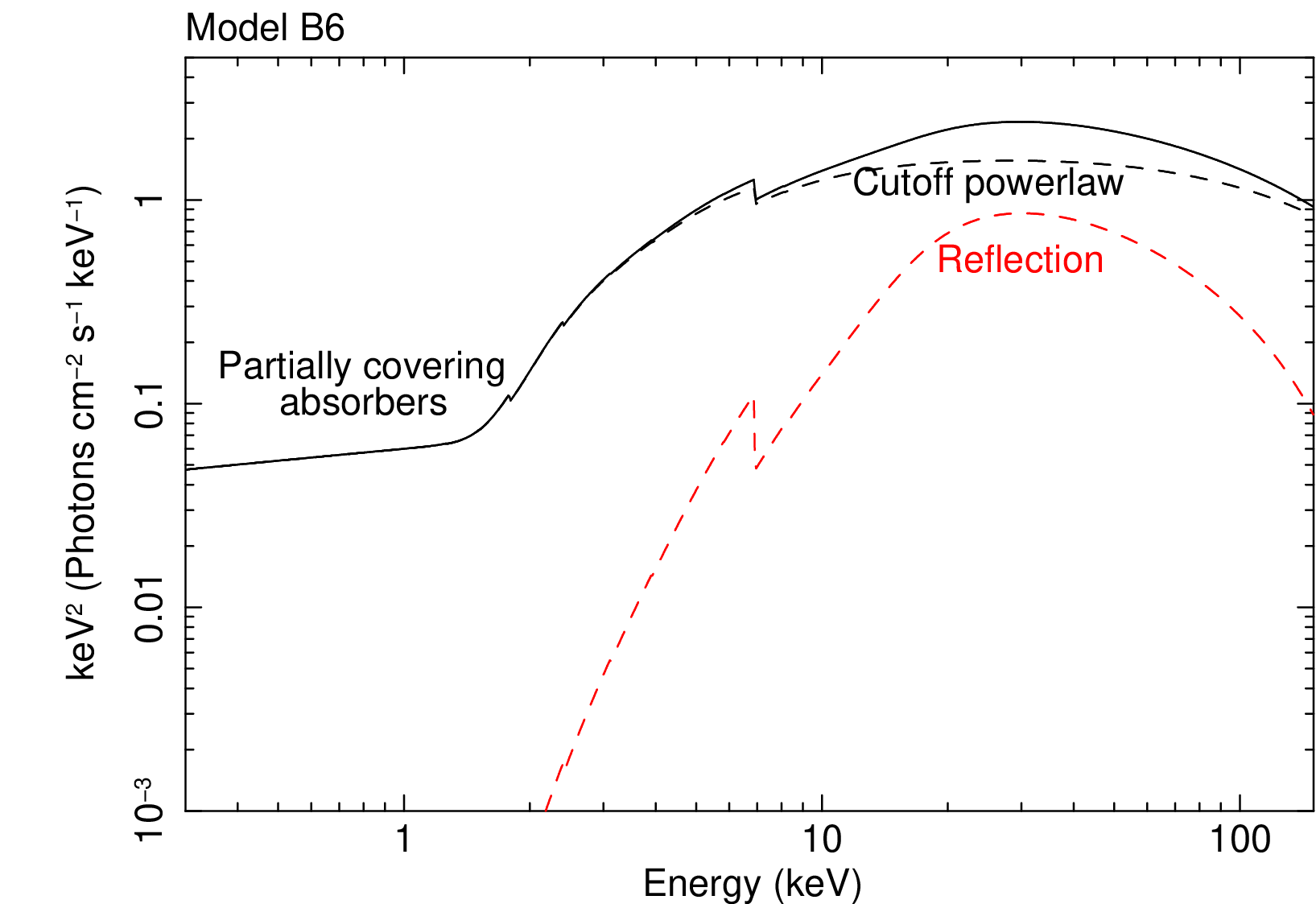}\end{minipage}
\par\bigskip
\begin{minipage}[!b]{.48\textwidth}
\centering
\includegraphics[height=9cm,angle=270]{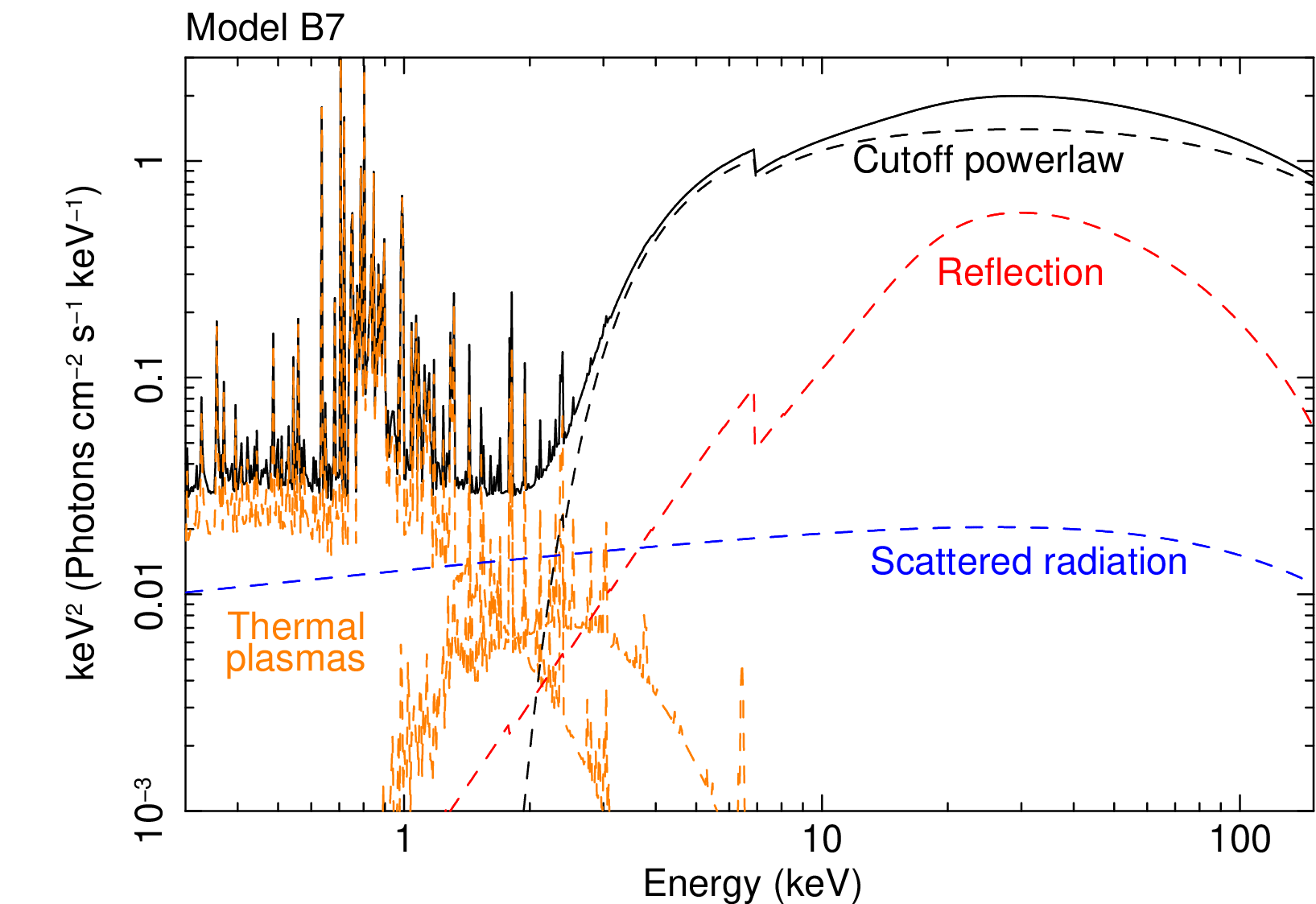}\end{minipage}
\begin{minipage}[!b]{.48\textwidth}
\centering
\includegraphics[height=9cm,angle=270]{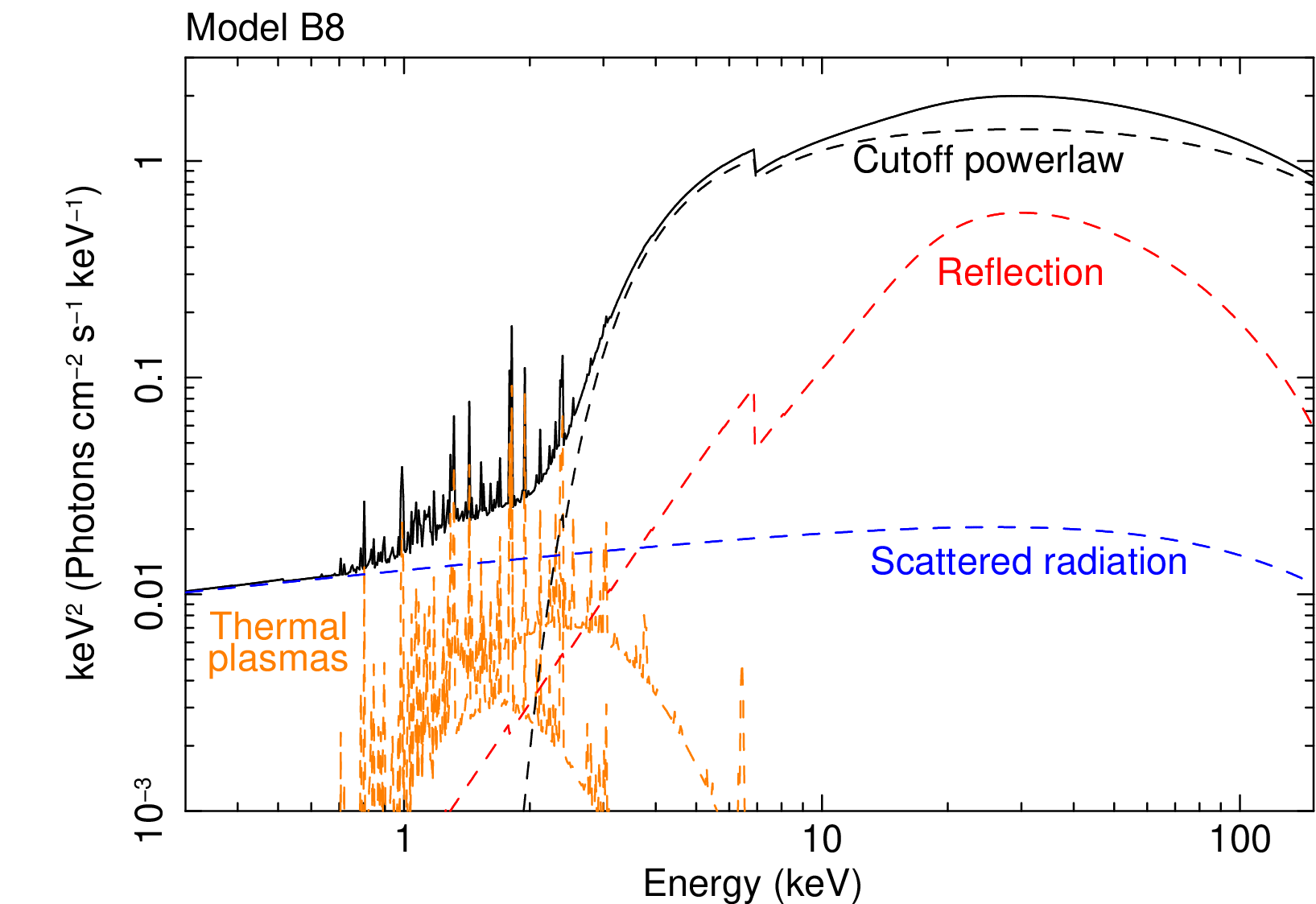}\end{minipage}
\par\bigskip
\begin{minipage}[!b]{.48\textwidth}
\centering
\includegraphics[height=9cm,angle=270]{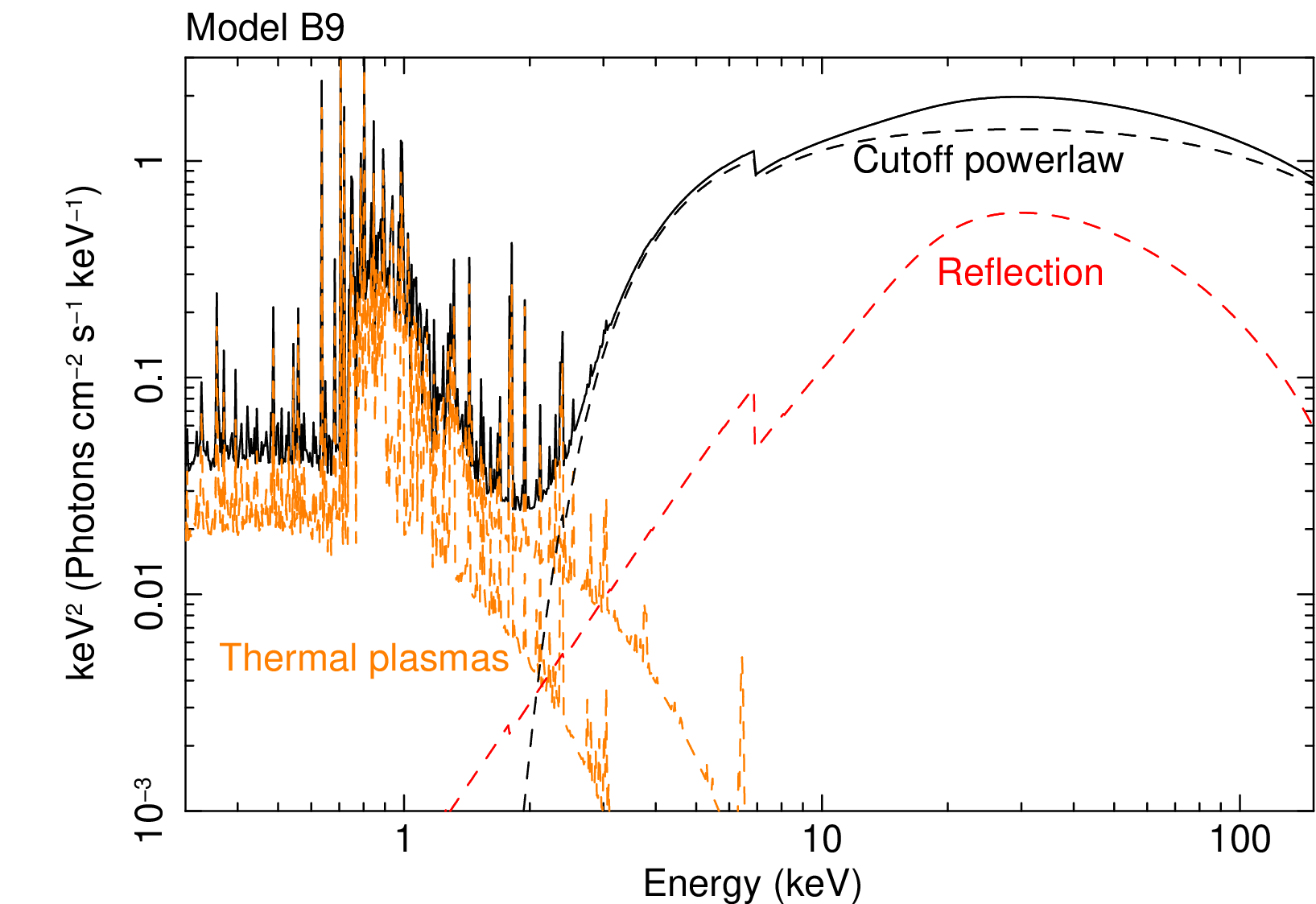}\end{minipage}
\begin{minipage}[!b]{.48\textwidth}
\centering
\includegraphics[height=9cm,angle=270]{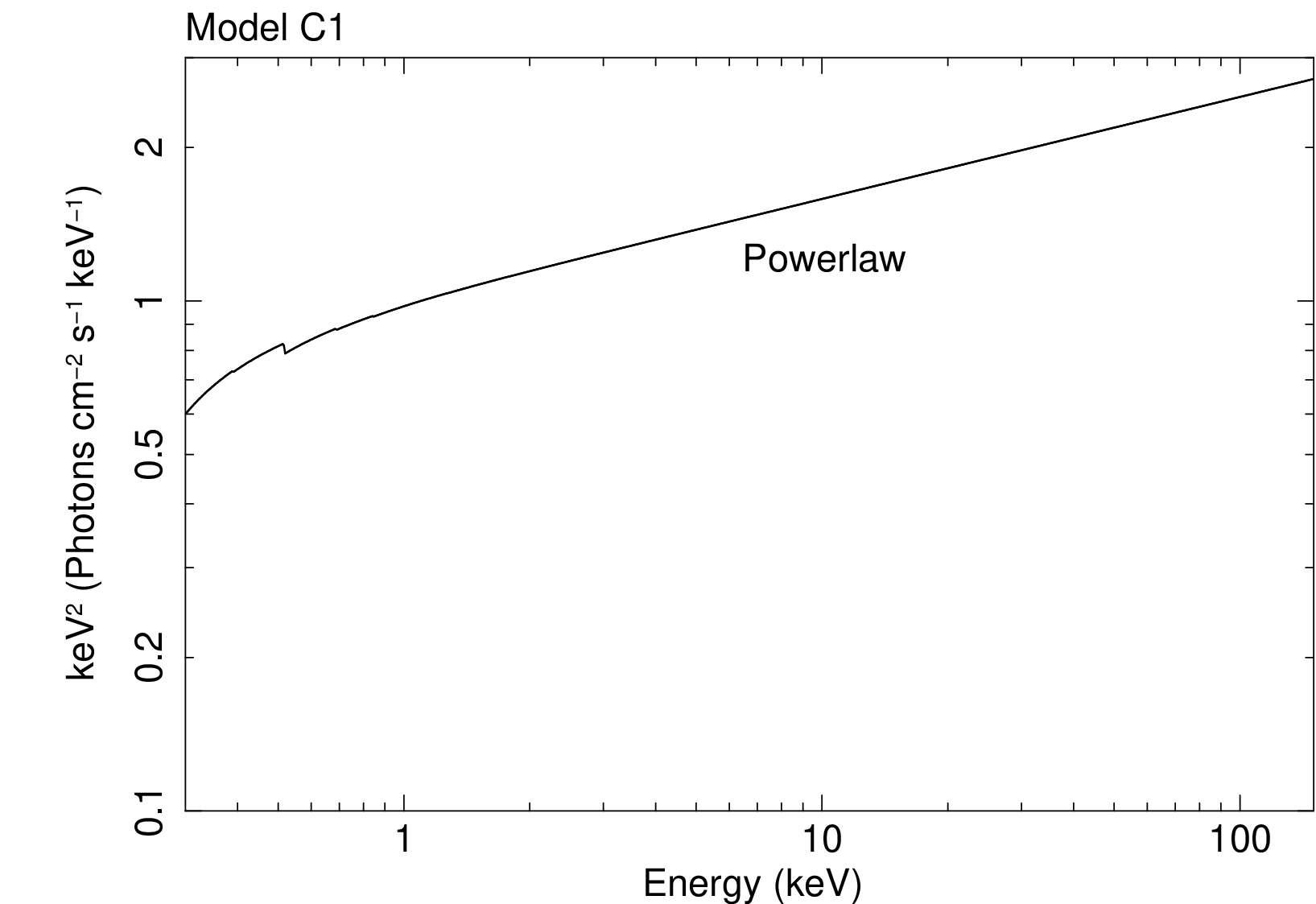}\end{minipage}
 \begin{minipage}[t]{1\textwidth}
  \caption{Models used for the X-ray spectral analysis (B5 to C1). See Table\,\ref{tab:modelcomponents} and $\S$\ref{sect:spectralModels} for details.}
\label{fig:spectra_model3}
 \end{minipage}
\end{figure*}

\begin{figure*}[h!]
\centering
 %% 1st image
 %% 2nd image
%
%
%
\begin{minipage}[!b]{.48\textwidth}
\centering
\includegraphics[height=9cm,angle=270]{modelC2.ps}\end{minipage}
 %% 2nd image
\begin{minipage}[!b]{.48\textwidth}
\centering
\includegraphics[height=9cm,angle=270]{modelC3.ps}\end{minipage}
\par\bigskip
\begin{minipage}[!b]{.48\textwidth}
\centering
\includegraphics[height=9cm,angle=270]{modelC4.ps}\end{minipage}
\begin{minipage}[!b]{.48\textwidth}
\centering
\includegraphics[height=9cm,angle=270]{modelC5.ps}\end{minipage}
\par\bigskip
\begin{minipage}[!b]{.48\textwidth}
\centering
\includegraphics[height=9cm,angle=270]{modelC6.ps}\end{minipage}
\begin{minipage}[!b]{.48\textwidth}
\centering
\includegraphics[height=9cm,angle=270]{modelC7.ps}\end{minipage}
%
% %% caption
 \begin{minipage}[t]{1\textwidth}
  \caption{Models used for the X-ray spectral analysis (C2 to C7). See Table\,\ref{tab:modelcomponents} and $\S$\ref{sect:spectralModels} for details.}
\label{fig:spectra_model4}
 \end{minipage}
\end{figure*}
\clearpage
\begin{figure*}[t!]
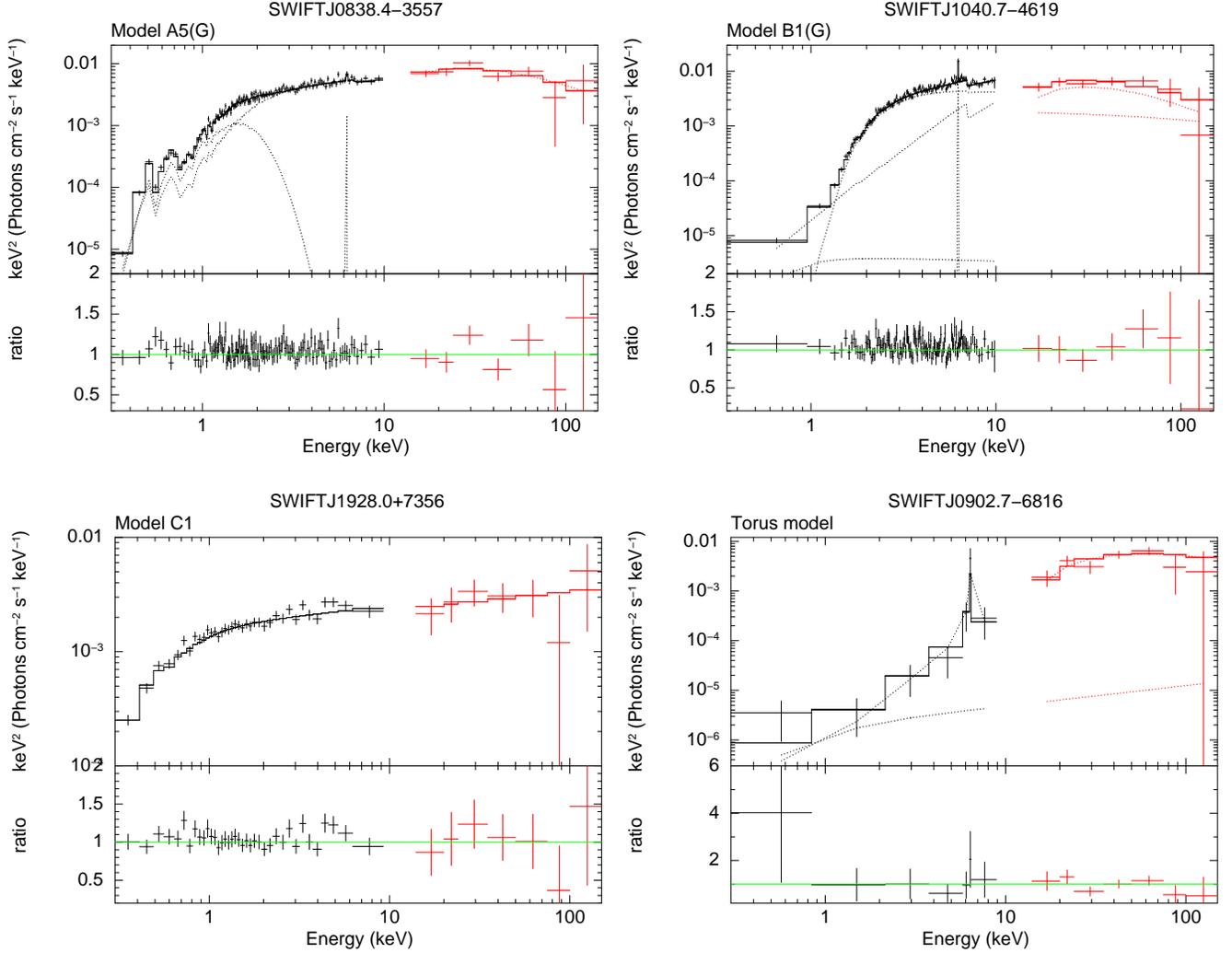

\centering
 %% 1st image
 %% 2nd image
\begin{minipage}[!b]{.48\textwidth}
\centering
\includegraphics[height=9cm,angle=270]{modelA.ps}\end{minipage}
\begin{minipage}[!b]{.48\textwidth}
\centering
\includegraphics[height=9cm,angle=270]{modelB.ps}\end{minipage}
\par\bigskip
\begin{minipage}[!b]{.48\textwidth}
\centering
\includegraphics[height=9cm,angle=270]{modelC.ps}\end{minipage}
\begin{minipage}[!b]{.48\textwidth}
\centering
\includegraphics[height=9cm,angle=270]{model_torus.ps}\end{minipage}
% %% caption
 \begin{minipage}[t]{1\textwidth}
  \caption{Example of spectra fitted with different models. {\it Top left panel}: {\it XMM-Newton} EPIC/PN (black) and {\it Swift}/BAT (red) spectrum of SWIFT\,J0838.4$-$3557 fitted with model A5, which considers the X-ray continuum and a blackbody component absorbed by a partially covering ionized absorber, plus a Gaussian line; {\it Top right panel}: {\it XMM-Newton} EPIC/PN (black) and {\it Swift}/BAT (red) spectrum of SWIFT\,J1040.7$-$4619 fitted with model B1, which includes an obscured X-ray continuum, reflection, a scattered cutoff power-law component, and a Gaussian line. {\it Bottom left panel} {\it Swift}/XRT (black) and BAT (red) spectrum of the blazar SWIFT\,J1928.0$+$7356 fitted with model C1 (i.e. power-law continuum plus absorption); {\it Bottom right panel}: {\it Swift}/XRT (black) and BAT (red) spectrum of the CT AGN SWIFT\,J0902.7$-$6816 fitted with a self-consistent torus model plus a scattered component. The bottom panels of the four figures show the ratio between the models and the data.}
\label{fig:spectra}
 \end{minipage}
\end{figure*}

In the next sections, we report in detail the spectral components ($\S$\ref{sect:modelcomponents}) and the models ($\S$\ref{sect:fitUnob}, \ref{sect:fitOb} and \ref{sect:fitBlaz}) we used for the broad-band X-ray spectral fitting. The histograms showing the number of times each best-fit model was used are illustrated in Fig.\,\ref{fig:distribution_models} for unobscured AGN (top panel), obscured AGN (middle panel), and blazars (bottom panel). The main spectral parameters obtained by the broad-band X-ray spectral fitting are reported in Table\,\ref{tbl-3}. In the top panel of Fig.\,\ref{fig:ChiDOF} we show the values of the $\chi^2$/C-stat obtained with the best-fit model versus the number of degrees of freedom (DOF) for the sources in our sample, while in the bottom panel we show the ratio between $\chi^2$ (or C-stat) and the DOF versus the number of counts. The median value of the ratio between $\chi^{2}$ and the DOF is $0.996\pm0.008$, confirming the satisfactory quality of the spectral fitting.

\subsection{Model components}\label{sect:modelcomponents}

In the following we describe the different components used for the X-ray spectral fitting and the free parameters of each model.

\subsubsection{X-ray continuum}
For the X-ray continuum of the non-blazar AGN we used a power-law component with a high-energy cutoff (\textsc{cutoffpl} in \textsc{xspec}). The free parameters of this model are the photon index ($\Gamma$), the energy of the cutoff ($E_{\rm C}$) and the normalization ($no_{\rm\,cut}$). To take into account reprocessing of the primary X-ray continuum by circumnuclear material we used the \textsc{pexrav} model \citep{Magdziarz:1995pi}, which assumes reflection by a semi-infinite slab. The inclination angle $i$ was fixed to $30$\,degrees for all objects, in order to have a value of the reflection parameter ($R=\Omega/2\pi$, where $\Omega$ is the covering factor of the reflecting material) independent of any assumption on the geometry of obscured and unobscured AGN. The metallicity was fixed to solar, and thus the sole free parameter of this model is $R$.
For blazars we used a simple power-law model (\textsc{pow}) or, when required by the fit, a broken (\textsc{bkn}) or a double-broken (\textsc{bkn2}) power-law. For the power-law model the free parameters are the photon index ($\Gamma$) and the normalization ($no_{\rm\,pow}$). 

\noindent The broken power-law model considers a continuum that changes its slope at an energy $E_{\rm\,brk}$. The two different photon indices are $\Gamma_{1}$ and $\Gamma_{2}$ for $E<E_{\rm\,brk}$ and $E>E_{\rm\,brk}$, respectively. The free parameters of this model are $E_{\rm\,brk}$, $\Gamma_{1}$, $\Gamma_{2}$ and the normalization ($no_{\rm\,bkn}$). In the double broken power-law model the continuum changes slope twice: at $E_{\rm\,brk}^1$ and $E_{\rm\,brk}^2$. The photon indices are $\Gamma_{1}$, $\Gamma_{2}$ and $\Gamma_{3}$ for $E<E_{\rm\,brk}^1$, $E_{\rm\,brk}^1\leq E<E_{\rm\,brk}^2$ and $E\geq E_{\rm\,brk}^2$, respectively. The free parameters are $E_{\rm\,brk}^1$, $E_{\rm\,brk}^2$,  $\Gamma_{1}$, $\Gamma_{2}$, $\Gamma_{3}$ and the normalization ($no_{\rm\,bkn2}$). No reflection component was considered for blazars, since most of the X-ray emission arises from the jets, which washes out any signature of reprocessed radiation. In Table\,\ref{tbl-3} we report the values of $\Gamma_{1}$ for the blazars for which a broken power-law continuum was used, while the values of $\Gamma_{2}$, $\Gamma_{3}$, $E_{\rm\,brk}^1$ and $E_{\rm\,brk}^2$ are reported in Table\,\ref{tbl-bknpo}.

\subsubsection{Absorption}
Absorption of the X-ray radiation by neutral material occurs due to the combined effect of photoelectric absorption and Compton scattering.
Photoelectric absorption was taken into account using the \textsc{zphabs} model, with the redshift fixed to the systemic redshift of each source. Compton scattering was considered using the \textsc{cabs} model. The only free parameter for these two models is the column density, which was tied to have the same value in all fits [i.e., $N_{\rm\,H}(\textsc{cabs})$=$N_{\rm\,H}(\textsc{zphabs})$=$N_{\rm\,H}$]. Whenever the column density could not be constrained because the source was completely unobscured, it was fixed to $N_{\rm H}/\rm cm^{-2}=0$. The redshift was fixed to $z=0$ for the sources for which no spectroscopic redshift was available. When required by the data we used a partial covering neutral absorber model \textsc{zpcfabs}, whose free parameters are $N_{\rm H}$ and the covering fraction ($f_{\rm\,cov}$). The values of $N_{\rm H}$ from the best-fit models are listed in Table\,\ref{tbl-3}

Absorption by ionized gas (also referred to as ``warm absorption") was taken into account using the \textsc{zxipcf} model \citep{Reeves:2008qy}, which uses a grid of XSTAR absorption models \citep{Bautista:2001jk,Kallman:2001ty}. The free parameters of this model are the column density  ($N_{\rm H}^{\rm W}$), the ionization parameter ($\xi$) and the covering factor ($f_{\rm\,cov}^{\rm\,W}$) of the warm absorber. The ionisation parameter is defined as $\xi = L_{\rm\,ion}/nr^2$, where $n$ is the density of the absorber, $L_{\rm\,ion}$ is the ionizing luminosity of the source in the range 5\,eV--300\,keV, and $r$ is the distance between the ionizing source and the absorbing material. The values of $N_{\rm H}^{\rm W}$, $\xi$ and $f_{\rm\,cov}^{\rm\,W}$ obtained by our spectral analysis are listed in Table\,\ref{tbl-wa}.

\subsubsection{Soft excess}
An excess over the X-ray primary emission below $\sim 1-2$\,keV (the ``soft excess") has been found in both obscured and unobscured sources, although it is widely believed to have a very different physical origin in the two cases. For unobscured objects the soft excess might be due to any of three potential mechanisms: i) blurred relativistic reflection (e.g., \citealp{Crummy:2006mi,Fabian:2009db,Vasudevan:2014ul}); ii) Comptonization of the seed optical/UV photons in plasma colder than that responsible for the primary X-ray component (e.g., \citealp{Mehdipour:2011if,Done:2012bf,Boissay:2014vf,Boissay:2016fp}); iii) smeared absorption by ionized material (e.g., \citealp{Gierlinski:2004by}). For obscured objects this feature could have one or several of the following origins: i) emission from a thermal plasma possibly related to star formation (e.g., \citealp{Iwasawa:2011dk}); ii) radiative-recombination continuum created by gas photoionized by the AGN (e.g., \citealp{Bianchi:2006zp,Guainazzi:2007fv}); iii) scattering of the primary X-ray emission in Compton-thin circumnuclear material (e.g., \citealp{Ueda:2007th}). 

Given the different physical origin, we adopted different models to reproduce the soft excess in unobscured and obscured AGN. For unobscured sources we used a blackbody component (\textsc{bbody}), with the free parameters being the temperature ($kT_{\rm\,bb}$) and the normalization. This is not a physical model, and it provides only a phenomenological representation of this feature; given the uncertain origin of the soft excess we deem this to be the best approach. For obscured AGN a second cutoff power-law component was added to the model, with the values of the photon index, the cutoff energy and the normalizations fixed to those of the primary X-ray emission. A multiplicative constant ($f_{\rm\,scatt}$), of typically a few percents of the primary X-ray emission ($\S$\ref{sect:SoftExcess_obs}), was added to this second cutoff power-law to renormalise the flux, as a free parameter. For obscured sources we also added, when necessary, a collisionally ionized plasma (\textsc{apec}). The free parameters of the \textsc{apec} model are the temperature ($kT_{\rm\,therm.}$) and the normalization. It should be noted that an unobscured scattered component could also be due to a partially covering absorber, in particular for values of $f_{\rm\,scatt}\geq 5-10\%$.

\subsubsection{Fe\,K$\alpha$ emission lines}
The fluorescent iron K$\alpha$ emission line has been observed almost ubiquitously in the X-ray spectra of AGN (e.g., \citealp{Mushotzky:1993bf,Shu:2010qv}), and is composed of a narrow (e.g., \citealp{Shu:2011ul,Iwasawa:2012wq,Ricci:2013fk,Ricci:2013vn,Ricci:2014fr}) and a broad (e.g., \citealp{Mushotzky:1995kl,Nandra:1997kq,Patrick:2012eu}) component. The narrow Fe K$\alpha$ line could originate in the molecular torus (e.g., \citealp{Nandra:2006rw,Ricci:2014dq}), in the broad-line region (e.g., \citealp{Bianchi:2008ua}), in an intermediate region between these two \citep{Gandhi:2015db}, and from very extended ($> 10-100$\,pc) material (e.g., \citealp{Young:2001jw,Arevalo:2014nx,Bauer:2015oq}).

The broad component is likely to be due to relativistic reflection from the innermost region of the accretion flow (e.g., \citealp{Fabian:2000fj,Brenneman:2009cs,Reynolds:2014tg}), although at least for some objects it might be due to the distortion of the X-ray continuum created by highly-ionized absorbing material along the line-of-sight (e.g., \citealp{Turner:2009ek} and references therein).

In this work we do {\it not} attempt to reproduce the broad Fe\,K$\alpha$ emission line in the {\it Swift}/BAT AGN under study, and we limit our analysis to the more prominent and more common narrow component. This was done by adding a Gaussian emission line profile ($\textsc{zgauss}$ in \textsc{xspec}) to all high-quality spectra (i.e., {\it XMM-Newton} EPIC/PN, {\it Chandra}, {\it Suzaku}/XIS) unless the line could not be constrained. A Gaussian line was also taken into account for {\it Swift}/XRT spectra if residuals at $\sim 6.4$\,keV were found fitting the continuum. The parameters of this component are the peak energy, the width ($\sigma$) and the normalization ($n_{\rm Gauss}$) of the line. The energy of the line was fixed to 6.4\,keV if it could not be constrained, while the width was fixed to 10\,eV (i.e., lower than the energy resolution of the X-ray instruments we used) if the line was not resolved. The values of the energy, equivalent width (EW), width and normalization of the lines obtained are listed in Table\,\ref{tbl:kalphalines}. The properties of the Fe\,K$\alpha$ line, and its relation with the physical characteristics of the accreting SMBH, will be discussed in detail in a forthcoming publication.

\subsection{Spectral models}\label{sect:spectralModels}
In the following we list the different models that we adopted for the X-ray spectral fitting of our sources. In Table\,\ref{tab:modelcomponents} we summarize the different spectral models, while in Figs.\,\ref{fig:spectra_model1}--\ref{fig:spectra_model4} we illustrate the models we used, highlighting the different components. In Fig.\,\ref{fig:spectra} we show, as an example of the typical fitting quality, four broad-band X-ray spectra of different types of AGN (unobscured, obscured, blazar and CT).

\subsubsection{Unobscured sources}\label{sect:fitUnob}
A total of 352 AGN were fitted using this set of models (i.e., model A1 to A8, top panel of Fig.\,\ref{fig:distribution_models}).

\smallskip

\noindent{\bf Model\,\,A1}:

\smallskip
\noindent\textsc{tbabs$_{\rm\,Gal}$$\times$$C_{\rm\,BAT}\times$zphabs$\times$cabs$\times$pexrav}.
\smallskip

\noindent This model includes primary X-ray emission and reflection, both of them obscured by the same column density. Used for 101 AGN.
\smallskip

\noindent{\bf Model\,\,A2}:

\smallskip
\noindent\textsc{tbabs$_{\rm\,Gal}$$\times$$C_{\rm\,BAT}\times$zphabs$\times$cabs$\times$(pexrav + bb)}.
\smallskip

\noindent This model is the same as model A1, with the addition of a blackbody component to take into account the presence of a soft excess below 1 keV. Used for 170 sources.
\smallskip

\noindent {\bf Model\,\,A3}: 
\smallskip

\noindent\textsc{tbabs$_{\rm\,Gal}\times$$C_{\rm\,BAT}\times$$\times$zphabs$\times$cabs$\times$zxipcf$\times$pexrav}.
\smallskip

\noindent Same as model A1 plus a layer of partially covering ionized material. Used for 35 AGN.
\smallskip

\noindent{\bf Model\,\,A4}:
\smallskip

\noindent\textsc{tbabs$_{\rm\,Gal}$$\times$$C_{\rm\,BAT}\times$zphabs$\times$cabs$\times$zxipcf$\times$ zxipcf$\times$pexrav}.
\smallskip

\noindent Same as Model A3 plus another layer of partially covering ionized material. Used for 6 objects.
\smallskip

\noindent{\bf Model\,\,A5}: 
\smallskip

\noindent\textsc{tbabs$_{\rm\,Gal}$$\times$$C_{\rm\,BAT}\times$zphabs$\times$cabs$\times$zxipcf$\times$(pexrav + bb)}.
\smallskip

\noindent Same as model A2 plus a partially covering ionized absorber. Used for 33 X-ray spectra.
\smallskip

\noindent{\bf Model\,\,A6}: 
\smallskip

\noindent\textsc{tbabs$_{\rm\,Gal}$$\times$$C_{\rm\,BAT}\times$zphabs$\times$cabs$\times$zxipcf$\times$ zxipcf$\times$(pexrav + bb)}.
\smallskip

\noindent Same as model A5 plus another partially covering ionized absorber. Used for 5 sources.\smallskip

\begin{figure*}[t!]
\centering
\begin{minipage}[!b]{.48\textwidth}
\centering
\includegraphics[width=8.5cm]{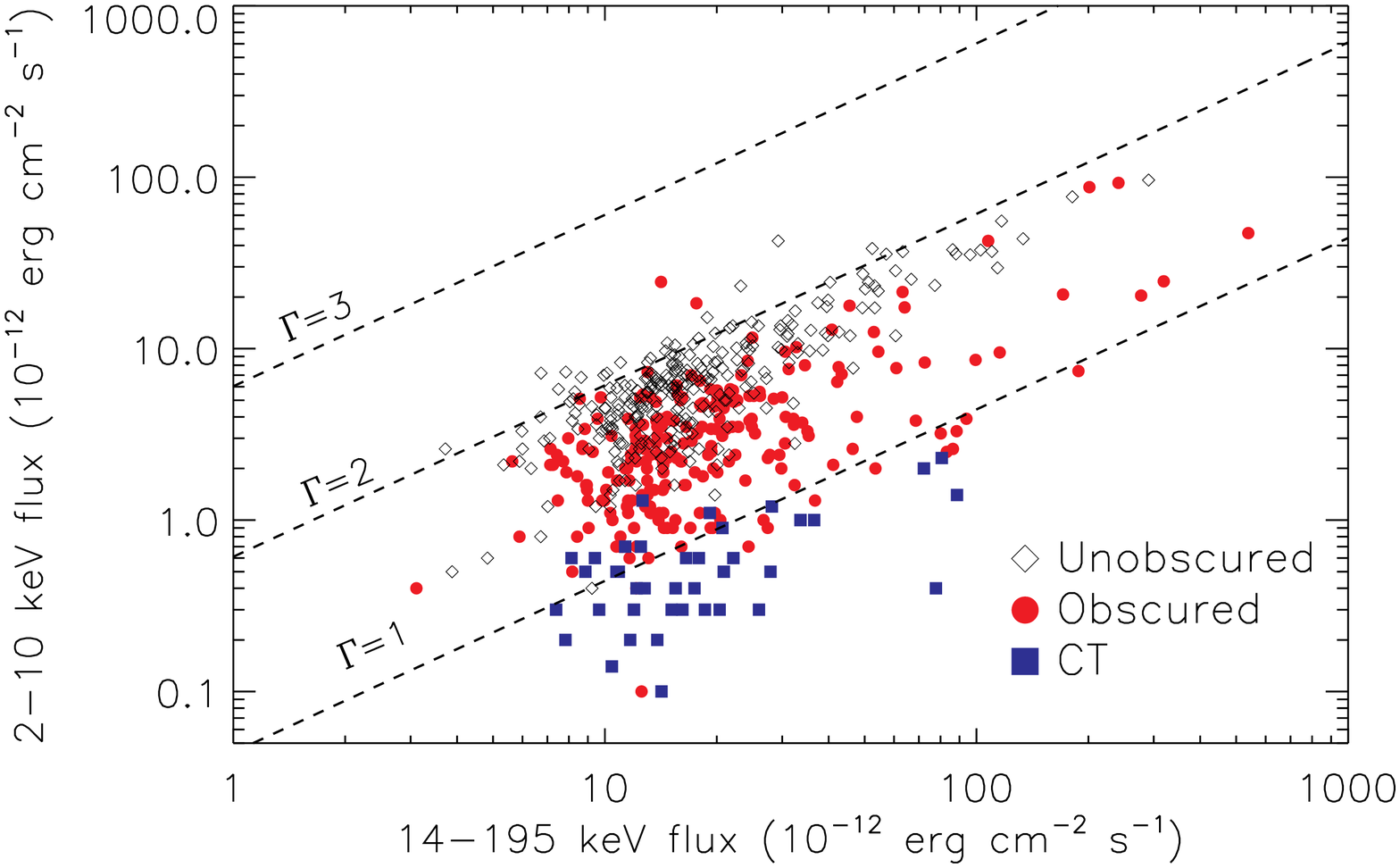}\end{minipage}
\begin{minipage}[!b]{.48\textwidth}
\centering
\includegraphics[width=8.5cm]{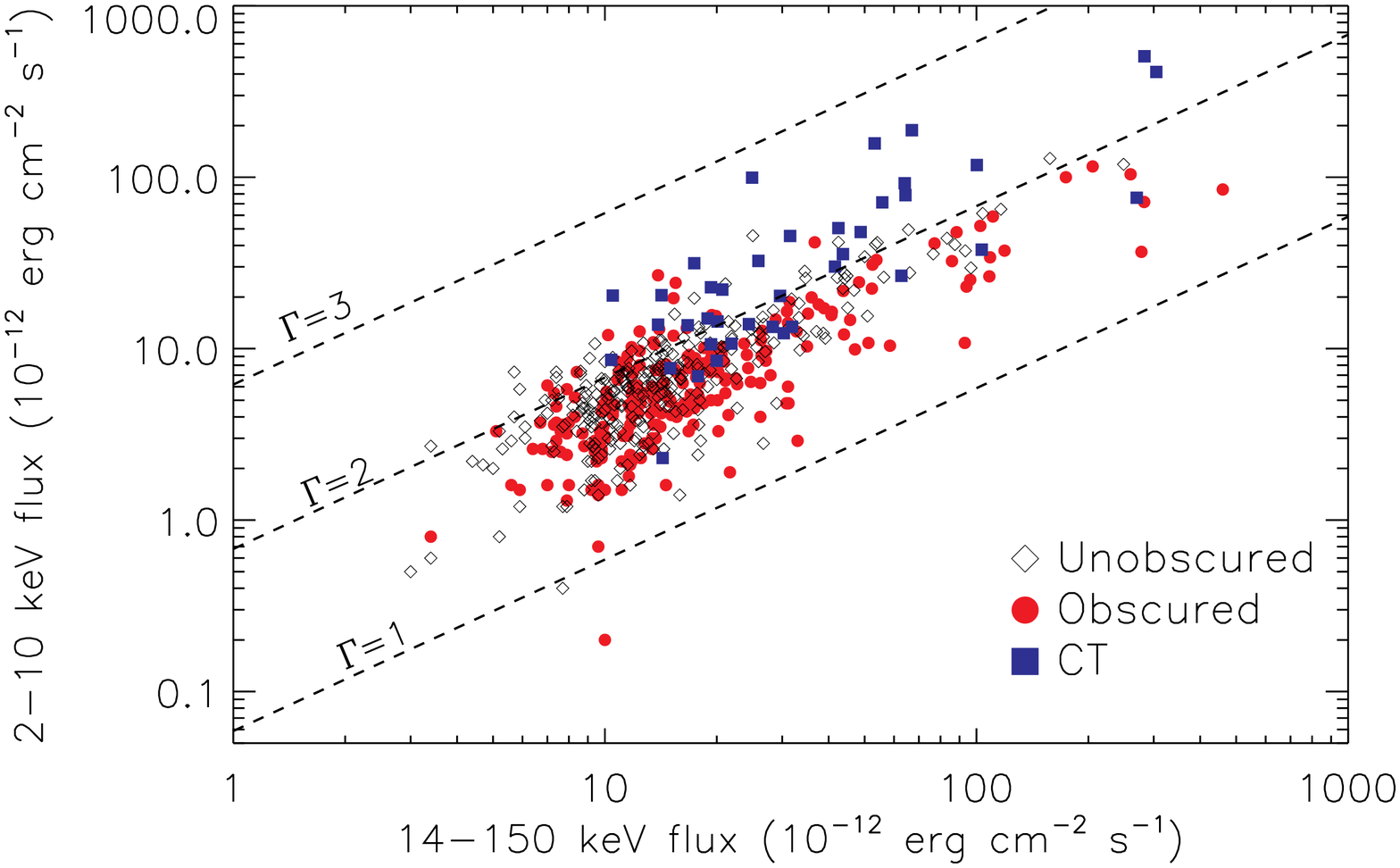}\end{minipage}
% %% caption
 \begin{minipage}[t]{1\textwidth}
  \caption{{\it Left panel:} observed 2--10\,keV flux versus the observed 14--195\,keV {\it Swift}/BAT flux reported in \cite{Baumgartner:2013uq} for the non-blazar AGN of our sample. {\it Right panel:} absorption-corrected 2--10\,keV flux versus the absorption-corrected flux in the 14--150\,keV range. In both panels the dashed lines represent the fluxes expected for values of the photon index of $\Gamma=1$, 2 and 3, assuming that the X-ray continuum is produced by a simple power law component. The non-blazar AGN are divided into unobscured, obscured and CT. }
\label{fig:f210corrvsf14150corr}
 \end{minipage}
\end{figure*}

\noindent{\bf Model\,\,A7}:
\smallskip

\noindent\textsc{tbabs$_{\rm\,Gal}$$\times$$C_{\rm\,BAT}\times$(zphabs$\times$cabs$\times$pexrav + apec + apec)}\smallskip

\noindent This model was used only for SWIFT\,J0955.5$+$6907 (M81) and adds emission from two components of collisionally ionized plasma to model A1. These two components, with different temperatures, are included to take into account the emission from hot gas, possibly associated to star formation and point sources \citep{Page:2003zr}.
\smallskip

\noindent{\bf Model\,\,A8}:
\smallskip

\noindent\textsc{tbabs$_{\rm\,Gal}$$\times$$C_{\rm\,BAT}\times$zpcfabs$\times$zxipcf$\times$cabs$\times$(pexrav + bb)}\smallskip

\noindent This model was used only for SWIFT\,J1210.5$+$3924 (NGC\,4151) and adds a partially covering neutral absorber to model A5. These two components, with different temperatures are added to take into account the emission from hot gas, possibly associated to star formation and point sources \citep{Page:2003zr}.
\smallskip

\subsubsection{Obscured sources}\label{sect:fitOb}
A total of 386 objects were fitted using this set of models (i.e., model B1 to B9, middle panel of Fig.\,\ref{fig:distribution_models}).
For obscured sources we separated the primary X-ray emission from the reflection in order to leave the latter unobscured. Reprocessed X-ray radiation was  taken into account by tying the values of the normalization of the power-law and of the cutoff energy to those of the primary X-ray emission, while leaving the reflection component free to vary. The value of $R$ was set to have only negative values, in order to consider only X-ray reflection in the model. For the sources for which reprocessed and primary X-ray emission were decomposed, the cross-calibration constant was added only to the primary X-ray emission. This reflects a scenario in which most of the X-ray variability is due to the obscured primary X-ray continuum, while unobscured reflected radiation, produced in the torus and/or in the BLR, does not vary significantly on the timescales probed here (e.g., \citealp{Arevalo:2014nx}).

\smallskip
\noindent{\bf Model\,\,B1}:\smallskip

\noindent\textsc{tbabs$_{\rm\,Gal}$$\times$($C_{\rm\,BAT}\times$zphabs$\times$cabs$\times$cutoffpl + pexrav + $f_{\rm\,scatt}$$\times$cutoffpl)}.
\smallskip

\noindent This model considers an absorbed primary X-ray emission, an unobscured reflection component, and a scattered component. Used for 272 sources.
\smallskip

\noindent{\bf Model\,\,B2}: \smallskip

\noindent\textsc{tbabs$_{\rm\,Gal}$$\times$($C_{\rm\,BAT}\times$zphabs$\times$cabs$\times$cutoffpl + pexrav +}
\textsc{$f_{\rm\,scatt}$$\times$cutoffpl + apec)}. 
\smallskip

\noindent Same as model B1 plus a collisionally ionized plasma component. Used for 87 AGN.
\smallskip

\noindent{\bf Model\,\,B3}:\smallskip

\noindent\textsc{tbabs$_{\rm\,Gal}$$\times$($C_{\rm\,BAT}\times$zphabs$\times$cabs$\times$cutoffpl + pexrav +}
\textsc{ $f_{\rm\,scatt}$$\times$cutoffpl + apec + apec)}. \smallskip

\noindent Same as model B2 plus a second collisionally ionized plasma. Used for 19 objects.\smallskip

\noindent{\bf Model\,\,B4}:\smallskip

\noindent\textsc{tbabs$_{\rm\,Gal}$$\times$($C_{\rm\,BAT}\times$zphabs$\times$cabs$\times$cutoffpl + pexrav + }
\textsc{$f_{\rm\,scatt}$$\times$cutoffpl + apec + apec + apec)}.\smallskip

\noindent Same as model B3 plus a third collisionally ionized plasma. Used only for SWIFT\,J1322.2$-$1641 (MCG\,$-$03$-$34$-$064).\smallskip

\noindent{\bf Model\,\,B5}:\smallskip

\noindent\textsc{tbabs$_{\rm\,Gal}$$\times$($C_{\rm\,BAT}\times$zphabs$\times$cabs$\times$zxipcf$\times$cutoffpl + pexrav }
\textsc{+ $f_{\rm\,scatt}$$\times$cutoffpl + apec)}.\smallskip

\noindent Same as model B2 plus a partially covering ionized absorber. Used only for SWIFT\,J0333.6$-$3607 (NGC\,1365).\smallskip

\noindent{\bf Model\,\,B6}:\smallskip

\noindent\textsc{tbabs$_{\rm\,Gal}$$\times$($C_{\rm\,BAT}\times$zpcfabs$\times$zpcfabs$\times$cabs$\times$pexrav)}.\smallskip

\noindent Considers a double partially covering absorber instead of fully covering material. This model was used only for SWIFT\,J0552.2$-$0727 (NGC\,2110), SWIFT\,J2124.6$+$5057 (4C\,50.55) and SWIFT\,J2223.9$-$0207 (3C 445). For the latter also a thermal plasma was added to the model. The column density of the \textsc{cabs} term was fixed to the sum of the values of $N_{\rm H}$ of the two partially covering absorbers, weighted over their covering factor.\smallskip

\noindent{\bf Model\,\,B7}:\smallskip

\noindent\textsc{tbabs$_{\rm\,Gal}$$\times$($C_{\rm\,BAT}\times$zphabs$\times$cabs$\times$cutoffpl + pexrav + }
\textsc{$f_{\rm\,scatt}$$\times$cutoffpl + zphabs$\times$apec + apec)}. \smallskip

\noindent Same as Model B3 plus neutral absorption for one of the two collissionally ionized plasma models. Used only for SWIFT\,J1206.2$+$5243 (NGC\,4102).\smallskip

\noindent{\bf Model\,\,B8}:\smallskip

\noindent\textsc{tbabs$_{\rm\,Gal}$$\times$($C_{\rm\,BAT}\times$zphabs$\times$cabs$\times$cutoffpl + pexrav +}
\textsc{$f_{\rm\,scatt}$$\times$cutoffpl + zphabs$\times$apec + zphabs$\times$apec)}.\smallskip

\noindent Same as model B3 plus two neutral absorption components, one for each of the collissionally ionized plasma components. Used only for SWIFT\,J1652.9$+$0223 (NGC\,6240).\smallskip

\noindent{\bf Model\,\,B9}:\smallskip

\noindent\textsc{tbabs$_{\rm\,Gal}$$\times$($C_{\rm\,BAT}\times$zphabs$\times$cabs$\times$cutoffpl + pexrav + } \textsc{apec + apec)}.\smallskip

\noindent Same as model B3, with the exception of the scattered component. This model was applied only for SWIFT\,J0319.7$+$4132 (NGC\,1275).\smallskip

\begin{figure*}[t!]
\centering
 %% 1st image
 %% 2nd image
\begin{minipage}[!b]{.48\textwidth}
\centering
\includegraphics[width=9cm]{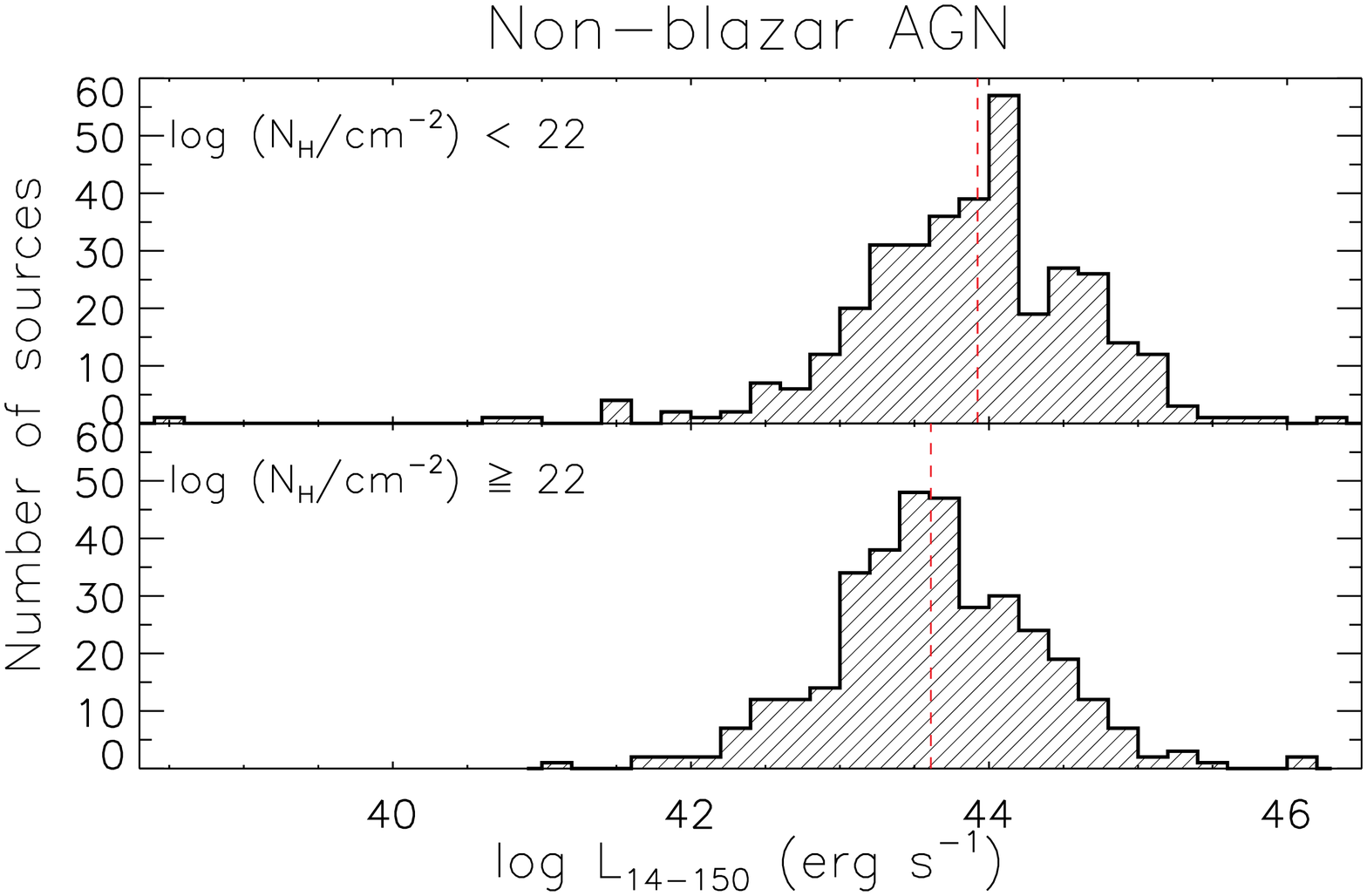}\end{minipage}
\begin{minipage}[!b]{.48\textwidth}
\centering
\includegraphics[width=9cm]{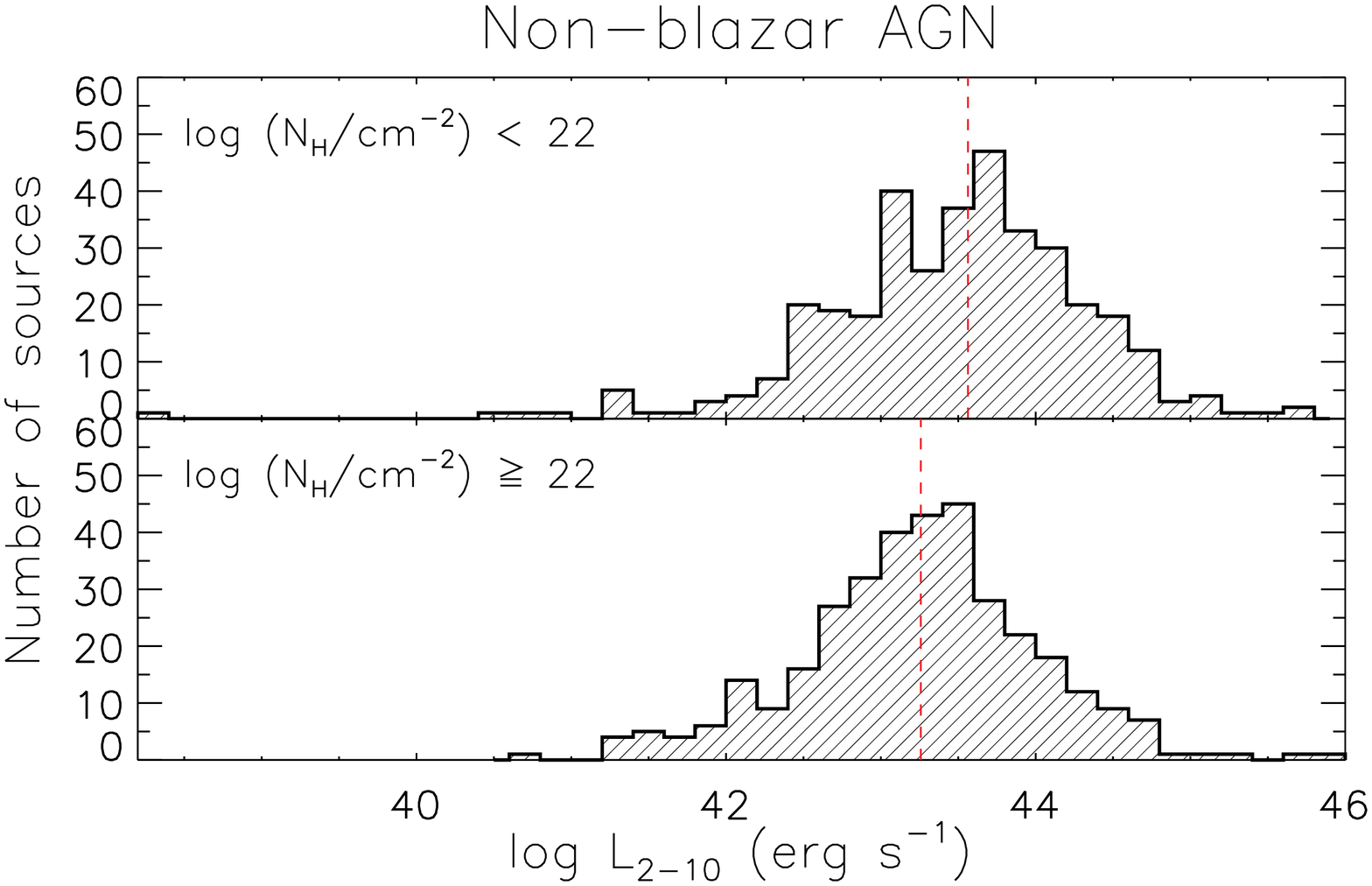}\end{minipage}
% %% caption
\begin{minipage}[!b]{.48\textwidth}
\centering
\par\medskip
\includegraphics[width=9cm]{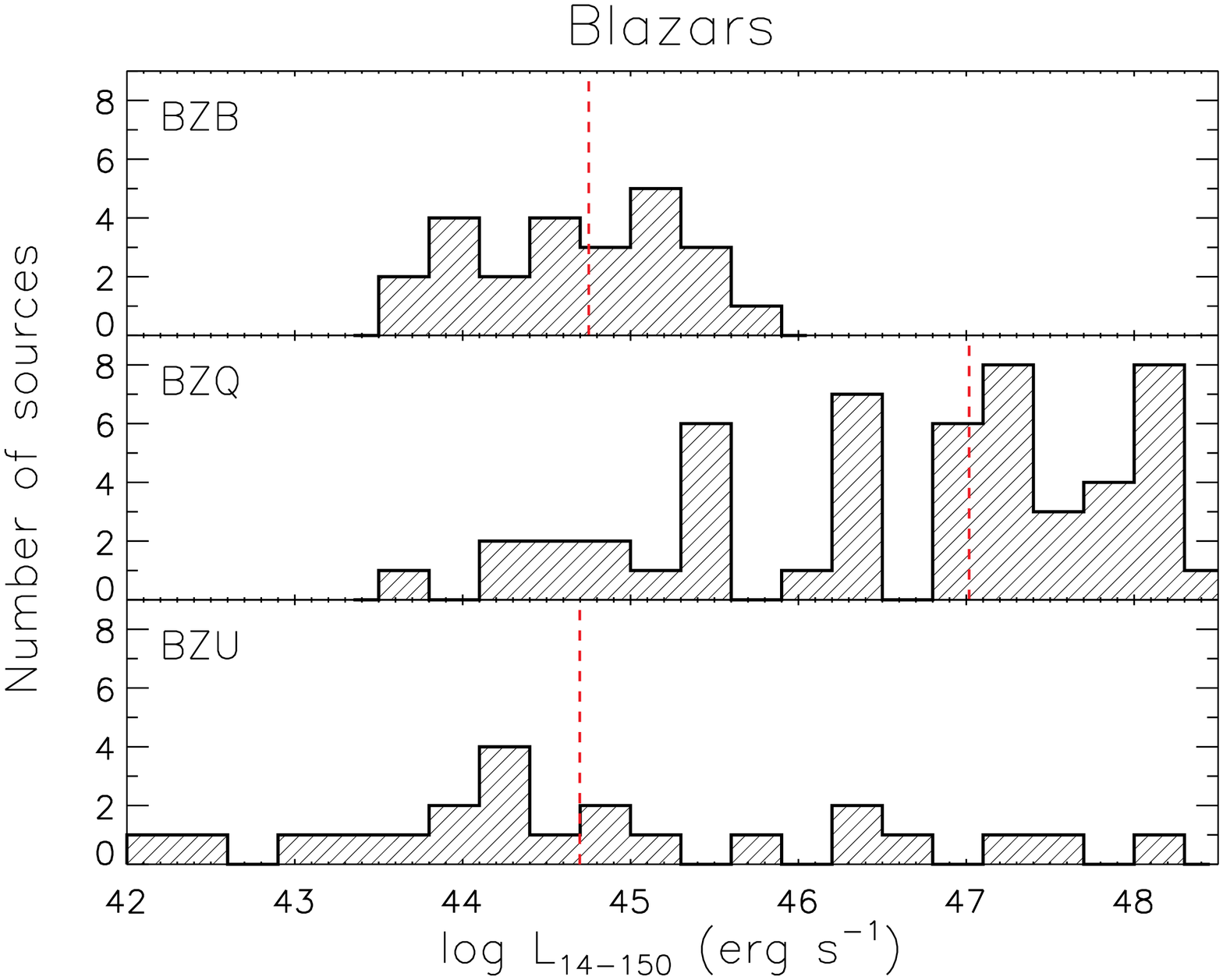}\end{minipage}
\begin{minipage}[!b]{.48\textwidth}
\centering
\par\medskip
\includegraphics[width=9cm]{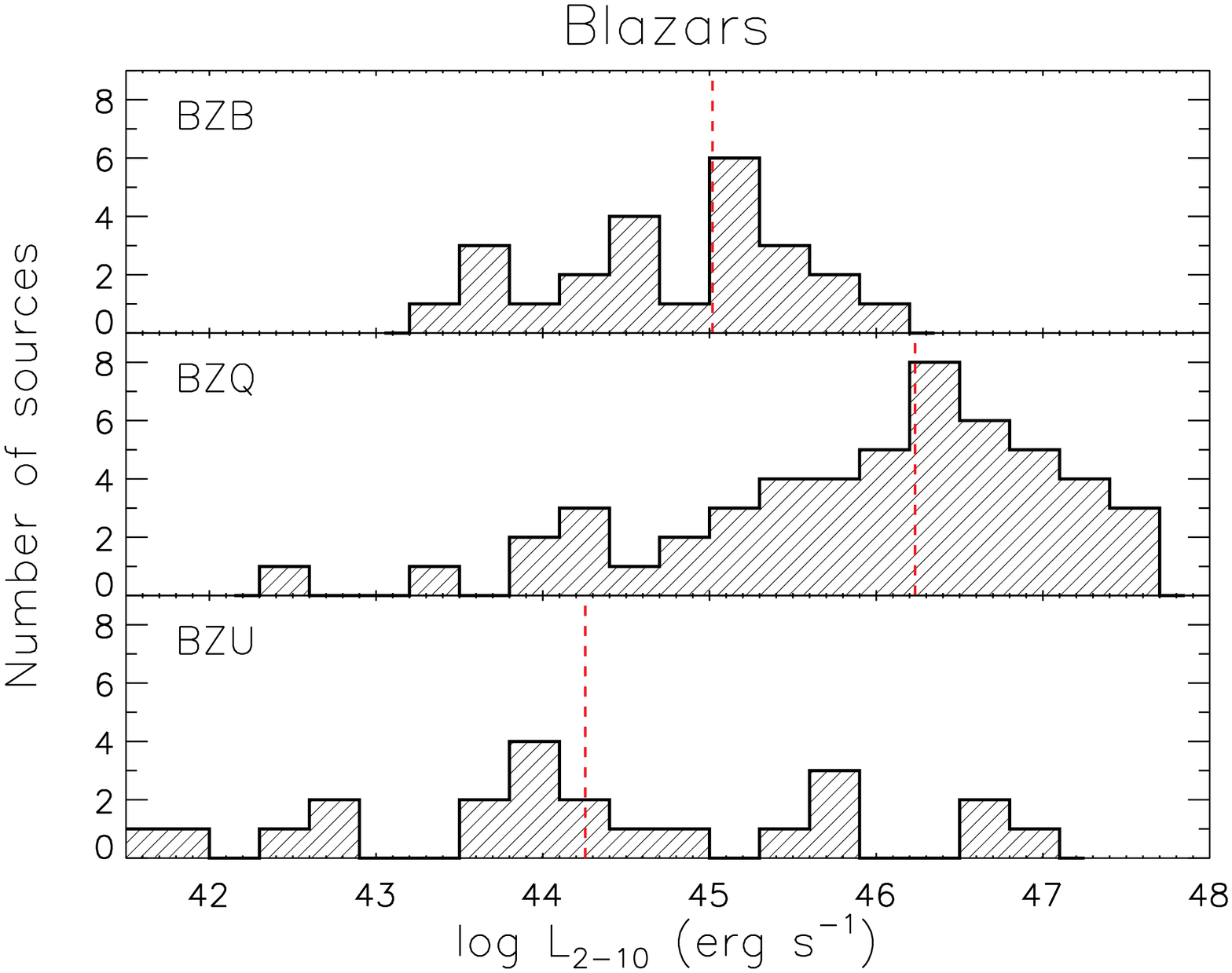}\end{minipage}
% %% caption
 \begin{minipage}[t]{1\textwidth}
  \caption{{\it Top panels:} Absorption-corrected 14--150\,keV (left panel) and 2--10\,keV (right panel) luminosity distributions of the non-blazar AGN of our sample. Non-blazar AGN are divided into sources with $N_{\rm H}<10^{22}\rm\,cm^{-2}$ and those with $N_{\rm H}\geq 10^{22}\rm\,cm^{-2}$. In both panels the red dashed vertical lines represent the median value of the luminosity. {\it Bottom panels:} Absorption-corrected 14--150\,keV (left panel) and 2--10\,keV (right panel) luminosity distributions of the blazars. The blazars are divided into BL Lacs (BZB), Flat Spectrum Radio Quasars (BZQ) and blazars of uncertain type (BZU). In both panels the red dashed vertical lines show the median value of the luminosity for the different types of blazars. The median values of the luminosities are listed in Table\,\ref{tab:median_nonblaz} and \ref{tab:median_blaz} for non-blazar AGN and blazars, respectively.}
\label{fig:Lumdistr_obsunobs}
 \end{minipage}
\end{figure*}

\begin{figure}[t!]
\centering
 %% 1st image
 %% 2nd image
%
\begin{minipage}[!b]{.48\textwidth}
\centering
\includegraphics[width=9cm]{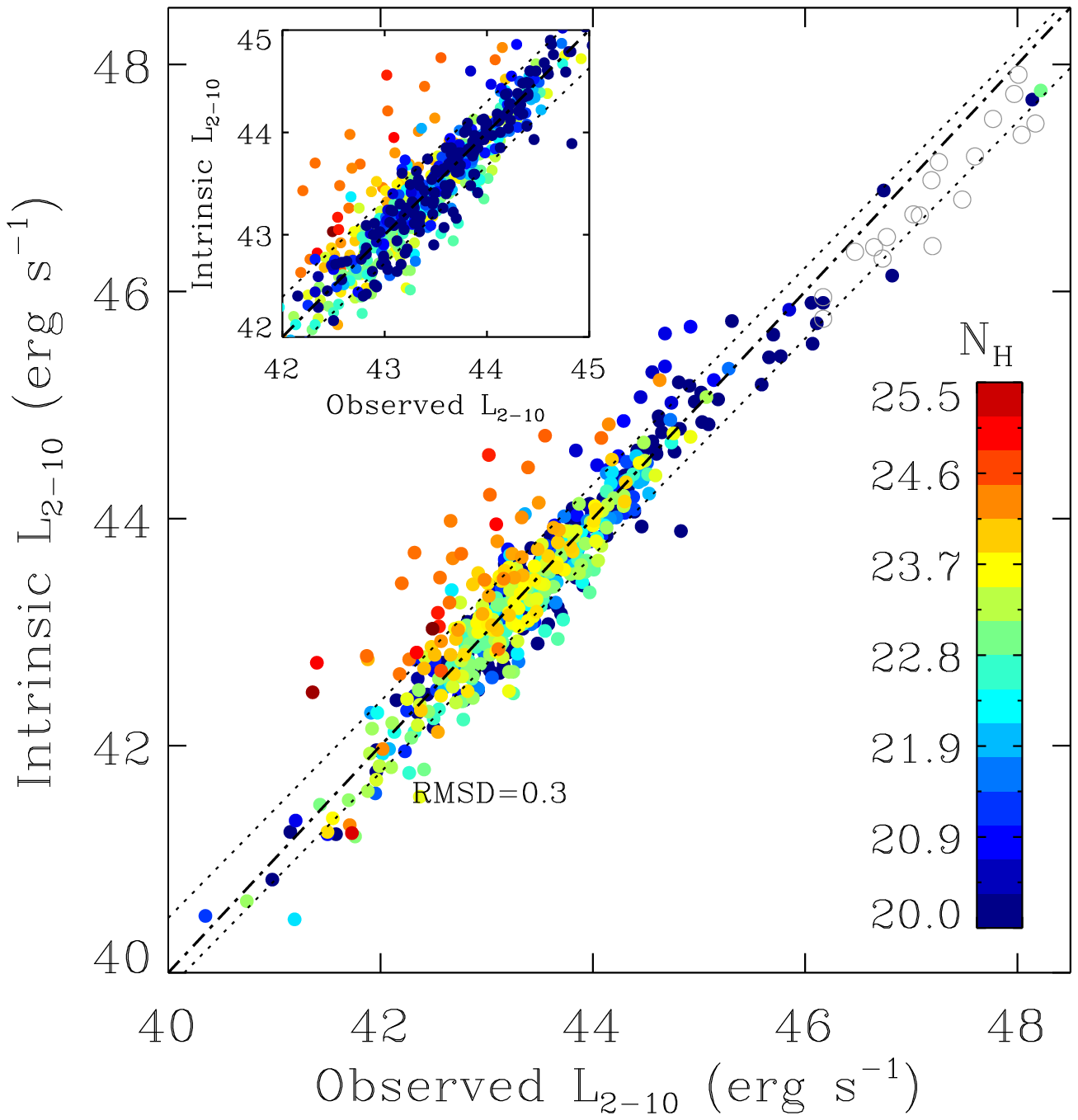}\end{minipage}
% %% caption
\par\smallskip
 \begin{minipage}[t]{0.48\textwidth}
  \caption{Intrinsic 2--10\,keV luminosity obtained by our spectral fitting versus the value calculated from the observed 14--195\,keV luminosity [$L_{2-10}(\rm obs)=0.37\times L_{14-195}$]. The top left panel shows a zoom in the $\log L_{2-10}=\log L_{2-10}(\rm obs)=42-45$ region, plotting the unobscured sources on top of the obscured. The dot-dashed line represents the $L_{2-10}=L_{2-10}(\rm obs)$ case, while the dotted lines show the scatter of the correlation. The blazars are illustrated as empty points, and were not used for the fit. The plot shows that for objects with column densities $\log(N_{\rm H}/\rm cm^{-2})\lesssim 23.7$ the two values of the 2--10\,keV luminosity are consistent (with a scatter of $\sim 0.3$\,dex). }
\label{fig:L210Ldr1}
 \end{minipage}
\end{figure}

\subsubsection{Blazars}\label{sect:fitBlaz}
A total of 97 objects were fitted using this set of models (i.e., C1 to C7, bottom panel of Fig.\,\ref{fig:distribution_models}).
For all the blazars, with the exception of the eight listed in $\S$\ref{sect:Xrayspecanalysis}, we considered a power-law for the primary X-ray emission and did not take into account reprocessed radiation. The models we applied are the following.

\smallskip

\noindent{\bf Model C1}:\smallskip

\noindent\textsc{tbabs$_{\rm\,Gal}$$\times$$C_{\rm\,BAT}\times$zphabs$\times$cabs$\times$pow}\smallskip

\noindent This model considers an absorbed power-law continuum. Used for 52 objects.\smallskip

\noindent{\bf Model C2}:\smallskip

\noindent\textsc{tbabs$_{\rm\,Gal}$$\times$$C_{\rm\,BAT}\times$(zphabs$\times$cabs$\times$pow+bb)}\smallskip

\noindent Same as model C1 plus a blackbody component. Used for ten blazars.\smallskip

\noindent{\bf Model C3}:\smallskip

\noindent \textsc{tbabs$_{\rm\,Gal}$$\times$$C_{\rm\,BAT}\times$zphabs$\times$cabs$\times$zxipcf$\times$pow}\smallskip

\noindent Same as model C1, including also an ionized absorption component. Used for three objects: SWIFT\,J0507.7$+$6732, SWIFT\,J1224.9$+$2122 and SWIFT\,J1625.9$+$4349 (87GB\,050246.4+673341, PG\,1222+216 and 87GB\,162418.8+435342, respectively).
\smallskip

\noindent{\bf Model C4}:\smallskip

\noindent\textsc{tbabs$_{\rm\,Gal}$$\times$$C_{\rm\,BAT}\times$zphabs$\times$cabs$\times$zxipcf$\times$zxipcf$\times$pow}\smallskip

\noindent Same as model C1 plus two ionized absorption components. Used for two objects: SWIFT\,J1557.8$-$7913 (PKS\,1549$-$79) and SWIFT\,J2346.8$+$5143 (2MASX\,J23470479+5142179).
\smallskip

\noindent{\bf Model C5}:\smallskip

\noindent\textsc{tbabs$_{\rm\,Gal}$$\times$$C_{\rm\,BAT}\times$(zphabs$\times$cabs$\times$pow+$f_{\rm\,scatt}\times$pow)}\smallskip

\noindent Same as model C1 plus a power-law component, which might be either unobscured jet emission or scattered emission as in obscured non-blazar AGN. Used for six blazars.
\smallskip

\noindent{\bf Model C6}:\smallskip

\noindent\textsc{tbabs$_{\rm\,Gal}$$\times$$C_{\rm\,BAT}\times$zphabs$\times$cabs$\times$bkn}\smallskip

\noindent In this model the primary X-ray emission is produced by a broken power-law. Used for 23 sources. \smallskip

\noindent{\bf Model C7}: \smallskip

\noindent\textsc{tbabs$_{\rm\,Gal}$$\times$$C_{\rm\,BAT}\times$(zphabs$\times$cabs$\times$bkn2)}\smallskip

\noindent In this model the primary X-ray component is described by a double broken power-law. Used only for SWIFT\,J1256.2$-$0551 (3C\,279).
\smallskip

\subsubsection{Other models}\label{sec:othermodels}
For two objects (SWIFT\,J0956.1+6942 and SWIFT\,J2234.8$-$2542) we used two additional models to reproduce their X-ray emission.
\smallskip

\noindent{\bf Model D1}:\smallskip

\noindent\textsc{tbabs$_{\rm\,Gal}$$\times$$C_{\rm\,BAT}\times$[zphabs$\times$cabs$\times$cutoffpl+ }\newline
\textsc{zphabs$\times$cabs$\times$(apec+apec)]}\smallskip

\noindent Just in one case, the nearby star forming galaxy SWIFT\,J0956.1+6942 (M\,82, $z=0.000677$, i.e. the closest extragalactic {\it Swift}/BAT source), it was necessary to use a model which consisted of a cutoff power-law X-ray continuum and two absorbed collisional plasmas. No AGN is present in this object (e.g., \citealp{Gandhi:2011qv}), and all the X-ray flux can be explained by star formation (e.g., \citealp{Ranalli:2008ve}). In this model the cutoff power-law represents the X-ray emission of the population of X-ray binaries, while the two collissionally ionized plasma models take into account the emission from hot gas. The results obtained by the spectral analysis for this object are summarised in Appendix\,\ref{app:m82}.
\smallskip

\noindent{\bf Model D2}:\smallskip

\noindent\textsc{tbabs$_{\rm\,Gal}\times$zphabs$\times$cabs$\times$pow }\smallskip

\noindent For SWIFT\,J2234.8$-$2542 we combined the results of the spectral analysis of the {\it BeppoSAX}  spectrum \citep{Malizia:2000yq} with the {\it Swift}/BAT spectrum. This was done fitting the {\it Swift}/BAT spectrum, fixing the column density  to the value reported by \cite{Malizia:2000yq}, with Model D2.

\subsubsection{Compton-thick AGN}\label{Sect:CTspecanalysis}

In order to self-consistently take into account absorbed and reprocessed X-ray radiation, the 75 sources that, after being fitted with the models reported above, were found to have column density values consistent with $N_{\rm H}\sim 10^{24}\rm\,cm^{-2}$ within their 90$\%$ uncertainties, were fitted with the spherical-toroidal model of \citet{Brightman:2011oq}. The free parameters of this model are the column density, which is the same for every line-of-sight intercepting the torus, the half-opening angle of the torus ($\theta_{\rm\,OA}$) and the inclination angle ($\theta_{\rm\,i}$). We left $\theta_{\rm\,OA}$ free to vary unless the parameter could not be constrained, in which case it was fixed to 60\,degrees. To reduce the degree of complexity of the models, the value of $\theta_{\rm\,i}$ was fixed to the maximum allowed value (87\,degrees) for all sources. The main properties of the CT AGN from our sample have been reported in a recently published paper \citep{Ricci:2015kx}, while the spectral parameters obtained are listed in Table\,\ref{tbl-torpar}.

\section{Results}\label{sect:results}

In the following we report the results obtained by the broad-band X-ray spectral analysis of the 836 AGN of our sample. In $\S$\ref{sect:Luminosities} we describe how we calculated the fluxes and luminosities, and discuss the luminosity distributions of different classes of AGN. In $\S$\ref{sect:Xraycontinuum} we discuss the properties of the X-ray continuum, and in particular of the photon index ($\S$\ref{sect:photonindex}), the cross-calibration constant ($\S$\ref{sect:crossbat}), the high-energy cutoff ($\S$\ref{sect:cutoff}), and the reflection component ($\S$\ref{sect:refl}). In $\S$\ref{sect:Absorption} we summarize the properties of the neutral ($\S$\ref{sect:NeutralAbsorption}) and ionized ($\S$\ref{sect:IonizedAbsorption}) absorbers, while in $\S$\ref{sect:SoftExcess} we discuss those of the soft excess for unobscured ($\S$\ref{sect:SoftExcess_unobs}) and obscured ($\S$\ref{sect:SoftExcess_obs}) non-blazar AGN. The median values of the parameters of non-blazar AGN and blazars are reported in Table\,\ref{tab:median_nonblaz} and Table\,\ref{tab:median_blaz}, respectively. Through the paper the errors on the median values are the median absolute deviations.

\subsection{Fluxes and Luminosities}\label{sect:Luminosities}

The absorption-corrected fluxes of the continuum emission (i.e., excluding the soft excess and Fe\,K$\alpha$ component) were measured in three energy bands: 2--10, 20--50 and 14--150\,keV ($F_{\mathrm{\,2-10}}$, $F_{\mathrm{\,20-50}}$ and $F_{\mathrm{\,14-150}}$, respectively). To be consistent with what was reported in the {\it Swift}/BAT catalogs we also report the fluxes in the 14--195\,keV band ($F_{\mathrm{\,14-195}}$), which were obtained by extrapolating the 14--150\,keV fluxes by assuming a power-law with a photon index of $\Gamma=1.8$, consistent with the typical value found for {\it Swift}/BAT AGN (see \S\ref{sect:photonindex}). In the left panel of Fig.\,\ref{fig:f210corrvsf14150corr} we show the observed 2--10 flux versus the 14--195\,keV flux, while in the right panel the absorption-corrected 2--10 versus the absorption-corrected 14--150\,keV flux. The plot shows that, once the emission has been corrected for absorption, most of the sources lie in the region predicted for a power-law continuum with a slope in the range $\Gamma=1-3$. In Table\,\ref{tbl:fluxes} we report the values of the observed 2--10 and 14--195\,keV fluxes ($F_{\mathrm{\,2-10}}^{\rm\,obs}$ and $F_{\mathrm{\,14-195}}^{\rm\,obs}$, respectively), and of the intrinsic 2--10, 20--50, 14--150 and 14--195\,keV fluxes for all the sources of our sample.

\begin{figure}[t!]
\centering
\includegraphics[width=9cm]{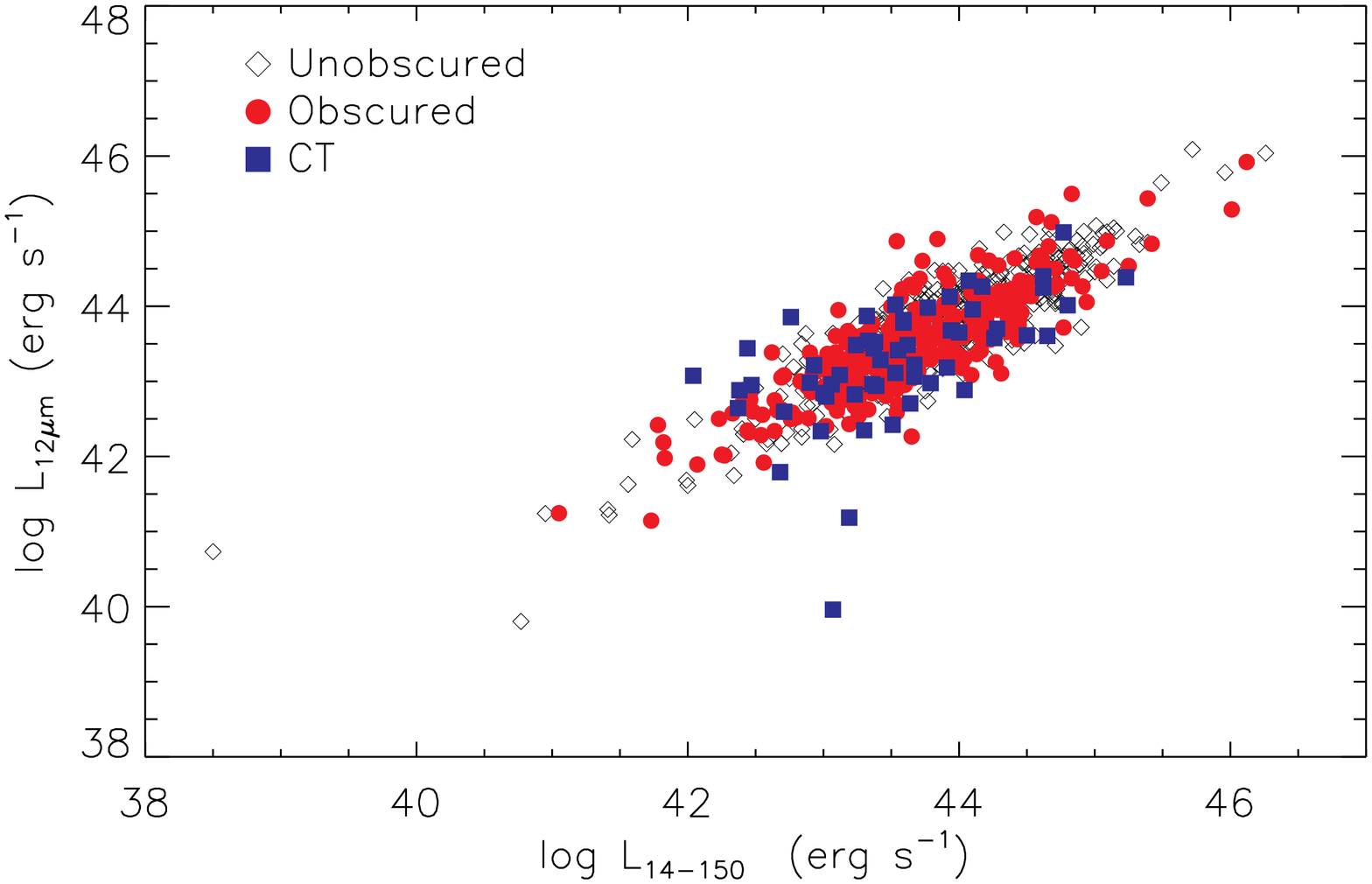}
\par\medskip
\includegraphics[width=9cm]{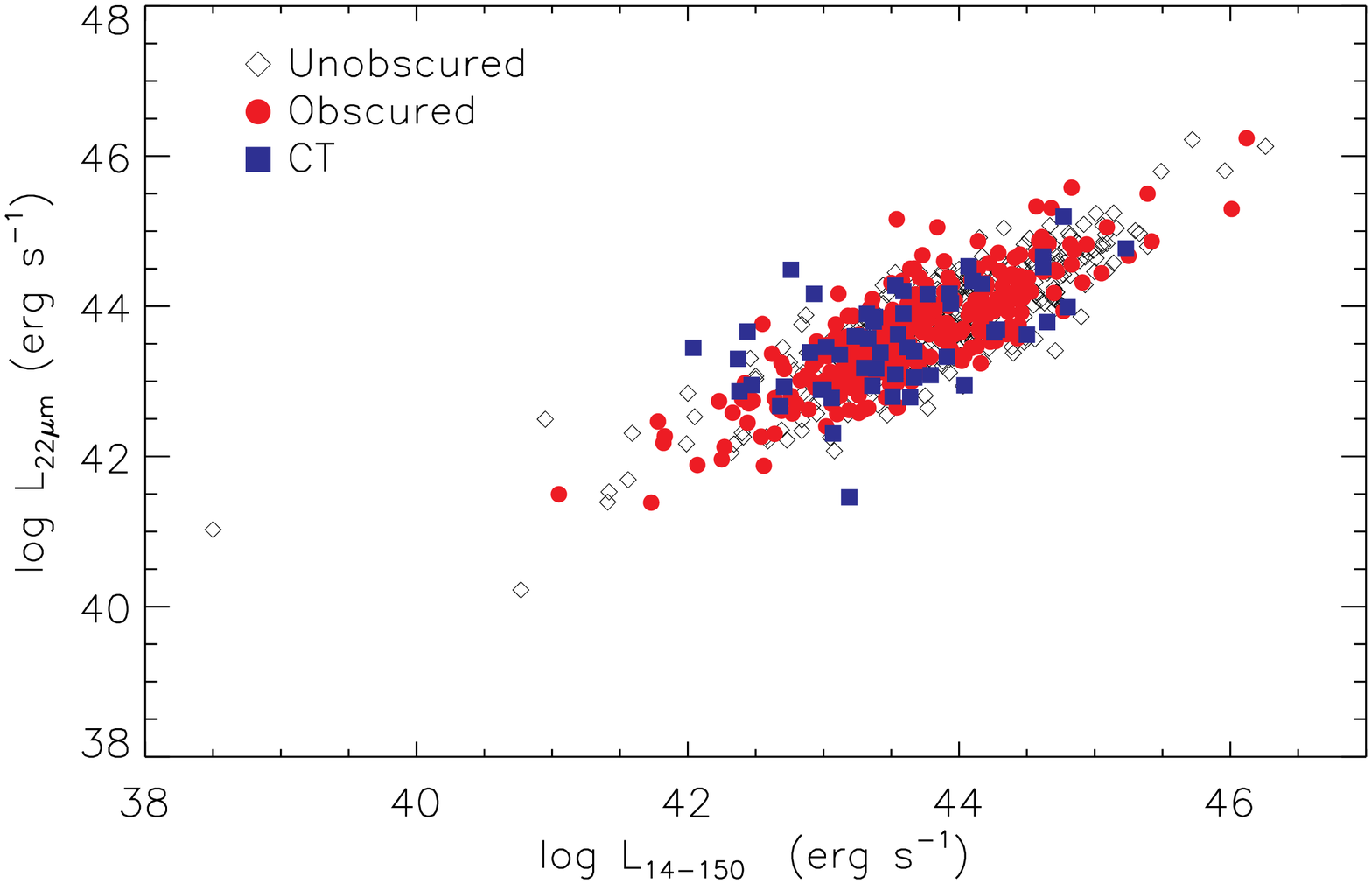}
% %% caption
%
  \caption{{\it Top panel:} Intrinsic 14--150\,keV luminosity versus the 12$\mu$m luminosity for unobscured (empty black diamonds), obscured (red circles) and CT (blue squares) non-blazar AGN. {\it Bottom panel:} Same as the top panel for the 22$\mu$m luminosities.}
\label{fig:LagnLMIR}
\end{figure}

\begin{figure*}[t!]
\centering
 %% 1st image
 %% 2nd image
\begin{minipage}[!b]{.48\textwidth}
\centering
\includegraphics[width=8.9cm]{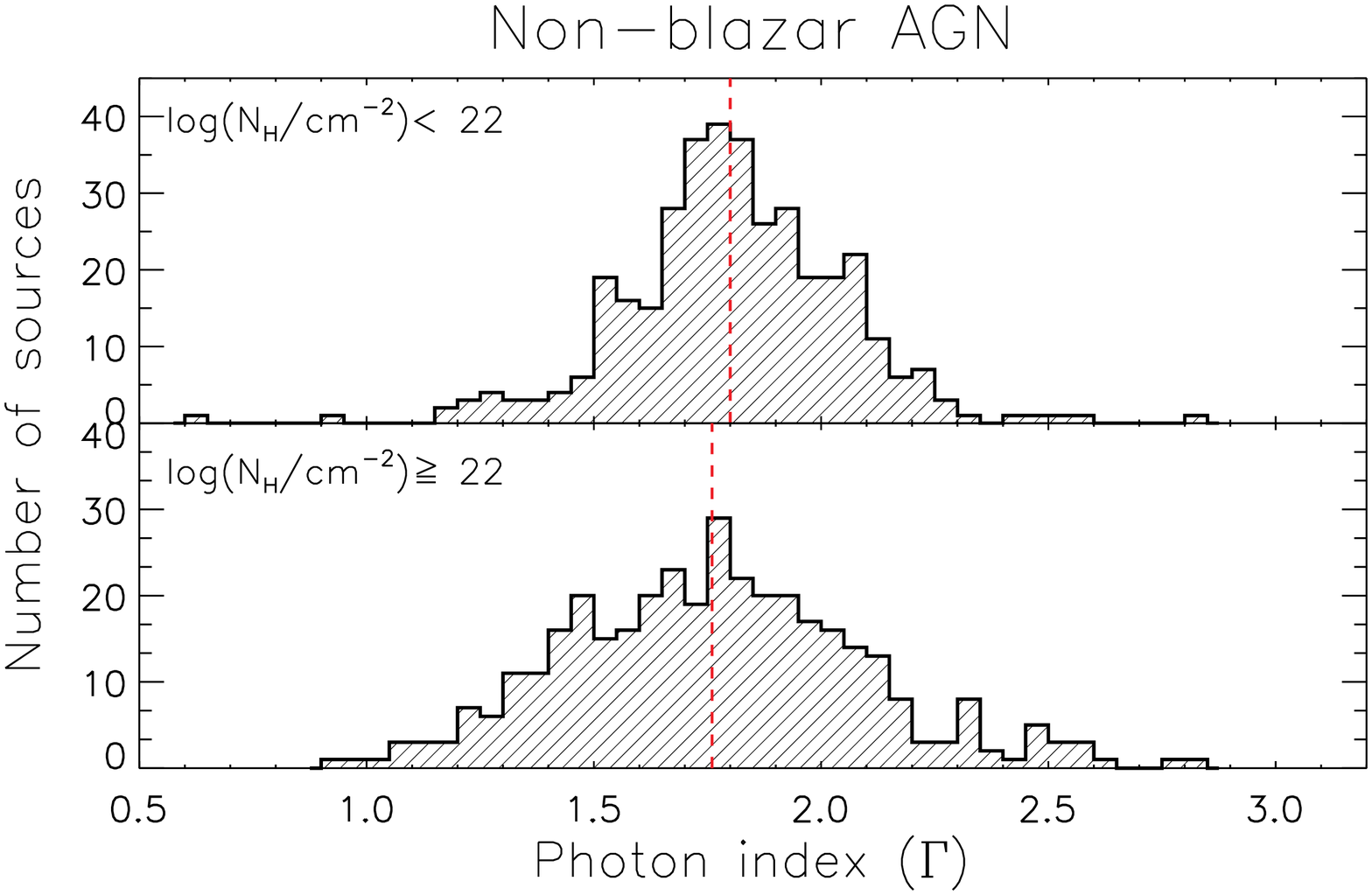}\end{minipage}
\begin{minipage}[!b]{.48\textwidth}
\centering
\includegraphics[width=8.9cm]{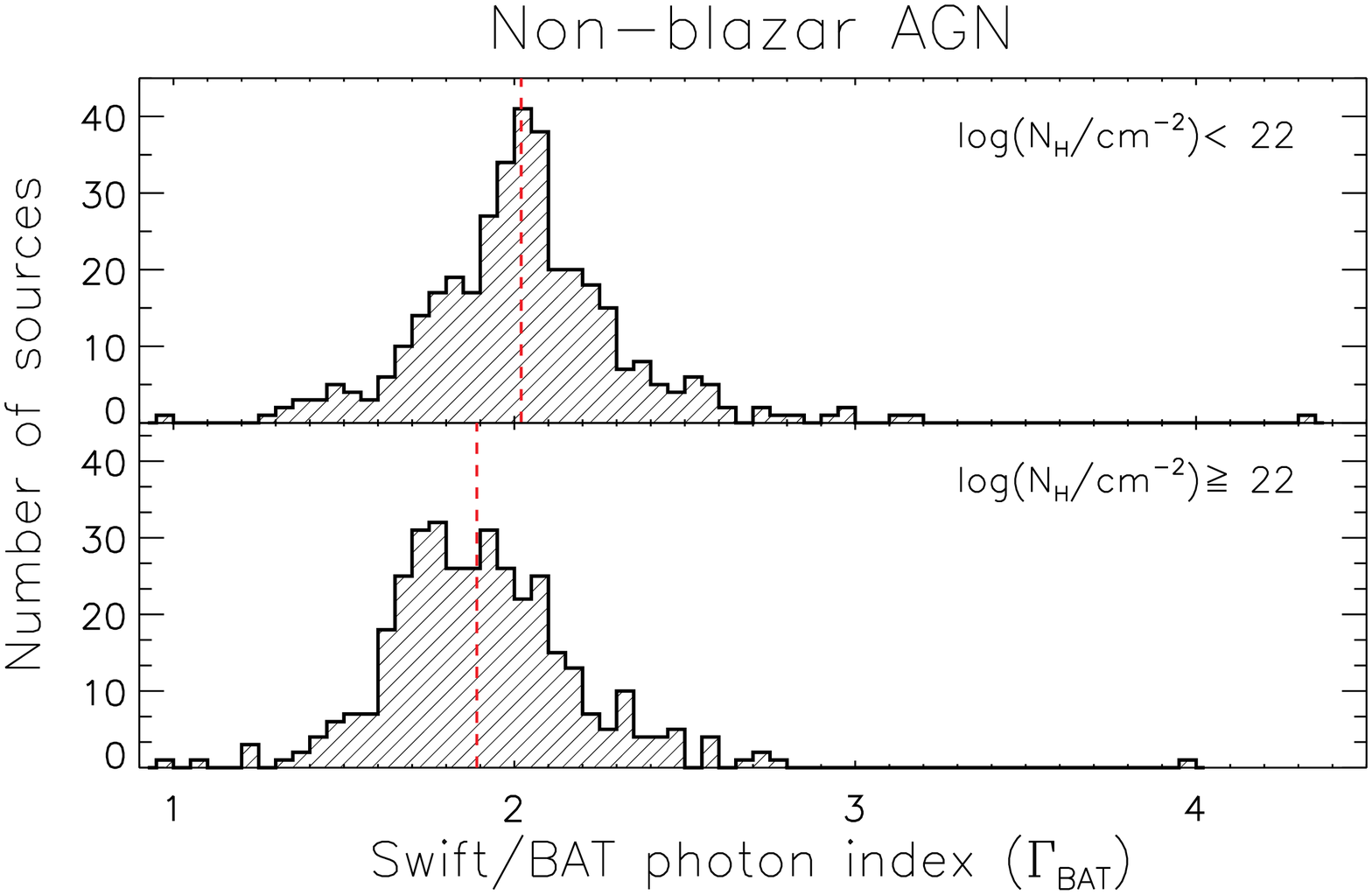}\end{minipage}
\begin{minipage}[!b]{.48\textwidth}
\centering
\par\medskip
\includegraphics[width=8.9cm]{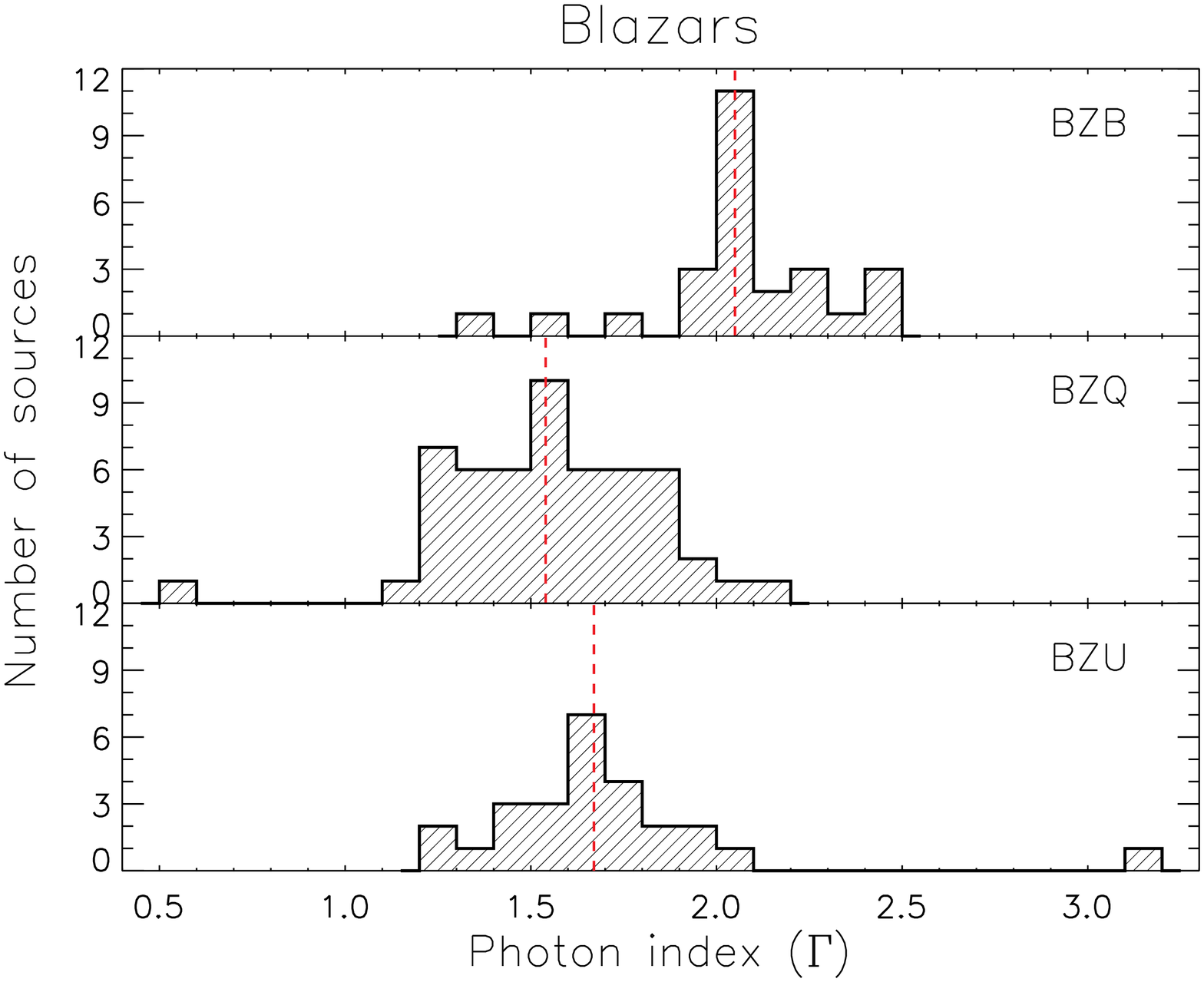}\end{minipage}
\begin{minipage}[!b]{.48\textwidth}
\centering
\par\medskip
\includegraphics[width=8.9cm]{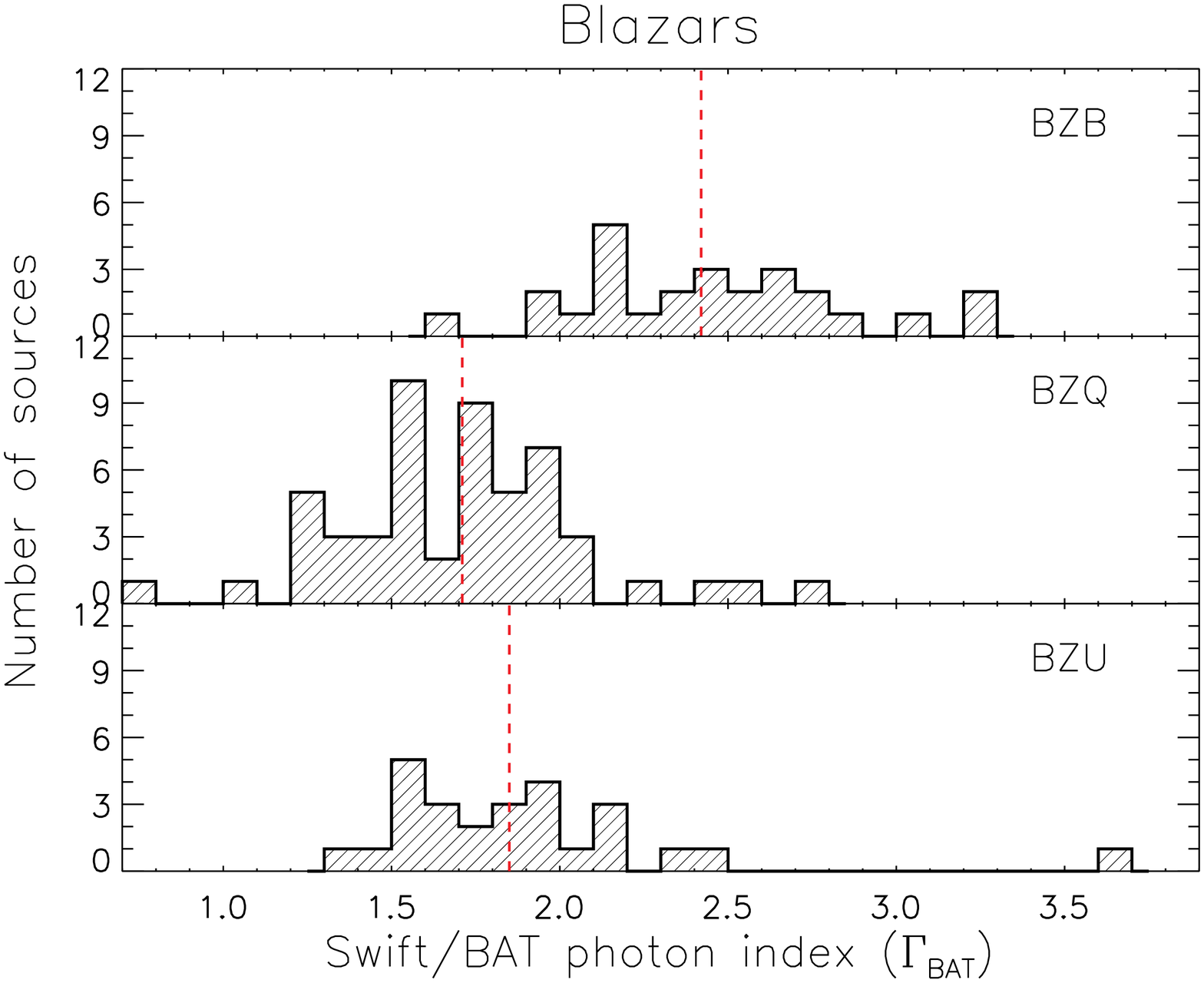}\end{minipage}
% %% caption
 \begin{minipage}[t]{1\textwidth}
  \caption{{\it Top left panel:} Distribution of the best-fitting photon indices obtained by the broad-band X-ray spectroscopy for non-blazar AGN with $N_{\rm H}<10^{22}\rm\,cm^{-2}$ (top panel) and $N_{\rm H}\geq 10^{22}\rm\,cm^{-2}$ (bottom panel). {\it Top right panel:} Distribution of the photon indices obtained fitting the {\it Swift}/BAT spectra of non-blazar AGN with $N_{\rm H}<10^{22}\rm\,cm^{-2}$ (top panel) and $N_{\rm H}\geq 10^{22}\rm\,cm^{-2}$ (bottom panel)  with a simple power-law model \citep{Baumgartner:2013uq}. {\it Bottom left panel:} same as the top left panel for blazars. The blazars are divided into BL Lacs (BZB), Flat Spectrum Radio Quasars (BZQ) and blazars of uncertain type (BZU). {\it Bottom right panel:} as the top right panel for blazars. In all panels the red dashed vertical lines show the median values of the distributions. The median photon indices are listed in Table\,\ref{tab:median_nonblaz} and \ref{tab:median_blaz} for non-blazar AGN and blazars, respectively.}
\label{fig:Gamma_hist}
 \end{minipage}
\end{figure*}

The absorption-corrected and $k$-corrected continuum luminosities were calculated, for the 803 sources for which spectroscopic redshifts were available, in the 2--10, 20--50, 14--150 and 14--195\,keV bands ($L_{\mathrm{\,2-10}}$, $L_{\mathrm{\,20-50}}$, $L_{\mathrm{\,14-150}}$ and $L_{\mathrm{\,14-195}}$, respectively) using the following relation:
\begin{equation}
L_{\mathrm{\,i}}=4\pi d_{\rm\,L}^{2}\frac{F_{\mathrm{\,i}}}{(1+z)^{2-\Gamma}},
\end{equation}
where $F_{\mathrm{\,i}}$ is either the 2--10, 20--50, 14--150 or 14--195\,keV flux, and $d_{\rm\,L}$ is the luminosity distance. We used redshift-independent distance for the 44\,objects at $z<0.01$ for which these measurements were available (see $\S$\ref{sect:redshift}). For blazars the $k$-correction was calculated using the broad-band $\Gamma$ for $L_{\mathrm{\,2-10}}$, and the photon index obtained by fitting the {\it Swift}/BAT spectra with a single power-law model ($\Gamma_{\rm BAT}$) for the $L_{\mathrm{\,20-50}}$, $L_{\mathrm{\,14-150}}$ and $L_{\mathrm{\,14-195}}$ luminosities. The observed (i..e., k-corrected but not absorption-corrected) 2--10\,keV and 14--195\,keV luminosities ($L_{2-10}^{\rm\,obs}$ and $L_{14-195}^{\rm\,obs}$, respectively) were calculated in a similar way. In Table\,\ref{tbl:luminosities} we report the observed 2--10\,keV and 14--195\,keV luminosities, and the intrinsic 2--10\,keV, 20--50\,keV, 14--150\,keV and 14--195\,keV luminosities for all the sources in our sample with spectroscopic redshifts.

\begin{figure*}[t!]
\centering
 %% 1st image
 %% 2nd image
\begin{minipage}[!b]{.48\textwidth}
\centering
\includegraphics[width=9cm]{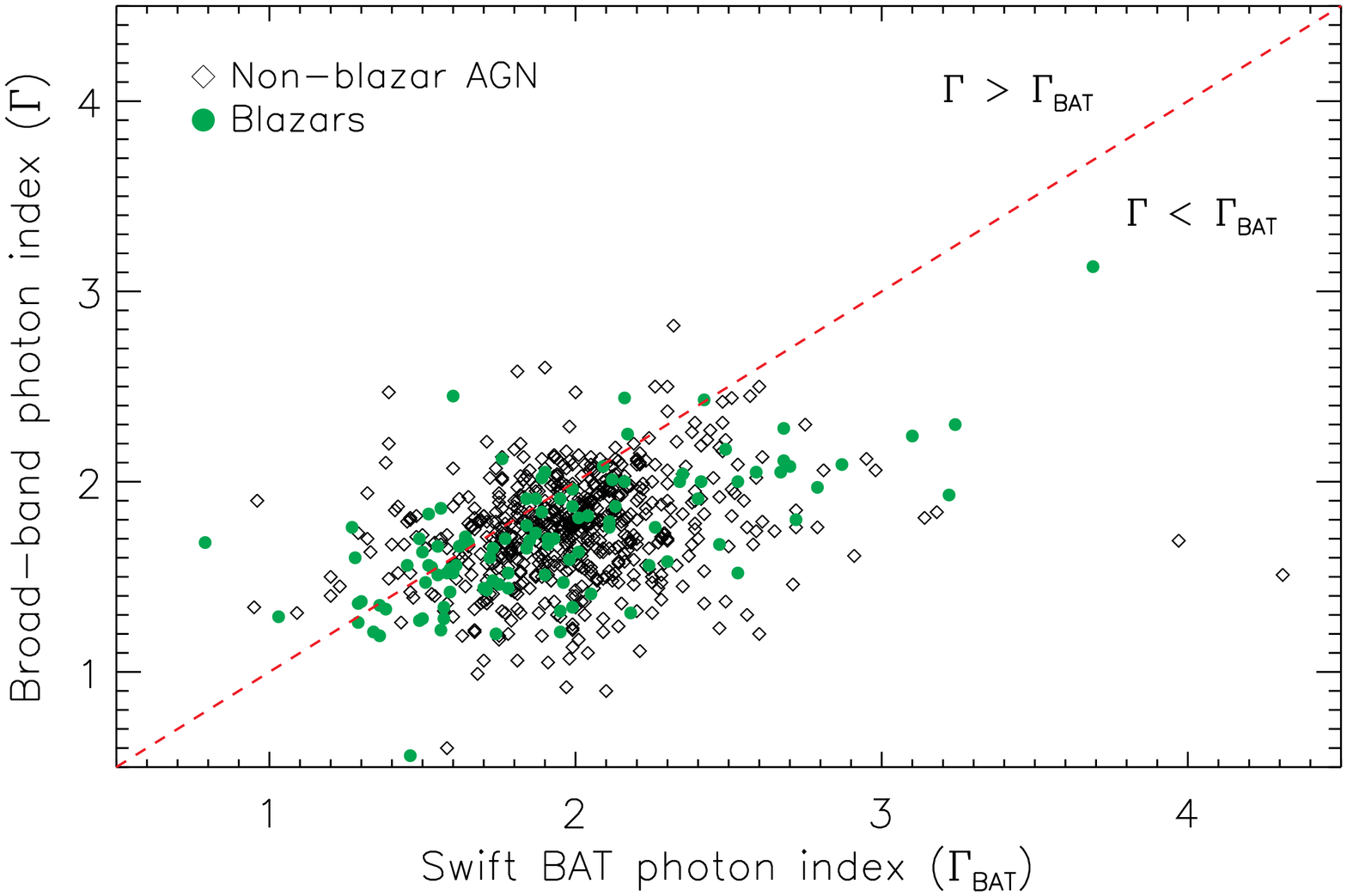}\end{minipage}
\begin{minipage}[!b]{.48\textwidth}
\centering
\includegraphics[width=9cm]{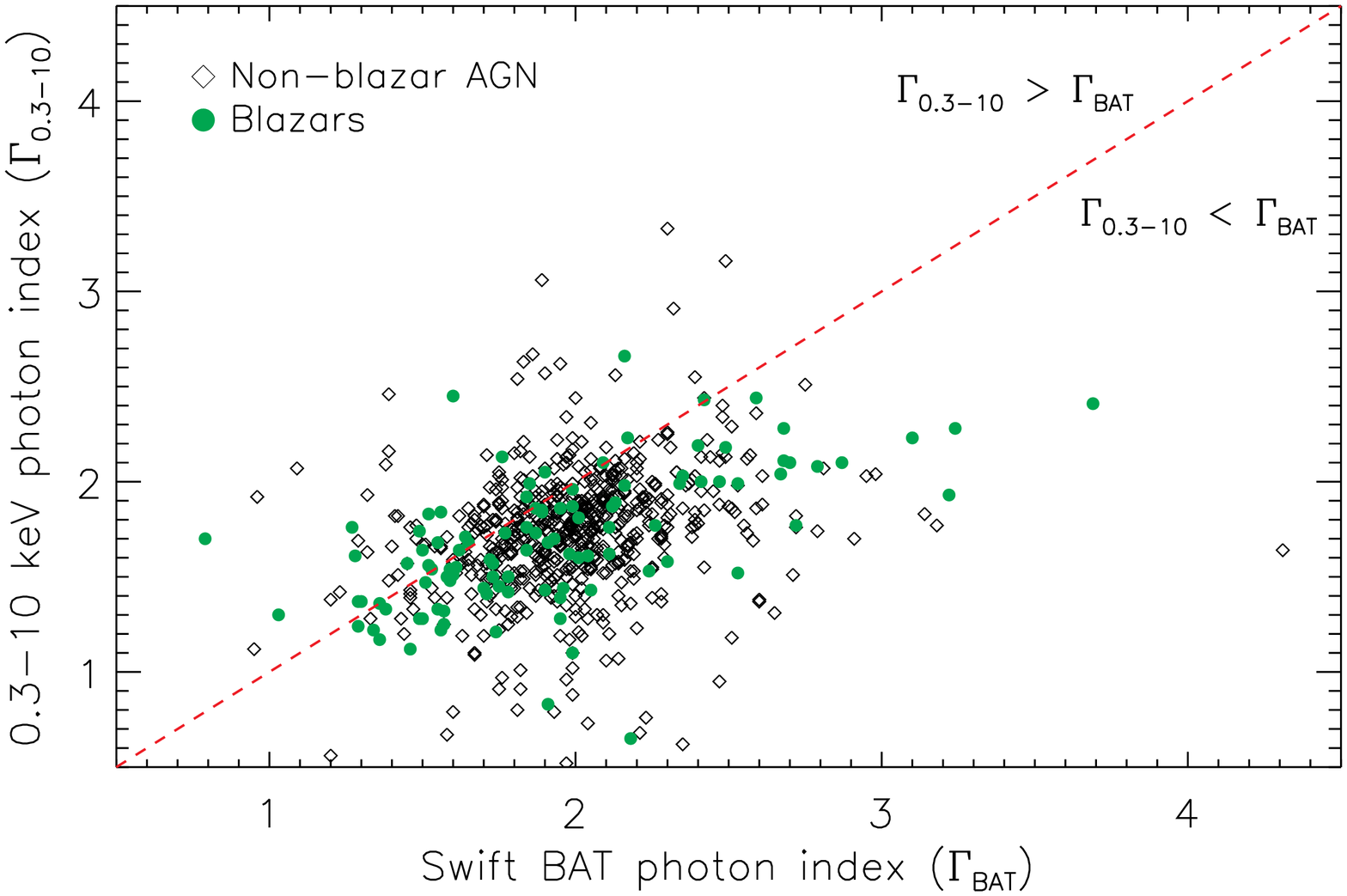}\end{minipage}
\par\medskip
\begin{minipage}[!b]{.48\textwidth}
\centering
\includegraphics[width=9cm]{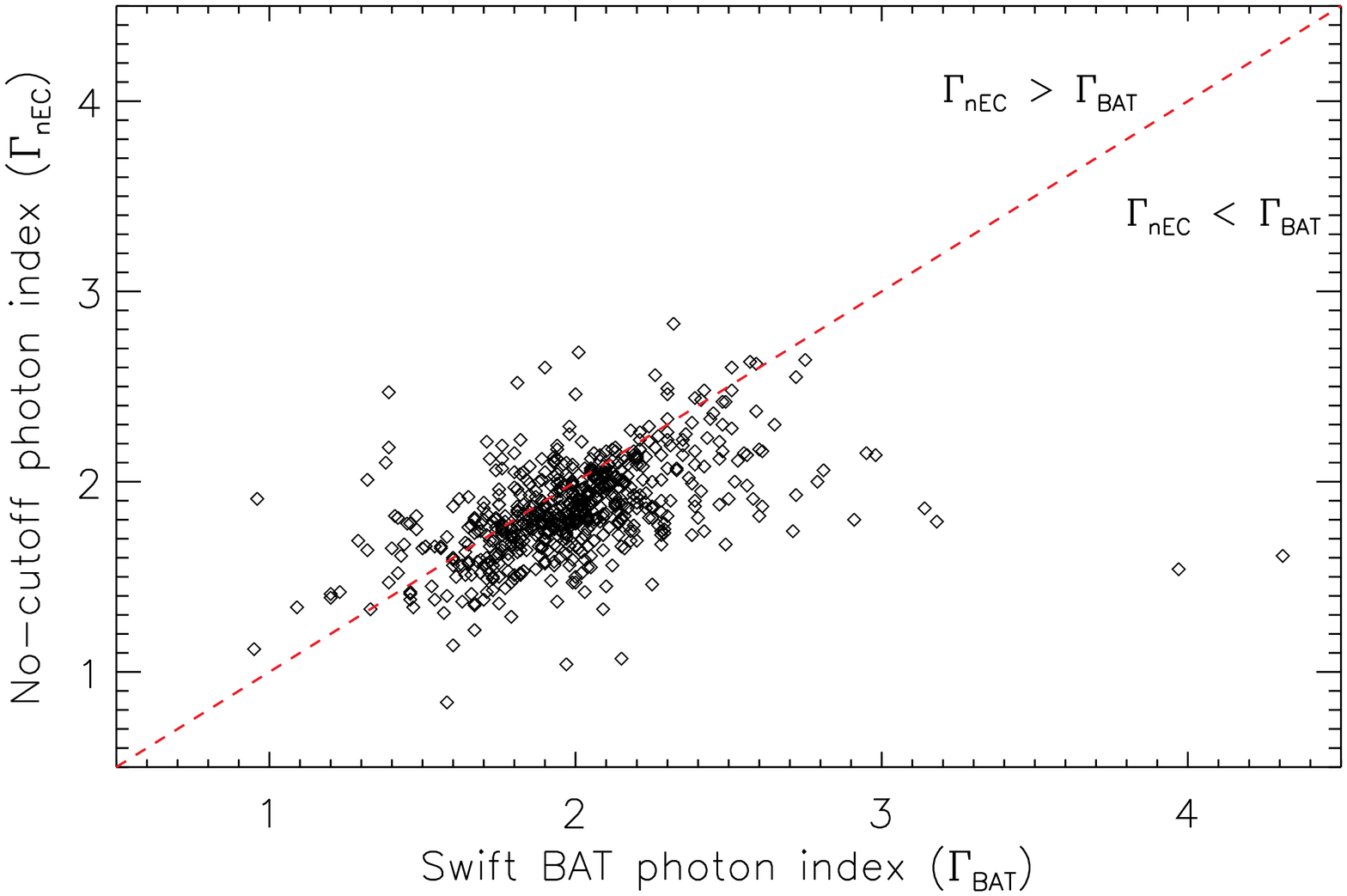}\end{minipage}
\begin{minipage}[!b]{.48\textwidth}
\centering
\includegraphics[width=9cm]{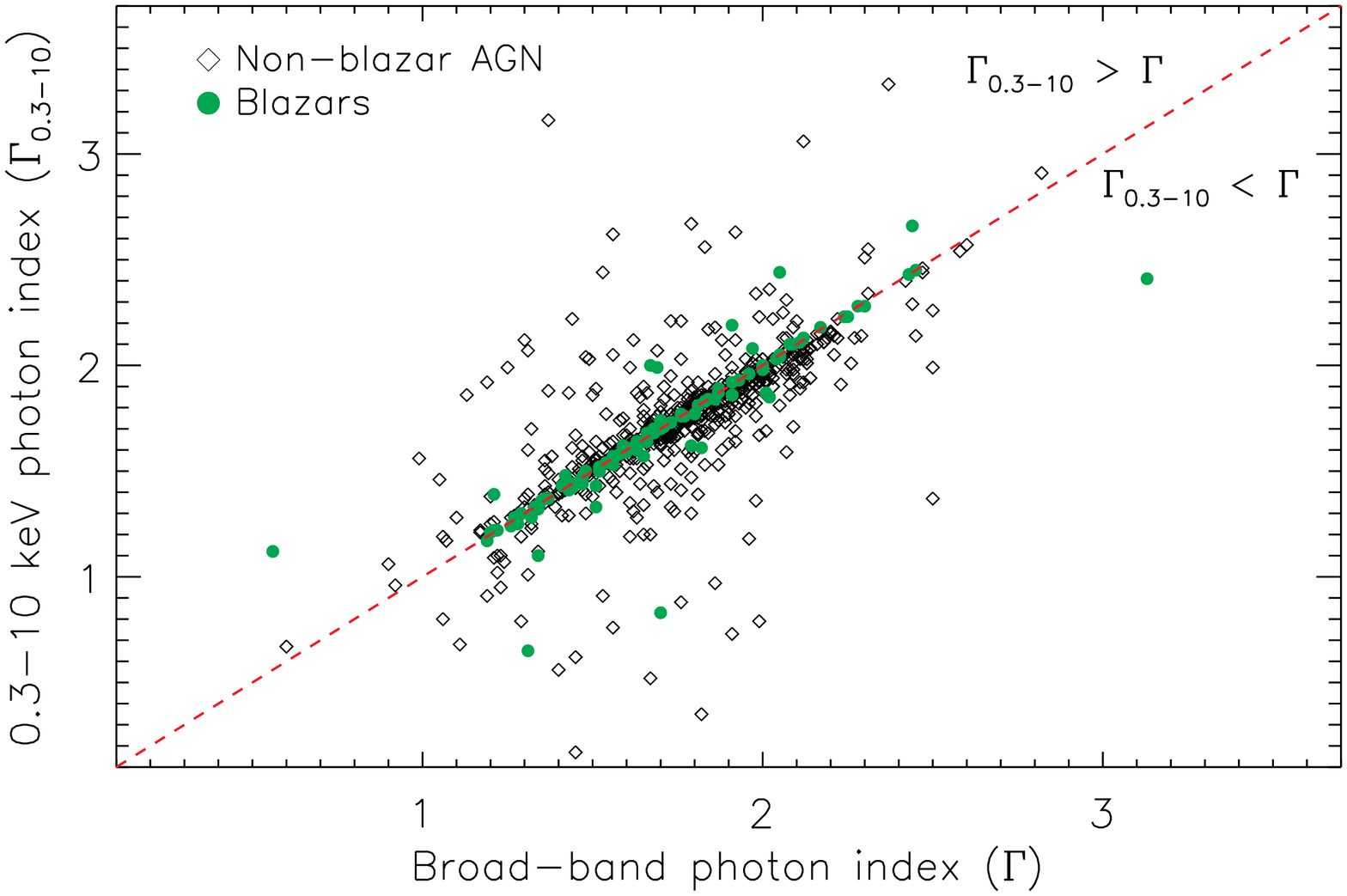}\end{minipage}
\par\medskip
\begin{minipage}[!b]{.48\textwidth}
\centering
\includegraphics[width=9cm]{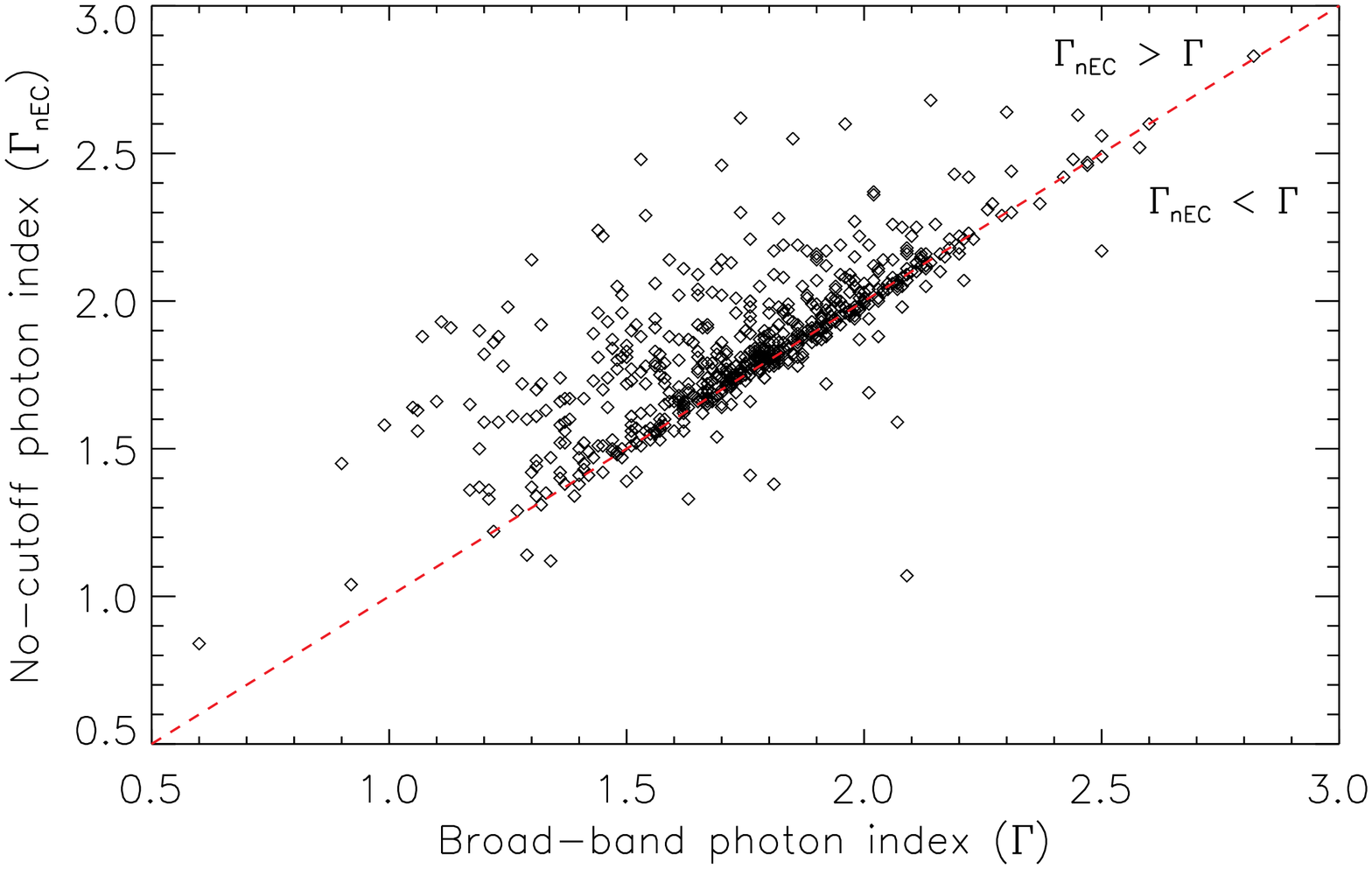}\end{minipage}
\begin{minipage}[!b]{.48\textwidth}
\centering
\includegraphics[width=9cm]{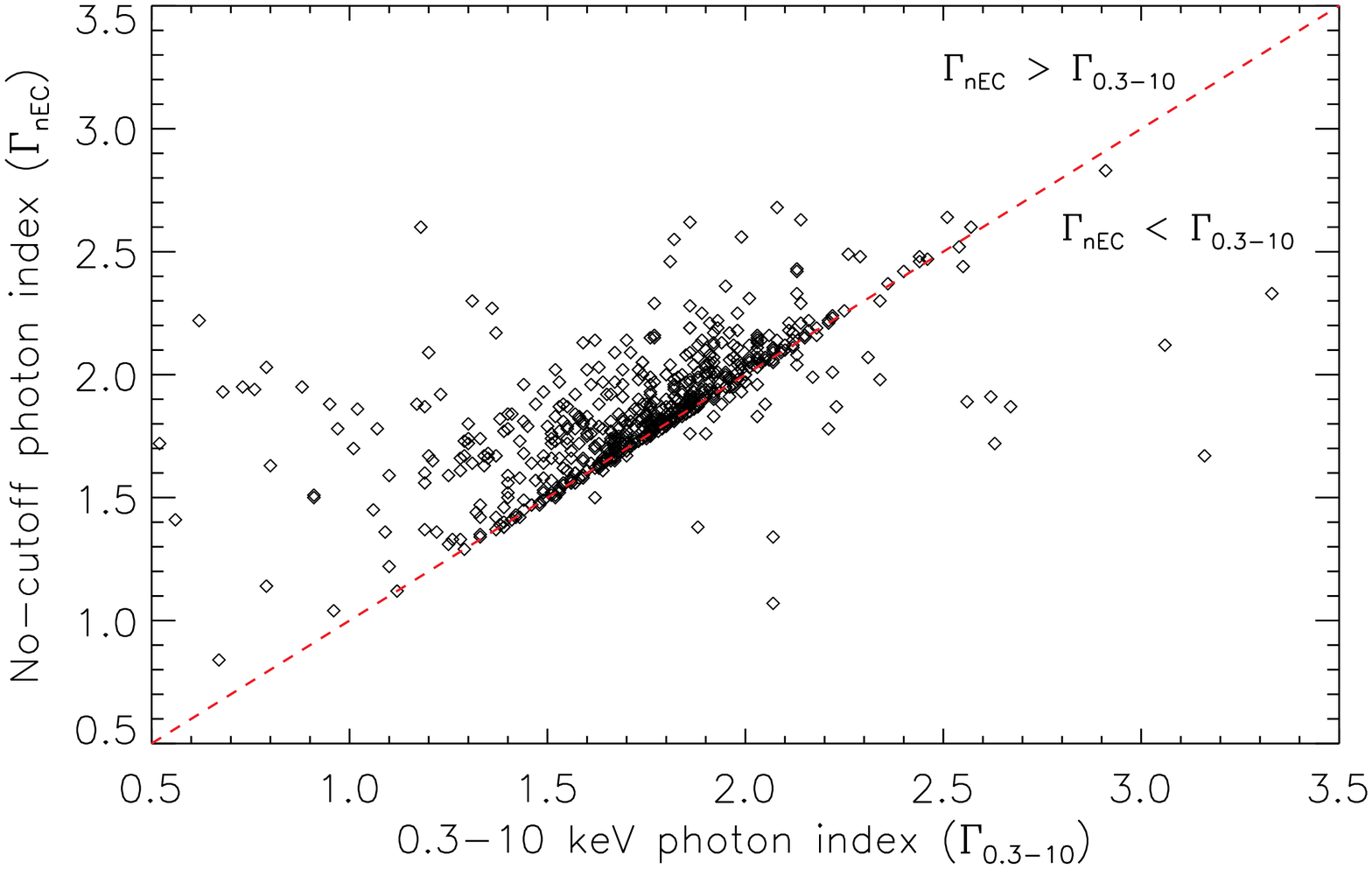}\end{minipage}
% %% caption
 \begin{minipage}[t]{1\textwidth}
  \caption{Scatter plot of the photon indices obtained using different approaches for objects with $\log (N_{\rm H}/\rm cm^{-2}) \leq 23.5$. The values of $\Gamma$ were inferred from the broad-band (0.3--150\,keV) X-ray spectral analysis as discussed in $\S$\ref{sect:Xrayspecanalysis}. $\Gamma_{\rm BAT}$ was obtained by fitting the 14--195\,keV {\it Swift}/BAT spectra with a simple power-law model, and were reported in \cite{Baumgartner:2013uq}. We calculated $\Gamma_{\rm nEc}$ by fixing the value of the high-energy cutoff to $E_{\rm C}=500\rm\,keV$, while $\Gamma_{0.3-10}$ was obtained using the best X-ray spectral model for each object, and fitting only the 0.3--10\,keV spectra fixing $E_{\rm C}=500\rm\,keV$ while setting the reflection parameter to $R=0$ (i.e. no reflection). The plot shows the values of $\Gamma$, $\Gamma_{\rm BAT}$ and $\Gamma_{0.3-10}$ for blazars and non-blazar AGN. For $\Gamma_{\rm nEc}$ we only show the non-blazar AGN, since no high-energy cutoff was considered when fitting blazars (with the few exceptions reported in $\S$\ref{sect:Xrayspecanalysis}).}
\label{fig:GVSg}
 \end{minipage}
\end{figure*}

The distributions of $L_{14-150}$ and $L_{2-10}$ for non-blazar AGN are shown in the top panels of Fig.\,\ref{fig:Lumdistr_obsunobs}. These panels show that the median luminosity of unobscured AGN is higher than that of obscured AGN. This is likely related to the intrinsically different luminosity functions of obscured and unobscured AGN (e.g., \citealp{Della-Ceca:2008bs,Burlon:2011uq,Aird:2015vn}), and is consistent with the observed decrease of the fraction of obscured sources with increasing luminosity (e.g., \citealp{Ueda:2003nx,Treister:2006xr,Burlon:2011uq,Ueda:2014ly}). Due to the strong beaming, blazars (right panel of Fig.\,\ref{fig:Lumdistr_obsunobs}) have on average higher luminosities than non-blazar AGN. As seen in many previous studies, among the blazars the Flat Spectrum Radio Quasars are significantly more luminous than the BL Lacs (e.g., \citealp{Fossati:1998hb,Ghisellini:2010la}).

In \cite{Koss:2017wo} we reported the observed 2--10\,keV luminosities obtained by transforming the observed 14--195\,keV luminosities into 2--10\,keV luminosities using a correction factor of 0.37 [i.e., $L_{2-10}(\rm obs)=0.37\times L_{14-195}$]. In Fig.\,\ref{fig:L210Ldr1} we show that the values of $L_{2-10}(\rm obs)$ for objects with $\log (N_{\rm H}/\rm cm^{-2})\lesssim 23.7$ are in good agreement with the 2--10\,keV luminosities inferred from our spectral analysis. The scatter of $\sim 0.3$\,dex in the plot is likely due to differences in the shape of the X-ray continuum, and to intrinsic flux variability of the primary X-ray emission. Above $\log (N_{\rm H}/\rm cm^{-2})\simeq 23.7$ Compton scattering becomes non-negligible, and even the hard X-ray emission is depleted (see Fig.\,1 of \citealp{Ricci:2015kx}), resulting in underestimated values of $L_{2-10}(\rm obs)$. It should be remarked that, while most CT AGN lie in the range $\Gamma=1.7-2.3$ (Fig.\,\ref{fig:f210corrvsf14150corr}), a few sources have absorption-corrected 2--10\,keV fluxes considerably higher than those expected using the 14--150\,keV flux, and their $L_{2-10}$ emission might be overestimated. In the top panel of Fig.\,\ref{fig:LagnLMIR} we illustrate the relation between the intrinsic 14--150\,keV luminosity and the 12$\mu$m luminosity (from {\it WISE} and high spatial resolution observations) for the objects of our sample. A clear correlation between the mid-IR and X-ray luminosity of non-blazar AGN has been found in the past years by a large number of studies (e.g., \citealp{Lutz:2004vn,Gandhi:2009uq,Asmus:2015ly,Stern:2015vn,Ichikawa:2012fk,Ichikawa:2017rw}), and has been interpreted as the signature of reprocessing of the accreting SMBH emission by dust in the torus. The plot shows that CT AGN follow the same trend as obscured and unobscured sources, thus confirming that we are able to accurately constrain the intrinsic hard X-ray luminosity for these objects. A similar trend is also observed considering the 22$\mu$m luminosities (bottom panel of Fig.\,\ref{fig:LagnLMIR}). We therefore recommend, for CT AGN, to use the 20--50\,keV and 14--150\,keV fluxes and luminosities, and the intrinsic 2--10\,keV fluxes and luminosities obtained from {\it Swift}/BAT luminosities assuming $\Gamma=1.8$ (see Table\,\ref{tbl:lumfluxCT}).

\begin{figure}[t!]
\centering
 %% 1st image
 %% 2nd image
\includegraphics[width=8.8cm]{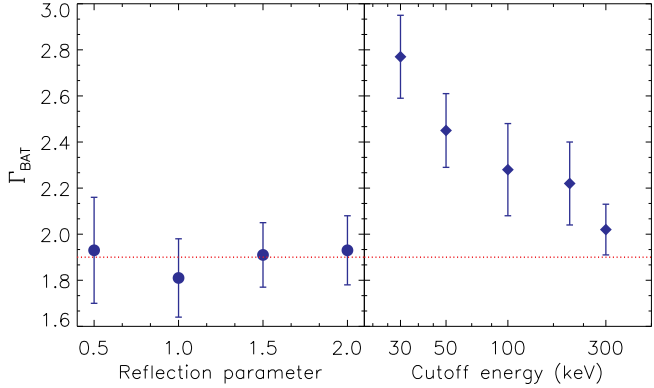}
% %% caption
  \caption{{\it Left panel:} Photon index obtained by fitting with a power-law model simulated {\it Swift}/BAT spectra [setting $\Gamma=1.9$ and $E_{\rm C}=500$\,keV and the 14--195\,keV flux to $(1-2)\times10^{-11}\rm\,erg\,cm^{-2}\,s^{-1}$] assuming different values of the reflection parameter. The red dotted line shows the correct value of photon index of the primary X-ray continuum. {\it Right panel:} same as left panel, setting  $\Gamma=1.9$ and $R=0.5$ and varying the values of the high-energy cutoff. The plot shows that decreasing values of the high-energy cutoff lead to the increase of the {\it Swift}/BAT photon index, while values of the reflection parameter do not typically affect significantly $\Gamma_{\rm BAT}$. Exposure times in the simulations were set to $9.5\times 10^{6}\rm\,s$, consistent with the exposure of $\sim 50\%$ of the sky for the 70-month {\it Swift}/BAT catalog.}
\label{fig:Gbat_sims}
\end{figure}

\subsection{X-ray continuum properties}\label{sect:Xraycontinuum}

\subsubsection{Photon index}\label{sect:photonindex}

The distribution of photon indices inferred from the broad-band spectral analysis is shown in the top left panel of Fig.\,\ref{fig:Gamma_hist} for unobscured and obscured AGN\footnote{The median of the uncertainties on the photon index for the whole sample of non-blazar AGN is $\Delta \Gamma=0.17$.}. We find that the median values (red dashed vertical lines) of $\Gamma$ for sources with $\log (N_{\rm H}/\rm cm^{-2})\geq 22$ and $\log (N_{\rm H}/\rm cm^{-2})< 22$ are consistent ($1.76\pm0.02$ and $1.80\pm0.02$, respectively). The total-non-blazar sample has a median of  $1.78\pm0.01$, consistent with the value found by \cite{Winter:2009qf} (see also \citealp{Alexander:2013ly}). Compton-thin obscured AGN have significantly lower photon indices ($\Gamma=1.70\pm0.02$) than unobscured objects, while CT AGN are found to typically have steeper slopes ($\Gamma=2.05\pm0.05$). The distribution of $\Gamma$ of sources with $\log (N_{\rm H}/\rm cm^{-2})\geq 22$ is broader than that of AGN with $\log (N_{\rm H}/\rm cm^{-2})< 22$, and a Kolmogorov-Smirnov test results in a p-value of $\sim 10^{-4}$, suggesting that the two populations are significantly different. The origin of this difference might be related to the larger fraction of sources with lower 0.3--10\,keV counts amongst the obscured sources, which increases the scatter of $\Gamma$, and to the population of CT AGN with $\Gamma>2$. 
A significant difference between sources with  $\log (N_{\rm H}/\rm cm^{-2})\geq 22$ and $\log (N_{\rm H}/\rm cm^{-2})< 22$ is found for the photon indices obtained by fitting the 14--195\,keV {\it Swift}/BAT spectrum with a simple power-law model ($\Gamma_{\rm BAT}$, top right panel of Fig.\,\ref{fig:Gamma_hist}), with unobscured AGN showing a steeper X-ray spectrum ($2.02\pm0.02$) than their $\log (N_{\rm H}/\rm cm^{-2})\geq 22$ counterparts ($1.89\pm0.02$). This is likely related to the fact that $\Gamma_{\rm BAT}$ is affected by obscuration, and that for $\log (N_{\rm H}/\rm\,cm^{-2})\gg 23$ the hard X-ray continuum is flattened by the presence of the reprocessed X-ray radiation emerging from the obscured primary continuum.

To further study the shape of the X-ray continuum we calculated the photon indices using two alternative approaches. First, we re-fitted all the non-blazar AGN of our sample using the same models described in $\S$\ref{sect:Xrayspecanalysis}, fixing the high-energy cutoff to the maximum value allowed ($E_{\rm C}=500$\,keV), in order to minimize the spectral curvature due to the cutoff. We report the values of the photon index obtained using this approach ($\Gamma_{\rm nEc}$) in Table\,\ref{tbl:otherGamma}. Second, we set $E_{\rm C}=500$\,keV, excluded the X-ray reprocessed emission\footnote{This was done fixing the reflection parameter to $R=0$ for the models A1--A8 and B6, and removing the reflection component for the models B1--B5 and B7--B9}, and fitted the data only in the 0.3--10\,keV band. This was done using the same best-fitted model obtained from the broad-band X-ray spectral analysis, ignoring, besides the high-energy cutoff and the reflection component, also the cross-calibration constant. The photon index obtained with this approach is affected by absorption for objects in which the line-of-sight column density is $N_{\rm H}\gg 10^{23}\rm\,cm^{-2}$, since Compton scattering would significantly deplete the X-ray emission and the relative influence of the reprocessed X-ray emission would be stronger. The values of the photon index obtained following this procedure ($\Gamma_{0.3-10}$) are also reported in Table\,\ref{tbl:otherGamma}.

\begin{figure}[t!]
\centering
 %% 1st image
 %% 2nd image
\begin{minipage}[!b]{.48\textwidth}
\centering
\includegraphics[width=8.8cm]{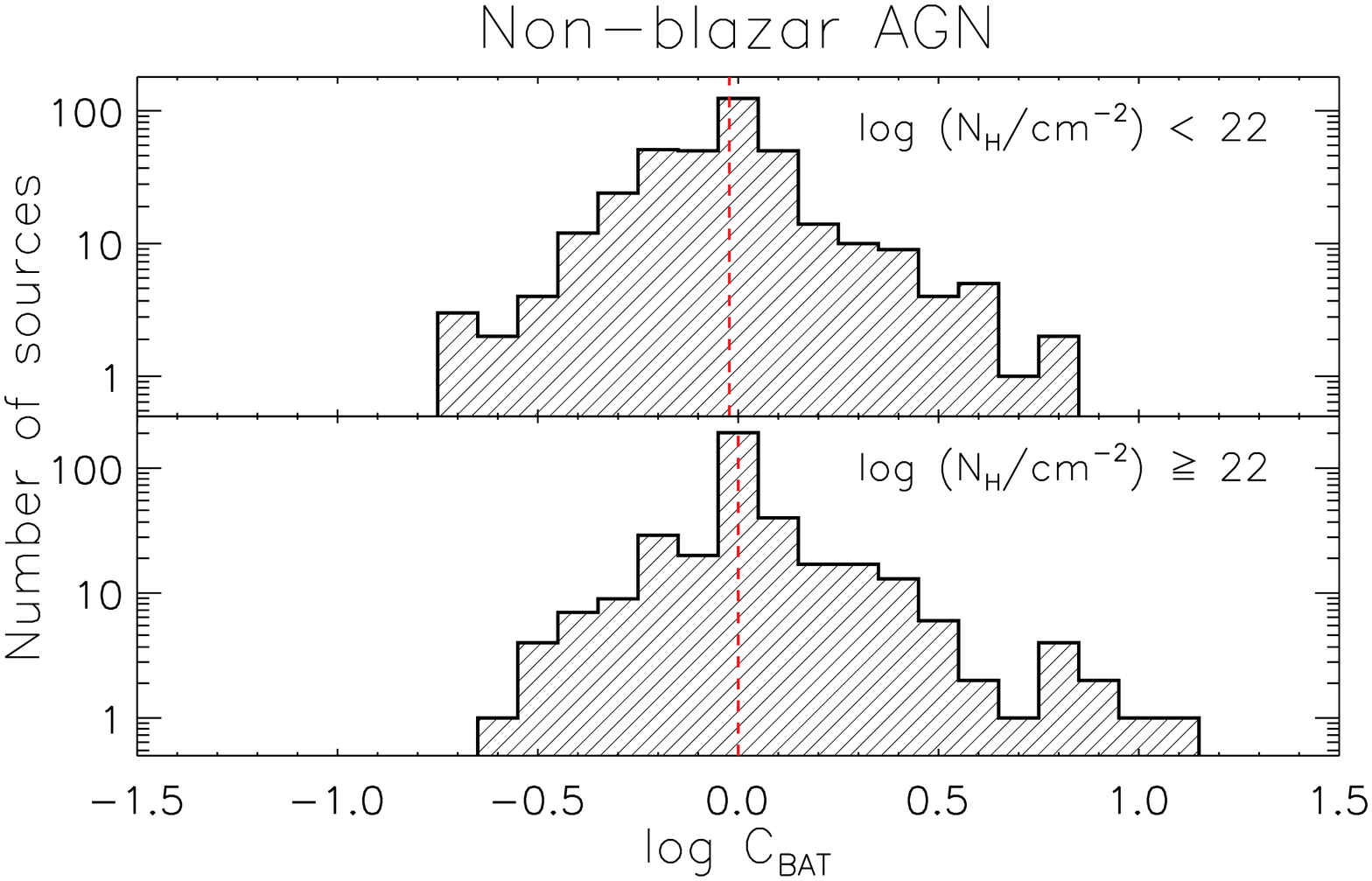}\end{minipage}
\begin{minipage}[!b]{.48\textwidth}
\centering
\par\medskip
\includegraphics[width=8.8cm]{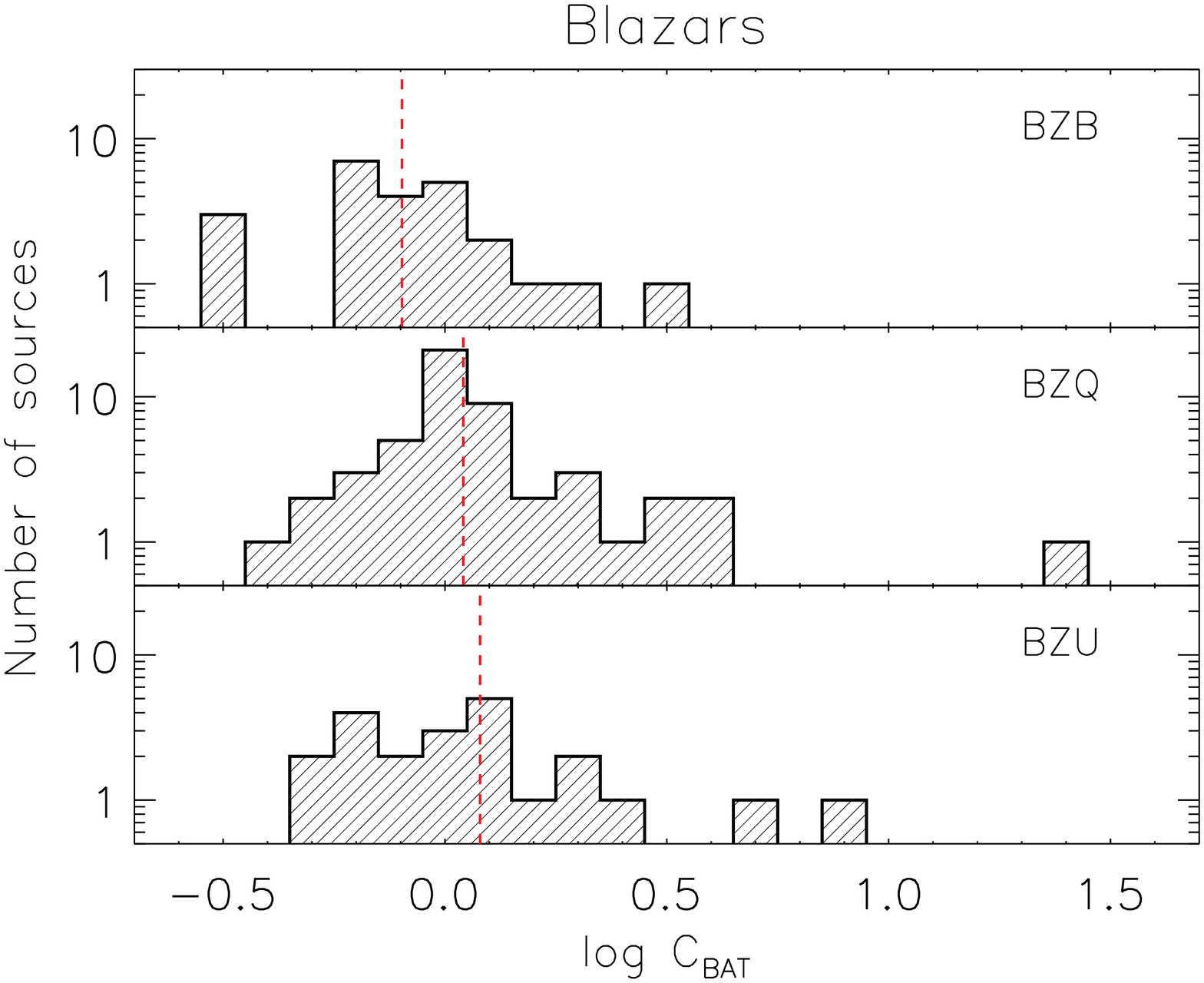}\end{minipage}
% %% caption
 \begin{minipage}[t]{0.48\textwidth}
  \caption{{\it Top panel:} Distribution of the cross-calibration ($C_{\rm\,BAT}$) constants obtained by our spectral analysis for the non-blazar AGN of our sample. The sample is divided into objects with $\log (N_{\rm H}/\rm cm^{-2})< 22$ (top panel) and $\log (N_{\rm H}/\rm cm^{-2})\geq 22$ (bottom panel). {\it Bottom panel:} same as left panel for the blazars. The blazars are divided into BL Lacs (BZB), Flat Spectrum Radio Quasars (BZQ) and blazars of uncertain type (BZU). The median values of $C_{\rm\,BAT}$ are listed in Table\,\ref{tab:median_nonblaz} and \ref{tab:median_blaz} for non-blazar and blazars, respectively. The dispersion in $C_{\rm\,BAT}$ shows the typical X-ray variability on the timescales probed here (days to several years). }
\label{fig:Cbatdistr}
 \end{minipage}
\end{figure}

In Fig.\,\ref{fig:GVSg} we compare the four different photon indices described above. We consider only objects with $\log (N_{\rm H}/\rm cm^{-2}) \leq 23.5$ since obscuration would most significantly affect $\Gamma_{0.3-10}$ above this value. The top left panel compares $\Gamma$ with $\Gamma_{\rm BAT}$ for blazars and non-blazar AGN. The plot shows that for non-blazar AGN the values of $\Gamma$ are typically lower than $\Gamma_{\rm BAT}$. This difference is likely due to the presence of a cutoff in the modelling of the primary X-ray continuum, as found by several studies of broad-band X-ray emission of AGN (e.g., \citealp{Ballantyne:2014if,Balokovic:2015mi,Lubinski:2016ao}). In agreement with the idea that a high-energy cutoff is almost ubiquitous in the X-ray spectra of AGN, we find that the large majority ($\sim 80\%$) of the objects have $\Gamma_{0.3-10} < \Gamma_{\rm BAT}$ (top right panel). Similarly, also $\Gamma_{\rm nEc}$ tends to be lower than $\Gamma_{\rm BAT}$ (central left panel). 

\begin{figure*}[t!]
\centering
 %% 1st image
 %% 2nd image
%
\begin{minipage}[!b]{.48\textwidth}
\centering
\includegraphics[width=9cm]{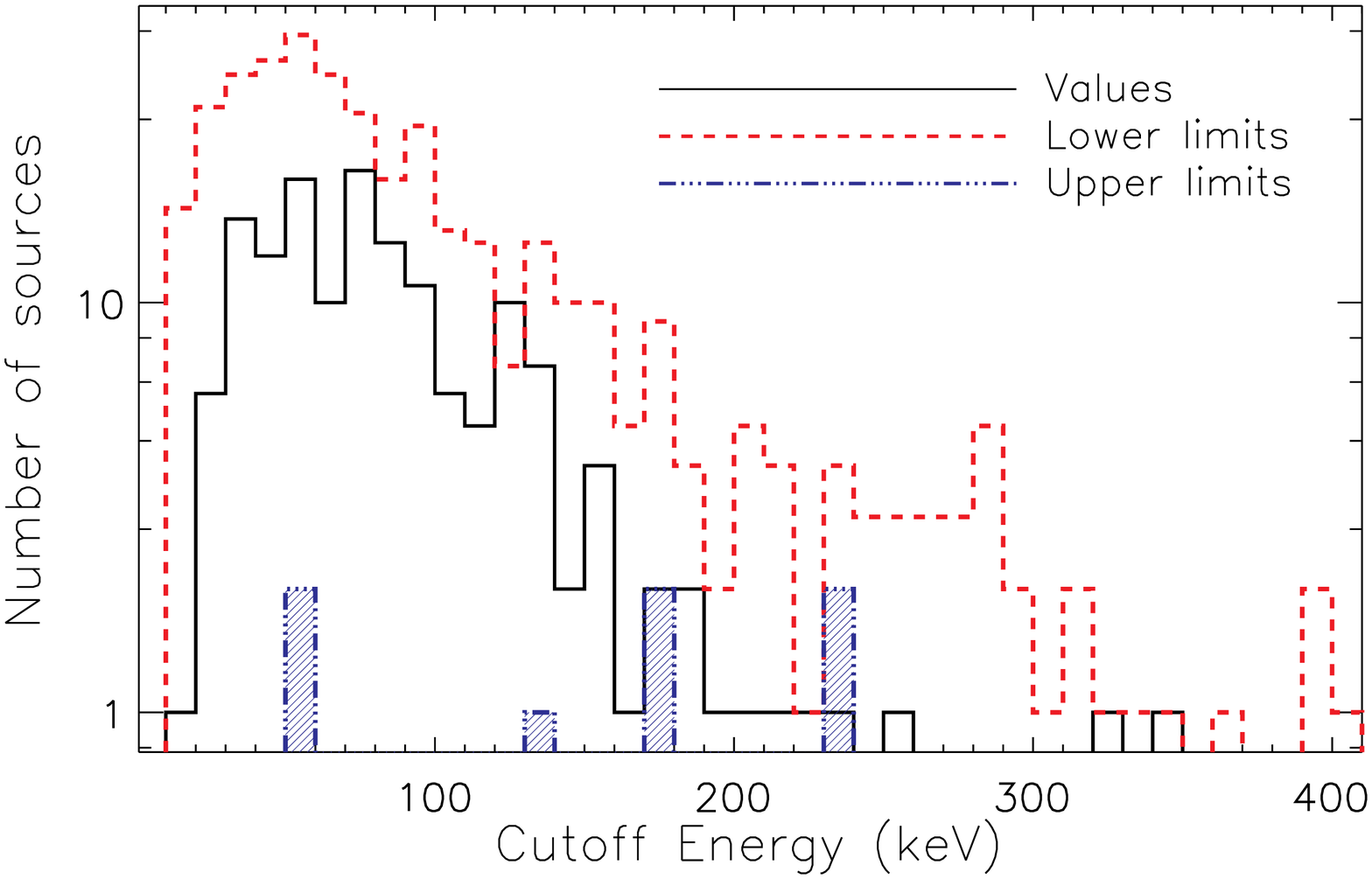}\end{minipage}
\begin{minipage}[!b]{.48\textwidth}
\centering
\includegraphics[width=9cm]{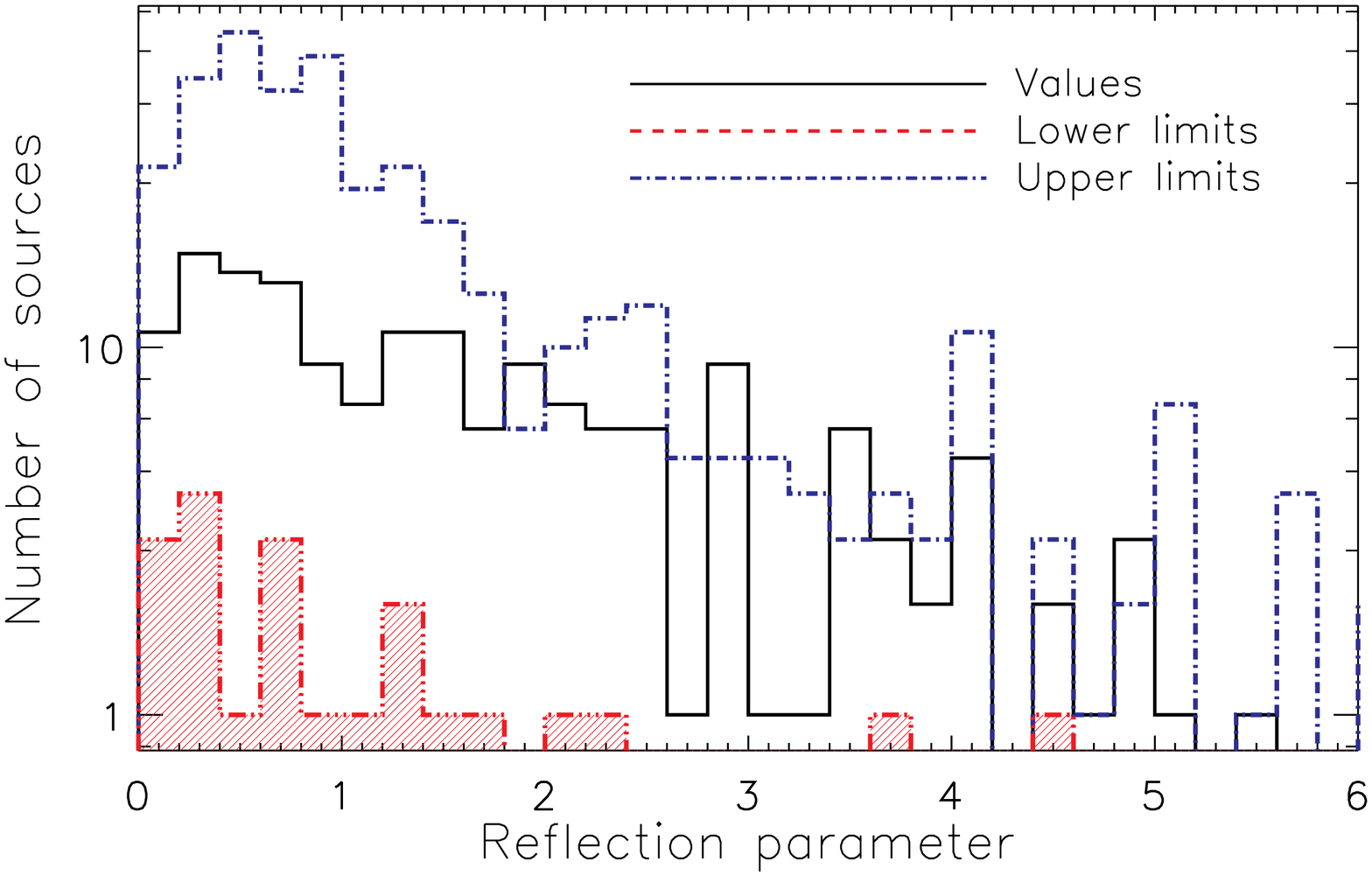}\end{minipage}
%
% %% caption
 \begin{minipage}[t]{1\textwidth}
  \caption{{\it Left panel:} Distribution of the cutoff energies for the non-blazar AGN. The plot shows the values (black continuos histogram), the lower limits (red dashed histogram) and the upper limits (blue dot-dot-dashed histogram). {\it Right panel:} Distribution of the reflection parameters for the non-blazar AGN. The plot shows the values (black continuos histogram), the lower limits (red dashed histogram) and the upper limits (blue dot-dashed histogram). The median values of $E_{\rm C}$ and $R$ are listed in Table\,\ref{tab:median_nonblaz}. }
\label{fig:hist_EC_R}
 \end{minipage}
\end{figure*}

To test how reflection influences the observed value of $\Gamma_{\rm BAT}$ we simulated {\it Swift}/BAT spectra using \textsc{pexrav}, setting $\Gamma=1.9$ and $E_{\rm C}=500$\,keV, the 14--195\,keV flux to $(1-2)\times10^{-11}\rm\,erg\,cm^{-2}\,s^{-1}$ (consistent with the typical flux of the BAT AGN of our sample) assuming different values of the reflection parameter. The spectrum was then fitted with a simple power-law model, similar to what has been done for the {\it Swift}/BAT 70-month catalog. The value of $\Gamma_{\rm BAT}$ does not increase significantly for larger values of the reflection parameter (left panel of Fig.\,\ref{fig:Gbat_sims}). We then carried out similar simulations, fixing the reflection parameter and varying the energy of the cutoff, and found that $\Gamma_{\rm BAT}$ increases significantly as the energy of the cutoff decreases (right panel of Fig.\,\ref{fig:Gbat_sims}). This clearly shows that the steepening of $\Gamma_{\rm BAT}$ with respect to $\Gamma$ is due to the presence of a cutoff.

We further find that, for most of our AGN (both blazars and non-blazar sources), $\Gamma_{0.3-10}\simeq \Gamma$ (central right panel of Fig.\,\ref{fig:GVSg}). For the blazars this is due to the fact that no cutoff or reflection were considered in the fitting, while for the non-blazar AGN this implies that we are able to recover the intrinsic value of $\Gamma$ for most objects.
The values of $\Gamma_{\rm nEc}$ are substantially higher than those of $\Gamma$ (bottom left panel) and $\Gamma_{0.3-10}$ (bottom right panel). In particular, only $\sim 21\%$ ($\sim 14\%$) of the objects have $\Gamma_{\rm nEc}$ $<\Gamma$ ($\Gamma_{\rm nEc}<\Gamma_{0.3-10}$). This is due to the fact that, to compensate for the spectral curvature due to the cutoff, the broad-band X-ray spectral fit carried out to obtain $\Gamma_{\rm nEc}$ results in larger values of $R$ and in steeper slopes.

The median value of both $\Gamma$ and $\Gamma_{\rm BAT}$ is lower for blazars ($\Gamma=1.68\pm0.04$, $\Gamma_{\rm BAT}=1.87\pm0.06$) than for the non-blazar AGN ($1.78\pm0.01$, $1.96\pm0.01$). The distributions of $\Gamma$ and $\Gamma_{\rm BAT}$ are illustrated in the left and right bottom panels of Fig.\,\ref{fig:Gamma_hist}, respectively.
The X-ray continuum of the Flat Spectrum Radio Quasars is very flat ($\Gamma=1.54\pm0.05$, $\Gamma_{\rm BAT}=1.71\pm0.06$), and it differs significantly from that of BL Lacs, which typically show steeper X-ray emission ($\Gamma=2.05\pm0.06$, $\Gamma_{\rm BAT}=2.42\pm0.10$). These differences in spectral shape are consistent with what has been found by \cite{Sambruna:2010ao} who, combining {\it Swift}/BAT with {\it Fermi}/LAT spectra,  argued that this behaviour (i.e., more luminous blazars are flatter in the hard X-ray band) is in agreement with the so called ``blazar sequence" (e.g., \citealp{Fossati:1998hb,Inoue:2009kq}).

\subsubsection{Cross-calibration constant}\label{sect:crossbat}

As mentioned in $\S$\ref{sect:Xrayspecanalysis}, a cross-calibration constant between the soft X-ray spectra and the 70-month averaged {\it Swift}/BAT spectra ($C_{\rm\,BAT}$) was added to all models. We find that both the obscured and unobscured sample have a median of $C_{\rm\,BAT}=1$, consistent with the idea that the 0.3--10\,keV observations are randomly sampling the variable flux of the X-ray source. 
In the top panel of Fig.\,\ref{fig:Cbatdistr} we illustrate the distribution of $\log C_{\rm\,BAT}$ for objects with $\log (N_{\rm H}/\rm cm^{-2}) < 22$ (top panel) and $\log (N_{\rm H}/\rm cm^{-2}) \geq 22$ (bottom panel). The standard deviation of $\log C_{\rm\,BAT}$ for the latter ($0.22$\,dex) is consistent with that of the former ($0.21$\,dex). 

The uncertainties on the values of $C_{\rm\,BAT}$ are typically higher for more obscured objects (with a median of $\Delta C_{\rm\,BAT}\sim 0.55$) than for those with $\log (N_{\rm H}/\rm cm^{-2}) < 22$ ($\Delta C_{\rm\,BAT}\sim 0.45$) and blazars ($\Delta C_{\rm\,BAT}\sim 0.25$). This difference is likely related to the complexity of the modelling when absorption is present, and to the fact that the soft X-ray spectra of the most obscured sources typically have a lower number of counts.

The standard deviation found for the whole sample of non-blazar AGN ($\sigma\simeq 0.22$\,dex) is similar to the dispersion obtained comparing $L_{2-10}(obs)$ to $L_{2-10}$ (0.3\,dex, Fig.\,\ref{fig:L210Ldr1}). The difference between these two values (0.22\,dex and 0.3\,dex) could be related to differences in the spectral shape of the X-ray continuum, which would increase the dispersion of $L_{2-10}(obs)$ versus $L_{2-10}$, as discussed in $\S$\ref{sect:Luminosities}. Studying a sample of 45 Compton-thin obscured {\it Swift}/BAT AGN with {\it Suzaku}, \cite{Kawamuro:2016fj} found a dispersion in $C_{\rm\,BAT}$ of $0.21$\,dex, which is consistent with the value obtained here. This confirms that with our spectral analysis we are able to quantify the intrinsic variability of the X-ray source, and that, on the timescales probed by our study (days to several years), the X-ray variability of non-blazar AGN is $\sim 0.2$\,dex.

While the median value of blazars is also consistent with one ($C_{\rm\,BAT}=1.00\pm0.35$, bottom panel of Fig.\,\ref{fig:Cbatdistr}), it has a larger scatter ($\sigma\simeq 0.3$\,dex) than for the non-blazar AGN, consistent with the stronger variability of these objects, observed across the entire electromagnetic spectrum (e.g., \citealp{Ulrich:1997qf}). The number of blazars with $C_{\rm\,BAT}> 2$ (10) is larger than that of objects of the same class with $C_{\rm\,BAT}<0.5$ (5). This is likely related to the fact that several soft X-ray observations of the blazars in our sample have been triggered by the object being in a bright state (e.g., \citealp{Stroh:2013jw}).

\subsubsection{High-energy cutoff}\label{sect:cutoff}

\begin{figure}[t!]
\centering
 %% 1st image
 %% 2nd image
%
\begin{minipage}[!b]{.48\textwidth}
\centering
\includegraphics[width=9cm]{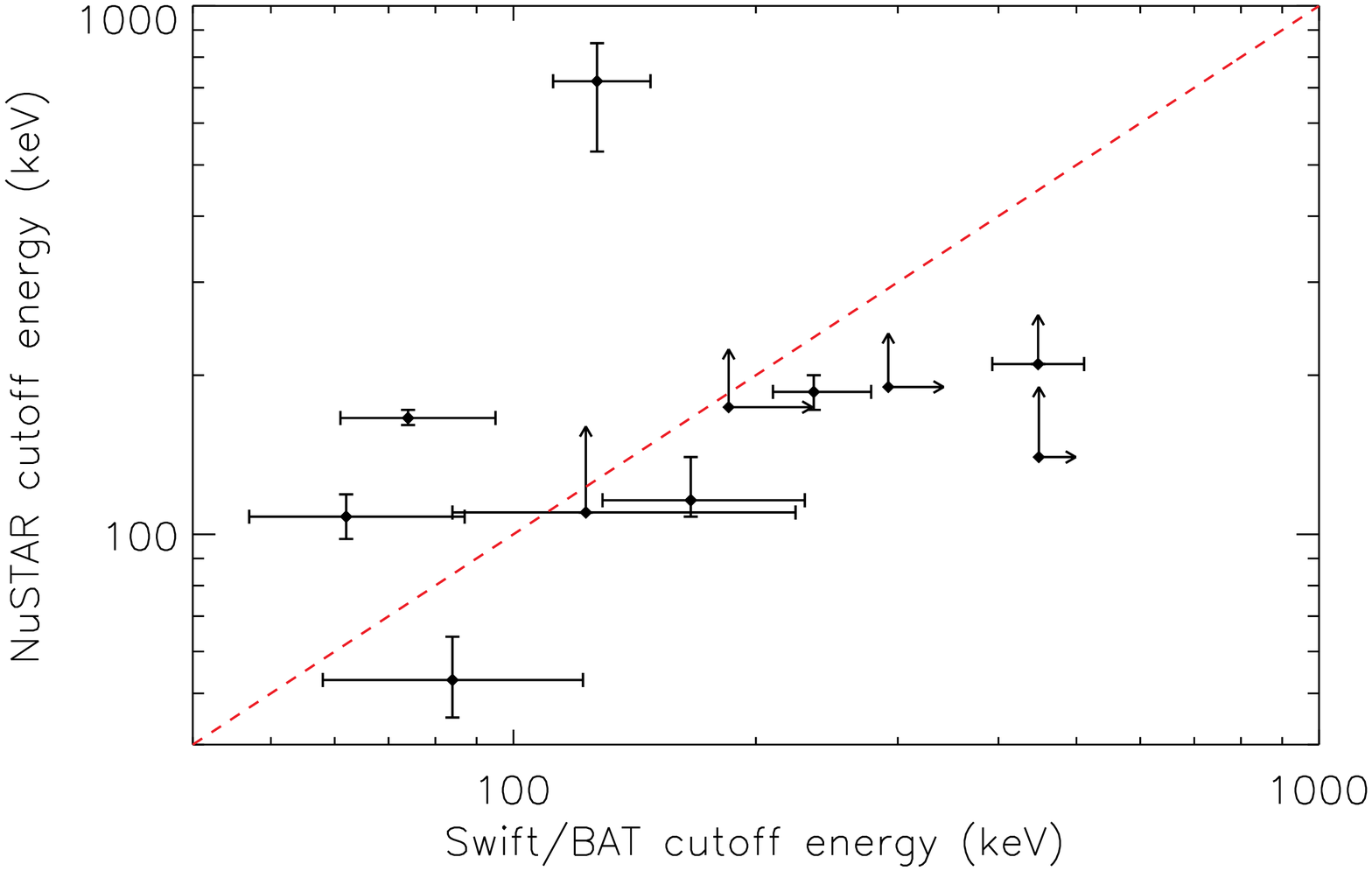}\end{minipage}
%
% %% caption
 \begin{minipage}[t]{.48\textwidth}
  \caption{Values of the cutoff energy obtained with {\it NuSTAR} (from \citealp{Marinucci:2016sp}) versus the values obtained by our analysis using {\it Swift}/BAT for the same objects. The objects reported in the plot are Ark\,120 \citep{Matt:2014fv}, Mrk\,335 \citep{Parker:2014zp}, NGC\,7213 \citep{Ursini:2015dk}, 3C\,390.3 \citep{Fabian:2015wq}, IC\,4329A \citep{Brenneman:2014mi}, NGC\,5506 \citep{Matt:2015fe}, SWIFT\,J2127.4+5654 \citep{Marinucci:2014cl}, MCG\,$-$05$-$23$-$016 \citep{Balokovic:2015mi}, MCG$-$06$-$30$-$015 \citep{Marinucci:2014fu}, NGC\,2110 \citep{Marinucci:2015eq} and SWIFT\,J1737.5$-$2908 \citep{Tortosa:2017kq}. The red dashed line illustrates the scenario in which the two energies are the same. The plot shows that the values of the high-energy cutoff are roughly in agreement, with the exception of NGC\,5506, for which the energy of the cutoff found by {\it NuSTAR} is significantly higher than that inferred using {\it Swift}/BAT.}
\label{fig:EC_nuBat}
 \end{minipage}
\end{figure}

The distribution of the cutoff energies for the non-blazar AGN of our sample are shown in the left panel of Fig.\,\ref{fig:hist_EC_R}. For 418 (7) sources only a lower (upper) limit of $E_{\rm C}$ could be obtained. As expected, most of the 161 sources for which it was possible to constrain $E_{\rm C}$ have values $E_{\rm C} \lesssim 100$\,keV, i.e. within the energy range covered by our spectral analysis, and a median value of $E_{\rm C}=76\pm6$\,keV. We found that AGN with $\log (N_{\rm H}/\rm cm^{-2})< 22$ and those with $\log (N_{\rm H}/\rm cm^{-2})\geq 22$ have consistent median cutoff energies ($80\pm7$\,keV and $74\pm11$\,keV, respectively). The eight CT AGN for which $E_{\rm C}$ could be constrained have a median of $43\pm15$\,keV, lower than Compton-thin and unobscured objects. This is probably due to the fact that \textsc{pexrav} fails to identify the curvature of the spectra of these heavily obscured objects as reprocessed emission, and associates it with an X-ray continuum with a low cutoff energy, and confirms the importance of using physical torus models to study CT AGN.
We find that the mean of our sample is $E_{\rm C}=90\pm5$\,keV and a standard deviation of $\sigma=61$\,keV. This is lower than the value found by \cite{Malizia:2014zt} studying the broad-band X-ray spectra of a sample of 41 type-I AGN ($E_{\rm C}=128$\,keV, $\sigma=46$\,keV). We compared the values of the cutoff energy obtained by our study with those found by the analysis of {\it NuSTAR} observations (\citealp{Tortosa:2017kq}, \citealp{Marinucci:2016sp} and references therein), and found that the values are roughly in agreement (Fig.\,\ref{fig:EC_nuBat}), with the exception of NGC\,5506, for which the energy of the cutoff found by {\it NuSTAR} ($720^{+130}_{-190}$\,keV, \citealp{Matt:2015fe}) is significantly higher than that inferred using {\it Swift}/BAT ($127^{+21}_{-15}$\,keV). For two of the objects reported by \cite{Marinucci:2016sp}, 3C\,382 \citep{Ballantyne:2014if} and Fairall\,9 \citep{Fabian:2015wq}, the data did not allow to constrain the energy of the cutoff.

Since for most of the sources of our sample we could only obtain lower and upper limits on the energy of the cutoff, the median and mean values reported above are not representative of the whole sample of {\it Swift}/BAT AGN. To better constrain the median of the sample, including also upper and lower limits, we performed 10,000 Montecarlo simulations for each value of $E_{\rm C}$. For each simulation we used the following approach: the values of $E_{\rm C}$ of the detections were substituted with values randomly selected from Gaussian distribution centered on the best-fit value, with the standard deviation given by its uncertainty; ii) the upper limits $U$ were substituted with a random value from a uniform distribution in the interval $[0,U]$; the lower limits $L$ were substituted with a value randomly selected from a uniform distribution in the interval $[$L$,1000]$. In the last step we assumed that the maximum value of the cutoff energy is 1,000\,keV. For each Montecarlo run we calculated the median of all values, and then used the mean of the 10,000 simulations. The values obtained are reported in Table\,\ref{tab:median_nonblaz}. We find that the whole sample has a median cutoff energy of $E_{\rm C}=381\pm16$\,keV, and the value of AGN with $\log (N_{\rm H}/\rm cm^{-2})< 22$ ($384\pm21$\,keV) is consistent with that of objects with $\log (N_{\rm H}/\rm cm^{-2})= 22-24$ ($386\pm25$\,keV), and with the one obtained for CT AGN ($341\pm70$\,keV). Considering a maximum cutoff energy of 500\,keV (800\,keV) we obtain, for the total sample, a median of $E_{\rm C}=244\pm8$\,keV ($332\pm12$\,keV). These values are in good agreement with what was obtained by \cite{Ballantyne:2014dp} ($E_{\rm C}=270^{+170}_{-80}$\,keV) fitting the X-ray luminosity function of local AGN in four energy bands.

To further test the typical values of the cutoff energy, and avoid issues related to the choice of the maximum cutoff energy, we also calculated the mean and median using the Kaplan-Meier estimator and including the lower limits (see \S5 of \citealp{Shimizu:2016hc} for details). We found that, for the whole sample, the median (mean) is $200\pm 29$\,keV ($319\pm23$\,keV), while for unobscured AGN is $210\pm 36$\,keV ($331\pm 29$\,keV). For objects with $\log (N_{\rm H}/\rm cm^{-2})= 22-24$ the median (mean) is $188\pm 27$\,keV ($262\pm22$\,keV), and for CT AGN is $449\pm64$\,keV ($272\pm 51$\,keV).

\subsubsection{Reflection}\label{sect:refl}

The distribution of the reflection parameters ($R$) are shown in the right panel of Fig.\,\ref{fig:hist_EC_R}. We could constrain $R$ for 170 objects, while for 490 we obtain an upper limit and for 23 a lower limit. The median value of $R$ among the sources for which the parameter could be constrained is $R=1.3\pm0.1$. It should be noted that this value does not correspond to the {\it intrinsic} median of hard X-ray selected AGN, since it is typically easier to infer this parameter when its value is large. We found that 51 objects have values of $R> 2$, albeit most of them have large uncertainties (the median fractional error is $\sim 70\%$), and only 16 are consistent with $R\geq 2$ within their respective uncertainties. Most of the objects (35) with $R\geq 2$ are unobscured AGN. This might lend support to the idea that this enhanced fraction of reprocessed flux with respect to the primary X-ray emission is caused by relativistic reflection, as predicted by the light-bending scenario (e.g., \citealp{Miniutti:2004si}).

We took into account the upper and lower limits on $R$ following the same approach used for $E_{\rm C}$ ($\S$\ref{sect:cutoff}). In this case however we allowed the values to vary in the range $[L,10]$, i.e. we assumed a maximum value of $R=10$. We find that the median of the whole sample is $R=0.53\pm0.09$. AGN with $\log (N_{\rm H}/\rm cm^{-2})< 22$ typically have larger reflection parameters ($R=0.83\pm0.14$) than those with $\log (N_{\rm H}/\rm cm^{-2})\geq 22$ ($R=0.37\pm0.11$). We also find that CT AGN have significantly lower intensity of the reprocessed X-ray continuum relative to the primary X-ray luminosity ($R=0.15\pm0.12$). The decrease of the reflection component with increasing obscuration would be in agreement with the idea that most of the reprocessing in AGN is due to the accretion disk, so that objects observed pole-on are able to see more of the reprocessed radiation than those observed edge-on. Our results are in disagreement with what was found by \cite{Ricci:2011zr} stacking {\it INTEGRAL} IBIS/ISGRI spectra, and with the results of \citet{Vasudevan:2013ys} and \citet{Esposito:2016kq} obtained by stacking {\it Swift}/BAT spectra. All these works in fact showed that the stacked spectra of obscured AGN tend to have more reflection than those of their less obscured counterparts. Similar results were also obtained by stacking {\it XMM-Newton} data by \cite{Corral:2011le}. While the origin of this difference is still not clear, a possible explanation could be that the stacking of spectra with different column densities would artificially produce the curvature observed in the averaged spectrum.

\subsection{Absorption properties}\label{sect:Absorption}

\subsubsection{Neutral absorption}\label{sect:NeutralAbsorption}

\begin{figure}[t!]
\centering
 %% 1st image
 %% 2nd image
\begin{minipage}[!b]{.48\textwidth}
\centering
\includegraphics[width=8.8cm]{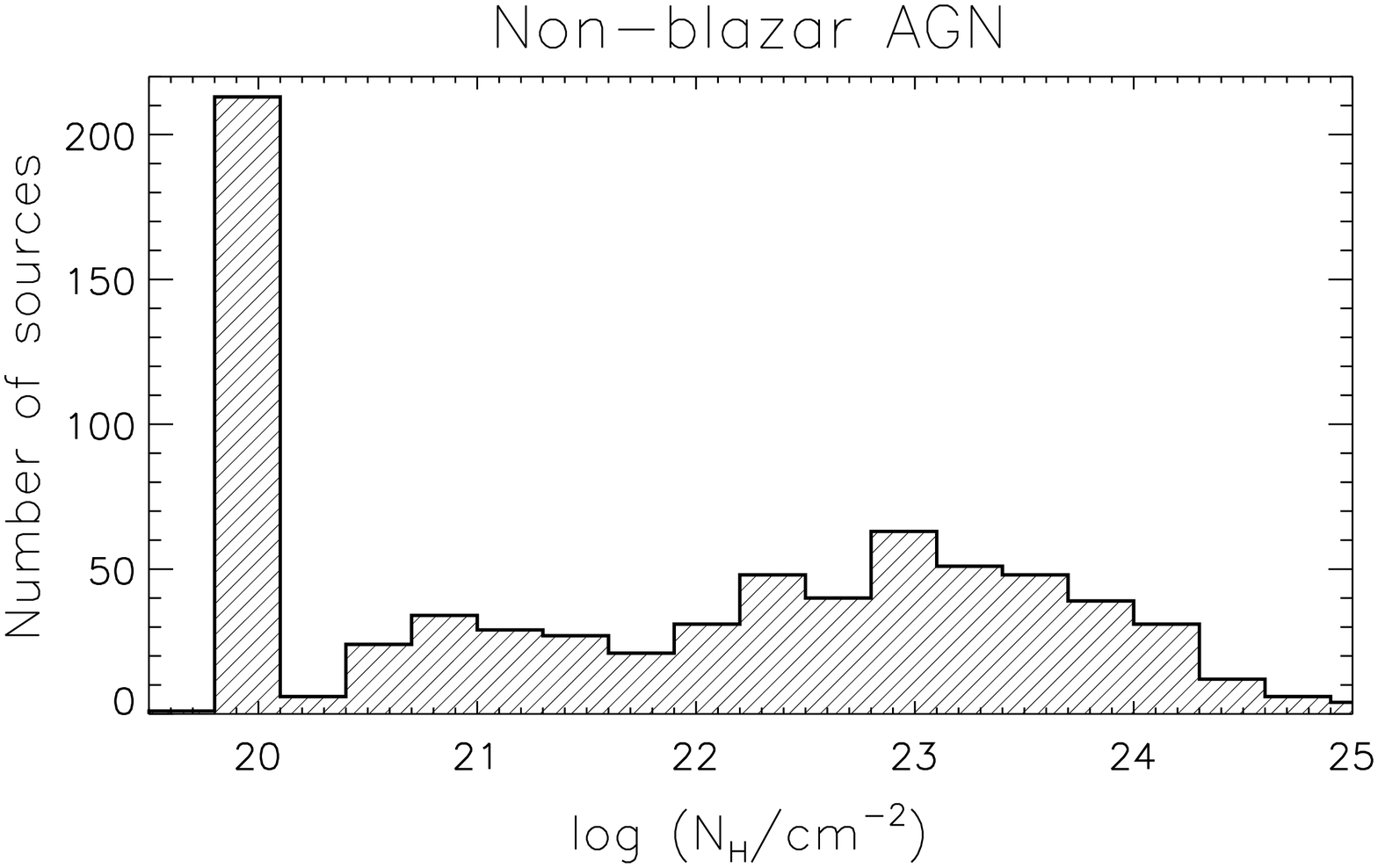}\end{minipage}
\par\smallskip
\begin{minipage}[!b]{.48\textwidth}
\centering
\includegraphics[width=8.8cm]{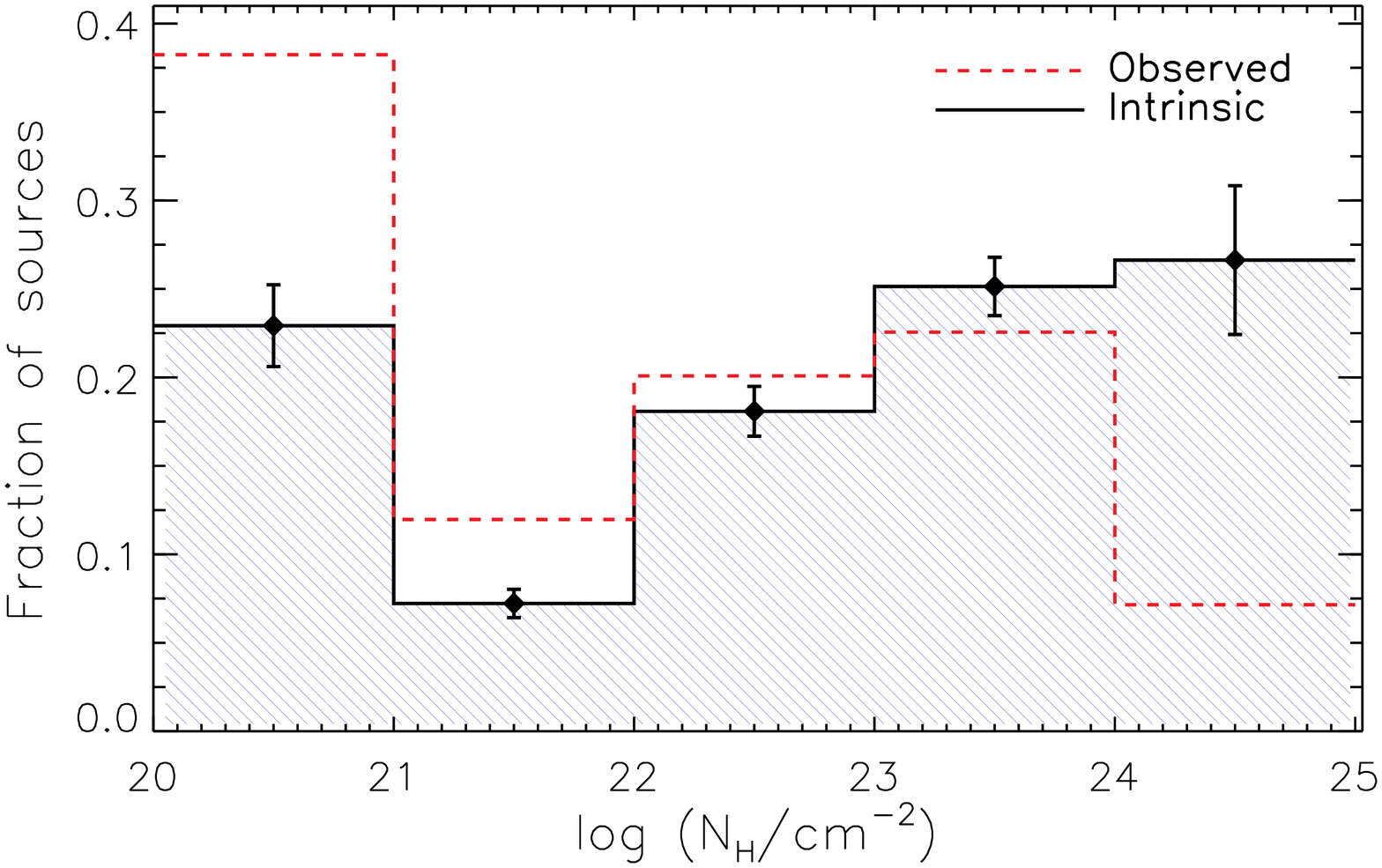}\end{minipage}
\par\smallskip
\begin{minipage}[!b]{.48\textwidth}
\centering
\par\medskip
\includegraphics[width=8.8cm]{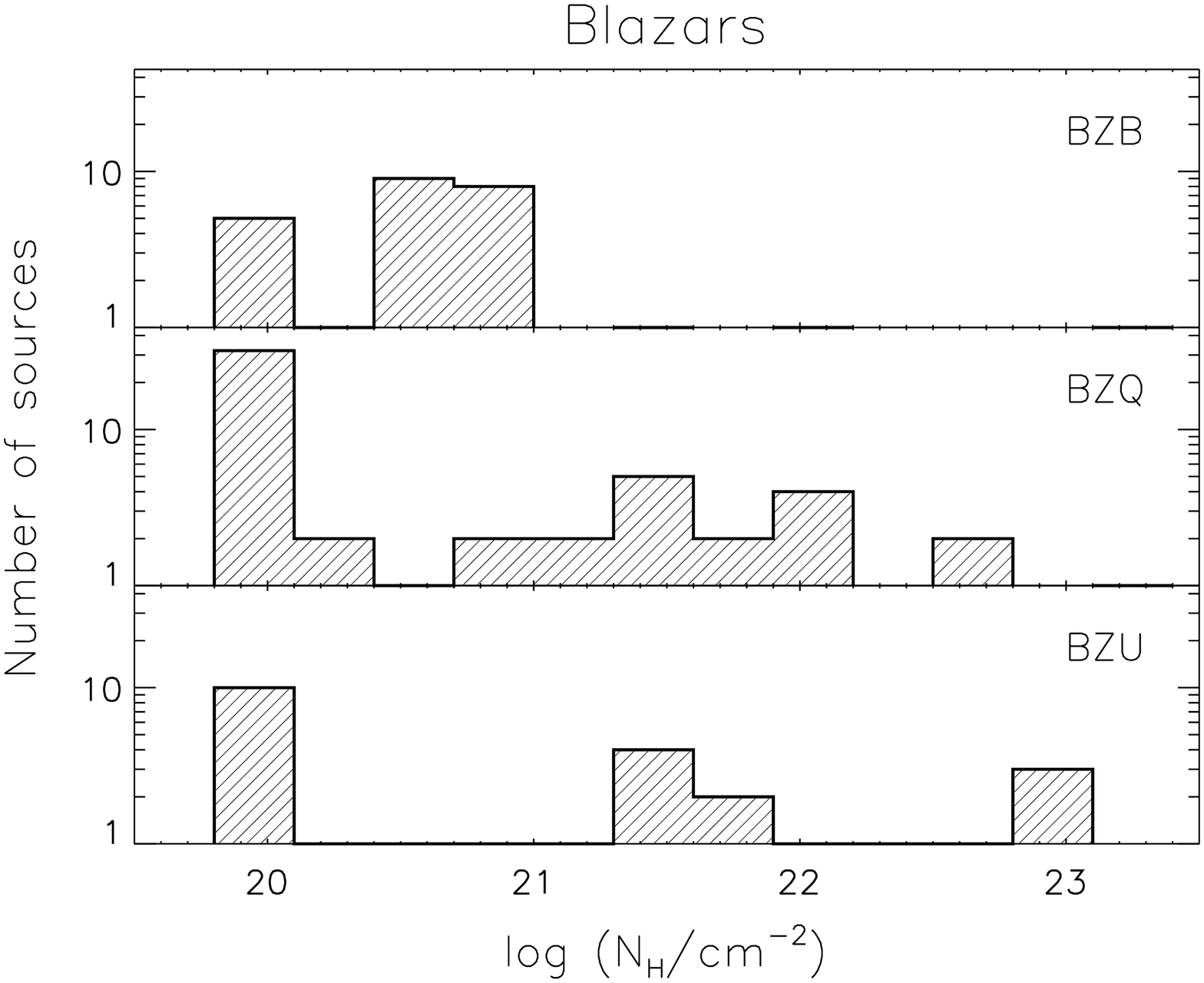}\end{minipage}
% %% caption
 \begin{minipage}[t]{.48\textwidth}
  \caption{{\it Top panel:} Distribution of the line-of-sight column density for the non-blazar AGN. Sources which do not show any sign of obscuration were arbitrarily assigned $\log (N_{\rm H}/\rm cm^{-2})= 20$ for visual clarity. {\it Central panel:} Observed (red dashed line) and intrinsic (black continuous line, from \citealp{Ricci:2015kx}) column density distribution of non-blazar AGN. {\it Bottom panel:} Same as the top panel for blazars. The sources were divided into BL Lacs (BZB), Flat Spectrum Radio Quasars (BZQ) and blazars of uncertain type (BZU). The median values of $N_{\rm H}$ are listed in Table\,\ref{tab:median_nonblaz} and \ref{tab:median_blaz} for non-blazar AGN and blazars, respectively.}
\label{fig:NHdistr}
 \end{minipage}
\end{figure}

\begin{figure*}[t!]
\centering
 %% 1st image
 %% 2nd image
\includegraphics[width=0.48\textwidth,angle=0]{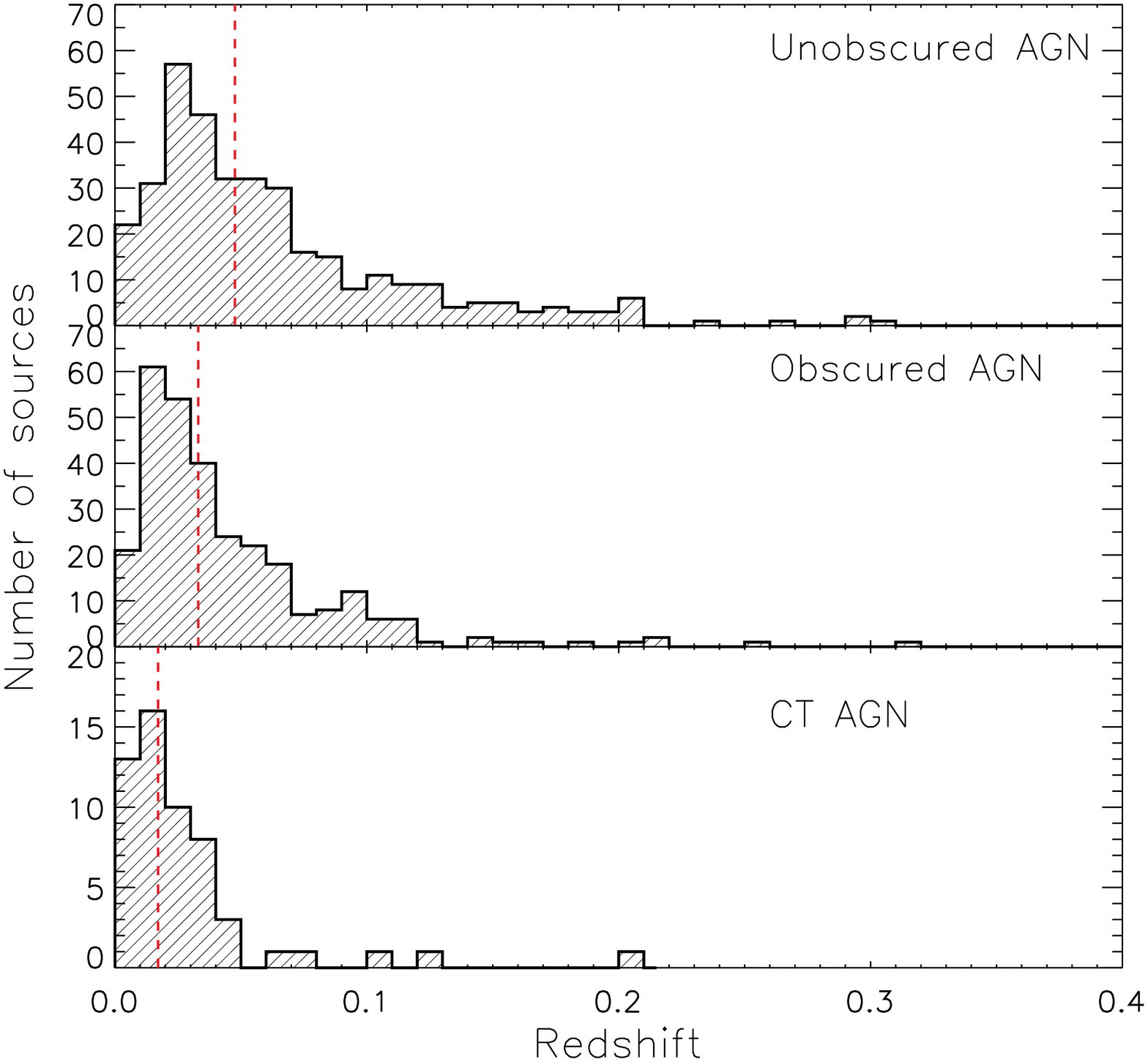}
\includegraphics[width=0.48\textwidth,angle=0]{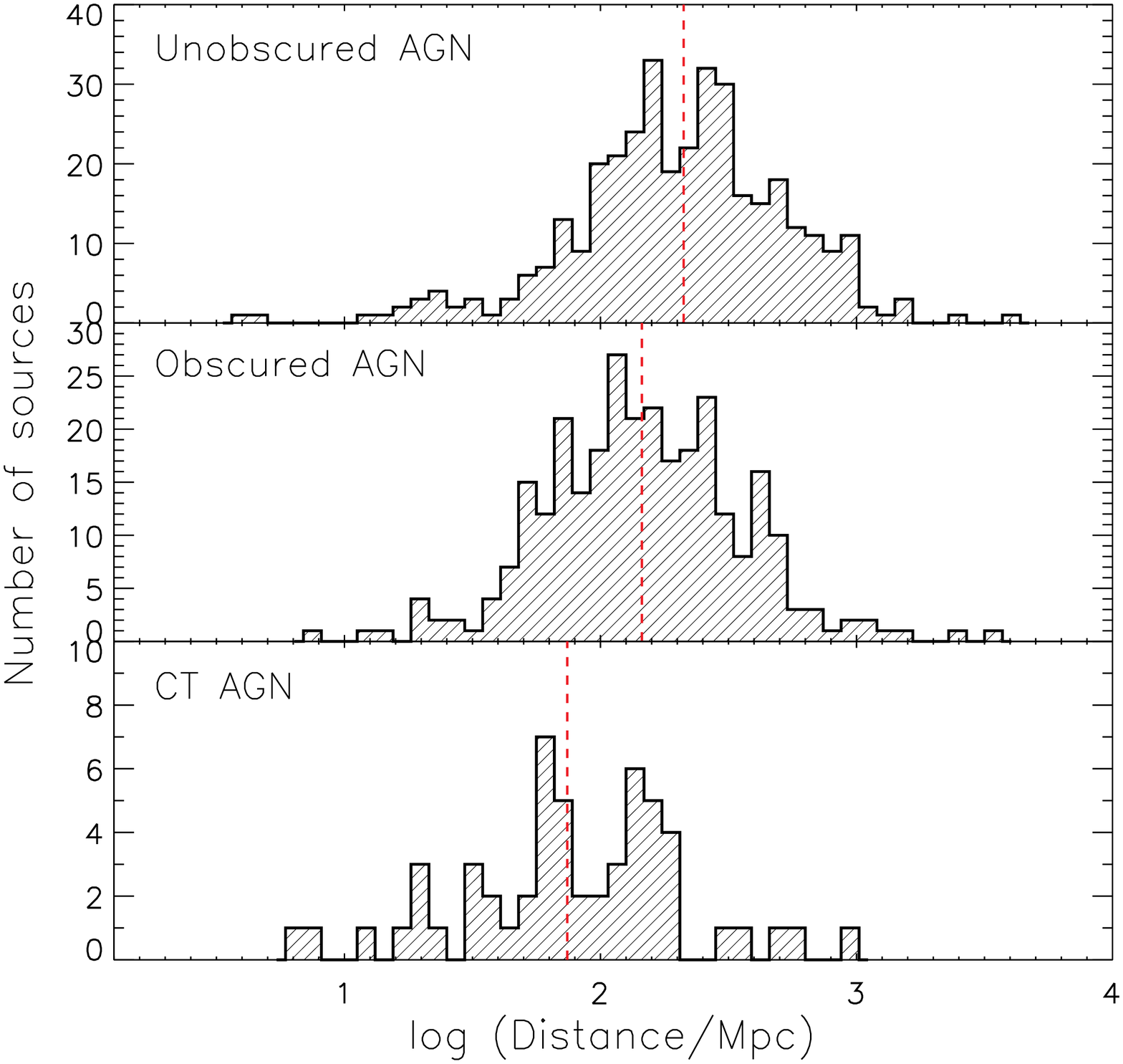}
% %% caption
% \begin{minipage}[t]{1\textwidth}
  \caption{Distribution of the spectroscopic redshifts (left panels) and distances (right panels) of non-blazar AGN divided according to their line-of-sight column density: unobscured (top panels; $N_{\rm H}<10^{22}\rm\,cm^{-2}$), obscured Compton-thin (middle panels; $10^{22}<N_{\rm H}<10^{24}\rm\,cm^{-2}$) and CT (bottom panels; $N_{\rm H}\geq10^{24}\rm\,cm^{-2}$). The red dashed lines show the median values of the redshift for the three subsamples. The plots are zoomed in the $z=0-0.4$ range, similarly to Fig.\,\ref{fig:hist_redshift}. The fact that obscured Compton-thin AGN (middle panel) have a lower median redshift and distance ($z=0.033$, $D=145.2$\,Mpc) than the unobscured AGN ($z=0.047$, $D=211.3$\,Mpc) is related to the difference in their luminosity distributions (see $\S$\ref{sect:Luminosities} for discussion). The lower redshift and distance of CT AGN ($z=0.017$, $D=74.2$\,Mpc) is instead due to the influence of obscuration, which allows to detect only the nearest objects of this class (see Fig.\,3 of \citealp{Ricci:2015kx}).}
\label{fig:redshifts_obsunobsct}
% \end{minipage}
\end{figure*}

A total of 366 non-blazar AGN in our sample have $\log (N_{\rm H}/\rm cm^{-2})\geq 22$, while 365 have $\log (N_{\rm H}/\rm cm^{-2})< 22$ (top and middle panels of Fig.\,\ref{fig:NHdistr}). Among the blazars, only 13 sources are obscured, while the remaining 92 objects are unobscured (bottom panel of Fig.\,\ref{fig:NHdistr}). The difference between the column density distributions of blazars and non-blazar AGN could be either related to: i) the very strong radiation field of the former; ii) the fact that in blazars a significant fraction of the X-ray emission is emitted by the jet, which implies that the region producing X-ray radiation is more extended and hence more difficult to significantly obscure in blazars than in non-blazar AGN; iii) the fact that blazars are observed pole-on and therefore it is less likely for the X-ray source to be obscured by the torus.

In Fig.\,\ref{fig:redshifts_obsunobsct} we show the redshift (left panel) and distance (right panel) distribution of non-blazar AGN, divided according to their line-of-sight column density into unobscured (top panel), obscured Compton-thin (middle panel) and CT (bottom panel). The fact that obscured Compton-thin AGN (middle panel) have a lower median redshift ($z=0.033$) than the unobscured AGN ($z=0.047$) is related to the difference in their luminosity distributions (see $\S$\ref{sect:Luminosities} for discussion), while the lower redshift of CT AGN ($z=0.017$) is instead due to the influence of obscuration, which allows to detect only the nearest objects of this class (see Fig.\,3 of \citealp{Ricci:2015kx}).

\begin{figure}[t!]
\centering
 %% 1st image
 %% 2nd image
%
%\centering
\includegraphics[width=9cm]{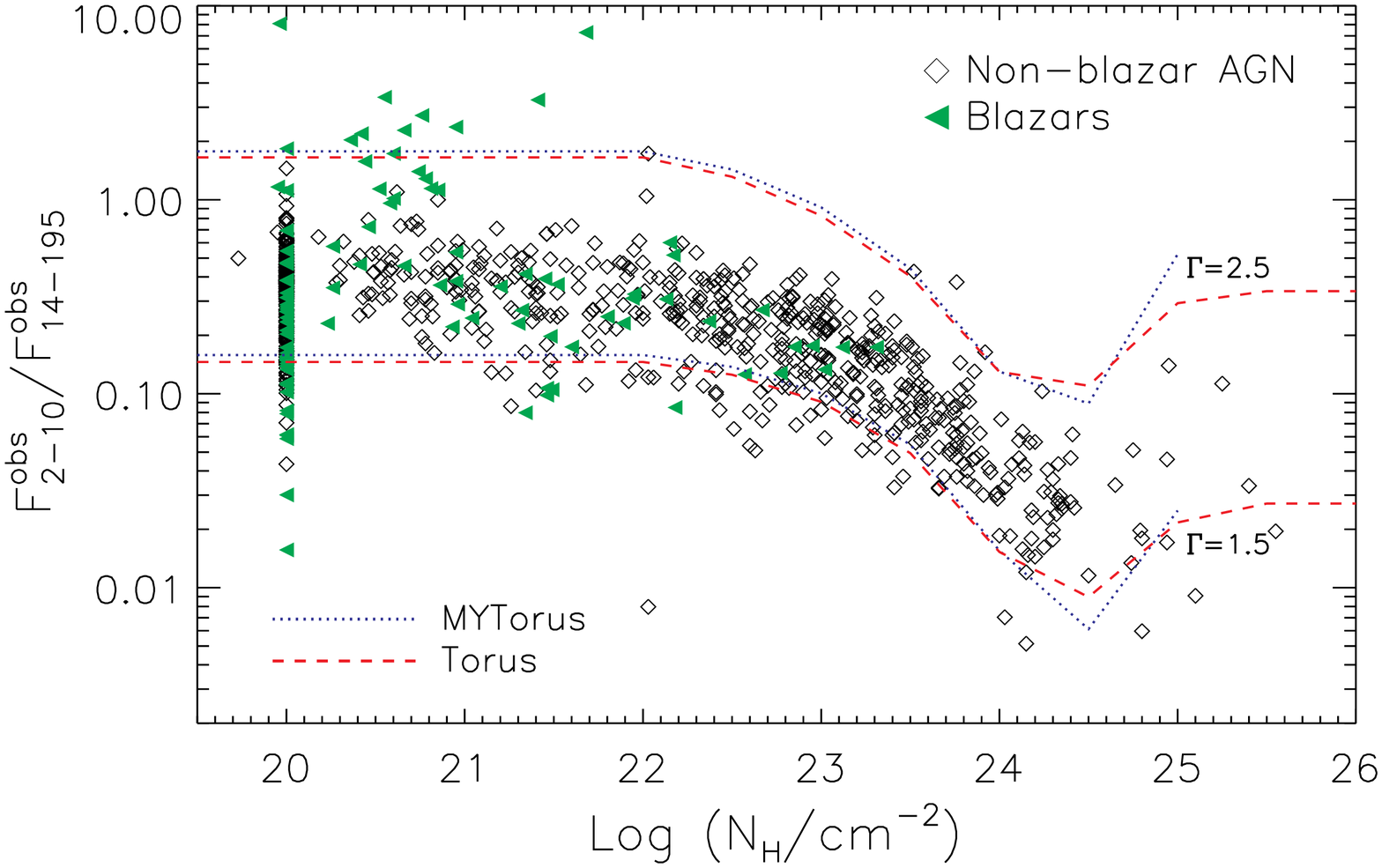}%\end{minipage}
% %% caption
% \begin{minipage}[t]{1\textwidth}
  \caption{Ratio of the observed 2--10\,keV and 14--195\,keV flux versus the column density inferred from the broad-band X-ray spectral analysis for the non-blazar AGN (black diamonds) and the blazars (green triangles). The plot shows the clear decrease of the $F_{\,2-10}^{\rm\,obs}/F_{\,14-150}^{\rm\,obs}$ flux ratio for $\log (N_{\rm H}/\rm cm^{-2})>22$ due to the stronger effect of absorption below 10\,keV. The black dotted and red dashed lines represent the expected flux ratios (for $\Gamma=1.5$ and $\Gamma=2.5$) obtained considering the \textsc{MYTorus} model and the torus model of \cite{Brightman:2011oq}, respectively.}
\label{fig:FluxratioVSnh}
% \end{minipage}
\end{figure}

In Fig.\,\ref{fig:FluxratioVSnh} we illustrate the ratio between the observed fluxes in the 2--10\,keV and 14--195\,keV bands ($F_{2-10}^{\rm\,obs}/F_{14-195}^{\rm\,obs}$) versus the column density. Due to the very different impact of absorption in these two bands, we expect that, as the line-of-sight column density increases, $F_{2-10}^{\rm\,obs}/F_{14-195}^{\rm\,obs}$ would decrease. For $N_{\rm H}\geq 10^{23.5}\rm\,cm^{-2}$ absorption plays a significant role also in the 14--195\,keV band, so that the trend is expected to flatten. The plot shows a clear decrease of $F_{2-10}^{\rm\,obs}/F_{14-195}^{\rm\,obs}$ for $\log (N_{\rm H}/\rm cm^{-2})\gtrsim 22$. In particular $\sim 90\%$ of the non-blazar AGN for which $F_{2-10}^{\rm\,obs}/F_{14-195}^{\rm\,obs}<0.1$ have $N_{\rm H}>10^{23}\rm\,cm^{-2}$. For flux ratios of $F_{2-10}^{\rm\,obs}/F_{14-195}^{\rm\,obs}<0.03$, $\sim 94\%$ of the AGN are CT. 
However, we identify a few exceptions to this general trend. Two AGN with $\log (N_{\rm H}/\rm cm^{-2})\geq 24$ have $F_{2-10}^{\rm\,obs}/F_{14-195}^{\rm\,obs}>0.11$, both of which are well known CT AGN: ESO\,138$-$G001 ($F_{2-10}^{\rm\,obs}/F_{14-195}^{\rm\,obs}=0.113$; e.g., \citealp{Piconcelli:2011zt}) and NGC\,1068 ($F_{2-10}^{\rm\,obs}/F_{14-195}^{\rm\,obs}=0.14$; e.g., \citealp{Bauer:2015oq}). These two objects are amongst the most obscured of our sample, and the 14--195\,keV flux is also strongly affected by absorption, which naturally leads to a higher value of $F_{2-10}^{\rm\,obs}/F_{14-195}^{\rm\,obs}$ with respect to the transmission-dominated CT AGN. In Fig.\,\ref{fig:FluxratioVSnh} we also illustrate that for most sources $F_{2-10}^{\rm\,obs}/F_{14-195}^{\rm\,obs}$ is consistent with the expected flux ratios. These theoretical values were obtained using two different spectral models: the torus model of \citeauthor{Brightman:2011oq} (\citeyear{Brightman:2011oq}, red dashed lines) and \textsc{MYTorus} (\citealp{Murphy:2009uq}, black dotted lines). The expected $F_{2-10}^{\rm\,obs}/F_{14-195}^{\rm\,obs}$ were calculated considering the maximum value of the inclination angle allowed by the models, a half-opening angle of $60^{\circ}$ for the torus model of \cite{Brightman:2011oq}, and two different values of the photon index ($\Gamma=1.5$ and $\Gamma=2.5$). We added to the models a scattered power-law component with $f_{\rm\,scatt}=1\%$, consistent with the typical value found for {\it Swift}/BAT AGN (see $\S$\ref{sect:SoftExcess_obs}).
The only Compton-thin object with a very low flux ratio ($F_{2-10}^{\rm\,obs}/F_{14-195}^{\rm\,obs}<0.01$) is UGC\,12243. This source requires a very large cross-calibration constant ($C_{\rm\,BAT}=11.8^{+22}_{-3.4}$) which could be due to extreme variability, with the source being in a very low-flux state at the time of the {\it XMM-Newton} observation.

\begin{figure}[t!]
\centering
 %% 1st image
 %% 2nd image
%
\begin{minipage}[!b]{.48\textwidth}
\centering
\includegraphics[width=9cm]{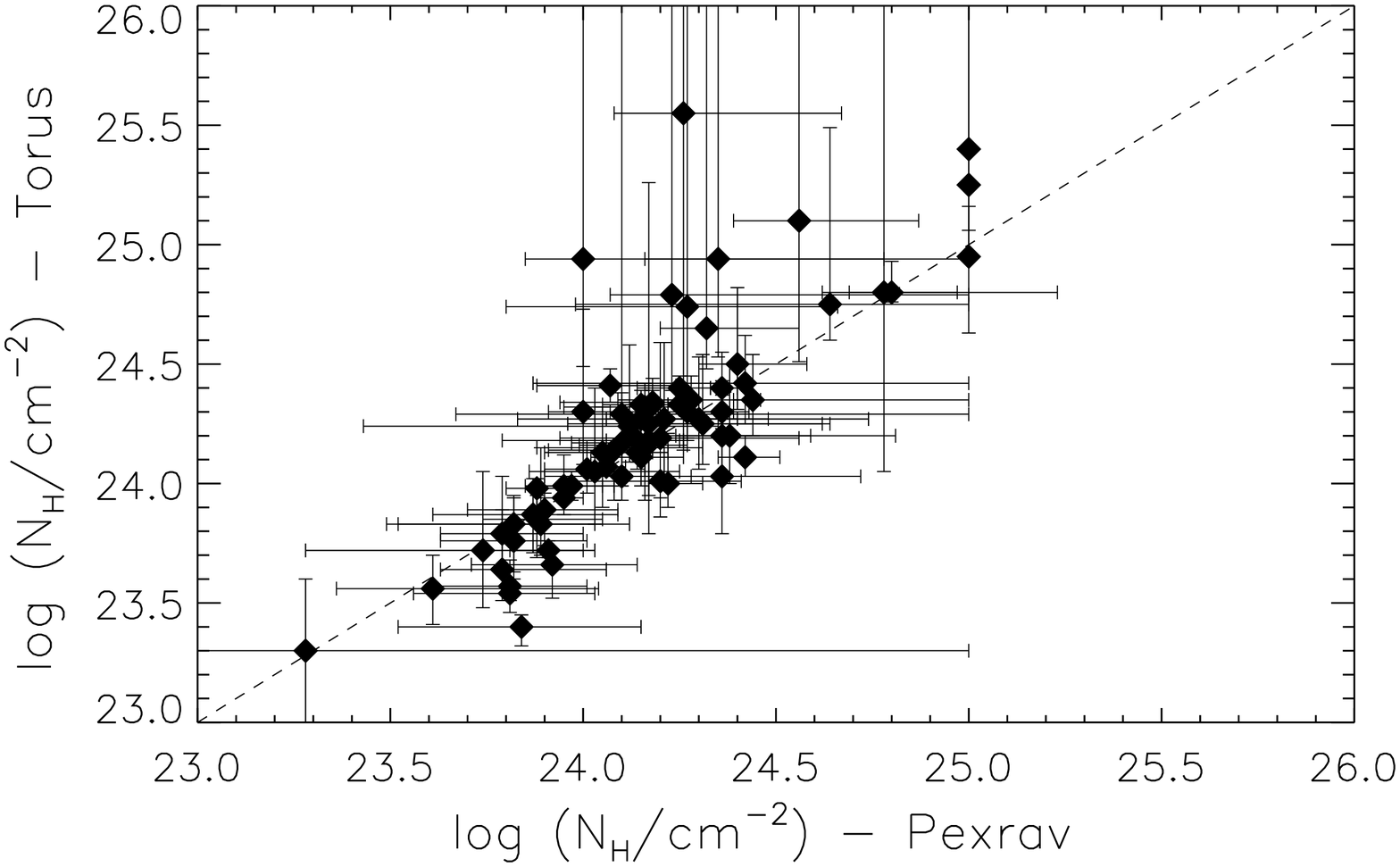}\end{minipage}
\par\smallskip
\begin{minipage}[!b]{.48\textwidth}
\centering
\includegraphics[width=9cm]{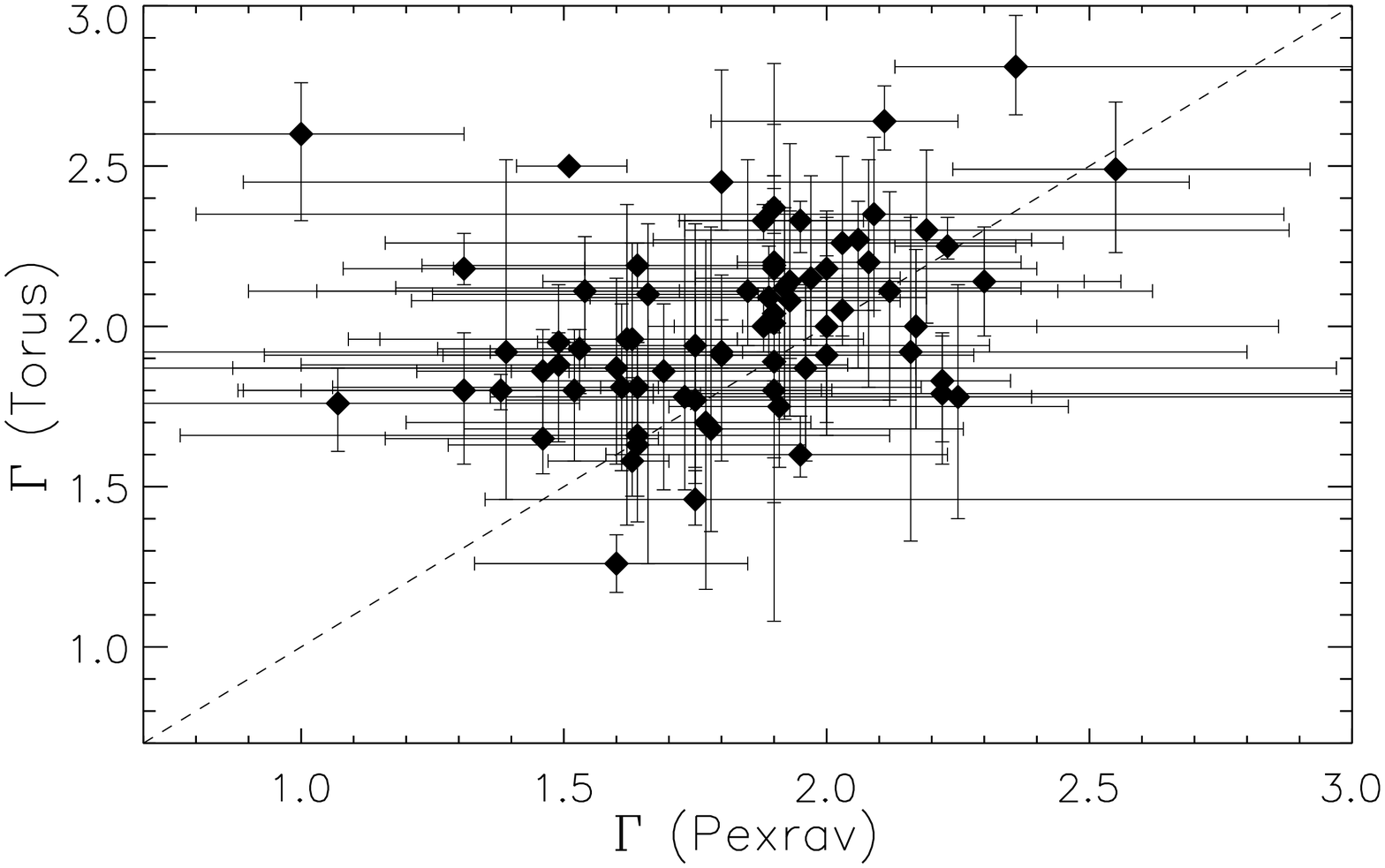}\end{minipage}
\par\smallskip
\begin{minipage}[!b]{.48\textwidth}
\centering
\includegraphics[width=9cm]{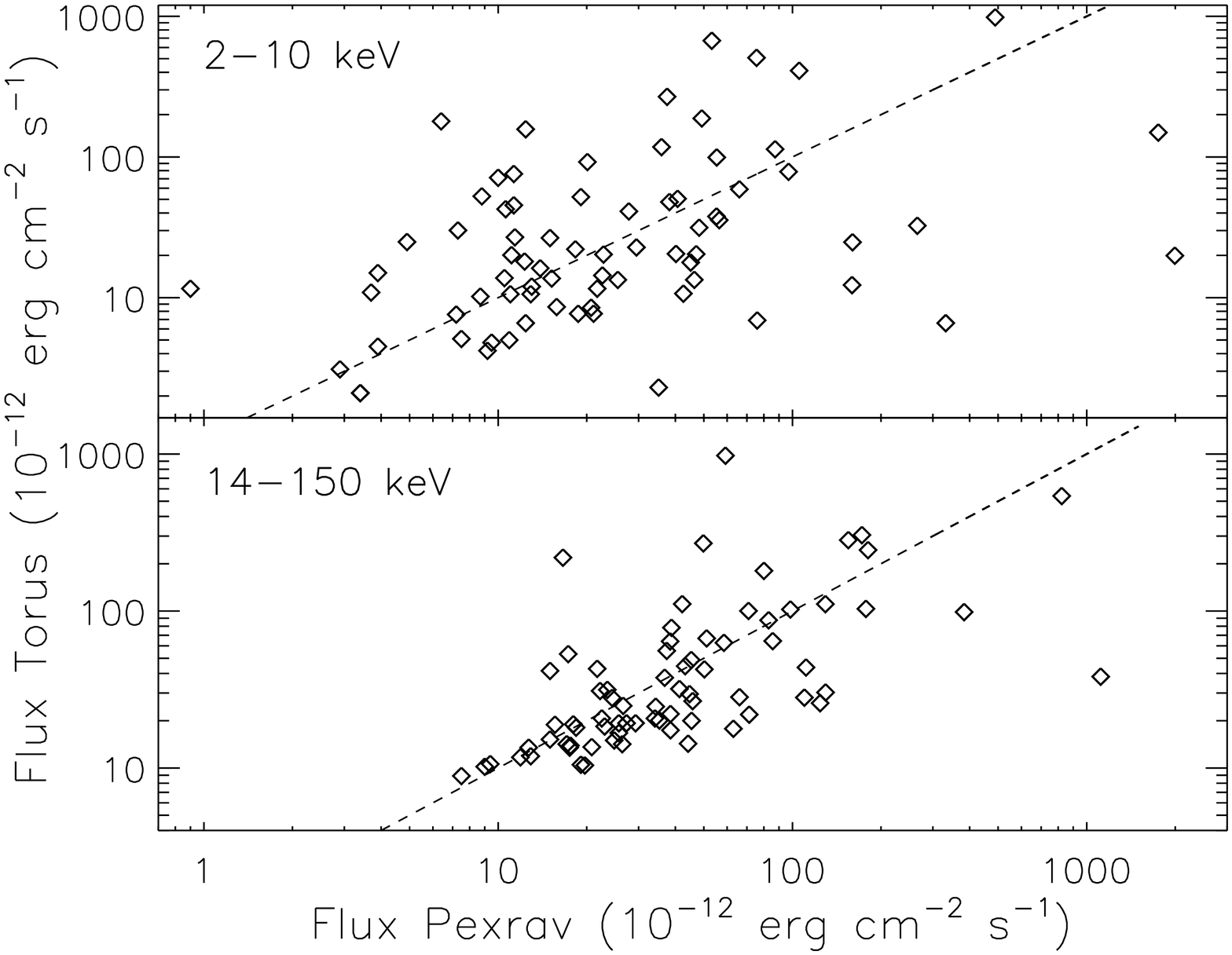}\end{minipage}
% %% caption
 \begin{minipage}[t]{.48\textwidth}
  \caption{{\it Top panel}: Scatter plot of the values of the column density obtained by fitting the 75 most obscured objects of our sample (i.e., all those with column densities consistent with $10^{24}\rm\,cm^{-2}$ within their 90\% confidence intervals) using \textsc{pexrav} and the torus model of \cite{Brightman:2011oq}. The dashed line shows the 1:1 relation between the two values of $N_{\rm H}$. {\it Central panel}: same as top panel for the photon index of the primary X-ray continuum ($\Gamma$). {\it Bottom panel}: same as top panel for the absorption-corrected flux in the 2--10\,keV and 14--150\,keV energy ranges. The plots show that \textsc{pexrav} tends to underestimate $N_{\rm H}$ [for $\log (N_{\rm H}/\rm cm^{-2})\gtrsim 24.3$] and $\Gamma$ with respect to torus models.}\label{fig:NHGpexVSNHGtor}\end{minipage}
\end{figure}

While for most sources we consider only a single layer of neutral obscuring material, the unabsorbed power-law component used in our analysis to reproduce the scattered emission allows to account also for partially covering obscuration of the X-ray source. Typically, the values of $f_{\rm\,scatt}$ of optically-selected AGN are of the order $\sim1-5$\% (e.g., \citealp{Bianchi:2007bq}), so that values considerably larger than this might imply the contribution of some leaked primary X-ray continuum. In our sample we find that a total of 22 (40) non-blazar AGN have $f_{\rm\,scatt} \geq 10\%$ ($\geq 5\%$). Alternative explanations for the significant contribution of an unobscured component at low energies include the presence of  strong star formation or a jet component dominating the X-ray emission below $\sim 2-3$\,keV. This has been found to be the case for radio galaxies \citep{Hardcastle:2006rq,Hardcastle:2009ca}, which very often show additional unobscured power-law emission. In agreement with this, we find that several of the objects with $f_{\rm\,scatt} \geq 10\%$ are radio loud, such as for example Cygnus A ($f_{\rm\,scatt}=13.3\%$) and 4C +21.55 ($30.6\%$). Three out of the six blazars for which a scattered component was added to the X-ray spectrum also show $f_{\rm\,scatt} \geq 5\%$. For three objects, SWIFT\,J0552.2$-$0727 (NGC\,2110)\footnote{For NGC\,2110 the two absorbers have column densities of $N_{\rm H}^{1}=17.9^{+1.6}_{-1.5}\times10^{22}\rm\,cm^{-2}$ and $N_{\rm H}^{2}=2.70^{+0.03}_{-0.03}\times10^{22}\rm\,cm^{-2}$, and cover $f_{\rm\,cov}^{1}=34.5^{+0.9}_{-0.9}\%$ and $f_{\rm\,cov}^{2}=95.3^{+0.3}_{-0.4}\%$ of the X-ray source, respectively. An additional, fully covering, absorber with a column density of $N_{\rm H}=1.0^{+0.1}_{-0.1}\times10^{21}\rm\,cm^{-2}$ is required to well reproduce the data.}, SWIFT\,J2124.6$+$5057 (4C\,50.55)\footnote{For 4C\,50.55 the absorbers have column densities of $N_{\rm H}^{1}=2.2^{+0.1}_{-0.2}\times10^{22}\rm\,cm^{-2}$ and $N_{\rm H}^{2}=20^{+3}_{-4}\times10^{22}\rm\,cm^{-2}$, covering $f_{\rm\,cov}^{1}=92^{+1}_{-1}\%$ and $f_{\rm\,cov}^{2}=43^{+2}_{-6}\%$ of the X-ray source, respectively }, and SWIFT\,J2223.9$-$0207 (3C\,445)\footnote{For 3C\,445 the two absorbers have column densities of $N_{\rm H}^{1}=34^{+7}_{-6}\times10^{22}\rm\,cm^{-2}$ and $N_{\rm H}^{2}=6.7^{+1.4}_{-1.2}\times10^{22}\rm\,cm^{-2}$, covering $f_{\rm\,cov}^{1}=83^{+4}_{-5}\%$ and $f_{\rm\,cov}^{2}=98.3^{+0.5}_{-0.4}\%$ of the X-ray source, respectively.}, we find that two layers of partially covering neutral material are needed to reproduce the X-ray spectrum. For these sources the values of the column density reported in Table\,\ref{tbl-3} are the sum of the different components multiplied by the covering factor (as noted in $\S$\ref{sect:spectralModels}).

\begin{figure}[t!]
\centering
%\centering
\includegraphics[width=9cm]{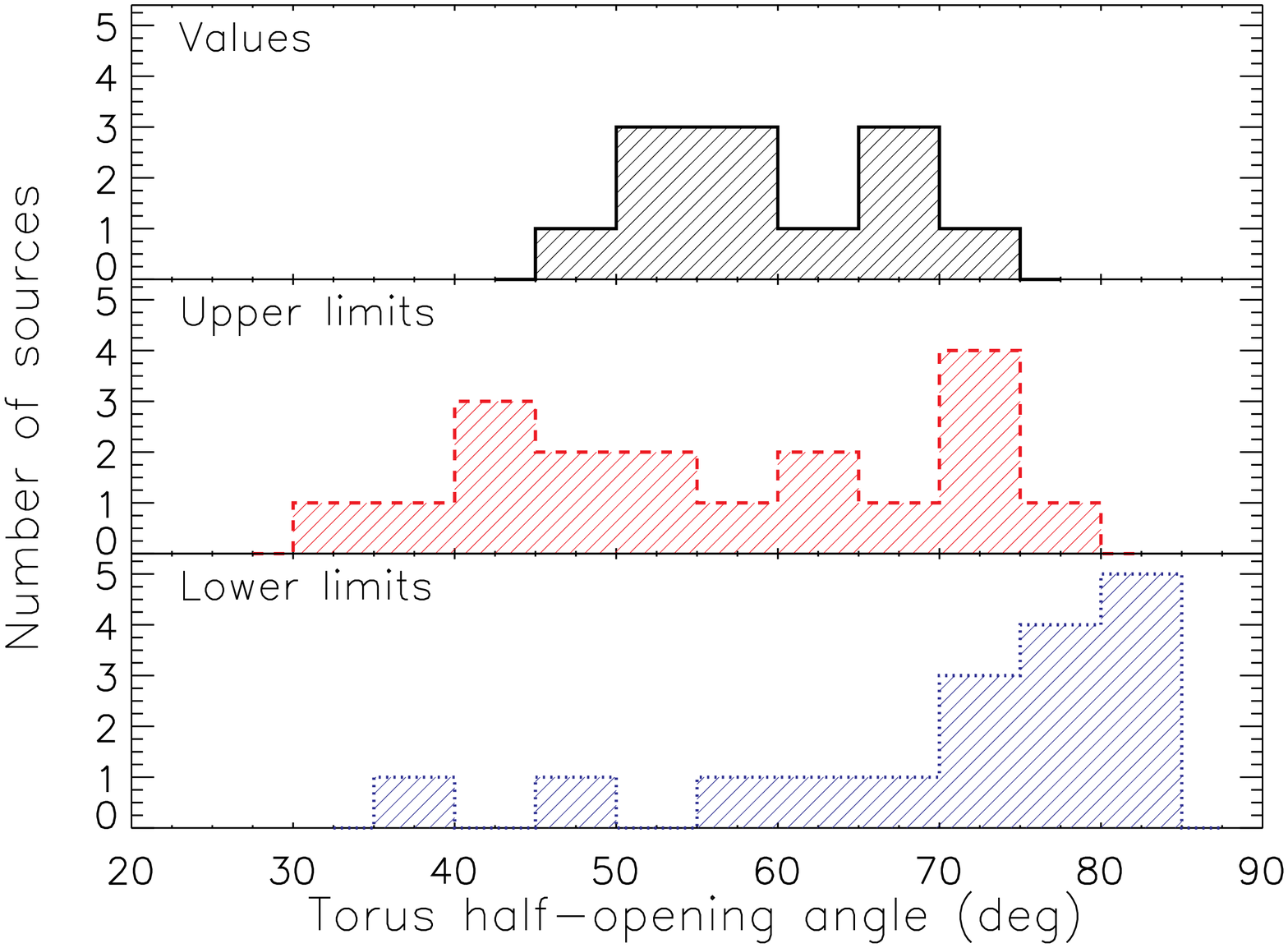}%\end{minipage}
% %% caption
  \caption{Distribution of the half-opening angle of the torus obtained by fitting with the torus model of \cite{Brightman:2011oq} the broad-band X-ray spectra of the objects with column densities consistent with $\log (N_{\rm H}/\rm cm^{-2})=24$ within their 90\% uncertainties. The plot shows the values (top panels), upper limits (middle panel) and lower limits (bottom panel). }
\label{fig:thetaOAdistr}
% \end{minipage}
\end{figure}

To better constrain the column density, the 75 objects with values of $N_{\rm H}$ consistent with $10^{24}\rm\,cm^{-2}$ within their 90\% confidence interval were fitted with the torus model of \cite{Brightman:2011oq} (see Table\,\ref{tbl-torpar} for the parameters obtained by the spectral fitting). With this approach we found that 55 {\it Swift}/BAT AGN are CT \citep{Ricci:2015kx}, of these 26 were identified as CT candidates for the first time. A similar study was recently carried out by \cite{Akylas:2016si}, who found 53 CT AGN, confirming the CT nature of most of our candidates. The two objects reported by \cite{Akylas:2016si} as CT but not listed in \cite{Ricci:2015kx} are NGC\,4941 and NGC\,3081. Both of these sources have column densities that are either consistent with being CT, or are heavily obscured. From our analysis we find that NGC\,4941 has a line-of-sight column density of $\log (N_{\rm H}/\rm cm^{-2})=23.91$ and the 90\% confidence interval is  $\log (N_{\rm H}/\rm cm^{-2})=23.81-24.00$. NGC\,3081 is also found to be heavily obscured [$\log (N_{\rm H}/\rm cm^{-2})=23.91$], with a 90\% confidence interval of $\log (N_{\rm H}/\rm cm^{-2})=23.87-23.95$. The fact that these two AGN are heavily obscured is supported also by the very large EW of their Fe\,K$\alpha$ features: $340^{+87}_{-17}\rm\,eV$ and $304^{+33}_{-18}\rm\,eV$ for NGC\,4941 and NGC\,3081, respectively. In \cite{Ricci:2015kx} we did not report SWIFT\,J0025.8$+$6818 (2MASX\,J00253292+6821442) as CT, since the analysis of the combined {\it Swift}/BAT and the XRT spectra resulted in a column density consistent with CT within the 90\% confidence interval, but with the best-fit value below the threshold ($7.8^{+5.4}_{-4.5}\times 10^{23}\rm\,cm^{-2}$ and $6.8^{+7.4}_{-1.7}\times 10^{23}\rm\,cm^{-2}$ for the phenomenological and torus model, respectively). However, the analysis of the combined {\it Chandra} and {\it Swift}/BAT spectra, carried out with the torus model, confirms that this source is CT, with a column density of $1.4^{+0.7}_{-0.9}\times10^{24}\rm\,cm^{-2}$. The large EW of the Fe\,K$\alpha$ ($836_{-778}^{+595}$\,eV) also strongly supports the idea that this source is CT.

\begin{figure}[t!]
\centering
\begin{minipage}[!b]{.48\textwidth}
\centering
\includegraphics[width=9cm]{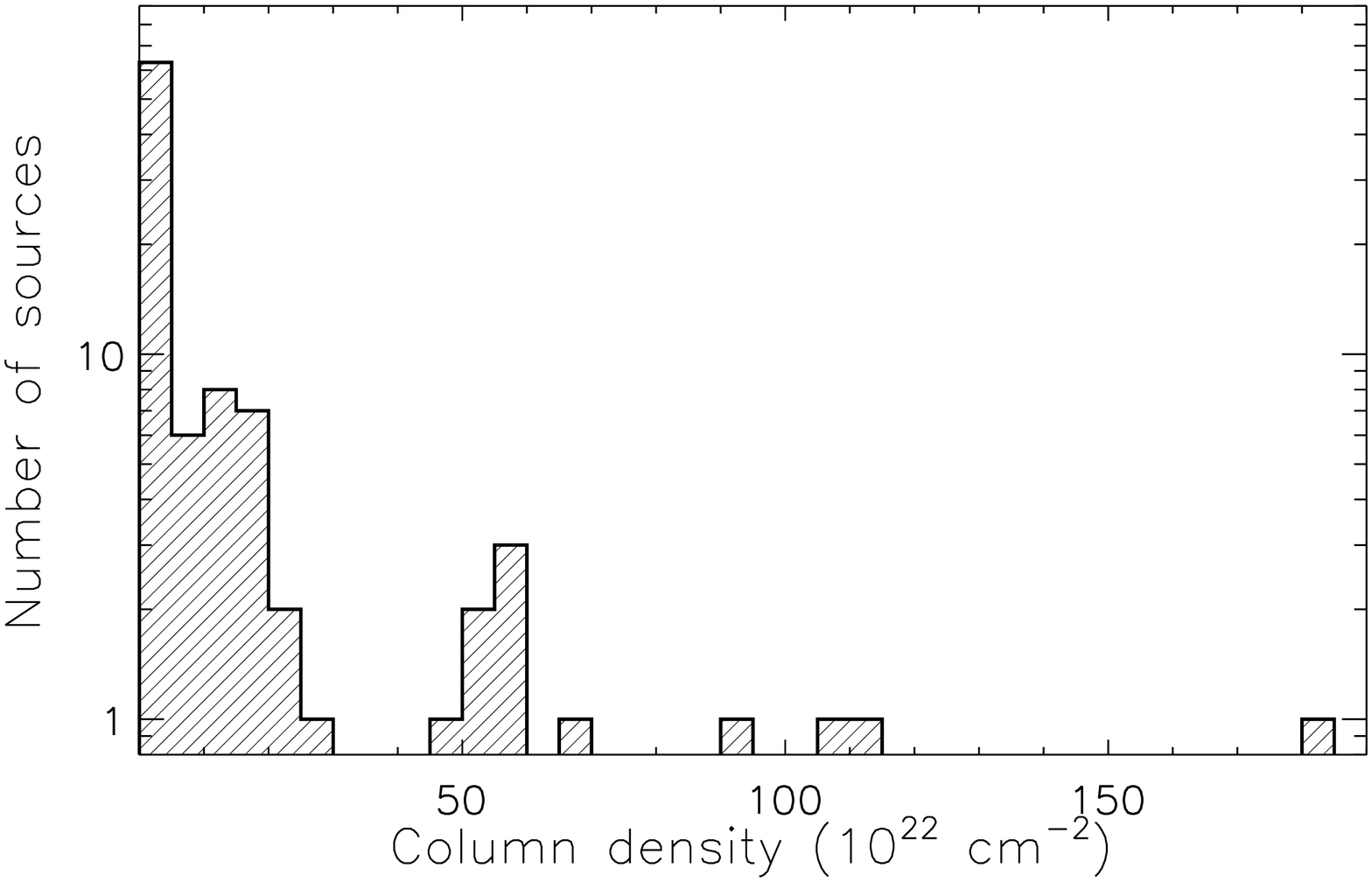}\end{minipage}
\par\medskip
\begin{minipage}[!b]{.48\textwidth}
\centering
\includegraphics[width=9cm]{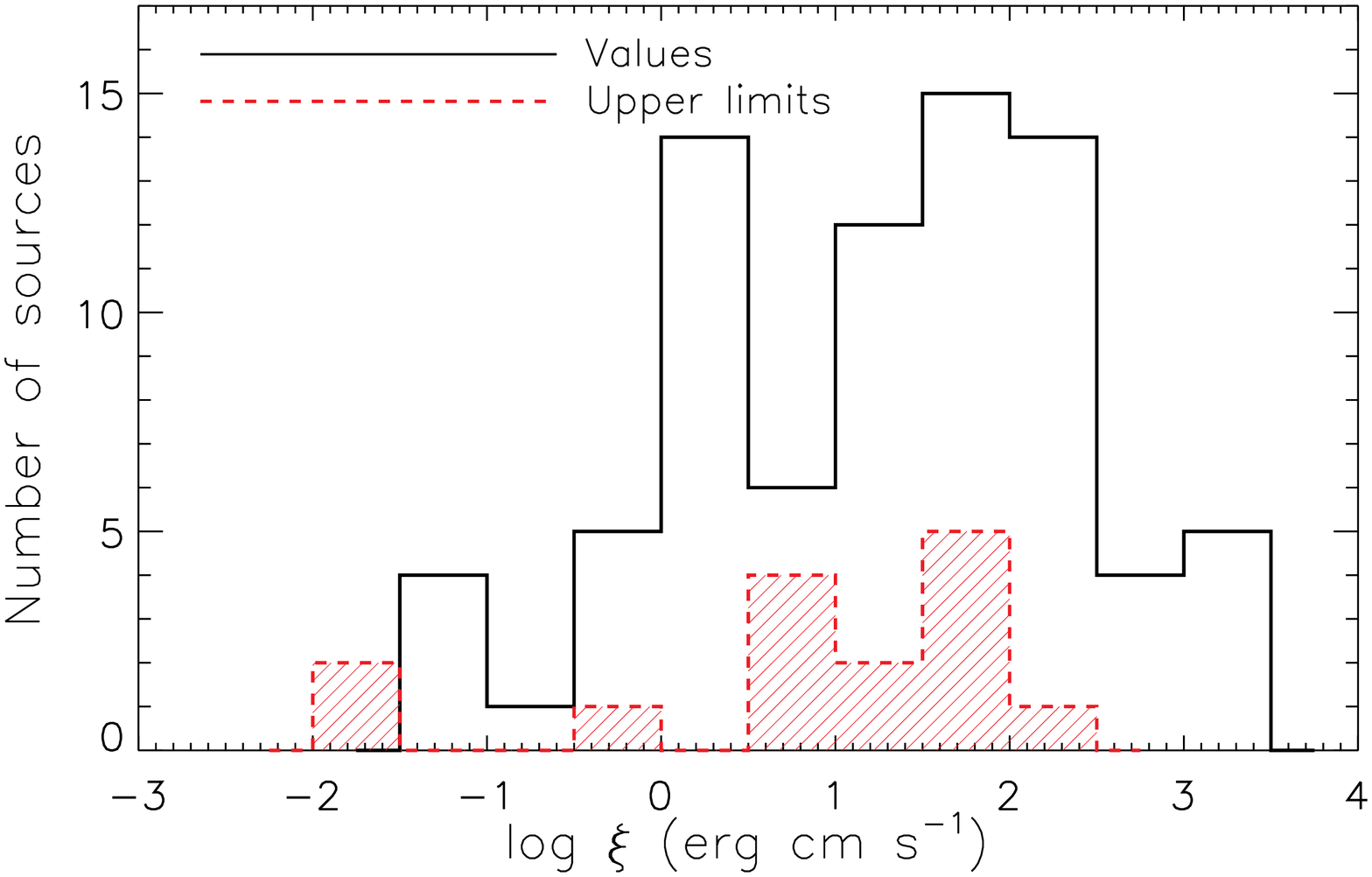}\end{minipage}
\par\medskip
\begin{minipage}[!b]{.48\textwidth}
\centering
\includegraphics[width=9cm]{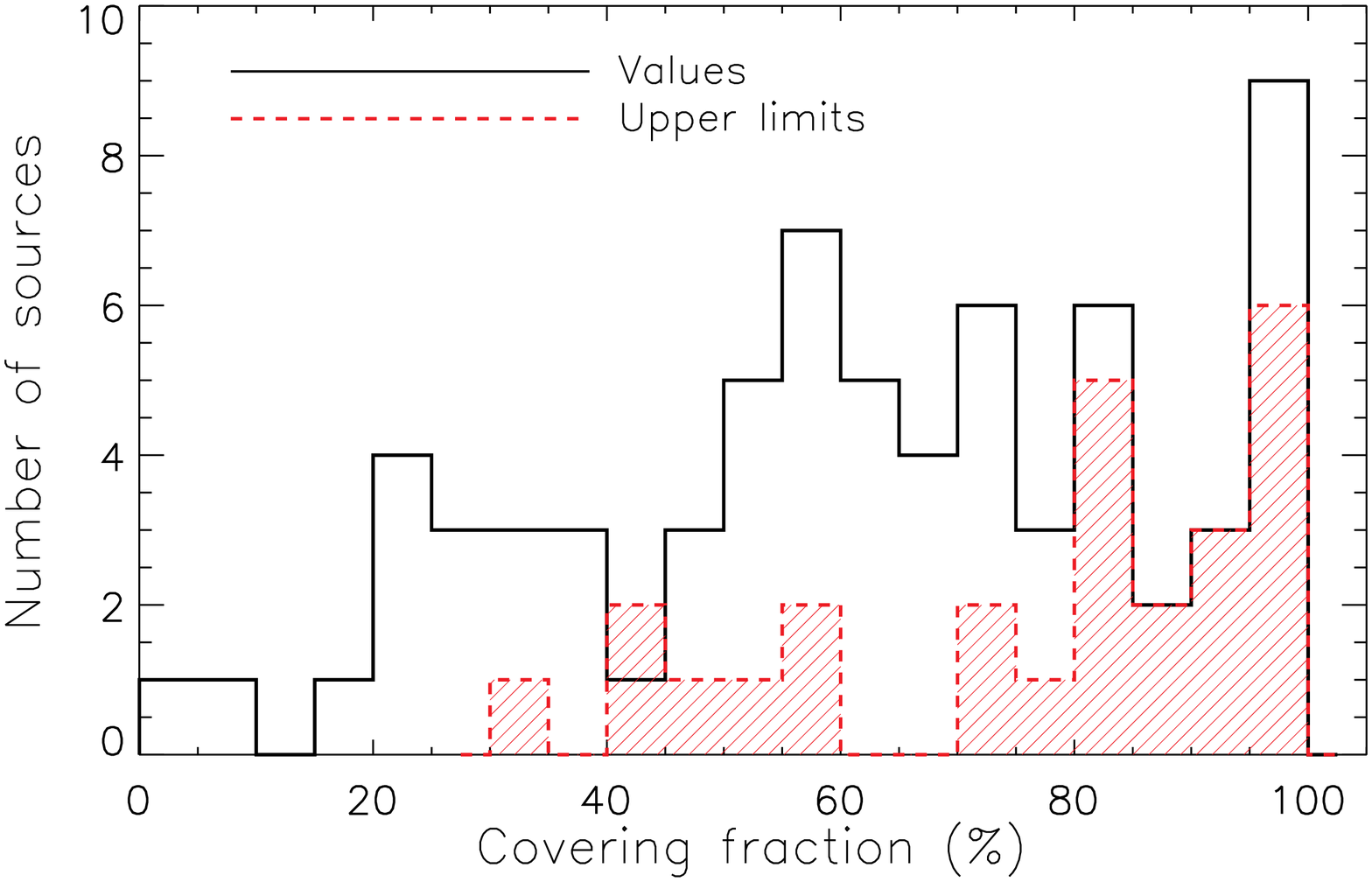}\end{minipage}
% %% caption
 \begin{minipage}[t]{0.48\textwidth}
  \caption{Distribution of the column density (top panel), ionization parameter (middle panel) and covering factor (bottom panel) of the ionized absorbers for the sources of our sample (see $\S$\ref{sect:IonizedAbsorption} for details). For the ionization parameter and the covering factor we also show the values of the upper limits (red dashed lines) obtained. The median values of the parameters are listed in Table\,\ref{tab:median_nonblaz}.}
\label{fig:warmabs}
 \end{minipage}
\end{figure}

In the top panel of Fig.\,\ref{fig:NHGpexVSNHGtor} we compare the column density obtained using the \textsc{pexrav} model versus that inferred from the torus model. There is a good agreement between the two values of $N_{\rm H}$ up to $\log (N_{\rm H}/\rm cm^{-2})\simeq 24.3$. Around $\log (N_{\rm H}/\rm cm^{-2})\simeq 24.5$ the column densities obtained with the torus model become typically larger (5 out of 7 objects). Performing a linear fit of the form $\log N_{\rm H}({\rm Torus})= \alpha \times \log N_{\rm H}({\rm Pexrav}) + \beta$ we found a slope of $\alpha=1.17\pm0.10$. The fact that $\alpha >1$ is due to the difficulty of constraining column densities with \textsc{pexrav} for $\log (N_{\rm H}/\rm cm^{-2}) \gtrsim 24.5$, since for these levels of obscuration most of the primary X-ray emission is depleted by absorption and the source is reflection dominated. Using torus models, on the other hand, it is possible to employ the shape of the reprocessed emission, as well as that of the absorbed primary X-ray emission, to infer the column density. In the central panel of Fig.\,\ref{fig:NHGpexVSNHGtor} we show the values of $\Gamma$ obtained by the two models. The plot shows that the slopes inferred by the torus model tend to be steeper than those obtained with the phenomenological model (i.e., using \textsc{pexrav}). In particular $\sim 72\%$ of the sources have $\Gamma\left({\rm Torus}\right) > \Gamma\left({\rm Pexrav}\right)$. This shows the importance of self-consistently taking into account absorbed and reprocessed X-ray emission in the most obscured AGN. The absorption-corrected fluxes obtained with \textsc{pexrav} and the torus model in the 2--10\,keV and 14--150\,keV energy ranges are illustrated in the bottom panel of Fig.\,\ref{fig:NHGpexVSNHGtor}. The plot shows that the dispersion between the fluxes obtained with the two models is lower in the 14--150\,keV than in the 2--10\,keV band. This is a straightforward consequence of the fact that the corrections are smaller in the 14--150\,keV band.

The distribution of the half-opening angle of the torus obtained by our analysis is illustrated in Fig.\,\ref{fig:thetaOAdistr}. For most of the objects (28) we could not constrain $\theta_{\rm\,OA}$, while for 18 (17) we are only able to infer an upper (lower) limit. For the 12 objects for which we could constrain $\theta_{\rm\,OA}$ we find a median value of $58\pm3$\,degrees.
The detailed study of the obscuration properties of {\it Swift}/BAT AGN is reported in \citep{Ricci:2017fq}, where we discuss how absorption is related to the physical characteristics of the accreting SMBH.

\subsubsection{Ionized absorption}\label{sect:IonizedAbsorption}

Ionized absorption has been found to be a common characteristic of unobscured AGN. Early studies carried out with {\it ASCA} found evidence of O\,VII and O\,VIII absorption edges in $\sim 50\%$ of the sources (e.g., \citealp{Reynolds:1997bs,George:1998hp}). More recent studies, carried out using {\it Suzaku}, {\it XMM-Newton}, and {\it Chandra}, have confirmed the presence of these warm absorbers, showing that they are related to outflows (e.g., \citealp{Kaspi:2000sw,Kaastra:2000lo}) with velocities of the order of 100--1,000\,$\rm\,km\,s^{-1}$ (e.g., \citealp{Tombesi:2013yo} and references therein). The analysis of high-quality {\it Suzaku} and {\it XMM-Newton} spectra has additionally pointed out that some objects also show highly ionized outflows with velocities exceeding 10,000\,$\rm\,km\,s^{-1}$, the so-called ultra-fast outflows (e.g., \citealp{Tombesi:2010df,Tombesi:2010zh,Gofford:2013jx}).

Evidence of ionized absorption was found in 86 AGN in our sample. Of these, five are blazars, one is an obscured AGN (NGC\,1365), and the remaining 80 are unobscured AGN.  This implies that, on average, the covering factor of the warm absorbers in non-blazar unobscured AGN is $\gtrsim 22\%$. This can be used only as a lower limit, since we might be missing ionized absorbers in objects too faint for high-quality X-ray spectra. Eleven non-blazar AGN and two blazars require two ionized absorbers to well reproduce the 0.3--10\,keV spectra. 

\begin{figure}[t!]
\centering
 %% 1st image
 %% 2nd image
%\begin{minipage}[!b]{.48\textwidth}
\centering
\includegraphics[width=9cm]{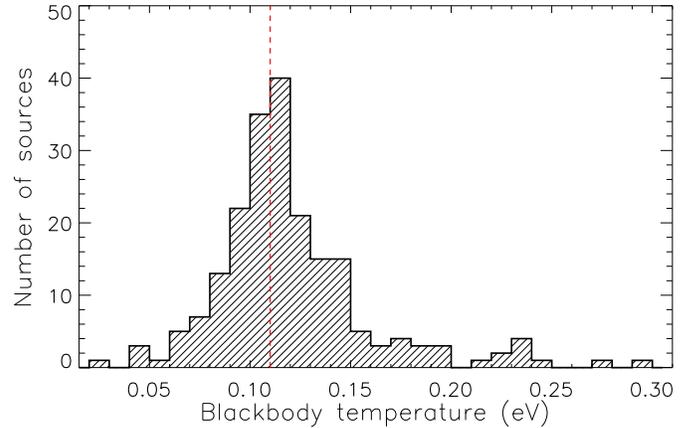}%\end{minipage}
% %% caption
% \begin{minipage}[t]{1\textwidth}
  \caption{Distribution of the blackbody temperatures obtained for unobscured non-blazar AGN. The red dashed vertical line shows the median value (see Table\,\ref{tab:median_nonblaz}).}
\label{fig:BBhist}
% \end{minipage}
\end{figure}

In Fig.\,\ref{fig:warmabs} we show the distribution of the column density (top panel), the ionization parameter (middle panel) and the covering factor (lower panel) of the warm absorbers. We find that the median column density is $N_{\rm H}^{\rm W}=2.8\times 10^{22}\rm\,cm^{-2}$. The median value of the ionization parameter (covering factor), for the objects for which we could constrain the values, is $\log [\xi/(\rm\,erg\,cm\,s^{-1})]=1.45$ ($f_{\rm\,cov}^{\rm\,W}=63\%$). These values are in good agreement with what was found by previous studies of warm absorbers (e.g., \citealp{Tombesi:2013yo} and references therein). A more detailed study of the properties of ionized absorption, and on the relation between the warm absorbers and the physical properties of the accreting SMBH will be discussed in a forthcoming paper.

\subsection{Soft excess}\label{sect:SoftExcess}

\subsubsection{Unobscured AGN}\label{sect:SoftExcess_unobs}

A total of 209 unobscured AGN (i.e., $\sim57\%$) in our sample show evidence for soft excesses. This is in agreement with what has been found by \citeauthor{Winter:2008fk} (\citeyear{Winter:2008fk,Winter:2009qf}; $\sim 40-50\%$) studying X-ray observations of smaller samples of {\it Swift}/BAT AGN, and with previous works carried out with {\it ASCA} ($\sim 40\%$, \citealp{Reeves:2000ya}). The presence of a soft excess had been found to be ubiquitous in optically-selected quasars (e.g., \citealp{Piconcelli:2005bj}). \cite{Scott:2012bc} estimated that, by correcting the results obtained for a large sample of type-I AGN \citep{Scott:2011le} to take into account detectability, the true percentage of sources with a soft excess is $75\pm23\%$. Similarly, studying a sample of 48 Seyfert 1-1.5 galaxies observed by {\it Suzaku} and {\it XMM-Newton}, \cite{Winter:2012xi} found that 94\% of the objects show a soft excess.

The soft excess was reproduced here using a simple phenomenological model, i.e. a blackbody with a variable temperature and normalization. We found that the median temperature of our sample is $kT_{\rm\,bb}=0.110\pm0.003$\,eV. As shown in Fig.\,\ref{fig:BBhist}, the distribution of $kT_{\rm\,bb}$ is very narrow ($\sigma=0.04\rm\,eV$), which is also in agreement with previous works (e.g., \citealp{Gierlinski:2004by}, \citealp{Winter:2012xi}).

\begin{figure}[t!]
\centering
\includegraphics[width=9cm]{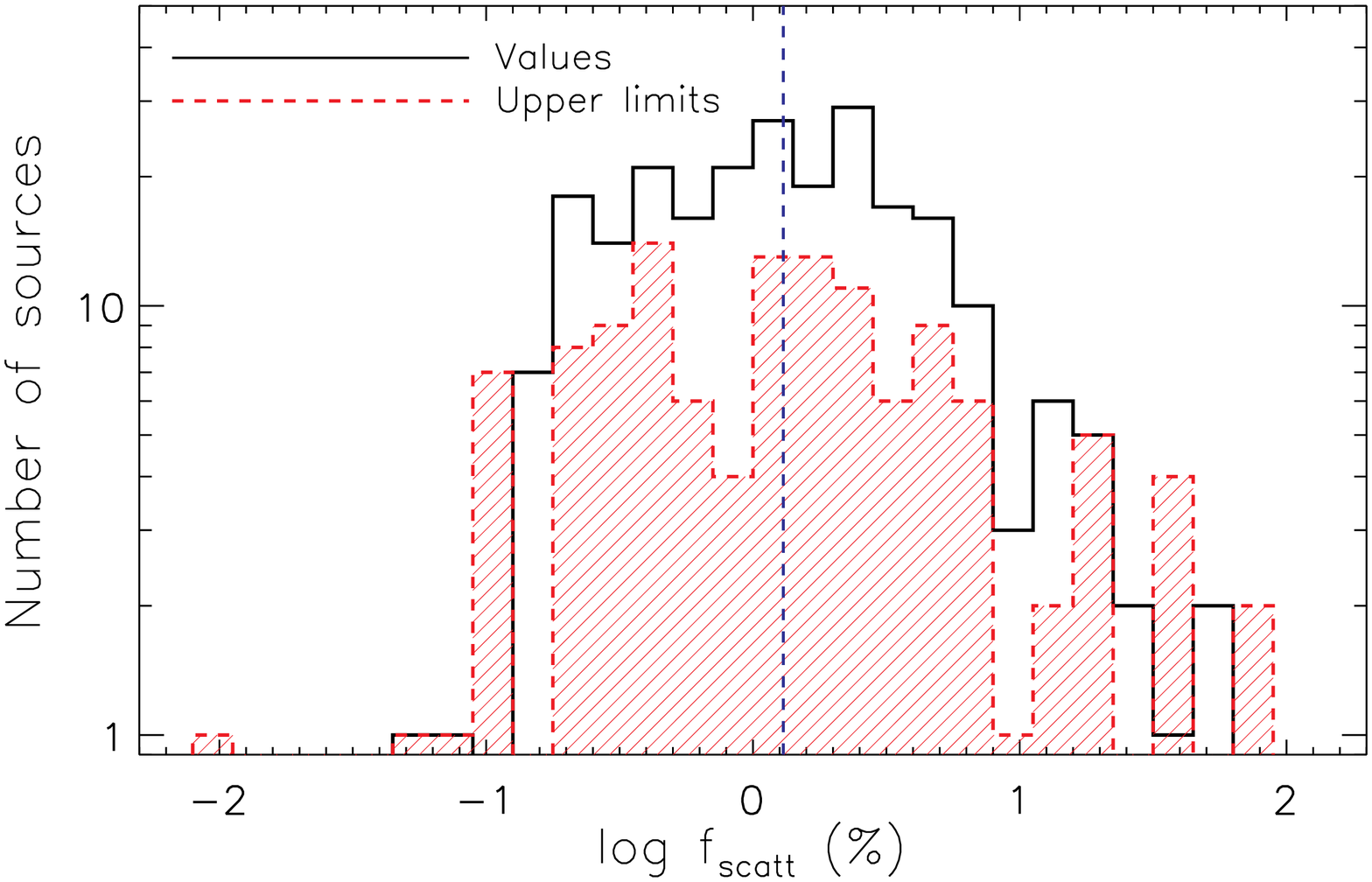}
% %% caption
  \caption{Distribution of the fraction of scattered X-ray radiation inferred for the obscured non-blazar AGN. The black line represents the measured values, while the red dashed line represents the upper limits. The vertical blue dashed line shows the median value of the distribution (see Table\,\ref{tab:median_nonblaz}).}
\label{fig:scattfractionHist}
\end{figure}

\subsubsection{Obscured AGN}\label{sect:SoftExcess_obs}

The soft excess in obscured AGN could have several origins. It could be related to Thomson scattering of the primary X-ray radiation in ionized gas, possibly located in the photo-ionized region (e.g.,  \citealp{Noguchi:2010lk}). In support of this idea, \cite{Ueda:2015mz} recently demonstrated that AGN with low scattering fractions also show smaller ratios of the extinction-corrected [O{\sc iii}\,5007] emission line to the intrinsic 2--10 keV luminosity. In addition to this, star formation could contribute significantly to the flux below $\sim2$\,keV. X-ray emission in star-forming regions is due to a population of X-ray binaries and to collisionally ionized plasma (e.g., \citealp{Ranalli:2008ve}). A significant contribution might also come from radiative-recombination, which would create emission lines and a continuum below $\sim2$\,keV (e.g., \citealp{Marinucci:2011lp}).

Here we do not attempt to model photoionized emission, and we reproduced the soft excess in obscured AGN as a combination of a scattered continuum and one or more collissionally ionized plasmas. A scattered component was added to the X-ray spectral model of 388 sources. The value of $f_{\rm\,scatt}$ could be constrained for 251 objects, while for 137 only an upper limit was obtained (Fig.\,\ref{fig:scattfractionHist}). The median value of the scattered fraction for the objects for which this value could be constrained is $f_{\rm\,scatt}=1.4\pm0.6\%$, with 101 objects having $f_{\rm\,scatt} < 1\%$. Following the same approach described in $\S$\ref{sect:cutoff}, we took into account also the lower limits in the calculation of the median value of $f_{\rm\,scatt}$, and found a median of $f_{\rm\,scatt}=1.0\pm0.4\%$ for the total sample (see Table\,\ref{tab:median_nonblaz}).
Optically-selected AGN have been shown to have significantly higher values of the fraction of scattered radiation ($f_{\rm\,scatt}\sim1-5$\%; e.g., \citealp{Bianchi:2007bq}). Smaller values of $f_{\rm\,scatt}$ imply either a small opening angle of the torus or a smaller amount of gas responsible for the scattering. Studying {\it Suzaku} observations of {\it Swift}/BAT type-II AGN,  \cite{Ueda:2007th} found the first evidence for AGN with $f_{\rm\,scatt} < 0.5\%$, and concluded that these objects are ``buried AGN", i.e. objects that have a torus with a very large covering factor (see also \citealp{Eguchi:2009xq,Eguchi:2011tk}). It has been argued by \cite{Honig:2014fv} that object with $f_{\rm\,scatt} < 0.5\%$ reside in highly inclined galaxies or merger systems, which would increase the probability that the scattered emission is obscured by the host galaxy. \cite{Ueda:2015mz} showed that, while in some cases it is possible that the host galaxy is responsible for the lower scattered fraction, more than half of the objects with $f_{\rm\,scatt} < 0.5\%$ in their sample are free from absorption by interstellar matter along the disk of the host galaxies. In our sample, 93 AGN ($\sim 25\%$ of the objects) have $f_{\rm\,scatt} < 0.5\%$. The scattered emission, and the relation between $f_{\rm\,scatt}$ and the physical properties of the accreting system, will be the subject of a forthcoming work (Ricci et al., in prep.).

\begin{figure}[t!]
\centering
\includegraphics[width=9cm]{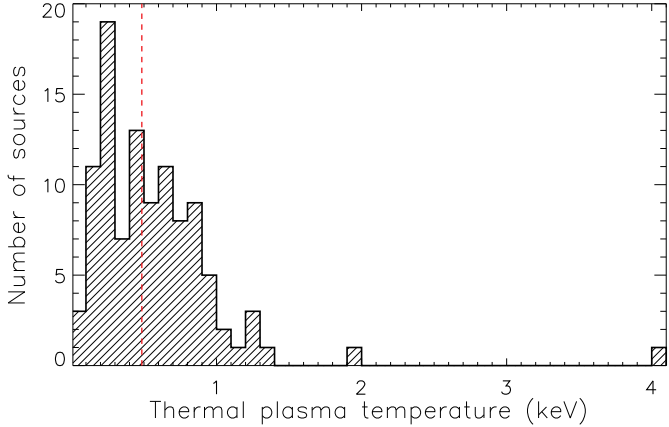}
% %% caption
  \caption{Distribution of the temperatures of the thermal plasma. When more than one component was present, we averaged the values of the temperature of each object. The vertical red dashed line shows the median value of the distribution (see Table\,\ref{tab:median_nonblaz}).}
\label{fig:Apec_hist}
\end{figure}

A thermal plasma has been used for 112 objects in total (Fig.\,\ref{fig:Apec_hist}). For 88 object we applied a single thermal component, while for 23 two components were necessary, and in one case (MCG\,$-$03$-$34$-$064) we used three thermal plasma components with different temperatures. The median value of the temperature of this component is $kT_{\rm\,therm.}=0.49\pm0.06$\,keV.

\section{Summary and conclusion}\label{sect:Summary}

We described here our detailed broad-band X-ray spectral analysis of the AGN reported in the {\it Swift}/BAT 70-month catalog. Soft X-ray (0.3--10\,keV) spectra are available for 836 of the 838 sources detected by {\it Swift}/BAT in its first 70 months of observations, implying a completeness of $\sim 99.8\%$. 
Our sample consists of the 836 {\it Swift}/BAT objects with soft X-ray observations, of which 731 are non-beamed AGN and 105 are blazars (26 BL Lacs, 53 Flat Spectrum Radio Quasars and 26 of uncertain type). A variety of X-ray spectral models was applied to the 0.3--150\,keV spectra. The median values of the spectral parameters of non-blazar AGN and blazars obtained by our spectral analysis are listed in Table\,\ref{tab:median_nonblaz} and Table\,\ref{tab:median_blaz}, respectively. The sample and the X-ray observation log are listed in Tables\,\ref{tbl-list} and \ref{tbl-2}, respectively. In Table\,\ref{tbl-3} we report the main spectral parameters obtained by our analysis for all sources.
In the following we summarize our main findings.

\begin{itemize}
\item We report the values of the intrinsic (i.e., absorption-corrected and $k$-corrected) luminosities for all objects of our sample in the 2--10\,keV, 20--50\,keV, 14--150\,keV and 14--195\,keV bands. We find that unobscured AGN have typically higher {\it intrinsic} luminosities than obscured AGN, in agreement with the decrease of the fraction of obscured sources with increasing luminosity. Among the blazars, Flat Spectrum Radio Quasars are significantly more luminous, in all the bands discussed here, than BL Lacs, while blazars of uncertain type typically have the lowest luminosities (see $\S$\ref{sect:Luminosities} and Fig.\,\ref{fig:Lumdistr_obsunobs}).
\item The median value of the photon index obtained by the broad-band X-ray spectral analysis of the non-blazar AGN is $\Gamma=1.78\pm0.01$. Obscured and unobscured objects have consistent spectral slopes, with medians of $\Gamma=1.76\pm0.02$ and $\Gamma=1.80\pm0.02$, respectively (see $\S$\ref{sect:photonindex} and Fig.\,\ref{fig:Gamma_hist}).  The values of $\Gamma$ and of the photon index obtained by fitting the 0.3--10\,keV spectrum alone ($\Gamma_{0.3-10}$) are typically lower than the {\it Swift}/BAT photon index ($\Gamma_{\rm BAT}$, see Fig.\,\ref{fig:GVSg}). The steepening of the X-ray continuum in the hard X-ray band suggests that the presence of a high-energy cutoff is almost ubiquitous in non-blazar AGN.
\item Flat Spectrum Radio Quasars have typically flatter slopes ($\Gamma=1.54\pm0.05$, $\Gamma_{\rm BAT}=1.71\pm0.06$) than BL Lacs ($\Gamma=2.05\pm0.06$, $\Gamma_{\rm BAT}=2.42\pm0.10$). See $\S$\ref{sect:photonindex} and Fig.\,\ref{fig:Gamma_hist}. 
\item A cross-calibration constant was added to all models. We found that both non-blazar AGN and blazars have a median value of $C_{\rm\,BAT}=1$ (see $\S$\ref{sect:crossbat} and Fig.\,\ref{fig:Cbatdistr}). From the dispersion in $C_{\rm\,BAT}$ of non-blazar AGN we conclude that the typical variability of these objects, on the timescales probed by our study (days to several years), is $\sim 0.2$\,dex.
\item The cutoff energy of non-blazar AGN, considering also the upper and lower limits (assuming a maximum value of $E_{\rm C}=1000$\,keV), has a median of $E_{\rm C}=381\pm16$\,keV, and obscured and unobscured AGN have consistent median $E_{\rm C}$ (see $\S$\ref{sect:cutoff} and the left panel of Fig.\,\ref{fig:hist_EC_R}). Ignoring the upper and lower limit the median values is significantly lower ($E_{\rm C}=76\pm6$\,keV). This is due to the fact that, at the typical signal-to-noise ratio of our data, we can mostly constrain values of $E_{\rm C}\lesssim 100\rm\,keV$. Using the Kaplan-Meier estimator and including the lower limits we found that, for the whole sample, the median (mean) is $200\pm 29$\,keV ($319\pm23$\,keV).
\item We find that the reflection parameter of non-blazar AGN, taking into account also upper and lower limits, has a median of $R=0.53\pm0.09$ (see $\S$\ref{sect:refl} and the right panel of Fig.\,\ref{fig:hist_EC_R}). Unobscured objects have a larger median value ($R=0.83\pm0.14$) than Compton-thin obscured AGN ($R=0.38\pm0.11$) and CT sources ($R=0.15\pm0.12$). The decrease of the reflection component with increasing obscuration is in agreement with the idea that most of the reprocessed X-ray radiation originates in the accretion disk.
For the objects for which we could constrain $R$ we find a median value of $R=1.2\pm 0.2$. This is larger in unobscured AGN ($R=1.4\pm 0.3$) than in Compton-thin obscured AGN ($R=0.6\pm 0.2$) and in CT sources ($R=0.25\pm 0.10$). 
\item We find that the non-blazar sample is almost equally divided into AGN with $N_{\rm H}<10^{22}\rm\,cm^{-2}$ (365) and those with $N_{\rm H}\geq 10^{22}\rm\,cm^{-2}$ (366). Among the latter 56 are Compton-thick ($N_{\rm H} \geq 10^{24}\rm\,cm^{-2}$), and 75 have column densities consistent with $N_{\rm H}\sim 10^{24}\rm\,cm^{-2}$ (see \S\ref{sect:NeutralAbsorption} and Fig.\,\ref{fig:NHdistr}). 
\item Evidence of ionized absorption is found in 86 AGN (\S\ref{sect:IonizedAbsorption} and Fig.\,\ref{fig:warmabs}), most of which are unobscured. This allowed us to conclude that the covering factor of the ionized material in AGN with $N_{\rm H}<10^{22}\rm\,cm^{-2}$ is $\gtrsim 22\%$.
\item The soft excess in unobscured AGN could be well reproduced by a blackbody model. We found that the range of temperatures is very narrow, with a median of $kT_{\rm\,bb}=0.11$ (see $\S$\ref{sect:SoftExcess_unobs} and Fig.\,\ref{fig:BBhist}).
\item The typical fraction of scattered radiation is $f_{\rm\,scatt}=1.0\pm0.5\%$, lower than what is usually inferred for optically-selected AGN (see $\S$\ref{sect:SoftExcess_obs} and Fig.\,\ref{fig:scattfractionHist}). A total of 93 objects ($\sim 25\%$) have values of $f_{\rm\,scatt}<0.5\%$, which might imply that they are either surrounded by a torus with a very large covering factor, or that the amount of material responsible for the Thomson scattering in these objects is small. A total of 22 (40) objects have values of $f_{\rm\,scatt}\geq 10\%$ ($\geq 5\%$), which implies that either the obscuring material partially covers the X-ray source or that jet emission contributes significantly to the X-ray flux at $\lesssim 2$\,keV.
\end{itemize}

In the tables in the Appendix we report the parameters of the broken power-laws (Table\,\ref{tbl-bknpo}), of the warm absorbers (Table\,\ref{tbl-wa}), of the Gaussian lines (Table\,\ref{tbl:kalphalines}), as well as the observed and intrinsic fluxes (Table\,\ref{tbl:fluxes}) and luminosities (Tables\,\ref{tbl:luminosities} and \ref{tbl:lumfluxCT}). In Table\,\ref{tbl:otherGamma} we list the values of $\Gamma_{\rm\,nEC}$ and $\Gamma_{0.3-10}$, while in Table\,\ref{tbl-torpar} we report the parameters obtained by fitting with a torus model the broad-band X-ray spectra of the 75 AGN with values of $N_{\rm H}$ consistent with $10^{24}\rm\,cm^{-2}$ within their 90\% confidence interval.

This work is part of a large effort aimed at shedding light on the multi-wavelength properties of the least biased sample of local AGN available. A series of forthcoming publications will investigate, in detail, the relations between several of the properties measured here and those of the accreting SMBHs and host galaxies. These include, among others, the absorption properties (\citealp{Ricci:2017fq}, Ricci et al. in prep.); Fe\,K$\alpha$ line; the fraction of scattered radiation; and the relation between mergers and SMBH accretion.

\tabletypesize{\normalsize}
\begin{deluxetable*}{lccccc} 
\tablecaption{Median values of the parameters obtained by the broad-band X-ray spectral analysis of the non-blazar AGN of our sample. \label{tab:median_nonblaz}}
\tablewidth{0pt}
\tablehead{
\colhead{(1)} & \colhead{(2)} & \colhead{(3)} & \colhead{(4)} & \colhead{(5)} & \colhead{(6)}  \\
\noalign{\smallskip}
\noalign{\smallskip}
\noalign{\smallskip}
& \multicolumn{5}{c}{\bf \large Non-blazar AGN}  	\\
\noalign{\smallskip}
 \cline{2-6}
\noalign{\smallskip}
 \colhead{ } & \colhead{All} & \colhead{$<22$} & \colhead{$\geq22$}  & \colhead{$22-24$} & \colhead{$\geq 24$}
}
\startdata
\noalign{\smallskip}
$\log L_{\rm\,2-10}$$^{a}$ & 43.41   &  43.56    &  43.26  & 43.24	& 43.46	  \\
\noalign{\smallskip}
$\log L_{\rm\,20-50}$$^{b}$  &  43.38 &   43.58  &   43.22  & 43.23	& 43.12	   \\
\noalign{\smallskip}
$\log L_{\rm\,14-150}$$^{c}$  & 43.74  &  43.92   &   43.62  & 43.63	& 43.53	   \\
\noalign{\smallskip}
$\Gamma$$^{d}$ &  $1.78\pm0.01$ & $1.80\pm0.02$  &  $1.76\pm0.02$     &	$1.70\pm0.02$ &	$2.05\pm 0.05$    \\
\noalign{\smallskip}
$\Gamma_{\rm BAT}$$^{e}$ & $1.96\pm0.01$ &  $2.02\pm 0.02$  & $1.89\pm0.02$     & $1.90\pm0.02$	&	$1.75\pm0.04$    \\
\noalign{\smallskip}
$C_{\rm\,BAT}$$^{f}$   &  $1.00\pm0.05$ &   $1.00\pm0.05$  & $1.00\pm0.08$   &$1.00\pm0.09$	&	$1.00\pm0.12$    \\
\noalign{\smallskip}
$E_{\rm C}$$^{g}$ &  $381\pm16$  &  $384\pm21$   & $380\pm24$     & $386\pm25$ 	& $341\pm70$ 	  \\
\noalign{\smallskip}
$E_{\rm C}$$^{h}$ &  $200\pm29$  &  $210\pm36$   & $189\pm26$     & $188\pm27$ 	& $449\pm64$ 	  \\
\noalign{\smallskip}
$E_{\rm C}$$^{i}$ &   $76\pm6$ &  $80\pm7$   & $74\pm11$     & $74\pm11$	&  $43\pm15$	  \\
\noalign{\smallskip}
$R$$^{j}$  &  $0.53\pm0.09$ & $0.83\pm0.14$    & $0.37\pm0.11$    & $0.38\pm0.11$	&	$0.15\pm 0.12$   \\
\noalign{\smallskip}
$R$$^{k}$  &  $1.3\pm0.1$ & $1.8\pm0.2$    & $0.6\pm0.2$    & $0.6\pm0.2$	&	$0.25\pm0.10$   \\
\noalign{\smallskip}
$N_{\rm H}$$^{l}$   & $10^{22}$  &  $10^{20}$   &  $1.5\times 10^{23}$    &  $10^{23}$	&	$2\times10^{24}$  \\
\noalign{\smallskip}
$N_{\rm H}^{\rm W}$$^{m}$	 & $2.8\times 10^{22}$  & $2.8\times 10^{22}$    & \nodata    &	\nodata &	\nodata   \\
\noalign{\smallskip}
$f_{\rm\,C}^{\rm\,W}$$^{n}$	 & $63\pm4$  &   $63\pm4$    & \nodata    &\nodata	&	\nodata   \\
\noalign{\smallskip}
$\log \xi$$^{o}$  &  $1.45\pm0.16$ &  $1.45\pm0.16$   & \nodata    &	\nodata &	\nodata   \\
\noalign{\smallskip}
$kT_{\rm\,bb}$$^{p}$  &  $0.110 \pm 0.003$ &   $0.110 \pm 0.003$  & \nodata    & \nodata	& \nodata	   \\
\noalign{\smallskip}
$f_{\rm\,scatt}$$^{q}$  & $1.0 \pm 0.5$  &  \nodata   &  $1.0 \pm 0.4$    &   $1.1 \pm 0.5$	& $0.6 \pm 0.5$	  \\
\noalign{\smallskip}
$f_{\rm\,scatt}$$^{r}$  & $1.4 \pm 0.6$  &  \nodata   &  $1.4 \pm 0.6$    & $1.5 \pm 0.5$	& $0.4 \pm 0.3 $	  \\
\noalign{\smallskip}
$kT_{\rm\,therm.}$$^{s}$  &  $0.50\pm0.05$ &   \nodata  &  $0.50\pm0.05$ & $0.49\pm0.06$   	&	$0.56\pm0.07$
\enddata

\tablecomments{The table lists the median values of the parameters (1) obtained by the broad-band X-ray spectral analysis for: (2) the whole sample of non-blazar AGN, (3) objects with $N_{\rm H}<10^{22}\rm\,cm^{-2}$, (4) sources with $N_{\rm H}\geq 10^{22}\rm\,cm^{-2}$, (5) AGN with $N_{\rm H}= 10^{22-24}\rm\,cm^{-2}$, and (6) objects with $N_{\rm H}\geq  10^{24}\rm\,cm^{-2}$.}
\tablenotetext{a}{Absorption-corrected and $k$-corrected 2--10\,keV luminosity [$\log (\rm\,erg\,s^{-1})$]. See $\S$\ref{sect:Luminosities}. }
\tablenotetext{b}{Absorption-corrected and $k$-corrected 20--50\,keV luminosity [$\log (\rm\,erg\,s^{-1})$]. See $\S$\ref{sect:Luminosities}. }
\tablenotetext{c}{Absorption-corrected and $k$-corrected 14--150\,keV luminosity [$\log (\rm\,erg\,s^{-1})$]. See $\S$\ref{sect:Luminosities}. }
\tablenotetext{d}{Photon index obtained by fitting the 0.3--150\,keV spectrum ($\S$\ref{sect:photonindex}).}
\tablenotetext{e}{Photon index obtained by fitting {\it Swift}/BAT spectrum with a power-law model ($\S$\ref{sect:photonindex}).}
\tablenotetext{f}{Cross-calibration constant between the 0.3--10\,keV and the {\it Swift}/BAT spectra ($\S$\ref{sect:crossbat}).}
\tablenotetext{g}{Cutoff energy (keV) obtained considering also upper and lower limits ($\S$\ref{sect:cutoff}).}
\tablenotetext{h}{Cutoff energy (keV) obtained considering the lower limits and applying the Kaplan-Meier estimator ($\S$\ref{sect:cutoff}).}
\tablenotetext{i}{Cutoff energy (keV) obtained using only the sources for which the parameter could be inferred ($\S$\ref{sect:cutoff}).}
\tablenotetext{j}{Reflection parameter obtained considering also upper and lower limits ($\S$\ref{sect:refl}).}
\tablenotetext{k}{Reflection parameter obtained using only the sources for which the parameter could be inferred ($\S$\ref{sect:refl}).}
\tablenotetext{l}{Column density of the neutral material ($\rm\,cm^{-2}$). See \S\ref{sect:NeutralAbsorption}.}
\tablenotetext{m}{Column density of the ionized absorber ($\rm\,cm^{-2}$). See \S\ref{sect:IonizedAbsorption}.}
\tablenotetext{n}{Covering factor of the ionized absorber (\%). See \S\ref{sect:IonizedAbsorption}.}
\tablenotetext{o}{Ionization parameter of the warm absorber ($\rm\,erg\,cm\,s^{-1}$). See \S\ref{sect:IonizedAbsorption}.}
\tablenotetext{p}{Temperature of the blackbody component (keV). See $\S$\ref{sect:SoftExcess_unobs}.}
\tablenotetext{q}{Fraction of scattered radiation (\%) obtained considering also upper and lower limits ($\S$\ref{sect:SoftExcess_obs}).}
\tablenotetext{r}{Fraction of scattered radiation (\%) obtained using only the sources for which the parameter could be inferred ($\S$\ref{sect:SoftExcess_obs}).}
\tablenotetext{s}{Temperature of the thermal plasma (keV). See $\S$\ref{sect:SoftExcess_obs}.}

\end{deluxetable*}

\tabletypesize{\normalsize}
\begin{deluxetable*}{lcccc} 
\tablecaption{Median values of the parameters obtained by our spectral analysis\label{tab:median_blaz}.}
\tablewidth{0pt}
\tablehead{
\colhead{(1)} & \colhead{(2)} &    \colhead{(3)} & \colhead{(4)}  &  \colhead{(5)} \\
\noalign{\smallskip}
\noalign{\smallskip}
\noalign{\smallskip}
&  \multicolumn{4}{c}{\bf \large Blazars}	\\
\noalign{\smallskip}
 \cline{2-5}
\noalign{\smallskip}
\noalign{\smallskip}
 \colhead{ } & \colhead{All} &    \colhead{BZQ} & \colhead{BZB}  &  \colhead{BZU} 
}
\startdata
\noalign{\smallskip}
$\log L_{\rm\,2-10}$$^{a}$ & 45.37  &  46.23 & 45.09  &  44.25 \\
\noalign{\smallskip}
$\log L_{\rm\,20-50}$$^{b}$&  45.15 &  46.48 & 44.63   & 44.20  \\
\noalign{\smallskip}
$\log L_{\rm\,14-150}$$^{c}$ & 45.53   & 47.01  & 44.92  & 44.70   \\
\noalign{\smallskip}
$\Gamma$$^{d}$ &  $1.68\pm0.04$  &  $1.54\pm0.05$ &  $2.05\pm0.06$ & $1.67\pm0.09$      \\
\noalign{\smallskip}
$\Gamma_{\rm BAT}$$^{e}$ & $1.87\pm0.06$  &  $1.71\pm0.06$ & $2.42\pm0.10$   &  $1.85\pm0.11$    \\
\noalign{\smallskip}
$C_{\rm\,BAT}$$^{f}$  &  $1.00\pm0.35$ & $0.80\pm0.15$   & $1.10\pm0.63$  & $1.20\pm 0.45$  \\
\noalign{\smallskip}
$N_{\rm H}$$^{g}$  & $2.6\times10^{20}$  & $1.0\times10^{20}$   &  $4.0\times10^{20}$ & $1.6\times10^{21}$  \\
\enddata

\tablecomments{The table lists the median values of the parameters obtained by the broad-band X-ray spectral analysis (1) for the whole sample of blazars (2), for the Flat Spectrum Radio Quasars (3), the  BL Lacs (4), and the blazars of uncertain type (5).}

\tablenotetext{a}{Absorption-corrected and $k$-corrected 2--10\,keV luminosity [$\log(\rm\,erg\,s^{-1})$]. See $\S$\ref{sect:Luminosities}.}
\tablenotetext{b}{Absorption-corrected and $k$-corrected 20--50\,keV luminosity [$\log(\rm\,erg\,s^{-1})$]. See $\S$\ref{sect:Luminosities}.}
\tablenotetext{c}{Absorption-corrected and $k$-corrected 14--150\,keV luminosity [$\log(\rm\,erg\,s^{-1})$]. See $\S$\ref{sect:Luminosities}.}
\tablenotetext{d}{Photon index obtained by fitting the 0.3--150\,keV spectrum ($\S$\ref{sect:photonindex}).}
\tablenotetext{e}{Photon index obtained by fitting {\it Swift}/BAT spectrum with a power-law model ($\S$\ref{sect:photonindex}).}
\tablenotetext{f}{Cross-calibration constant between the 0.3--10\,keV and the {\it Swift}/BAT spectra ($\S$\ref{sect:photonindex}).}
\tablenotetext{g}{Column density of the neutral material ($\rm\,cm^{-2}$). See \S\ref{sect:NeutralAbsorption}.}

\end{deluxetable*}

\acknowledgments

This work is dedicated to the memory of our friend and collaborator Neil Gehrels.
We acknowledge the work that the {\it Swift}/BAT team has done to make this project possible. We are grateful to the anonymous referee for the prompt report that helped us improving the quality of the manuscript. We thank Taro Shimizu and Chin-Shin Chang for valuable discussion and for their help with the manuscript.
This work made use of data supplied by the UK Swift Science Data Centre at the University of Leicester. This work is sponsored by the Chinese Academy of Sciences (CAS), through a grant to the CAS South America Center for Astronomy (CASSACA) in Santiago, Chile. We acknowledge financial support from the CONICYT-Chile grants ``EMBIGGEN" Anillo ACT1101 (CR, ET), FONDECYT 1141218 (CR), FONDECYT 1160999 (ET), Basal-CATA PFB--06/2007 (CR, ET), the China-CONICYT fund (CR), the Swiss National Science Foundation (Grant PP00P2\_138979 and PP00P2\_166159, KS; grant 200021\_157021 KO), the Swiss National Science Foundation (SNSF) through the Ambizione fellowship grant PZ00P2\textunderscore154799/1 (MK), the National Key R\&D Program of China grant No. 2016YFA0400702 (LH), the National Science Foundation of China grants No. 11473002 and 1721303 (LH), the European Union's Seventh Framework programme under grant agreement 337595 (ERC Starting Grant, "CoSMass"; ID). Part of this work was carried out while CR was Fellow of the Japan Society for the Promotion of Science (JSPS) at Kyoto University. This work was partly supported by the Grant-in-Aid for Scientific Research 17K05384 (YU) from the Ministry of Education, Culture, Sports, Science and Technology of Japan (MEXT). This research has made use of the Tartarus (Version 3.1) database, created by Paul O'Neill and Kirpal Nandra at Imperial College London, and Jane Turner at NASA/GSFC. This publication made use of data products from the Wide-field Infrared Survey Explorer, which is a joint project of the University of California, Los Angeles, and the Jet Propulsion Laboratory/California Institute of Technology, funded by the National Aeronautics and Space Administration. This research made use of: the NASA/IPAC Extragalactic Database (NED), which is operated by the Jet Propulsion Laboratory, California Institute of Technology, under contract with the National Aeronautics and Space Administration; the SIMBAD database, operated at CDS, Strasbourg, France \citep{Wenger:2000pb}; the Aladin sky atlas, developed at CDS, Strasbourg Observatory, France \citep{Bonnarel:2000kk,Boch:2014qa}; Astropy, a community-developed core Python package for Astronomy \citep{Astropy-Collaboration:2013mi}; TOPCAT \citep{Taylor:2005if}.

{\it Facilities:}  \facility{AKARI}, \facility{ASCA}, \facility{BeppoSAX}, \facility{Chandra}, \facility{GALEX}, \facility{IRAS}, \facility{Suzaku}, \facility{Swift}, \facility{XMM-Newton}, \facility{WISE}.

\clearpage 

\bibliographystyle{apj} %% 
\bibliography{BASScatalog}

\appendix

\shorttitle{}
\shortauthors{}

\section{New soft X-ray counterparts}\label{app:new counterparts}

We found a different X-ray counterpart than that reported by the {\it Swift}/BAT 70 months catalog and by the studies reported above for only eight sources. In the following we report details about these objects. 

\smallskip
\noindent\textbf{SWIFT\,J0216.3+5128}.
The X-ray counterpart reported in the BAT catalog is 2MASX\,J02162987+5126246. The source is instead more likely associated to the X-ray bright Seyfert\,2 IGR\,J02164+5126/2MASS\,J02162672+5125251, which is clearly detected in the 10.0\,ks {\it XMM-Newton} EPIC/PN observation (see also \citealp{Masetti:2009wo}).

\smallskip
\noindent\textbf{SWIFT\,J0223.4+4551}.
The {\it Swift}/BAT source was identified with the triplet of galaxies V Zw 232, but the X-ray counterpart of this object is the central galaxy, 2MASX\,J02233309+4549162.

\smallskip
\noindent\textbf{SWIFT\,J0350.1$-$5019}.
The source is reported as being the counterpart of  2MASX\,J03502377$-$5018354. Besides  2MASX\,J03502377$-$5018354, three other X-ray sources are evident around the BAT position of the sources: 2MASX\,J03501198$-$5017165, 2MASX\,J03502377$-$5018354 and ESO\,201$-$4. The first two objects are too weak to be the counterpart of the {\it Swift}/BAT source, while the CT AGN ESO\,201$-$4 is at a flux level consistent with that of SWIFT\,J0350.1$-$5019.

\smallskip
\noindent\textbf{SWIFT\,J0528.1$-$3933}.
The source is identified with the interacting galaxy pair ESO\,306$-$IG001. From the analysis of the {\it XMM-Newton} image we found that the X-ray counterpart of this source is ESO\,306$-$IG001 NED01.

\smallskip
\noindent\textbf{SWIFTJ0654.6+0700}. 
This object was previously identified with 2MASS\,J06543368+0703024 (also in \citealp{Cusumano:2010qf}), while we find that the right counterpart is 2MASS\,J06543417+0703210 (see also \citealp{Parisi:2014bx}). The new identification is coincident with a bright {\it WISE} source.

\smallskip
\noindent\textbf{SWIFT\,J0744.0+2914}.
The 10.8\,ks {\it Chandra}/ACIS observation did not detect the counterpart reported in the 70-month BAT catalog, UGC\,03995A. The most likely counterpart of SWIFT\,J0744.0+2914 is the nearby galaxy UGC\,03995B, which is detected in the 0.3--10\,keV band and is interacting with UGC\,03995A \citep{Koss:2012fj}.

\smallskip
\noindent\textbf{SWIFT\,J0919.2+5528}.
The counterpart is not SBS\,0915+556, as reported in the 70-month catalog, but it is the CT AGN Mrk\,106, which is significantly brighter and closer to the position of the hard X-ray source than SBS\,0915+556.

\smallskip
\noindent\textbf{SWIFT\,J1238.6+0928}. 
The source was identified as being the hard X-ray counterpart of  the galaxy VCC\,1759. VCC\,1759 was not detected in the 20.1\,ks {\it XMM-Newton} EPIC/PN image, and the most likely counterpart for SWIFT\,J1238.6+0928 is the Seyfert\,2 2MASX\,J12384342+0927362 (see also \citealp{Malizia:2016tg}).

\smallskip
\noindent\textbf{SWIFTJ1354.5+1326}. 
The source was originally identified with 2MASX\,J13542913+1328068, which is part of an interacting galaxy pair with 2MASS\,J13542908+1327571 (at a distance of 10\arcsec). However, the former object is not detected in {\it WISE}, while the latter has an infrared brightness consistent with what is expected from its intrinsic X-ray flux. Therefore, it is more likely that the {\it Swift}/BAT source is associated with 2MASS\,J13542908+1327571.

\smallskip
\noindent\textbf{SWIFT\,J1535.8$-$5749}.
The 0.3--10\,keV counterpart of the source is IGR\,J15360$-$5750 \citep{Malizia:2010ez}, which is brighter and closer to the BAT position than the object reported in the BAT catalog, 1RXS\,J153552.8$-$575055.

\smallskip
\noindent\textbf{SWIFTJ1747.8+6837B}.
The source was identified with VII Zw 742, a system made of an interacting pair of galaxies (2MASX J17465994+6836392 and 2MASS J17465953+6836303) with a separation of $8.8\arcsec$. The two sources are blended in {\it WISE}, with the southern one being brighter.  The {\it Swift/XRT} observation of this source showed that the position of the X-ray counterpart coincides with that of 2MASS\,J17465953+6836303.

\smallskip
\noindent\textbf{SWIFTJ1856.1+1539}.
This object was originally identified with 2MASX\,J18560128+1538059. However, the right counterpart is more likely to be 2MASS\,J18560056+1537584, which has a bright, red {\it WISE} counterpart. This new identification is supported by what is reported by \cite{Rodriguez:2008hb}.

\smallskip
\noindent\textbf{SWIFT\,J2007.0$-$3433}.
The {\it Chandra}/ACIS 0.3--10\,keV image shows that the X-ray counterpart is not ESO\,399$-$20, but the nearby galaxy MCG\,$-$06$-$44$-$018, with which ESO\,399$-$20 is interacting.

\section{Dual AGN}\label{app:dualAGN}

Several objects in our sample are known to be dual AGN (e.g., \citealp{Koss:2012fj}), in the following we discuss the influence of the two sources on the {\it Swift}/BAT emission and list the cases in which the X-ray spectral properties of both sources are reported in our catalog. Those objects are listed as D1 and D2.

\smallskip
\noindent\textbf{SWIFT\,J0209.5$-$1010}.
This object is the combination of two AGN at $56\arcsec$: NGC\,833 and NGC\,835 with the former having a 2--10\,keV flux $\sim 1/3$ of the latter \citep{Koss:2011ve}. Due to the fact that both sources contribute to the 14--195\,keV flux we report the spectral characteristics of both NGC\,833 (SWIFT\,J0209.5-1010D1) and NGC\,835 (SWIFT\,J0209.5-1010D2).

\smallskip
\noindent\textbf{SWIFT\,J0324.9+4044}. This object is composed of two AGN at $12\arcsec$ from each other: IRAS\,03219+4031 and 2MASX\,J03251221+4042021. In the 2--10\,keV band IRAS\,03219+4031 is $\sim 12$ times brighter than 2MASX\,J03251221+4042021 \citep{Koss:2011ve}, and was therefore considered to be the counterpart of the {\it Swift}/BAT source.

\smallskip
\noindent\textbf{SWIFT\,J0602.2+2829}. This source is composed of IRAS\,05589+2828 and 2MASX\,J06021107+2828382 (at a distance of $20\arcsec$), with the former having a 2--10\,keV X-ray flux 1100 higher than the latter \citep{Koss:2011ve}.

\smallskip
\noindent\textbf{SWIFT\,J0945.6$-$1420}. This object is composed by two sources at $180\arcsec$ from each other: NGC\,2992 and NGC\,2993. Since the former AGN is $\sim 7$ times brighter in the 2--10\,keV band than the latter \citep{Koss:2011ve} it was considered as the counterpart of SWIFT\,J0945.6$-$1420.

\smallskip
\noindent\textbf{SWIFT\,J1023.5+1952}. This dual AGN is composed of NGC\,3227 and NGC\,3226, located at $130\arcsec$ from each other. In the 2--10\,keV band NGC\,3227 is $\sim 73$ times brighter than NGC\,3226 \citep{Koss:2011ve}, and was therefore assumed to be the counterpart of SWIFT\,J1023.5+1952.

\smallskip
\noindent\textbf{SWIFT\,J1136.0+2132}.
The counterpart of this source is a late-stage galaxy merger composed of two nuclei with a projected separation of 3.4\,kpc ($5\arcsec$): Mrk\,739E and Mrk\,739W \citep{Koss:2011ve}. The 2--10 observed flux of Mrk\,739W is $\sim 9\%$ of that of Mrk\,739E \citep{Koss:2011ve}, and the latter source was reported as the counterpart of the BAT AGN.

\smallskip
\noindent\textbf{SWIFT\,J1315.8+4420}. This source is composed of UGC\,8327 NED01 and UGC\,8327 NED02, located at $37\arcsec$ from each other. The 2--10\,keV flux of UGC\,8327 NED01 is $\sim 1140$ times higher than that of UGC\,8327 NED02 \citep{Koss:2011ve}, and this source was therefore considered to be the counterpart of the {\it Swift}/BAT object.

\smallskip
\noindent\textbf{SWIFT\,J1334.8$-$2328}.
The {\it XMM-Newton} EPIC/PN image reveals that, within the error box of the {\it Swift}/BAT position, both members of the interacting pair ESO\,509$-$IG066 \citep{Guainazzi:2005fk} are detected. These two sources are ESO\,509$-$IG066E/NED02 (SWIFT\,J1334.8$-$2328D1) and ESO 509$-$IG066W/NED01 (SWIFT\,J1334.8$-$2328D2), with the eastern component having a 2--10\,keV flux $\sim 60\%$ of the western component.

\smallskip
\noindent\textbf{SWIFT\,J1341.2+3023}. The source is composed of two objects at $59\arcsec$: Mrk\,268 and Mrk\,268SE. The counterpart of the {\it Swift}/BAT source is Mrk\,268, since it has a 2--10\,keV flux $>108$ times higher than that of its companion \citep{Koss:2011ve}.

\smallskip
\noindent\textbf{SWIFT\,J1355.9+1822}. This source is another known dual AGN, with the two nuclei (Mrk\,463E and Mrk\,463W, at $4\arcsec$) having a projected separation of 3.4\,kpc \citep{Bianchi:2008uq}. The {\it Chandra} study of \citep{Bianchi:2008uq} showed that both nuclei are obscured, and \citep{Koss:2011ve} reported that the 2--10\,keV observed flux of Mrk\,463W is $\sim 25\%$ that of Mrk\,463E.

\smallskip
\noindent\textbf{SWIFT\,J1652.9+0223}. The luminous infrared galaxy NGC\,6240 is known to host two Compton-thick AGN separated by 0.7\,kpc ($1.8\arcsec$): NGC\,6240N and NGC\,6240S \citep{Komossa:2003eu}. The two sources have comparable observed 2--10\,keV fluxes \citep{Koss:2011ve}, and \cite{Puccetti:2016dz} have recently shown that both sources contribute significantly to the flux above 10\,keV. The intrinsic 10--40\,keV luminosities are $7.1\times 10^{43}\rm\,erg\,s^{-1}$ and $2.7\times 10^{43}\rm\,erg\,s^{-1}$ for the southern and northern nucleus, respectively \citep{Puccetti:2016dz}. Since the {\it XMM-Newton} observation used here lacks the spatial resolution to resolve the two nuclei the system was considered to be one source.

\smallskip
\noindent\textbf{SWIFT\,J1816.0+4236}. The source is composed of UGC\,11185 NED01 and UGC\,11185 NED02 (at $28\arcsec$). Since UGC\,11185 NED02 has a 2--10\,keV flux $>23.3$ times larger than UGC\,11185 NED01 \citep{Koss:2011ve}, it was considered to be the counterpart of the BAT source.

\smallskip
\noindent\textbf{SWIFT\,J2028.5+2543}.
This source is a combination of two objects at $91\arcsec$ from each other: NGC\,6921 and MCG\,+04$-$48$-$002, with the former AGN contributing to $\sim 20\%$ of the 14--195\,keV flux. Since the two AGN contribute significantly to the {\it Swift}/BAT flux we reported the characteristics of both NGC\,6921 (SWIFT\,J2028.5+2543D1) and MCG\,+04$-$48$-$002 (SWIFT\,J2028.5+2543D2) in the catalog.

\smallskip
\noindent\textbf{SWIFT\,J2328.9+0328}.
This object is the combination of the unobscured AGN NGC\,7679 and the CT AGN NGC\,7682 (at $270\arcsec$). Since NGC\,7679 contributes to only $\sim 10\%$ of the 14--195\,keV flux, we adopted NGC\,7682 as the counterpart of the {\it Swift}/BAT source.

\section{The starburst galaxy SWIFT\,J0956.1+6942 (M\,82)}\label{app:m82}

The starburst galaxy M\,82 does not host an AGN, and its broad-band X-ray spectrum was fitted with the D1 model (\S\ref{sec:othermodels}; $\chi^{2}/\rm DOF=1294.9/1209$). We obtained a photon index of $\Gamma=1.31^{+0.43 }_{-0.55}$, a cutoff energy of $11^{+17}_{-6}$\,keV, and temperatures of the thermal plasma of $0.68^{+0.23}_{-0.32}$\,keV and $1.62^{+0.20}_{-0.13}$\,keV. The column density of the neutral material obscuring the cutoff power law component is $3.5^{+1.1}_{-1.4}\times 10^{22}\rm\,cm^{-2}$, while that of the material obscuring the thermal plasma is $5.0^{+6.8}_{-2.3}\times 10^{21}\rm\,cm^{-2}$. The cross calibration constant between the {\it XMM-Newton} and the {\it Swift}/BAT spectrum is consistent with unity ($C_{\rm\,BAT}=1.2^{+0.5}_{-0.3}$).

The observed fluxes in the 2--10 and 14--195\,keV bands are $F_{2-10}^{\rm\,obs}=11.2\times 10^{-12}\rm\,erg\,cm^{-2}\,s^{-1}$ and $F_{14-195}^{\rm\,obs}=6.4\times 10^{-12}\rm\,erg\,cm^{-2}\,s^{-1}$, respectively. The intrinsic fluxes in the 2--10, 20--50 and 14--150\,keV bands are $F_{2-10}=13.8\times 10^{-12}\rm\,erg\,cm^{-2}\,s^{-1}$, $F_{20-50}=2.5\times 10^{-12}\rm\,erg\,cm^{-2}\,s^{-1}$ and $F_{14-150}=5.0\times 10^{-12}\rm\,erg\,cm^{-2}\,s^{-1}$, respectively. The intrinsic luminosities, in the same energy ranges, are: $\log (L_{2-10}/\rm erg\,s^{-1})=40.17$, $\log (L_{20-50}/\rm erg\,s^{-1})=39.43$ and $\log (L_{14-150}/\rm erg\,s^{-1})=39.73$.

\clearpage
%\begin{turnpage}
\begin{landscape}
\LongTables
\tabletypesize{\normalsize}
\begin{deluxetable*}{llrrccc} 
\tablecaption{List of the {\it Swift}/BAT AGN from the 70-month catalog. \label{tbl-list}}
\tablewidth{0pt}
\tablehead{
\colhead{(1)} & \colhead{(2)} & \colhead{(3)} & \colhead{(4)} & \colhead{(5)} &  \colhead{(6)}  &  \colhead{(7)}  \\
\noalign{\smallskip}	
\colhead{Source} & \colhead{Counterpart} & \colhead{RA} & \colhead{DEC} & \colhead{Redshift} &   \colhead{Distance [Mpc]} & \colhead{BZCAT}  
}
\startdata

SWIFT\,J0001.0$-$0708            &            2MASX\,J00004876-0709117       &         0.2032   &    $-$7.1532		&	0.0375   				&        165.2      & \nodata			 \\ \noalign{\smallskip}			   
SWIFT\,J0001.6$-$7701         &   2MASX\,J00014596$-$7657144      &            0.4419   &        $-$76.9540 		&	0.0584  &        261.2	& \nodata		            \\ \noalign{\smallskip}				
SWIFT\,J0002.5+0323           &   NGC\,7811                       &            0.6103   &          3.3519   		&  	0.0255  &        111.3  	& \nodata		              \\ \noalign{\smallskip}				
SWIFT\,J0003.3+2737              &            2MASX\,J00032742+2739173       &         0.8643   &    27.6548			&	0.0397  				&        175.2      & \nodata			 \\ \noalign{\smallskip}			   
SWIFT\,J0005.0+7021                 &         2MASX\,J00040192+7019185           &       1.0082    &   70.3217	&	0.0960 &        440.7	& \nodata		              \\ \noalign{\smallskip}	                          
SWIFT\,J0006.2+2012           &   Mrk\,335                        &            1.5813   &         20.2029 			&	0.0258  &        112.7	& \nodata		              \\ \noalign{\smallskip}				
SWIFT\,J0009.4$-$0037               &         SDSS\,J000911.57$-$003654.7        &       2.2982    &   $-$0.6152	&	0.0733 &        331.3	& \nodata		          \\ \noalign{\smallskip} 			                          
SWIFT\,J0010.5+1057           &   Mrk\,1501                       &            2.6292   &         10.9749   		&  	0.0893  &        408.1  	&	BZQ 		             \\ \noalign{\smallskip}				
SWIFT\,J0017.1+8134             &          	[HB89]\,0014+813                    &      4.2853  &	     81.5856 	&	3.3660  				&      29188.0	  & BZQ		    	 \\ \noalign{\smallskip}                    
SWIFT\,J0021.2$-$1909               &         LEDA\,1348                         &       5.2814    &  $-$19.1682	&	0.0956 &        438.7	& \nodata		          \\ \noalign{\smallskip} 				                          

\enddata

\tablecomments{Table \ref{tbl-list} is published in its entirety in the electronic edition of the {\it Astrophysical Journal}. A portion is shown here for guidance. The table reports: (1) the {\it Swift}/BAT ID, (2) the name of the counterpart, (3, 4) the coordinates (in degrees, J2000), (5) the redshift, (6) the distance in Mpc, and (7) the BZCAT class.}

\tablenotetext{a}{For these sources only a photometric redshift (and its 1$\sigma$ confidence interval) is reported.}
\tablenotetext{b}{Sources for which redshift-independent measurements of the distance are reported.}

\end{deluxetable*}
%\end{turnpage}
\clearpage
\end{landscape}

\clearpage
%\begin{turnpage}
%\begin{landscape}
\LongTables
\tabletypesize{\normalsize}
\begin{deluxetable*}{lcccc} 
\tablecaption{Log of the soft X-ray observations.\label{tbl-2}}
\tablewidth{0pt}
\tablehead{
\colhead{(1)} & \colhead{(2)} & \colhead{(3)} & \colhead{(4)} & \colhead{(5)} \\
\noalign{\smallskip} 
\colhead{Source} & \colhead{Soft X-ray Facility/Instrument} & \colhead{Obs. ID} & \colhead{Exposure [ks]} & \colhead{Counts} 
}
\startdata
SWIFT\,J0001.0$-$0708    & {\it Swift}/XRT			& 40885001      & 8.5   &     	 	815     	 \\ \noalign{\smallskip}			   
SWIFT\,J0001.6$-$7701      	& {\it Swift}/XRT         	&	41138001		& 9.6   	  & 	346                                             \\ \noalign{\smallskip}				
SWIFT\,J0002.5+0323      	& {\it Swift}/XRT         	&	47107002		& 10.8   	   &    3035                                          \\ \noalign{\smallskip}				
SWIFT\,J0003.3+2737      & {\it Swift}/XRT			& 41139002      & 9.7   &         	100     	 \\ \noalign{\smallskip}			   
SWIFT\,J0005.0+7021     & {\it XMM$-$Newton} EPIC/PN			&	0550450101        			& 14.8  	&	  5343                                 \\ \noalign{\smallskip}	                          
SWIFT\,J0006.2+2012      	& {\it Swift}/XRT         	&	35755001		& 204.3   	  &     56302                                         \\ \noalign{\smallskip}				
SWIFT\,J0009.4$-$0037   & {\it Swift}/XRT						&	41140001					& 21.0   	&	  91                                 \\ \noalign{\smallskip} 			                          
SWIFT\,J0010.5+1057      	& {\it Swift}/XRT         	&	36363001		& 21.8   	  &     5148                                         \\ \noalign{\smallskip}				
SWIFT\,J0017.1+8134       	& {\it XMM$-$Newton} EPIC/PN	& 0112620201		&	19.1    &	15950    \\ \noalign{\smallskip}                    
SWIFT\,J0021.2$-$1909   & {\it Swift}/XRT						&	40886001					& 8.0   	&	  1515                                 \\ \noalign{\smallskip}

\enddata

\tablecomments{Table \ref{tbl-2} is published in its entirety in the electronic edition of the {\it Astrophysical Journal}. A portion is shown here for guidance
regarding its form and content. The table reports: (1) the {\it Swift}/BAT ID of the source, (2) the X-ray facility used, (3) the ID of the observations, (4) the exposure and (5) the number of counts.}

\end{deluxetable*}
%\end{turnpage}
\clearpage
%\end{landscape}

\clearpage
%\begin{turnpage}
\begin{landscape}
\LongTables
\begin{deluxetable*}{llrrcccccr} 
%\tabletypesize{\scriptsize}
\tabletypesize{\footnotesize}
\tablecaption{Parameters obtained by the analysis of the broad-band X-ray spectra. \label{tbl-3}}
\tablewidth{0pt}
\tablehead{
\colhead{(1)} & \colhead{(2)} & \colhead{(3)} & \colhead{(4)} & \colhead{(5)} &  \colhead{(6)} &    \colhead{(7)} & \colhead{(8)}  &  \colhead{(9)} &  \colhead{(10)} \\
\noalign{\smallskip} 
\colhead{Source} & \colhead{$\log N_{\rm H}$} & \colhead{$\Gamma$} & \colhead{$E_c$} & \colhead{$R$} &  \colhead{$C_{\rm\,BAT}$} &    \colhead{$f_{\rm\,scatt}$} & \colhead{$kT$}  &  \colhead{Model} &  \colhead{Statistic/DOF} \\
%}
\colhead{ } & \colhead{ [\scriptsize{$\rm\,cm^{-2}$}]} & \colhead{ } & \colhead{[\scriptsize{keV}] } & \colhead{ } &  \colhead{ } &  \colhead{[\scriptsize{\%}] } & \colhead{ [\scriptsize{keV}]}  &  \colhead{ } &  \colhead{ } 
}
\startdata
SWIFT\,J0001.0$-$0708       &		  	$22.19\,\,[22.11-22.26]$  	&	$ 1.64^{+0.60 }_{-0.18 }$	&	$\geq 29$				&		$\leq 1.7$				&		$0.8^{+0.6}_{-0.3}$			&	$\leq 1.1  $	 			&	\nodata	 													&	B1			&		 34.7/40    	\phantom{a}[$\chi^2$]	\\ \noalign{\smallskip}
SWIFT\,J0001.6$-$7701      	&		\nodata						&	$ 1.83^{+0.57 }_{-0.37 }$	&	$\geq 47$						&	$\leq 1.3$				&		$1.0^{+0.4}_{-0.8}$		     		&	\nodata						&	\nodata										&   A3		&		 14.8/17   \,\,[$\chi^2$]	\\ \noalign{\smallskip}
SWIFT\,J0002.5$+$0323      	&		\nodata						&	$ 2.23^{+0.06 }_{-0.06 }$	&	NC								&	$4.2^{+3.0}_{-0.7}$		&		$0.5^{+0.5}_{-0.2}$			     	&	\nodata						&	\nodata						 				&   A1		&		 131.5/120  \,\,[$\chi^2$] 	\\ \noalign{\smallskip}
SWIFT\,J0003.3$+$2737       &		  	$22.86\,\,[22.78-22.98]$  	&	$ 1.76^{+0.28 }_{-0.40 }$	&	NC						&		$1.1^{+2.2}_{-1.0}$		&		$1.0^{a}$					&	$\leq 0.7  $				&	\nodata	 						 							&	B1			&		 83.7/111    		 \phantom{a}[C]	\\ \noalign{\smallskip}
SWIFT\,J0005.0$+$7021      &	$22.61\,\,[22.56-22.68]$  		&	$1.46^{+0.29 }_{-0.20 }$		&			$47^{+47}_{-20}$	&		$\leq 2.4$				&	$1.0^{a}$					&	$\leq 0.7  $		 			&	\nodata	 																&		B1(G)		&		 243.3/241    		\phantom{a}[$\chi^2$] 	\\ \noalign{\smallskip}
SWIFT\,J0006.2$+$2012      	&		$20.48\,\,[20.26-20.51]$  	&	$ 2.82^{+0.08 }_{-0.03 }$	&	$\geq 185$						&	$0.8^{+1.0}_{-0.4}$		&		$ 3.7^{+1.4}_{-0.6}$ 				&	\nodata						&	$0.149^{+0.002 }_{-0.008 }$	 				&   A6(G)	&		 571.1/515  \,\,[$\chi^2$] 	\\ \noalign{\smallskip}
SWIFT\,J0009.4$-$0037      &	$23.61\,\,[23.36-24.04]$  		&	$1.64^{+0.48 }_{-0.87 }$		&			NC					&		$0.8^{+1.7}_{-0.4}$		&	$1.0^{a}$					&	$1.0^{+3.6}_{-0.9}  $	 		&	\nodata	 																&		B1(G)		&		 99.0/83    		\phantom{a}[C]   			 \\ \noalign{\smallskip}		
SWIFT\,J0010.5$+$1057      	&		$21.04\,\,[21.00-21.11]$  	&	$ 1.73^{+0.03 }_{-0.07 }$	&	$\geq 269$						&	$0.6^{+0.2}_{-0.2}$		&		$1.1^{+0.2}_{-0.2}$			     	&	\nodata						&	$0.122^{+0.007 }_{-0.007 }$	 				&  	A2(G)	&		 232.1/196  \,\,[$\chi^2$] 	\\ \noalign{\smallskip}
SWIFT\,J0017.1$+$8134       &		$21.95\,\,[21.85-22.04]$ 		&	$ 1.52^{+0.02 }_{-0.02 }$			&		\nodata				&		\nodata					&		$2.3^{+2.2}_{-1.2}$				&	\nodata	 							&	\nodata	 						&		C6				&		585.1/565   \phantom{a}[$\chi^2$] 	\\ \noalign{\smallskip}	
SWIFT\,J0021.2$-$1909      &	$21.98\,\,[21.92-22.08]$  		&	$1.70^{+0.22 }_{-0.13 }$		&			$86^{+211}_{-44}$	&		$\leq 0.2$				&	$0.7^{+0.1}_{-0.3}$			&	$2.0^{+1.2}_{-1.2}  $	 		&	\nodata	 																&		B1			&		 72.0/71    		\phantom{a}[$\chi^2$] 	\\ \noalign{\smallskip}

\enddata

\tablecomments{Table \ref{tbl-3} is published in its entirety in the electronic edition of the {\it Astrophysical Journal}. A portion is shown here for guidance
regarding its form and content. The table reports: (1) the {\it Swift}/BAT ID of the source, (2) the value of the logarithm of the column density and the 90\% confidence interval, (3) the photon index, (4) the energy of the cutoff, (5) the reflection parameter, (6) the cross-calibration between {\it Swift}/BAT and the soft-ray spectra, (7) the fraction of scattered emission observed in the soft X-ray band,  (8) the temperature of the blackbody component (for unobscured objects) or of the thermal plasma component (for obscured objects), (9) the model used (G is reported when a Gaussian line around $\sim$6--7\,keV was added, see $\S$\ref{sect:Xrayspecanalysis} for details) and (10) the value of the statistic and the number of degrees of freedom. Objects for which no column density was reported have $\log (N_{\rm\,H}/\rm cm^{-2})\lesssim 20$.}
\tablenotetext{NC}{Value not constrained.}
\tablenotetext{$^{*}$}{Value fixed.}
\tablenotetext{$a$}{The value of $C_{\rm\,BAT}$ was not constrained so that the constant was fixed to one.}
\end{deluxetable*}
%\end{turnpage}
\clearpage
\end{landscape}

\clearpage
%\begin{turnpage}
%\begin{landscape}
\tabletypesize{\normalsize}
\begin{deluxetable*}{lcc} 
\tablecaption{Parameters of the broken power-law continuum.\label{tbl-bknpo}}
\tablewidth{0pt}
\tablehead{
\colhead{(1)} & \colhead{(2)} & \colhead{(3)} \\
 \noalign{\smallskip}
\colhead{Source} & \colhead{$\Gamma_{2}$} & \colhead{$E_{\rm\,brk}$ [\scriptsize{keV}]} 
}
\startdata
SWIFT\,J0017.1$+$8134	&	$2.6^{+0.7}_{-0.5}$	 		&		$13^{+31}_{-10}$ \\ \noalign{\smallskip}
SWIFT\,J0036.0$+$5951	&	$2.96^{+0.13 }_{-0.11}$		&		$7^{+1}_{-1}$				\\ \noalign{\smallskip}
SWIFT\,J0122.0$-$2818		&	$-2.2^{+0.4 }_{-0.2 }$		&	$88^{+27}_{-16}$			\\ \noalign{\smallskip}
SWIFT\,J0122.9$+$3420	&	$2.52^{+0.51 }_{-0.40}$		&		$17^{+8}_{-15}$				\\ \noalign{\smallskip}
SWIFT\,J0142.0$+$3922		&	$1.74^{+0.09}_{-0.08}$		&	$2.5^{+0.6}_{-0.5}$			\\ \noalign{\smallskip}
SWIFT\,J0225.0$+$1847	&	$1.71^{+0.20 }_{-0.16 }$	&		$7.5^{+1.4}_{-0.4}$				\\ \noalign{\smallskip}
SWIFT\,J0404.0$-$3604	&	$1.95^{+0.14 }_{-0.09 }$	&		$3.9^{+1.4}_{-0.8}$				\\ \noalign{\smallskip}
SWIFT\,J0550.7$-$3212A	&	$3.58^{+1.35 }_{-0.66 }$	&		$18.8^{+2.0}_{-2.5}$				\\ \noalign{\smallskip}	
SWIFT\,J0710.3$+$5908	&	$2.85^{+1.37 }_{-0.57 }$	&		$32^{+10}_{-9}$					\\ \noalign{\smallskip}
SWIFT\,J0841.4$+$7052	&	$1.63^{+0.02 }_{-0.02 }$	&		$3.7^{+0.6}_{-0.5}$				\\ \noalign{\smallskip}

\tablecomments{Table \ref{tbl-bknpo} is published in its entirety in the electronic edition of the {\it Astrophysical Journal}. A portion is shown here for guidance
regarding its form and content. The table reports: (1) the {\it Swift}/BAT ID of the source, (2) the second photon index and (3) the energy of the break between $\Gamma_1$ and $\Gamma_2$. For SWIFT\,J1256.2$-$0551 two sets of values are reported since the source was fitted using model C7, which considers a double broken power-law. For this object the second line reports the value of $\Gamma_{3}$ and $E_{\rm\,brk}^2$.}

\end{deluxetable*}
%\end{turnpage}
\clearpage

\clearpage
%\begin{turnpage}
%\begin{landscape}
\tabletypesize{\normalsize}
\begin{deluxetable*}{lccc} 
\tablecaption{Parameters of the warm absorbers.\label{tbl-wa}}
\tablewidth{0pt}
\tablehead{
\colhead{(1)} & \colhead{(2)} & \colhead{(3)} &  \colhead{(4)}
\\ \noalign{\smallskip}
\colhead{Source} & \colhead{$N_{\rm H}^{\rm W}$  } & \colhead{$\log \xi$} &  \colhead{$f_{\rm\,cov}^{\rm\,W} $}\\
 \noalign{\smallskip}
 \colhead{ } & \colhead{[\scriptsize{$10^{22}\rm\,cm^{-2}$}]} & \colhead{[\scriptsize{$\rm erg\,cm\,s^{-1}$}] } &  \colhead{ [\scriptsize{$\%$}]}
}
\startdata
SWIFT\,J0001.6$-$7701		&		$10^{+5}_{-3}		$	&	$1.4^{+0.6}_{-0.1}		$	&	$96^{+2}_{-3}$	\\ \noalign{\smallskip}
SWIFT\,J0006.2$+$2012		&		$2.91^{+0.06}_{-0.04}$	&	$0.42^{+0.02}_{-0.02}	$	&	$95^{+2}_{-1}$	\\ \noalign{\smallskip}
SWIFT\,J0006.2$+$2012		&		$52^{+8}_{-9}		$	&	$3.18^{+0.06}_{-0.13}	$	&	$80^{+7}_{-9}$	\\ \noalign{\smallskip}
SWIFT\,J0113.8$-$1450		&		$0.44^{+0.18}_{-0.16}$	&	$0.4^{+0.4}_{-0.5}		$	&	$49^{+3}_{-6}$	\\ \noalign{\smallskip}
SWIFT\,J0154.9$-$2707		&	$3^{+5}_{-2}$			&	$\leq 1.6$				&	$74^{+6}_{-4}$	\\ \noalign{\smallskip}
SWIFT\,J0207.0$+$2931		&		$0.39^{+0.07}_{-0.06}$	&	$0.44^{+0.15}_{-0.23}	$	&	$71^{+8}_{-10}$	\\ \noalign{\smallskip}
SWIFT\,J0207.0$+$2931		&		$0.79^{+0.33}_{-0.16}$	&	$2.41^{+0.15}_{-0.14}	$	&	$\geq 84$	\\ \noalign{\smallskip}
SWIFT\,J0208.5$-$1738		&		$5^{+9}_{-2}		$	&	$1.3^{+0.9}_{-1.5}	$	&	$58^{+14}_{-21}$	\\ \noalign{\smallskip}
SWIFT\,J0222.3$+$2509		&	$0.5^{+1.3}_{-0.2}$	&	$\leq 1.5$	&	$68^{+10}_{-18}$ \\ \noalign{\smallskip}
SWIFT\,J0226.4$-$2821		&		$2.8^{+1.7}_{-1.1}	$	&	$0.9^{+0.3}_{-0.3}		$	&	$\geq 97$	\\ \noalign{\smallskip}

		\enddata

\tablecomments{Table \ref{tbl-wa} is published in its entirety in the electronic edition of the {\it Astrophysical Journal}. A portion is shown here for guidance
regarding its form and content. The table reports: (1) the {\it Swift}/BAT ID of the source, (2) the column density, (3) ionization parameter and (4) covering fraction of the ionized absorber. Sources reported more than once were fitted considering more than one layer of ionized absorbing material.}

\end{deluxetable*}

\clearpage
%\begin{turnpage}
%\begin{landscape}
\tabletypesize{\normalsize}
\begin{deluxetable*}{lcccc} 
\tablecaption{Parameters of the Gaussian lines\label{tbl:kalphalines}.}
\tablewidth{0pt}
\tablehead{
\colhead{(1)} & \colhead{(2)} & \colhead{(3)} &  \colhead{(4)} & \colhead{(5)} 
\\ \noalign{\smallskip}
\colhead{Source} & \colhead{$E_{\rm\,K\alpha}$} & \colhead{$EW$ } &  \colhead{$\sigma$ } & \colhead{Normalization} \\
\colhead{} & \colhead{ [\scriptsize{keV}]} & \colhead{[\scriptsize{eV}]} &  \colhead{ [\scriptsize{eV}]} & \colhead{ [\scriptsize{$10^{-6}\rm\,ph\,cm^{-2}\,s^{-1}$}]} 
}
\startdata
SWIFT\,J0005.0$+$7021      &	  		$6.43_{-0.07}^{+0.08}$	&	$80^{+25}_{-36}$		&		\nodata							&	$4.8_{-2.7}^{+2.7}$  \\ \noalign{\smallskip}  
SWIFT\,J0006.2$+$2012      &   $6.4^{\rm a}$					&	$110^{+76}_{-53}$			&		\nodata							&		 $6.9_{-2.8}^{+3.2}$   \\ \noalign{\smallskip}
SWIFT\,J0009.4$-$0037      &		  	$6.4^{\rm a}$				&	$\leq 676$				&		\nodata							&	$5.5_{-3.9}^{+8.1}$	  \\ \noalign{\smallskip}  
SWIFT\,J0010.5$+$1057      &   $6.12_{-0.38}^{+0.09}$		&	$162^{+116}_{-113}$			&		\nodata							&		 $17.8_{-9.5}^{+13.5}$	  \\ \noalign{\smallskip}
SWIFT\,J0025.8$+$6818	   &			$6.48_{-0.05}^{+0.06}$  &	 $836^{+595}_{-778}$	&		\nodata							&	 $11.9_{-6.8}^{+9.3}$  \\ \noalign{\smallskip}
SWIFT\,J0030.0$-$5904	&		$5.5_{-0.5}^{+0.7}$		 	&	$7300^{+4056}_{-5641}$			&	\nodata					&		$\leq 9.2$				\\ \noalign{\smallskip}
SWIFT\,J0036.3$+$4540      &   $6.52_{-0.09}^{+0.60}$		&	$585^{+232}_{-411}$			&		\nodata							&		 $37.4_{-27.4}^{+31.0}$	     \\ \noalign{\smallskip}
SWIFT\,J0042.9$+$3016B     &     $6.4^{\rm a}$					&	$\leq 721$					&		\nodata							&		$\leq 26.4$	  \\ \noalign{\smallskip}
SWIFT\,J0046.2$-$4008   &		$6.40_{-0.04}^{+0.04}$		&	$92^{+30}_{-10 }$				&	\nodata				&			$4.0_{-1.4}^{+1.4}$  		\\ \noalign{\smallskip}
SWIFT\,J0048.8$+$3155      &			$6.39_{-0.02}^{+0.03}$	&	 $53^{+7}_{-6}$			&		\nodata							&	 $24.7_{-6.0}^{+6.0}$  \\ \noalign{\smallskip}

		\enddata

\tablecomments{Table \ref{tbl:kalphalines} is published in its entirety in the electronic edition of the {\it Astrophysical Journal}. A portion is shown here for guidance
regarding its form and content. The table lists, for (1) the sources for which a Gaussian line was added to the broad-band spectral model: the (2) energy, (3) equivalent width, (4) width and (5) normalization of the line. The uncertainties on the equivalent width represent the $68\%$ confidence interval.}

\tablenotetext{a}{Value fixed.}
\tablenotetext{NC}{Not constrained.}

\end{deluxetable*}

\clearpage
%\begin{turnpage}
\begin{landscape}
\LongTables
\begin{deluxetable*}{lllccccc} 
%\tabletypesize{\scriptsize}
\tabletypesize{\footnotesize}
\tablecaption{Spectral parameters obtained with the torus model.\label{tbl-torpar}}
\tablewidth{0pt}
\tablehead{
\colhead{(1)} & \colhead{(2)} & \colhead{(3)} & \colhead{(4)} & \colhead{(5)} &  \colhead{(6)} &  \colhead{(7)} &  \colhead{(8)} \\
\noalign{\smallskip} 
\colhead{Source} & \colhead{$\theta_{\rm\,OA}$} & \colhead{$\log N_{\rm H}$} & \colhead{$\Gamma$} & \colhead{$f_{\rm\,scatt}$} &  \colhead{$kT$} &  \colhead{$C_{\rm\,BAT}$} &  \colhead{Statistic/DOF} \\
%}
\colhead{ } & \colhead{ [{\scriptsize Deg}]} & \colhead{ [{\scriptsize$\rm\,cm^{-2}$}]} & \colhead{ } & \colhead{ [{\scriptsize\%}]} &  \colhead{ [{\scriptsize keV}]} &  \colhead{ } &  \colhead{ } 
}
\startdata

SWIFT\,J0009.4$-$0037 	&	$60^{\rm a}$				&	23.56\,\,[23.41 -- 23.70]		&		$1.66^{+0.29}_{-0.27}$	&		$1.1^{+1.7}_{-0.7}$		&  \nodata												&	\nodata						&	102.1/86			\\ \noalign{\smallskip}
SWIFT\,J0025.8$+$6818 	&	$\geq 42$					&	24.14\,\,[23.72 -- 24.32]		&		$1.94^{+0.27}_{-0.56}$	&		$\leq 1.3$  			&  \nodata												&	$0.39^{+0.78}_{-0.27}$		&	46.5/49				\\ \noalign{\smallskip}
SWIFT\,J0030.0$-$5904	&	$60^{\rm a}$				&	24.03\,\,[23.79 -- 24.43]		&		$1.78^{+0.35}_{-0.38}$	&		$\leq 0.8  $			&	\nodata												&\nodata						&	22.3/17				\\ \noalign{\smallskip}
SWIFT\,J0105.5$-$4213	&	$60^{\rm a}$				&	24.18\,\,[23.95 -- 24.30]		&		$2.12^{+0.25}_{-0.41}$	&		$\leq 0.6  $			&	\nodata												&\nodata						&	32.6/38								\\ \noalign{\smallskip}
SWIFT\,J0106.8$+$0639 	&	$\leq 66$					&	23.54\,\,[23.46 -- 23.66]		&		$1.83^{+0.15}_{-0.19}$	&		$1.6^{+1.4}_{-0.8}$		&  \nodata												&	\nodata						&	99.2/104			\\ \noalign{\smallskip}
SWIFT\,J0111.4$-$3808 	&	$\geq 78$					&	24.33\,\,[24.32 -- 24.34]		&		$2.64^{+0.11}_{-0.09}$ 	&		$\leq 0.1  $			&	\nodata												&\nodata						&	380.7/295							\\ \noalign{\smallskip}
SWIFT\,J0122.8$+$5003	&	$58^{+12}_{-14}$			&	24.24\,\,[24.09 -- 24.58]		&		$2.81^{+0.16}_{-0.15}$	&		$\leq 0.1  $			&  \nodata												&\nodata						&	91.6/115							\\ \noalign{\smallskip}	
SWIFT\,J0128.9$-$6039	&	$60^{\rm a}$				&	24.13\,\,[23.90 -- 24.32]		&		$2.18^{+0.21}_{-0.36}$	&		$\leq 0.1  $			&  \nodata												&\nodata						&	25.7/25								\\ \noalign{\smallskip}
SWIFT\,J0130.0$-$4218	&	$\geq 36$					&	24.20\,\,[24.02 -- 24.55]		&		$2.45^{+0.35}_{-0.15}$	&		$\leq 0.02  $			&  \nodata												&\nodata						&	22.5/22								\\ \noalign{\smallskip}
SWIFT\,J0131.8$-$3307	&	$\leq 79$					&	23.89\,\,[23.74 -- 23.96]	 	&		$2.08^{+0.28}_{-0.36}$	&		$\leq 0.1$  			&	$0.85^{+0.21}_{-0.27}$								&	\nodata						&	49.4/42				\\ \noalign{\smallskip}

\enddata

\tablecomments{Table \ref{tbl-torpar} is published in its entirety in the electronic edition of the {\it Astrophysical Journal}. A portion is shown here for guidance
regarding its form and content. The table lists, for the objects with $\log (N_{\rm H}/\rm cm^{-2})$ within their 90\% confidence intervals, the (1) sources for which the torus model of \citet{Brightman:2011oq} was applied: (2) the half-opening angle of the torus, (3) the column density, (4) the photon index, (5) the fraction of scattered radiation, (6) the temperature of the thermal plasma, (7) the cross-calibration constant between the soft and hard X-ray spectra and (8) the value of the statistic and the DOF.}

\tablenotetext{a}{Value fixed.}

\end{deluxetable*}
%\end{turnpage}
\clearpage
\end{landscape}

\clearpage
%\begin{turnpage}
\begin{landscape}
\LongTables
\begin{deluxetable*}{lccccccc} 
\tabletypesize{\scriptsize}
\tablecaption{X-ray fluxes of the sources of our sample.\label{tbl:fluxes}}
\tablewidth{0pt}
\tablehead{
\colhead{} & \multicolumn2c{{\bf \large Observed}} & & \multicolumn4c{{\bf \large Intrinsic}} \\
\noalign{\smallskip} 
 \cline{2-3} \cline{5-8} \\
%\noalign{\smallskip}
\colhead{(1)} & \colhead{(2)} & \colhead{(3)}  & \colhead{} & \colhead{(4)} &\colhead{(5)}  & \colhead{(6)}  	&  \colhead{(7)}  \\
\noalign{\smallskip} 
\colhead{SWIFT ID} & \colhead{$F_{2-10}^{\rm\,obs}$} & \colhead{$F_{14-195}^{\rm\,obs}$}& \colhead{}  & \colhead{$F_{2-10}$} &\colhead{$F_{20-50}$}  & \colhead{$F_{14-150}$}  	&  \colhead{$F_{14-195}$} \\
\colhead{} & \colhead{ [\scriptsize{$10^{-12}\rm\,erg\,cm^{-2}\,s^{-1}$}]} & \colhead{[\scriptsize{$10^{-12}\rm\,erg\,cm^{-2}\,s^{-1}$}]} & \colhead{} & \colhead{[\scriptsize{$10^{-12}\rm\,erg\,cm^{-2}\,s^{-1}$}]} &\colhead{[\scriptsize{$10^{-12}\rm\,erg\,cm^{-2}\,s^{-1}$}]}  & \colhead{[\scriptsize{$10^{-12}\rm\,erg\,cm^{-2}\,s^{-1}$}]}  	&   \colhead{[\scriptsize{$10^{-12}\rm\,erg\,cm^{-2}\,s^{-1}$}]}   
%
%}
}
\startdata

   SWIFT\,J0001.0$-$0708  &       7.3  &      13.0  &      &   8.4  &       5.4  &      10.8  &      12.3 \\ \noalign{\smallskip}
   SWIFT\,J0001.6$-$7701  &       3.2  &      10.1  &     &    5.2  &       3.7  &       9.0  &      10.3 \\ \noalign{\smallskip}
   SWIFT\,J0002.5$+$0323  &       4.8  &      11.7  &     &    4.8  &       4.0  &       8.7  &       9.9 \\ \noalign{\smallskip}
   SWIFT\,J0003.3$+$2737  &       1.7  &      13.0  &    &     3.1  &       4.3  &      11.2  &      12.8 \\ \noalign{\smallskip}
   SWIFT\,J0005.0$+$7021  &       3.6  &      12.7  &    &     4.7  &       5.6  &      10.6  &      12.1 \\ \noalign{\smallskip}
   SWIFT\,J0006.2$+$2012  &       6.0  &      18.4  &    &    11.2  &       7.0  &      14.2  &      16.2 \\ \noalign{\smallskip}
   SWIFT\,J0009.4$-$0037  &       0.8  &       9.3  &      &   3.1  &       3.5  &      10.6  &      12.1 \\ \noalign{\smallskip}
   SWIFT\,J0010.5$+$1057  &       7.7  &      31.4  &    &     7.7  &      10.5  &      26.8  &      30.6 \\ \noalign{\smallskip}
   SWIFT\,J0017.1$+$8134  &       3.3  &      10.1  &    &     3.4  &       4.3  &       9.6  &      11.0 \\ \noalign{\smallskip}
   SWIFT\,J0021.2$-$1909  &      10.6  &      17.3  &    &    11.2  &       6.9  &      14.0  &      16.0 \\ \noalign{\smallskip}

\enddata

\tablecomments{Table \ref{tbl:fluxes} is published in its entirety in the electronic edition of the {\it Astrophysical Journal}. A portion is shown here for guidance
regarding its form and content. The table lists: the observed (2) 2--10\,keV and (3) 14--195\,keV fluxes;  the intrinsic (4) 2--10\,keV, (5) 20--50\,keV, (6) 14--150\,keV and (7) 14--195\,keV fluxes.}

\end{deluxetable*}
%\end{turnpage}
\clearpage
\end{landscape}

\clearpage
%\begin{turnpage}
%\begin{landscape}
\LongTables
\begin{deluxetable*}{lccccccc} 
\tabletypesize{\scriptsize}
\tablecaption{X-ray luminosities of the sources of our sample.\label{tbl:luminosities}}
\tablewidth{0pt}
\tablehead{
\colhead{} & \multicolumn2c{{\bf \large Observed}} & & \multicolumn4c{{\bf \large Intrinsic}} \\
\noalign{\smallskip} 
 \cline{2-3} \cline{5-8} \\
\colhead{(1)} & \colhead{(2)} & \colhead{(3)}  & \colhead{} & \colhead{(4)} &\colhead{(5)}  & \colhead{(6)}  	&  \colhead{(7)}  \\
\noalign{\smallskip} 
\colhead{SWIFT ID} & \colhead{$\log L_{2-10}^{\rm\,obs}$} & \colhead{$\log L_{14-195}^{\rm\,obs}$}  & \colhead{} &  \colhead{$\log L_{2-10}$} &  \colhead{$\log L_{20-50}$} &  \colhead{$\log L_{14-150}$} &  \colhead{$\log L_{14-195}$} \\
\colhead{} &  \colhead{[\scriptsize{$\rm\,erg\,s^{-1}$}]} &  \colhead{[\scriptsize{$\rm\,erg\,s^{-1}$}]}  & \colhead{} &  \colhead{[\scriptsize{$\rm\,erg\,s^{-1}$}]} &  \colhead{[\scriptsize{$\rm\,erg\,s^{-1}$}]} &  \colhead{[\scriptsize{$\rm\,erg\,s^{-1}$}]} &  \colhead{[\scriptsize{$\rm\,erg\,s^{-1}$}]} 
}
\startdata

   SWIFT\,J0001.0$-$0708  &     43.37  &     43.62  &     &  43.43  &     43.24  &     43.54  &     43.60 \\ \noalign{\smallskip}
   SWIFT\,J0001.6$-$7701  &     43.41  &     43.91  &     &  43.62  &     43.48  &     43.86  &     43.92 \\ \noalign{\smallskip}
   SWIFT\,J0002.5$+$0323  &     42.85  &     43.24  &    &   42.85  &     42.78  &     43.11  &     43.17 \\ \noalign{\smallskip}
   SWIFT\,J0003.3$+$2737  &     42.79  &     43.67  &     &  43.05  &     43.19  &     43.61  &     43.67 \\ \noalign{\smallskip}
   SWIFT\,J0005.0$+$7021  &     43.90  &     44.45  &    &   44.02  &     44.09  &     44.37  &     44.43 \\ \noalign{\smallskip}
   SWIFT\,J0006.2$+$2012  &     42.97  &     43.46  &    &   43.24  &     43.04  &     43.34  &     43.40 \\ \noalign{\smallskip}
   SWIFT\,J0009.4$-$0037  &     43.01  &     44.07  &    &   43.60  &     43.65  &     44.13  &     44.19 \\ \noalign{\smallskip}
   SWIFT\,J0010.5$+$1057  &     44.18  &     44.79  &    &   44.18  &     44.31  &     44.72  &     44.78 \\ \noalign{\smallskip}
   SWIFT\,J0017.1$+$8134  &     47.22  &     47.71  &    &   47.23  &     47.33  &     47.68  &     47.74 \\ \noalign{\smallskip}
   SWIFT\,J0021.2$-$1909  &     44.38  &     44.59  &    &   44.40  &     44.19  &     44.50  &     44.55 \\ \noalign{\smallskip}

\enddata

\tablecomments{Table \ref{tbl:luminosities} is published in its entirety in the electronic edition of the {\it Astrophysical Journal}. A portion is shown here for guidance
regarding its form and content. The table lists: the observed (2) 2--10\,keV and (3) 14--195\,keV luminosities;  the intrinsic (4) 2--10\,keV, (5) 20--50\,keV, (6) 14--150\,keV and (7) 14--195\,keV luminosities.}

\end{deluxetable*}
%\end{turnpage}
\clearpage
%\end{landscape}

\clearpage
%\begin{turnpage}
%\begin{landscape}
\LongTables
\begin{deluxetable*}{lcc} 
\tabletypesize{\scriptsize}
\tablecaption{Intrinsic 2--10\,keV fluxes and luminosities of the CT sources of our sample.\label{tbl:lumfluxCT}}
\tablewidth{0pt}
\tablehead{
\colhead{(1)} & \colhead{(2)} & \colhead{(3)}  \\
\noalign{\smallskip} 
\colhead{SWIFT ID} & \colhead{$F_{2-10}$} & \colhead{$\log L_{2-10}$}   \\
\colhead{} &  \colhead{[\scriptsize{$10^{-12}\rm\,erg\,cm^{-2}\,s^{-1}$}]} &  \colhead{[\scriptsize{$\rm\,erg\,s^{-1}$}]}  
}
\startdata
   SWIFT\,J0025.8$+$6818   &     13.1   &    42.63 \\ \noalign{\smallskip}
   SWIFT\,J0030.0$-$5904   &      7.5   &    43.16 \\ \noalign{\smallskip}
   SWIFT\,J0105.5$-$4213   &      7.1   &    43.17 \\ \noalign{\smallskip}
   SWIFT\,J0111.4$-$3808   &     28.4   &    42.95 \\ \noalign{\smallskip}
   SWIFT\,J0122.8$+$5003   &     10.6   &    43.00 \\ \noalign{\smallskip}
   SWIFT\,J0128.9$-$6039   &      5.9   &    44.86 \\ \noalign{\smallskip}
   SWIFT\,J0130.0$-$4218   &      7.4   &    43.05 \\ \noalign{\smallskip}
   SWIFT\,J0242.6$+$0000   &     76.4   &    42.39 \\ \noalign{\smallskip}
   SWIFT\,J0250.7$+$4142   &     12.0   &    42.75 \\ \noalign{\smallskip}
   SWIFT\,J0251.3$+$5441   &     18.9   &    42.99

\enddata

\tablecomments{Table \ref{tbl:lumfluxCT} is published in its entirety in the electronic edition of the {\it Astrophysical Journal}. A portion is shown here for guidance
regarding its form and content. The table lists: the intrinsic 2--10\,keV (2) fluxes and (3) luminosities calculated from the intrinsic 14--150\,keV values assuming $\Gamma=1.8$.}

\end{deluxetable*}
%\end{turnpage}
\clearpage
%\end{landscape}

\clearpage
%\begin{turnpage}
%\begin{landscape}
\tabletypesize{\normalsize}
\begin{deluxetable*}{lcc} 
\tablecaption{Values of $\Gamma_{\rm nEc}$ and $\Gamma_{0.3-10}$. \label{tbl:otherGamma}}
\tablewidth{0pt}
\tablehead{
\colhead{(1)} & \colhead{(2)} & \colhead{(3)} \\
\noalign{\smallskip} 
\colhead{Source} & \colhead{$\Gamma_{\rm nEc}$} & \colhead{$\Gamma_{0.3-10}$} 
}
\startdata

SWIFT\,J0001.0$-$0708          	 & 	$1.86_{-0.25}^{+0.26}$	& 		$1.85_{-0.32}^{+0.23}$	\\ \noalign{\smallskip}	
SWIFT\,J0001.6$-$7701    	&	$1.96_{-0.49}^{+0.60}$		&		$1.69_{-0.42}^{+0.90}$	\\ \noalign{\smallskip}		
SWIFT\,J0002.5$+$0323    	&	$2.22_{-0.06}^{+0.07}$		&		$2.16_{-0.05}^{+0.05}$	\\ \noalign{\smallskip}
SWIFT\,J0003.3$+$2737            &	$1.41_{-0.24}^{+0.53}$	&		$1.41_{-2.23}^{+0.36}$	\\ \noalign{\smallskip}  	
SWIFT\,J0005.0$+$7021        &	$1.93_{-0.05}^{+0.42}$	&		$1.49_{-0.10}^{+0.11}$	\\ \noalign{\smallskip}
SWIFT\,J0006.2$+$2012    	&	$2.83_{-0.12}^{+0.20}$		&		$2.91_{-0.11}^{+0.09}$	\\ \noalign{\smallskip}
SWIFT\,J0009.4$-$0037        &	$2.19_{-0.54}^{+0.86}$	&		$1.35_{-0.89}^{+0.94}$	\\ \noalign{\smallskip}	
SWIFT\,J0010.5$+$1057    	&	$1.73_{-0.02}^{+0.04}$		&		$1.73_{-0.06}^{+0.03}$	\\ \noalign{\smallskip}
SWIFT\,J0017.1$+$8134    & \nodata    	 			&		$1.52_{-0.03}^{+0.03}$	\\ \noalign{\smallskip}	
SWIFT\,J0021.2$-$1909        &	$1.69_{-0.13}^{+0.23}$	&		$1.70_{-0.24}^{+0.10}$	\\ \noalign{\smallskip}

		\enddata

\tablecomments{Table \ref{tbl:otherGamma} is published in its entirety in the electronic edition of the {\it Astrophysical Journal}. A portion is shown here for guidance
regarding its form and content. The table lists the values of (2) $\Gamma_{\rm nEc}$ and (3) $\Gamma_{0.3-10}$.}

\end{deluxetable*}

\end{document}